\documentclass[twocolumn,superscriptaddress,aps,prx,longbibliography]{revtex4-1}
\usepackage{amsmath,amssymb}
\usepackage{graphicx}
\usepackage{epsfig}
\usepackage{hyperref}
\usepackage{xcolor,colordvi}
\usepackage{float}

\newcommand{\be}{\begin{equation}}
\newcommand{\ee}{\end{equation}}
\newcommand{\bea}{\begin{eqnarray}}
\newcommand{\eea}{\end{eqnarray}}
\newcommand{\ba}{\begin{eqnarray*}}
\newcommand{\ea}{\end{eqnarray*}}

\newcommand{\aver}[1]{\langle {#1} \rangle}

\newcommand{\es}[1]{\begin{split}#1\end{split}}
\newcommand{\beq}{\begin{equation}}
\newcommand{\eeq}{\end{equation}}

\newcommand{\lp}{\left(}
\newcommand{\rp}{\right)}
\newcommand{\lsq}{\left[}
\newcommand{\rsq}{\right]}
\newcommand{\lbr}{\left\lbrace}
\newcommand{\rbr}{\right\rbrace}
\newcommand{\da}{\dagger}
\newcommand{\bma}{\begin{pmatrix}}
\newcommand{\ema}{\end{pmatrix}}

\newcommand{\mo}{{-1}}
\newcommand{\rw}{\rightarrow}
\newcommand{\oh}{\frac{1}{2}}

\newcommand{\w}{\omega}
\newcommand{\pt}{\partial _t}
\newcommand{\dt}{ dt \,}

\newcommand{\re}{\text{Re}}
\newcommand{\im}{\text{Im}}
\newcommand{\tr}{\text{tr}}

\newcommand{\abs}[1]{ \left\lvert #1	\right\rvert}

\usepackage{bbold}
\newcommand{\id}{\mathbb{1}}

\newcommand{\hc}{\text{hc}}

\newcommand{\mcV}{\mathcal{V}}

\newcommand{\lind}{\mathcal{L}}

\newcommand{\eff}{\text{eff}}

\newcommand{\intre}{\int_{-\infty}^\infty}
\newcommand{\bb}{\bold{b}}

\begin{document}

\title{Dynamical Mean-Field Theory for Markovian Open Quantum Many-Body Systems}

\author{Orazio Scarlatella}
\affiliation{JEIP, USR 3573 CNRS, Coll\'ege de France,   PSL  Research  University, 11,  place  Marcelin  Berthelot, 75231 Paris Cedex 05, France}
\affiliation{Clarendon Laboratory, University of Oxford, Parks Road, Oxford OX1 3PU, United Kingdom}
\affiliation{Institut de Physique Th\'{e}orique, Universit\'{e} Paris Saclay, CNRS, CEA, F-91191 Gif-sur-Yvette, France 
}
\author{Aashish A. Clerk}
\affiliation{Pritzker School of Molecular Engineering, University of Chicago, 5640 S. Ellis Ave., Chicago, IL, 60637, USA}
\author{Rosario Fazio}
\affiliation{The Abdus Salam International Centre for Theoretical Physics, Strada Costiera 11, I-34151 Trieste, Italy}
\affiliation{Dipartimento di Fisica, Universit\`a di Napoli ``Federico II'', Monte S. Angelo, I-80126 Napoli, Italy}
\thanks{On leave}
\author{Marco Schir\'o}
\thanks{On Leave from: Institut de Physique Th\'{e}orique, Universit\'{e} Paris Saclay, CNRS, CEA, F-91191 Gif-sur-Yvette, France}
\affiliation{JEIP, USR 3573 CNRS, Coll\'ege de France,   PSL  Research  University, 11,  place  Marcelin  Berthelot, 75231 Paris Cedex 05, France}

\date{\today}

\begin{abstract}
Open quantum many body systems describe a number of experimental platforms relevant for quantum simulations, ranging from arrays of superconducting circuits hosting correlated states of light, to ultracold atoms in optical lattices in presence of controlled dissipative processes. Their theoretical understanding is hampered by the exponential scaling of their Hilbert space and by their intrinsic nonequilibrium nature, limiting the applicability of many traditional approaches.
In this work we extend the nonequilibrium bosonic Dynamical Mean Field Theory (DMFT) to Markovian open quantum systems. Within DMFT, a Lindblad master equation describing a lattice of dissipative bosonic particles is mapped onto an impurity problem describing a single site embedded in its Markovian environment and coupled to a self-consistent field and to a non-Markovian bath, where the latter accounts for fluctuations beyond Gutzwiller mean-field theory due to the finite lattice connectivity.We develop non-perturbative approach to solve this bosonic impurity problem, which dresses the impurity, featuring Markovian dissipative channels, with the non-Markovian bath, in a self-consistent scheme based on a resummation of non-crossing diagrams. 
As a first application of our approach, we address the steady-state of a driven-dissipative Bose-Hubbard model with two-body losses and incoherent pump. We show that DMFT captures hopping-induced dissipative processes, completely missed in Gutzwiller mean-field theory, which crucially determine the properties of the normal phase, including the redistribution of steady-state populations, the suppression of local gain and the emergence of a stationary quantum-Zeno regime. We argue that these processes compete with coherent hopping to determine the phase transition towards a non-equilibrium superfluid, leading to a strong renormalization of the phase boundary at finite-connectivity.
We show that this transition occurs as a finite-frequency instability, leading to an oscillating in time order parameter, that we connect with a quantum many-body synchronization transition of an array of quantum van der Pol oscillators. 
\end{abstract}

\maketitle

\section{Introduction}

Developments in quantum science and quantum engineering have brought forth a variety of platforms to store, control and process information at genuine quantum levels. Examples include trapped atoms and ions~\cite{CiracZollerPRL95},  quantum cavity-QED systems~\cite{Haroche_RMP}, superconducting qubits\cite{Schoelkopf2008} or quantum optomechanical systems~\cite{aspelmeyer2014}. These architectures are not only of great relevance for quantum technologies but also for the quantum simulation of emergent collective many-body phenomena.
We now have several experimental quantum simulators worldwide running on a variety of architectures, from ultracold atoms in optical lattices~\cite{BlochDalibardNascimbeneNatPhys12}, Rydberg gases~\cite{BrowaeysNatPhys}, trapped ions~\cite{BlattRoosNatPhys12} and coupled cavity arrays~\cite{Houck2012}. Such simulators have led to significant advances in our understanding of quantum many-body phases and offer us an opportunity to address deep unanswered questions concerning the behavior of quantum matter in novel unexplored regimes, particularly far away from thermal equilibrium.

Many of these systems are typically characterized by excitations with a finite lifetime, due to unavoidable losses, dephasing and decoherence processes originating from their coupling to external environments and therefore feature an intrinsic nonequilibrium nature.
Arrays of circuit QED cavities, for example, are emerging as a unique platform for the quantum simulation of driven-dissipative quantum many body systems \cite{Houck2012,Salathe2015,puertasmartinez2019,CarusottoEtAlNatPhys2020}, where the balance of drive and loss processes and the presence of strong non-linearities, allows one to reach interesting non-equilibrium stationary states. Experiments have recently started to show a variety of dissipative quantum phases and phase transitions \cite{FinkEtAlPRX17,FitzpatrickEtAlPRX17,Fink2018,Raftery2014},
including the recent experimental realization of a dissipatively stabilised Mott insulator \cite{ma2019}.
On a different front, recent advances with ultracold atoms in optical lattices allow the engineering of controlled dissipative processes, such as correlated losses~\cite{Syassen1329,TomitaEtAlScienceAdv17} or heating due to spontaneous emission~\cite{LuschenEtAlPRX17,bouganne2020} and to study the resulting quantum many body dynamics over long time scales. 
These experimental settings offer fresh new inputs for quantum simulation of strongly correlated driven-dissipative quantum many-body systems, at the interface between quantum optics and condensed matter physics \cite{CarusottoCiutiRMP13,ritsch2013,SchmidtKochAnnPhy13,LeHurReview16,noh2017,Hartmann2016}.

From a theoretical standpoint these experimental platforms can be well described as open Markovian quantum systems, of either fermionic/bosonic particles or quantum spins, evolving through a Lindblad master equation for the many body density matrix $\rho$~\cite{breuerPetruccione2010}
\be\label{eq:masterEquation1}
\partial_t \rho = - i[H,\rho]+\sum_{\alpha} \left(L_{\alpha} \rho L_{\alpha}^{\dagger}
-\frac{1}{2}\left\{L^{\dagger}_{\alpha} L_{\alpha},\rho \right\}\right)\,.
\ee
The crucial aspect of this problem lies in the interplay between unitary dynamics and the dissipative evolution controlled by jump operators $L_{\alpha}$, out of which non trivial stationary states can be engineered~\cite{DiehlEtalNatPhys08,Verstraete2009,Siddiqi_quantum_bath_engineering,Leghtas2015}. Open Markovian quantum many body systems are predicted to display a broad range of new transient dynamical phenomena~\cite{Tomadin_prl10,PolettiEtAlPRL12,PolettiEtAlPRL13,LudwigMarquardtPRL13} and dissipative quantum phase transitions~\cite{SiebereHuberAltmanDiehlPRL13,MarinoDiehlPRL16,SchiroPRL16,MingantiEtAlPRA18,RotaEtAlPRL19,YoungEtAlPRX20}.
 
Solving the Lindblad equation for many-body systems is extremely challenging. Exact numerical solutions based on the diagonalization of the Lindbladian superoperator, or direct time evolution, are limited to very small systems, and only slightly larger systems can be addressed with quantum trajectories~\cite{DaleyAdvPhys2014}.  For one dimensional systems an efficient representation of the problem in the language of matrix product operator is possible~\cite{VerstraeteEtAlPRL04,ZwolakVidalPRL04} and has been successfully used in the recent past~\cite{KildaEtAlPRL19}, however its extension to higher dimensional systems poses problems, although some recent results have been obtained~\cite{KshetrimayumNatComm17,LandaEtAlPRL20,kilda2021stability}. As a result, a number of theoretical approaches have been developed in recent years to tackle driven-dissipative lattice systems~\cite{FinazziEtAlPRL15,WeimerPRL15,JinEtAlPRX16,SiebererRepProgPhys2016,Biondi_2017,VicentiniEtAPRA18,WeimerEtAlArxiv19,
VicentiniEtAlPRL19,YoshiokaHamazakiPRB19,NagySavonaPRL19,HartmannCarleoPRL19,LandaEtAlPRL20}. 
Driven-dissipative correlated bosons, such as those described by Bose-Hubbard and related models, are particularly challenging to tackle numerically due to the unbounded Hilbert space.

A powerful non-perturbative approach to quantum lattice models, based on the limit of large lattice connectivity $z$~\cite{Metzner1989,Georges_Kotliar_PRB92}, is provided by the dynamical mean field theory (DMFT)
of correlated electrons~\cite{Review_DMFT_96} and bosons~\cite{ByczukVollhardtPRB08,AndersEtAlPRL10,andersWernerNJP2011}.
When applied to equilibrium lattice bosons DMFT captures, non-perturbatively, the $1/z$ corrections to Gutzwiller mean-field theory through the solution of a quantum impurity model. Originally introduced for equilibrium problems, DMFT has been successfully applied to a variety of nonequilibrium 
fermionic problems with or without dissipation~\cite{aokiWernerRMP2014}, including Markovian fermions~\cite{panas2019densitywave}, and to study unitary dynamics of correlated bosons~\cite{strandWernerPRX2015}. 

In this work we introduce a new approach to Markovian bosonic quantum many-body systems. Specifically we extend the bosonic non-equilibrium DMFT~\cite{strandWernerPRX2015} to systems evolving under the Lindblad master equation~(\ref{eq:masterEquation1}) and combine it with a strong coupling impurity solver tailored for correlated dissipative processes described by many-body jump operators.
In the large connectivity limit DMFT maps the Lindblad dynamics~(\ref{eq:masterEquation1}) onto a dissipative quantum impurity model describing a single interacting bosonic mode, subject to many-body Markovian dissipation,  coupled to a coherent field and a non-Markovian environment both to be determined self-consistently.  The non-Markovian DMFT bath describes the effect of hopping processes from neighboring sites, which are completely missed by Gutzwiller mean-field theory. We will show that these processes are particularly interesting in open quantum systems since they not only introduce coherent effects but also provide new dissipative channels.

Solving the resulting quantum impurity model in presence of multiple many-body jump operators remains highly non-trivial and state of the art impurity solvers for non-equilibrium DMFT cannot be efficiently used in this setting. Here we build upon our recent work on Markovian impurity models~\cite{Scarlatella2019} to develop and benchmark a DMFT solver for driven-dissipative bosonic systems. This approach is based on a super-operator hybridization expansion and resummation of an infinite class of diagrams known as non-crossing approximation (NCA).
As first application of the DMFT/NCA method we study a Bose-Hubbard model subject to two-particle losses and single particle incoherent pumping.  This model is directly relevant for recent experiments with ultracold bosonic atoms in optical lattices under controlled dissipation~\cite{TomitaEtAlScienceAdv17,bouganne2020} as well as for arrays of superconducting circuits~\cite{ma2019}. 
We predict the emergence of a dissipative phase transition from a normal to a superfluid phase, where above a critical hopping or pump strength the system spontaneously develops a coherent field oscillating in time, and discuss the effect of finite-connectivity corrections
to Gutzwiller mean-field. We highlight how this transition can be naturally interpreted as the onset of many body synchronization in an array of quantum Van der Pol oscillators, a phenomenon which recently attracted lots of attention~\cite{lee2013,lorch2016,walter2014,walter2015,roulet2018a,roulet2018,dutta2019,giorgi2012,manzano2013,qiao2018,SonarEtalPRL18,
jaseem2019,tindall2020}. 
We show that DMFT allows to predict results which are qualitatively missed by Gutzwiller mean-field theory, including the redistribution of steady-state population and the suppression of gain due to hopping processes, a stationary quantum Zeno regime and a new competition between coherent and dissipative processes towards symmetry breaking. These results reflect in large quantitative corrections to the Gutzwiller mean-field phase diagram.

\begin{figure*}[t]
\includegraphics[width=14cm]{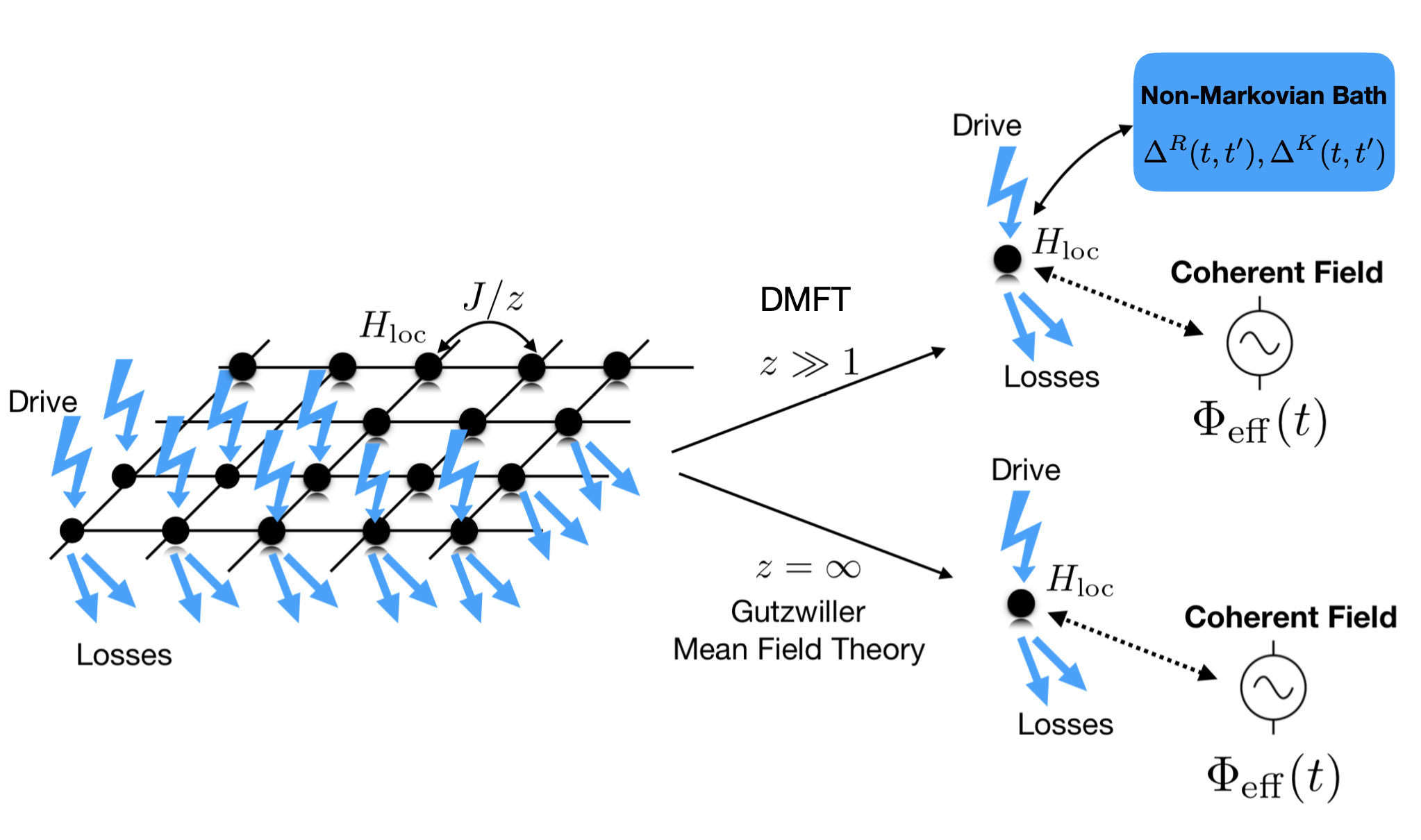}
\caption{Sketch of the DMFT mapping for open Markovian bosonic quantum systems. A lattice of interacting, driven-dissipative bosons coupled by hopping (left, see Eq.~(\ref{eq:BH}-\ref{eq:masterEquation2}) in the main text) is mapped in the large connectivity limit $z\gg 1$ onto a single site problem, with interactions, local Markovian drive and losses, coupled to a self-consistent environment (right, top panel). This includes a coherent drive and a non-Markovian bath (characterized by two independent hybridization functions $\Delta^R(t,t')$ and $\Delta^K(t,t')$) which accounts for fluctuations due to  neighboring sites. In the infinite coordination limit, $z=\infty$, one recovers the Gutzwiller mean-field theory (right, bottom panel), where the effect of the lattice is simply encoded in a self-consistent field.
}
\label{fig:sketch}
\end{figure*}

The paper is organized as follows. In the next section \ref{sec:summary} we summarize the main concepts and results of this work, which will be presented in more detail in the rest of the paper. In Sec.~\ref{sec:markovLattice} we introduce DMFT in more detail and discuss its physical content. In Sec. \ref{sec:impuritysolver} we discuss two methods to solve the quantum impurity problem: a strong coupling perturbative approach and a more powerful self-consistent NCA method. In Sec. \ref{sec:results} we present our results for an interacting Bose-Hubbard driven-dissipative lattice problem. Sec. \ref{sec:conclusions} is devoted to conclusions. In the Appendixes we provide technical details on the derivation of DMFT (Appendix~\ref{app:derivDMFT}), a non trivial consistency check on the NCA approximation (Appendix~\ref{app:bench}) and various analytical results quoted in the main text~(Appendixes~\ref{app:semiclassics}-\ref{app:ncaGreenFunctions}). 

\section{Summary of Main Results}
\label{sec:summary}

In this section we present a summary of the main results of this work, which will be discussed more in detail in the rest of the paper. In particular in Sec.~\ref{sec:summ_dmft} we discuss the formulation of DMFT for Markovian Bosons and of the NCA impurity solver, in Sec.~\ref{sec:prior} we discuss our theoretical approach in relation to prior work on open quantum many-body systems,  while in Sec.~\ref{sec:model} we present the application to a driven-dissipative Bose-Hubbard model with two-body losses and single-particle incoherent pump.

\subsection{Dynamical Mean-Field Theory for Open Markovian Bosons}
\label{sec:summ_dmft}

The class of models that we consider describe driven-dissipative bosonic particles on a lattice with coordination number $z$.
The many-body density matrix of the system evolves according to a Lindblad master equation 
\be\label{eq:masterEquation2}
\partial_t \rho = - i[H,\rho]+\sum_{i\mu} \left(L_{i\mu} \rho L_{i\mu}^{\dagger}
-\frac{1}{2}\left\{L^{\dagger}_{i\mu} L_{i\mu},\rho \right\}\right)\,.
\ee
with a set of local jump operators for each lattice site $L_{i \mu}$, accounting for dissipative processes, and with coherent evolution governed by the Hamiltonian
\be \label{eq:BH}
H=-\frac{J}{z}\sum_{\langle ij\rangle}\left(b^{\dagger}_i b_j+\hc\right)+\sum_i H_{loc}[b^{\dagger}_i,b_i]\,.
\ee
Here $[b_i,b^{\dagger}_i]=1$ are bosonic modes localized around the lattice site $i$, coupled together by a nearest-neighbours hopping term $J$. $H_{loc}[b^{\dagger}_i,b_i]$ is a generic local Hamiltonian, which can contain arbitrary local interactions. In order for the problem to remain well defined in the limit of infinite connectivity $z\rightarrow\infty$, to which we will compare our DMFT results, we scale the hopping with a $1/z$ factor which is the correct scaling for bosons~\cite{fisherFisherPRB1989,ByczukVollhardtPRB08,AndersEtAlPRL10}(see also Sec.~\ref{sec:mean_field_conn} for further details). 

The DMFT approach considers the master equation~(\ref{eq:masterEquation2}) in the limit of large, but finite, lattice connectivity $z\gg 1$. In fact, when the number of neighbors $z$ of a given lattice site is large, statistical and quantum fluctuations induced by the neighboring sites get small and can be treated in an approximate way, while the local, on-site physics is accounted for exactly.  
As we discuss in detail in Sec.~\ref{sec:markovLattice}, in the $z\gg 1$ limit the Lindblad lattice problem formally maps onto a quantum impurity model describing an interacting Markovian single-site, characterized by the same local Hamiltonian $H_{\rm loc}$ and local jump operators $L_{i\mu}$ entering Eq.~(\ref{eq:masterEquation1}), coupled to a time-dependent field $\Phi_{\rm eff}(t)$ acting as a coherent drive and a non-Markovian quantum bath (Fig.~\ref{fig:sketch}, top panel). These take into account the effect of the neighboring sites and have to be determined self-consistently through the calculation of impurity properties. As a result of the non-equilibrium nature of the problem, the non-Markovian environment is described in terms of two independent bath hybridization functions, the retarded $\Delta^R(t,t')$ and Keldysh $\Delta^K(t,t')$ which in a stationary-state encode information about spectrum and occupation of the single-particle excitations. In a generic non-equilibrium condition these are independent and not related by the fluctuation-dissipation theorem.

To appreciate the physical content of DMFT it is instructive to compare it with the widely used Gutzwiller mean-field theory. As we will show in Sec. \ref{sec:mean_field_conn} the latter coincides with the $z\rightarrow\infty$ solution of the many-body master equation and corresponds to DMFT when the non-Markovian bath is set to zero. Gutzwiller mean-field theory amounts to decouple the hopping term in the Hamiltonian~(\ref{eq:BH}), or equivalently assumes a product-state density matrix over different lattice sites, thus reducing the full many-body problem to a single site coupled to a self-consistent coherent field (Fig.~\ref{fig:sketch}, bottom panel). An obvious shortcoming of the Gutzwiller approach is that it cannot capture any coherent or dissipative processes arising from the coupling to neighboring sites, unless the system is in a broken-symmetry phase with a non-vanishing local order parameter, leading to a finite self-consistent field. The result is a particularly poor description of strongly interacting, yet incoherent, normal phases such as bosonic Mott insulators or many-body quantum Zeno phases that we discuss in this work, whose local properties within Gutzwller are completely independent on the hopping and identical to those of an isolated site. In this perspective DMFT accounts non-perturbatively, through the solution of a quantum impurity model with a non-Markovian bath $\Delta\sim 1/z$, for finite-connectivity corrections to Gutzwiller mean field theory, thus capturing fluctuations induced by the neighboring sites even in absence of an order parameter.

Although simplified with respect to the full master equation, the solution of DMFT equations and in particular of the quantum impurity problem sketched in Figure 1 still poses tremendous challenges. In particular the presence a Markovian environment containing arbitrary, possibly non-linear,  jump operators, in addition to local interactions and the non-Markovian DMFT bath  makes this problem hard to be solved efficiently with state of the art approaches for  non-equilibrium DMFT.
A major result of the present work is the development and benchmark of an impurity solver for Markovian bosonic DMFT based on the super-operator hybridization expansion formulated in Ref.~\onlinecite{Scarlatella2019} and applied there to a simple non-interacting fermionic resonant level model.  As we will discuss more in detail in Sec.~\ref{sec:NCA} this approach fully captures the underlying Markovian dynamics of the impurity problem in Fig 1 and accounts for the non-Markovian bath $\Delta$ through the resummation of an infinite class of diagrams in the impurity-bath coupling known as non-crossing approximation (NCA).

As we will discuss further on in the paper, the self-consistent nature of the non-crossing approximation (NCA) we use, as opposed to bare perturbative expansions to which we will compare our results, is crucial to fully capture the non-trivial correlations associated to the DMFT bath. 
We give a more complete picture of the DMFT formalism, including a discussion of the basic equations and of impurity solvers in Sec.~\ref{sec:markovLattice} and Sec.~\ref{sec:impuritysolver}.

\subsection{Relation to Prior Works}\label{sec:prior}

Here we wish to relate our approach with respect to previous works on nonequilibrium dissipative many-body systems.
 In the context of fermionic nonequilibrium DMFT, dissipation at single particle level (i.e. tunneling to external metallic leads) has been included before in several works, focusing for example on steady-state transport~\cite{joura2008steadystate,li2015electricfield,arrigoni2013,titvinidze2018charge,mattheis2018control,murakami2018nonequilibrium}, Floquet driven systems~\cite{tsuji2008correlated,murakami2017nonequilibrium,qin2017spectral} or photodoping~\cite{li2021nonequilibrium}.  We note that this type of dissipation is straightforward to handle within DMFT and pose no additional methodological challenges since it can be included within any impurity solver used for non-equilibrium DMFT in absence of dissipation. On the other hand many-body dissipative processes, such as those we focus here in the Lindblad context or those modeling the coupling between fermions and bosonic baths,  are more challenging to handle since they induce effective interactions. Up to date these have been included in non-equilibrium DMFT studies of dissipative problems only through perturbative expansions~\cite{eckstein2013photoinduced,golez2015dynamics,
Chen2016,bittner2018coupled,PeronaciEtAlPRB2020}.
In this respect our work goes beyond those studies in that all Markovian dissipative couplings (single and many-body) are treated on the same footing and encoded in the local Lindbladian of the impurity model, which opens up the possibilities for non-perturbative treatments of those couplings. This strategy is similar in spirit to what done for Markovian fermionic systems in Ref.~\onlinecite{panas2019densitywave}, where a discretization of the DMFT bath was used to solve the impurity problem with exact diagonalization. Here instead we use the NCA impurity solver which works directly in the thermodynamic limit of an infinite bath and does not introduce any discretization, which would be particularly severe for bosonic problems such as the one we focus here.

In the context of Markovian quantum many-body systems there have been recent methodological developments to
include correlations beyond mean-field theory. Although a precise comparison with DMFT is beyond the scope of this work, it is worth to discuss some of those methods here. The Cluster Mean-Field Theory~\cite{JinEtAlPRX16}, developed for driven-dissipative quantum spin models, is similar to DMFT in that it adds short-range correlations on top of a Gutzwiller mean-field, although this is achieved through the exact solution of a finite-size cluster, rather than through an infinite (non-interacting) bath. The Corner-Space Renormalization Method~\cite{FinazziEtAlPRL15} performs a diagonalization of the Lindbladian in a corner of the full Hilbert space, whose size is iteratively increased. As opposed to DMFT  which works in the thermodynamic limit, this is a finite-size method, which can however produce converged results for sizes much larger than brute force methods~\cite{rota2019quantum}. Both those approaches focus naturally on static correlations encoded in the stationary state density matrix while DMFT is constructed around the frequency-resolved Green's functions.

\subsection{Application to a Driven-Dissipative Bose Hubbard Lattice}
\label{sec:model}

In this work, we apply DMFT to a lattice model of driven-dissipative interacting bosons by specifying the local Hamiltonian and local jump operators entering Equations~(\ref{eq:masterEquation2}-\ref{eq:BH}). We consider for the former
\beq
\label{eq:locHam}
H_{\rm loc}=\omega_0  b^\da_i b_i  +\frac{U}{2}(  b^\da_i b_i )^2
\eeq
i.e. a characteristic frequency $\omega_0$ and on-site Kerr non linearity of strength $U$ while for the latter
we consider two kinds ($\mu=1,2$) of jump operators for each lattice site $i$, 
\bea\label{eqn:jumps}
L_{i2}=\sqrt{\eta}\,b_i b_i\\ \label{eqn:jumps2}
L_{i1}=\sqrt{r}\sqrt{ \eta}\,b_i^{\dagger}
\eea
We emphasize the correlated nature of the dissipative process encoded by $L_{i2}$ which acts only on states with multiple bosonic occupancy. This term will play a key role for our results. Interestingly this kind of loss process can be realized both with ultracold atoms~\cite{Syassen1329,TomitaEtAlScienceAdv17} as well with superconducting circuits~\cite{lescanne2020exponential}.
The resulting lattice model, Eq.~(\ref{eq:BH}-\ref{eqn:jumps2}), therefore describes a driven-dissipative realisation of the Bose-Hubbard model~\cite{fisherFisherPRB1989} whose properties in presence of incoherent drive and dissipation has been the subject of tremendous attention recently~\cite{HartmannEtAlNatPhys06,AngelakisEtAlPRA07,HartmannPRL10,Tomadin_prl10,JinEtAlPRL13,LeBoiteEtAlPRL13,LebreuillyEtAlPRA17,
foss-feig_emergent_2017,biondi_nonequilibrium_2017,VicentiniEtAPRA18,RotaEtAlPRL19,ScarlatellaFazioSchiroPRB19}. The specific form of dissipation we consider in Eq.~(\ref{eqn:jumps})  is rather unexplored in a many-body context, although few results are available in the literature that we will recall briefly here.

The many body master equation~(\ref{eq:masterEquation1},\ref{eq:locHam}-\ref{eqn:jumps2}) has a global $U(1)$ symmetry, associated with the invariance of the Liouvillian with respect to the transformation $b_i \rw b_i e^{i \theta }$,  and is time translational invariant (TTI). 
In the limit of a large number of bosons per site one expects a semiclassical description to work. The model reduces then to a discretized version of  Gross-Pitaevskii equation, largely studied in the context of exciton-polariton condensation~\cite{SiebereHuberAltmanDiehlPRL13}, which predicts a coherent phase of bosons for any $r>0$, independently of $J/U$. This phase, which spontaneously breaks both the $U(1)$ and TTI symmetry corresponds to a nonequilibrium superfluid. 
In the opposite regime of uncoupled sites, $J/U=0$, the steady-state density matrix is known analytically from Refs. \cite{dodonovMizrahiJPAMathGen1997,dykman1978} and describes an incoherent state: it is a statistical mixture of Fock states with $\aver{b_i} = 0$, as might be expected given the lack of any coherent driving. 
How those two different phases are connected upon increasing $J/U$, in the quantum regime of few bosons per site and finite lattice connectivity, is one of the main focus of this paper.

\begin{figure}[t]
\begin{center}
\epsfig{figure=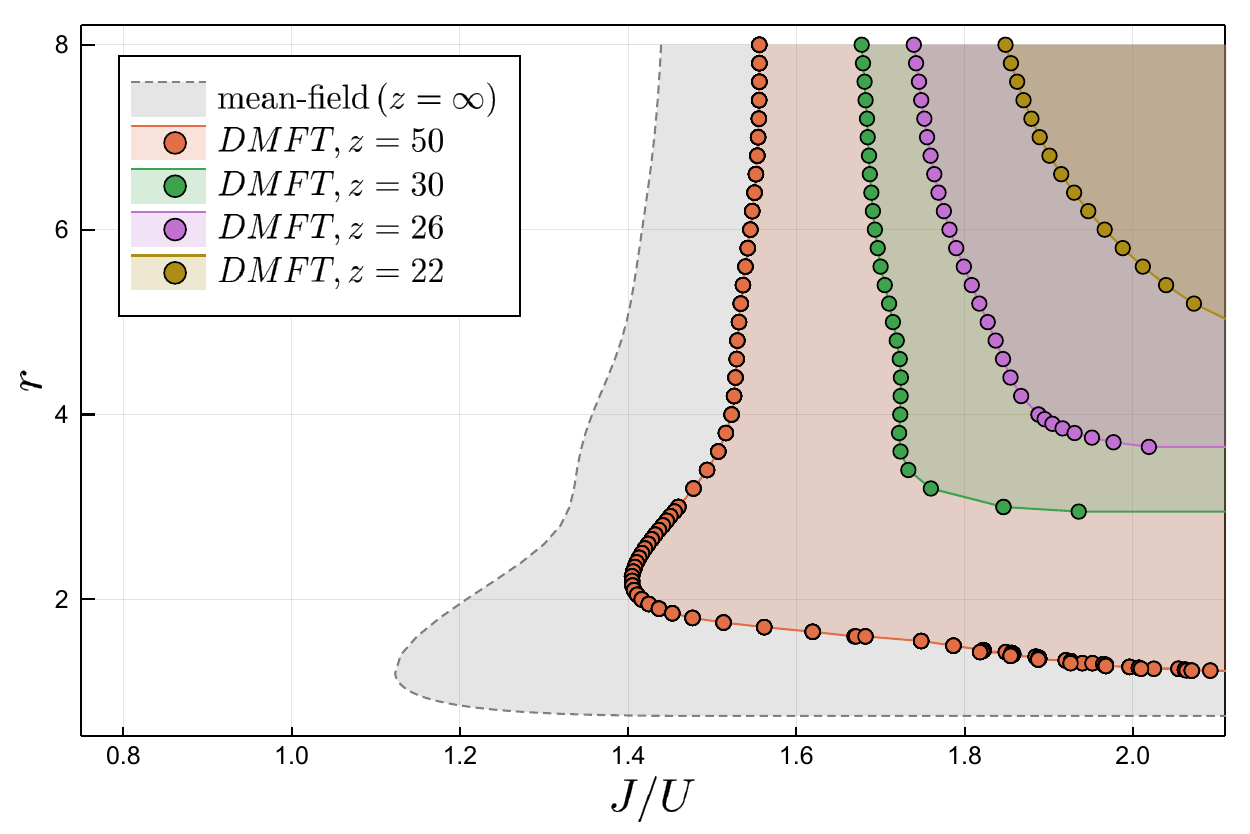,width=1\linewidth}
\caption{Phase Diagram of the Bose-Hubbard model (see Eq.(\ref{eq:BH}-\ref{eqn:jumps})) obtained with DMFT and NCA, as a function of pump/loss ratio $r$ and hopping/interaction ratio $J/U$, for different values of the lattice connectivity $z$, compared to the Gutzwiller mean-field one ($z=\infty$). A critical line $r_c(J)$ separates a normal low-hopping phase from a broken symmetry phase where the system develops a local order parameter oscillating in time at finite frequency. Decreasing the lattice connectivity, i.e. increasing quantum fluctuations due to the finite number of neighbors, the ordered phase shrinks and the entire phase boundary is reshaped. In particular the re-entrant behavior of the normal phase found in mean-field disappears at small values of $z$. Parameters: $\eta/U=0.02$, $dt=0.004$, $t_{ max}=10$, $dim_H=10$ (see Sec.~\ref{sec:results}).}
\label{fig:phaseDiag}
\end{center}
\end{figure}

In Figure~\ref{fig:phaseDiag} we plot the DMFT phase diagram for our Bose-Hubbard model as a function of $r$ (the dimensionless pump-to-loss ratio) and  $J/U$, for different values of the lattice connectivity $z$, together with the Gutzwiller mean field phase boundary corresponding to the $z=\infty$ limit~\footnote{We are not limited to the coordination numbers we spanned in Fig. \ref{fig:phaseDiag} and we can run DMFT equations for smaller values of $z$, but computing the phase diagram becomes particularly challenging, as it moves towards higher pump-to-loss $r$ values, as shown in Fig. \ref{fig:phaseDiag}, which are numerically hard to access requiring bigger Hilbert space sizes. 
}. For a given fixed value of $z$ we find a critical line $r_c(J)$ separating a small-hopping normal phase where the bosons remain incoherent, $\langle b_i\rangle=0$,  from a large hopping phase where the system develops a local order parameter breaking the $U(1)$ symmetry of the master equation. A first striking result that clearly appears from Figure~\ref{fig:phaseDiag} is that upon decreasing the connectivity, i.e. increasing the strength of fluctuations on top of the Gutzwiller mean field result, the phase diagram changes substantially. In particular the broken symmetry phase shrinks and moves toward larger values of pump and hopping.  Interestingly, the DMFT corrections to the phase diagram turn out to be substantially larger than for equilibrium lattice bosons~\cite{AndersEtAlPRL10,andersWernerNJP2011,strandWernerPRX2015}. 
The effect of finite-connectivity fluctuations is however not only quantitative. As we are going to discuss below, and more extensively in Sec.~\ref{sec:results}, these corrections arise from a qualitatively new physics that is captured by the DMFT/NCA description of the normal phase and completely missed by Gutzwiller mean-field theory.


As we will discuss in more details in Sec.~\ref{sec:spectFunctionNorm}-\ref{sec:occupation}, the normal phase of our model 
come with unique nonequilibrium properties, inherited from the local many-body physics of the single site problem~\cite{scarlatellaClerkSchiro2018}.  Above a pump threshold $r_{\rm ndos}$ the system develops a negative density of states (NDOS) at positive frequencies, a signature of incipient gain, i.e. energy emission in response to a weak coherent drive. Upon further increasing the pump to loss ratio above $r_{\rm inv}>r_{\rm ndos}$ the steady-state reduced density matrix shows population inversion, namely higher energy states become more occupied than lower energy ones.  Within Gutzwiller mean-field theory, which describes the normal phase as a product state of single sites, those scales are independent from the hopping $J$. DMFT instead shows that fluctuations due to finite connectivity reshape completely the spectral and distribution properties of the normal phase, leading in particular to a suppression of NDoS and population inversion. This arises from hopping-induced losses, a hallmark of the interplay between coherent and dissipative dynamics in our model,  which are the key physics captured by DMFT through the non-Markovian bath.

In Sec.~\ref{sec:DMFT_phaseDiag} we show that an interplay of NDoS and sufficiently strong hopping $J$ controls the true normal phase instability for values of the pump above $r_{\rm c}(J)$, when the system develops full phase coherence and enters the superfluid phase.
In particular we find that the unstable mode is modulated in time and that the system displays a finite frequency phase transition corresponding to an order parameter which develops undamped oscillations, thus breaking TTI~\cite{bucaJaksch2018,IeminiEtAlPRL18,ScarlatellaFazioSchiroPRB19}.
The large reduction of the normal phase at finite connectivity can be interpreted as an effect due to hopping-induced losses arising from the non-Markovian DMFT bath, which is able to wipe out the NDoS and absorb part of the energy emitted by the system, as we discuss more in detail in Sec.~\ref{sec:desynch}.This mechanism for the destruction of an ordered phase is of genuine nonequilibrium origin and cannot be understood in terms of an effective heating. Indeed, as we show in Sec.~\ref{sec:Teff} while an effective thermalisation is captured by DMFT through an effective temperature, this remains comparable to the Gutzwiller mean-field result up to small values of the connectivity and therefore cannot explain by itself the substantial reshape of the phase diagram.

The finite-frequency transition towards an oscillating nonequilibrium superfluid shares similarities with phenomena such as lasing, limit cycles and synchronization. As we show more in detail in Sec.~\ref{sec:synch}, the driven-dissipative Bose-Hubbard~(\ref{eq:BH}-\ref{eqn:jumps}) reduces in the semiclassical limit to an array of coupled classical van der Pol (vdP) oscillators~\cite{strogatz1991,matthews1991,cross2004}, which admits a stable limit cycle phase, a coherent phase with an order parameter oscillating in time at finite frequency, for any finite pump $r>0$ and any coupling $J$. 
In the quantum regime of few bosons per site, the picture qualitatively changes and a transition arises as a function of hopping $J$ depicted in Figure~\ref{fig:phaseDiag}. This can be interpreted, in light of this analogy, as a many-body quantum synchronization~\cite{lee2013,lorch2016,walter2014,giorgi2012,manzano2013,walter2015,roulet2018a,roulet2018,qiao2018,SonarEtalPRL18,
Davis_Tilley_2018,jaseem2019,tindall2020,dutta2019} where above a certain coupling $J$ all quantum VdP oscillators enter into a collective limit cycle phase.

 \begin{figure}[t]
\begin{center}
\epsfig{figure=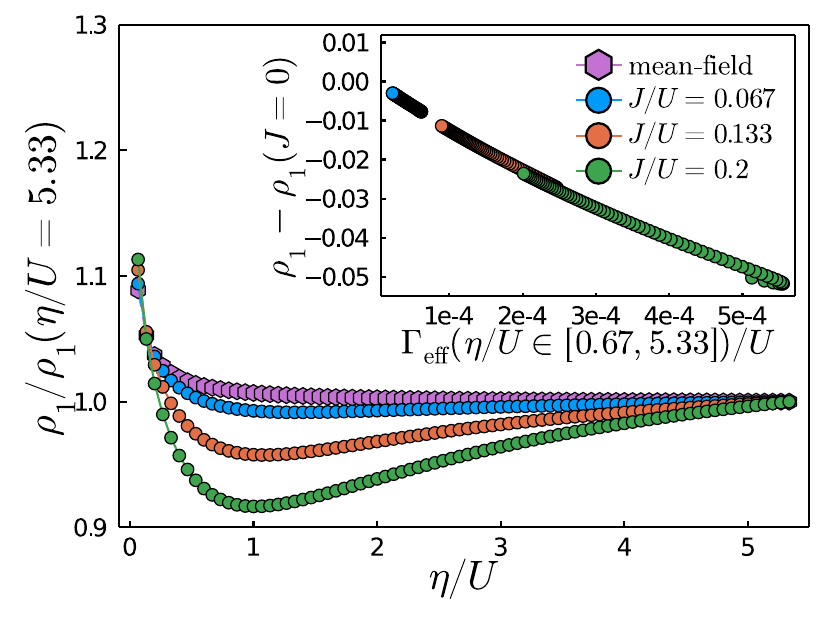,width=0.9\linewidth}
\caption{Emergence of Quantum Zeno Regime at fixed drive rate $f/U = r \eta/U =0.013$ and increasing two-particle loss rate $\eta$. Parameters: $t_{\rm max} =20$, $dt=0.002$, ${\rm dimH}=6$.  Probability of having one boson per site in the steady-state $\rho_1$ as a function of $\eta/U$, rescaled by its value at $\eta/U =2.67$ for convenience. It follows a non-monotonic behaviour with a minimum at $\eta=U$, which is more pronounced the larger the hopping $J$. 
This behaviour is a manifestation of quantum Zeno physics, controlled by the hopping-induced loss rate $\Gamma_{\rm eff }$ as we show clearly in the inset, which is instead completely missed by Gutzwiller.}
\label{fig:zeno_highlight}
\end{center}
\end{figure}

Finally, in order to highlight the role of hopping-induced dissipative processes and the qualitative differences between DMFT/NCA and Gutzwiller, in Sec.~\ref{sec:quantumzeno} we consider the limit of large two body losses $\eta\gg J$ for our Bose-Hubbard model. We note that this regime is experimentally accessible with ultracold gases in presence of inelastic collisions~\cite{Syassen1329,TomitaEtAlScienceAdv17}.
For large two-body losses, and in absence of any external pump, perturbation theory shows that one can restrict the Bose-Hubbard dynamics to a subspace made by hard-core bosonic states, the dark states of the local dissipator~\cite{garcia-ripoll2009}. Within this quantum Zeno subspace~ \cite{misra1977a,beige2000} the dominant dissipative processes left are those among neighboring sites, controlled by a hopping-induced loss rate $\Gamma_{\rm eff}= \frac{2 (J/z)^2}{U^2 + \eta^2} \eta $. This scale was shown to affect the transient dynamics~\cite{Syassen1329,garcia-ripoll2009,rossini2020b}, featuring a power-law decay towards the trivial zero-density steady-state.
Here we use DMFT/NCA to show that in presence of an additional small residual pump $f\ll \eta$ one can stabilize a Quantum Zeno stationary state in which physical properties are controlled by the scale $\Gamma_{\rm eff}$ (see Sec.~\ref{sec:quantumzeno}  for a detailed discussion).
An example is provided
in Figure~\ref{fig:zeno_highlight}, where we plot the DMFT/NCA steady-state occupation probability of the $n=1$ bosonic state versus $\eta/U$, for different values of $J/U$ and compare it with Gutzwiller results. The latter shows, independently of $J$, a very weak dependence from $\eta$ which would be completely absent if not for the small residual pump. DMFT results instead show a clear non-monotonous behavior as $J/U$ increases, with a minimum at $\eta\sim U$. This is a signature of the emergence of the Zeno scale $\Gamma_{\rm eff}$ controlling the physics, as we clearly reveal in the inset, where a scaling collapse is shown. We emphasize that the emergence of a Quantum Zeno stationary state represents a stringent test for the ability of DMFT of capturing hopping-induced losses, which are instead completely missed by Gutzwiller mean-field theory.

We give a more complete picture of our results for the Bose-Hubbard model in Sec.~\ref{sec:results}.

\section{Dynamical Mean-Field Theory For Markovian Bosons} 
\label{sec:markovLattice}

In this section we present the formalism of DMFT for Markovian bosons, including the basic self-consistency equations, its formal relation with Gutzwiller mean field theory and its physical interpretation, leaving its derivation to Appendix \ref{app:derivDMFT}. The starting point is to cast the Lindblad master equation \eqref{eq:masterEquation2} in the language of non-equilibrium Keldysh field theory, as discussed extensively in the literature~\cite{SiebererRepProgPhys2016}. The result is an action written in terms of bosonic coherent fields $\bar{b}_{i\alpha},b_{i\alpha}$ on each lattice site $i$ and on the upper/lower Keldysh contours, $\alpha=\pm$, which takes the form
\be\label{eqn:actionFromME}
\mathcal{S}=\intre dt \sum_{i\alpha}\alpha \bar{b}_{i\alpha} \, i\partial_t b_{i\alpha} - \intre dt \, i \lind 
\ee
where the Lindbladian $\lind$ is given by
\beq\label{eqn:actionFromME2}
\begin{split}
\lind &=  -i \lp H_+ - H_- \rp + \\
&+  \sum_{\mu,i} \gamma_{\mu} \lp  L_{i \, \mu +} \bar{L}_{i \, \mu -}   - \oh  \bar{L}_{i \, \mu +} L_{i \, \mu +}  - \oh L_{i \, \mu -} \bar{L}_{i \, \mu -}    \rp 
\end{split}
\eeq
with $H_{\alpha},L_{i\mu\alpha}$ are the expectation values of the Hamiltonian~(\ref{eq:BH}) and of the jump operators~(\ref{eqn:jumps}), expressed in terms of creation and annihilation operators belonging to the $\alpha$ contour, taken on bosonic coherent states. The full solution of the Keldysh action in Eq.(\ref{eqn:actionFromME}) is of course not possible in general, due to the coupling between many interacting modes and the presence of interaction, drive and dissipation. 

The key idea of DMFT is to write down an effective Keldysh action for the boson field of a single site of the lattice, obtained after integrating out all its neighbors~\cite{strandWernerPRX2015}. As we show in Appendix \ref{app:derivDMFT}, in the limit of large lattice connectivity, $z\gg 1$, this effective action has the closed form
\beq\label{eqn:S_eff}
\es{
\mathcal{S}_{\eff}&[\bb^\da_{\alpha},\bb_{\alpha}]=S_{loc}[\bb_{\alpha}^\da,\bb_{\alpha}]+\int dt \sum_{\alpha=\pm}\alpha\;\bold{\Phi}_{\eff\,\alpha}^\da (t) \bb_{\alpha}(t)+\\
& 
-\frac{1}{2}\int dtdt\sum_{\alpha,\beta=\pm}\alpha\beta\,\bb^\da_{\alpha}(t)\bold{\Delta}^{\alpha\beta}(t,t')\bb_{\beta}(t') 
}
\eeq
where $\alpha,\beta$ are Keldysh contour indices and we have dropped the site index from the local boson field for simplicity (we assume translational invariance) and grouped together creation/annihilation fields into a Nambu field 
\begin{eqnarray}\label{eqn:Nambu}
\bb^\da_{\alpha} =
\lp
\begin{array}{l}
\bar{b}_{\alpha} \; b_{\alpha}
\end{array}\rp
\,\;\;
\bb_{\alpha}=\lp
\begin{array}{l}
b_{\alpha}\\
\bar{b}_{\alpha}
\end{array}\rp
\end{eqnarray}
The above local Keldysh action describes a driven-dissipative quantum impurity model ~\cite{Scarlatella2019}. The first term in Eq.~(\ref{eqn:S_eff}), $S_{loc}[\bb_{\alpha}^\da,\bb_{\alpha}]$ is the local, on-site, contribution of the original lattice action~(\ref{eqn:actionFromME}-\ref{eqn:actionFromME2}) and therefore includes interactions, as well as Markovian incoherent drive and dissipation leading to off-diagonal terms in Keldysh space.
The second and third terms describe the feedback of the rest of lattice onto the site through its neighbors, in terms of an effective coherent drive  $\bold{\Phi}_{\eff\,\alpha}^\da (t) $ and an effective non-Markovian bath with hybridization function $\bold{\Delta}^{\alpha\beta}(t,t')$. Both these quantities have to be determined self-consistently, in particular the effective coherent drive reads
\bea
\label{eq:phieff}
\bold{\Phi}_{\eff\,\alpha}^\da =  J \bold{\Phi}^\da_{\alpha} (t) + \int \dt '\sum_{\beta=\pm}\beta \bold{\Phi}_{\beta}^\da(t')\bold{\Delta}^{\beta\alpha} (t',t) 
\eea
and has two contributions, the first coming from the average of the bosonic field as in Gutzwiller mean field theory
\be\label{eq:phi}
\bold{\Phi}^\da_{\alpha}  = \aver{\bb_{\alpha}^\da}_{S_\eff}
\ee 
and the second one coming from the non-Markovian bath, a non-trivial finite $z$ correction accounting for the feedback of neighboring sites on the local effective field~\cite{AndersEtAlPRL10,andersWernerNJP2011,strandWernerPRX2015}. This latter term, whose origin will be discussed more in detail in Appendix \ref{app:derivDMFT}, plays a key role within DMFT, in particular for what concerns the corrections to phase diagram as we discuss in Sec.~\ref{sec:DMFT_phaseDiag}-\ref{sec:desynch}.

The self-consistency relation for the Green's function depends on the specific choice of the lattice. In the following we will use the simplified relation for the Bethe lattice \cite{strandWernerPRX2015} 
\begin{align}
\label{eq:selfConsDelta}
\bold{\Delta}^{\alpha\beta}(t,t') =\frac{J^2}{z} \,  \bold{G}^{\alpha\beta}(t,t') 
\end{align}
which directly relates the hybridization function of the non-Markovian bath to the impurity connected Green's function
\be\label{eqn:GreenFunction}
\bold{G}^{\alpha\beta}(t,t') = -i\langle \bb_{\alpha}(t) \bb^\da_{\beta} (t')\rangle_{S_\eff}+i\bold{\Phi}_{\alpha}(t)\bold{\Phi}_{\beta}^\da(t') 
\ee
The DMFT solution of the original Markovian lattice problem thus requires one to solve the Keldysh action~(\ref{eqn:S_eff}), computing in particular the impurity Green's function~(\ref{eqn:GreenFunction}) and the average of the bosonic field~(\ref{eq:phi}), for given values of the non-Markovian bath $\Delta$ and effective field 
$\Phi$, and to iterate~(\ref{eq:phieff}-\ref{eq:selfConsDelta}) until self-consistency.

\subsection{Limit of infinite coordination number: Gutzwiller Mean Field Theory}
\label{sec:mean_field_conn}

It is instructive at this point to take explicitly the limit of infinite coordination number $z\rightarrow\infty$. In this limit,  the DMFT effective action \eqref{eqn:S_eff_lowest} becomes completely local in time
\beq\label{eqn:S_eff_lowest}
\es{
\mathcal{S}_{\eff}&[\bb^\da_{\alpha},\bb_{\alpha}]\xrightarrow{z=\infty}
S_{loc}[\bb_{\alpha}^\da,\bb_{\alpha}]+\int \sum_{\alpha}\alpha dt  \bold{\Phi}_{\eff\alpha}^\da (t) \bb_{\alpha}(t)
}
\eeq
 since the non-Markovian bath scales as $1/z$ (See Eq. ~\ref{eq:selfConsDelta}), and as such can be unfolded back into a master equation for a single-site density matrix $\rho$, which satisfies 
$$
\partial_t \rho(t)=i[ \bb^\da \bold{\Phi}(t),\rho]+\mathcal{L} \rho(t) 
$$  
where $\mathcal{L}$ is the local part of the Lindbladian and the feedback from the neighboring sites is carried by $\bold{\Phi}(t)=\tr [ \bold{b} \rho (t) ]$.  This corresponds to a factorized Gutzwiller-like ansatz for the lattice many body density matrix 
$$\rho_{\text{latt}}(t)=\prod_i \rho_i(t),$$
where $i$ is the site index and $\rho_i(t)\equiv \rho(t)$ because of translational invariance. In other words we have explicitly shown that, as for equilibrium or closed systems~\cite{AndersEtAlPRL10,strandWernerPRX2015} also for driven-dissipative lattice systems the infinite connectivity limit of bosons coincides with Gutzwiller mean-field theory. We note that when $\bold{\Phi} = 0 $, this mean-field describes completely uncoupled sites, while DMFT ($z<\infty$) captures the feedback from neighboring sites through the self-consistent bath $\bold{\Delta}$. In the following section we are going to add some physical intuition on how the DMFT action \eqref{eqn:S_eff} describes the effect of neighboring sites through a fictitious non-Markovian bath.

\subsection{DMFT Effective Action in the Classical/Quantum basis}
\label{sec:DMFTRAK}

We now give a physical interpretation to the various terms entering the DMFT effective action in Eq.~(\ref{eqn:S_eff}), in particular to the non-Markovian term.  It is useful to introduce the so called classical and quantum components of the bosonic field
\bea 
\bb_{\rm cl/q}(t)= \frac{\bb_+\pm \bb_-}{\sqrt{2}}\\
\bb^\da_{\rm cl/q}(t)= \frac{\bb^\da_+\pm \bb^\da_-}{\sqrt{2}}
\eea
in terms of which we can re-write the Keldysh action as 
\beq\label{eqn:S_effRAK}
\es{
\mathcal{S}_{\eff}&=S_{loc}[\bb_{\rm cl/q}^\da,\bb_{\rm cl/q}]
+
\int dt \bold{\Phi}_{\eff\,\rm cl}^\da\bb_{\rm q}
+\\
& 
-\frac{1}{2}\int  dtdt'
\left(
\bb^{\da}_{\rm q}(t)\bold{\Delta}^{R}(t,t')\bb_{\rm cl}(t')+hc\right)
+\\
&
-\frac{1}{2}\int  dtdt'\bb^{\da}_{\rm q}(t)\bold{\Delta}^{K}(t,t')\bb^{\da}_{\rm q}(t')
}
\eeq
In this basis only two independent combinations  of the non-Markovian kernels $\bold{\Delta}^{\alpha\beta}$ enter, namely the retarded component $\bold{\Delta}^R(t,t')=\theta(t-t') \lp \bold{\Delta}^{-+} (t,t') -\bold{\Delta}^{+-} (t,t')  \rp$,  which couples the classical and quantum components of the field and encodes the spectral properties of the bath resulting in a 
frequency dependent damping for the bosonic mode and the Keldysh component  $\bold{\Delta}^K=  \bold{\Delta}^{-+} (t,t') + \bold{\Delta}^{+-} (t,t')  $, encoding the occupation of the bath and resulting in a frequency dependent noise for the bosonic mode. It is worth stressing that the above Keldysh action contains quadratic, non-Markovian terms in the classical/quantum fields as well as non-linearities and higher powers of the classical/quantum fields included in the local part of the action $S_{loc}[\bb_{\rm cl/q}^\da,\bb_{\rm cl/q}]$. While the structure of Eq.~(\ref{eqn:S_effRAK}) is a generic feature of DMFT,  the local part of the action depends on the particular form of local interaction and jump operators. 

Finally, we can express also the impurity Green's functions Eq~(\ref{eqn:GreenFunction}) in this basis to obtain the retarded Green's function and the Keldysh one
\bea\label{eqn:GRAK}
\bold{G}^{R}(t,t') = -i\langle \bb_{\rm cl}(t) \bb^\da_{\rm q} (t')\rangle_{S_\eff}+i\bold{\Phi}_{\rm cl}(t)\bold{\Phi}_{\rm q}^\da(t') \\
\bold{G}^{K}(t,t') = -i\langle \bb_{\rm cl}(t) \bb^\da_{\rm cl} (t')\rangle_{S_\eff}+i\bold{\Phi}_{\rm cl}(t)\bold{\Phi}_{\rm cl}^\da(t')  \nonumber
\eea
Those correlation functions contain crucial physical information about the local physics of the driven-dissipative lattice problem.
The retarded Green's function in particular encodes information about the local excitation spectrum of the system and it is known to be a crucial probe for the transition from delocalization to Mottness in strongly correlated systems at equilibrium~\cite{Review_DMFT_96}. 
For open Markovian quantum systems the retarded Green's function contains, much like for closed equilibrium systems, information on the structure of the excitations on top of the stationary state~\cite{scarlatellaClerkSchiro2018} and it directly probes dissipative phase transitions where those excitations become unstable. Its poles correspond to eigenvalues of the Lindbladian in the single particle sector, which come with a characteristic frequency and lifetime, and their (possibly complex) weight. 
The retarded Green's function has also a directly physical meaning:  it describes the linear response of the expectation $\aver{b(t)}$ to a weak, classical field $h(t')$, which couples linearly to $b^\da$.  In the case where $b$ describes a photonic cavity mode, $G^R(t)$ can be directly measured by weakly coupling the cavity to an input-output waveguide and measuring the reflection of a weak probe tone (see e.g.~\cite{LemondePRL2013,lavitanClerkNJP2016}).

The Keldysh Green's function on the other hand contains information about how the finite frequency excitations above the stationary state are populated. In thermal equilibrium those two functions are not independent, but constrained to satisfy the fluctuation-dissipation theorem \cite{kamenev2011field}.

\subsection{Computing Lattice Quantities}
\label{sec:lattice}
Solving the DMFT effective action and computing the impurity Green's functions~(\ref{eqn:GRAK}) gives direct information on the local properties of the driven-dissipative lattice problem. Furthermore one can access non-local quantities, such as momentum distribution or non local correlation functions, through the lattice Green's functions at momentum $\textbf{k}$
\be\label{eq:latticeGreens}
\bold{G}^{\alpha\beta}_{\textbf{k}}(t,t')= -i\langle \bb_{\textbf{k}\alpha}(t) \bb^\da_{\textbf{k}\beta}(t')\rangle
+i\bold{\Phi}_{\textbf{k}\alpha}(t)\bold{\Phi}_{\textbf{k}\beta}^\da(t')\,.
\ee
These satisfies a Dyson equation with a lattice self-energy $\bold{\Sigma}^{\alpha\beta}(t,t')$, that within DMFT is momentum independent \cite{Georges_Kotliar_PRB92,aokiWernerRMP2014}, 
\beq \label{eq:latticeDyson}
\bold{G}^{\alpha\beta}_{\textbf{k}}(t,t')=\bold{g}^{\alpha\beta}_{\textbf{k}}(t,t')+
\sum_{\gamma\delta}\bold{g}^{\alpha\gamma}_{\textbf{k}}\otimes \bold{\Sigma}^{\gamma\delta}\otimes \bold{G}^{\delta\beta}_{\textbf{k}}(t,t')
\eeq
and coincides with the self-energy of the impurity problem
\beq
\label{eqchdmft:impurityGreens}
\bold{G}^{\alpha\beta}(t,t')=\bold{g}^{\alpha\beta}(t,t')+
\sum_{\gamma\delta}\bold{g}^{\alpha\gamma}\otimes \bold{\Sigma}^{\gamma\delta}\otimes \bold{G}^{\delta\beta}(t,t')
\eeq
where in the above equations $\otimes$ indicates time convolutions, $\bold{g}^{\alpha\beta}(t,t')$ are the Green's functions of the quantum impurity problem with no interactions, but including the non-Markovian bath $\bold{\Delta}$ and $\bold{g}^{\alpha\beta}_{\textbf{k}}(t,t')$ are the non-interacting lattice Green's functions.

\section{Quantum Impurity Solvers}
\label{sec:impuritysolver}

The main challenge behind our DMFT approach is to solve the Markovian quantum impurity model described by the Keldysh action ~\eqref{eqn:S_eff}, computing in particular the Green's functions.  
We stress that this remains a difficult task due to the presence of interactions on the impurity site, non-linear jump operators (such as our two-body losses) and the non-Markovian DMFT bath. While several impurity solvers have been developed in recent years for non-equilibrium DMFT~\cite{aokiWernerRMP2014}, none of them can be efficiently applied in our case (See Sec.~\ref{sec:prior} for a detailed discussion). To make progress we take explicit advantage of the Markovian structure of the impurity, which allows to treat non-linear jump operators as dissipative couplings of a local Lindbladian. 
This unleashes the possibility of developing strong-coupling impurity solvers for bosonic Markovian problems, which treat exactly the local Lindblad problem and include the effect of the non-Markovian DMFT bath through perturbative or non-perturbative schemes. We note that for non-equilibrium closed systems these strong-coupling methods represent the current state of the art of DMFT impurity solvers~\cite{aokiWernerRMP2014}. Here we develop two such schemes for bosonic Markovian systems, the Hubbard-I approximation and the more powerful Non-Crossing-Approximation, that we both present below. We comment in the Sec.~\ref{sec:conclusions} on possible methodological extensions.

\subsection{Hubbard-I Approximation}\label{sec:HubbardI}

The simplest approximation to solve the impurity problem~(\ref{eqn:S_eff}) is based on perturbation theory in the non-Markovian bath kernel $\bold{\Delta}$, and its lowest order is known as the Hubbard-I approximation \cite{Review_DMFT_96,strandWernerPRA2015}. As we will see this approach already gives a hopping dependence of correlation functions which goes beyond Gutzwiller mean-field theory, but misses important correlations due to the non-Markovian bath. Our DMFT approach will be based on the more powerful non-crossing approximation solver which we will introduce in the next section, but we will use Hubbard-I results for comparison and to motivate the need of a more powerful solver. 

For simplicity, we formulate Hubbard-I in the normal phase,  where $\bold{\Phi}= 0$ and anomalous correlation functions vanish, thus we can restrict to the first Nambu component and refer to it with non-bold symbols, e.g. $G^{\alpha\beta}= \bold{G}^{\alpha\beta}_{11} $ where $\alpha,\beta=\pm$ are Keldysh indexes.  We also focus on the stationary state regime, where Green's functions depend on time differences and we can move to the frequency domain i.e. $G^{\alpha\beta}(\omega)$, which is the case we will consider in our application in Sec. \ref{sec:results}. 

The impurity Green's function obeys a Dyson equation, see Eq.~(\ref{eqchdmft:impurityGreens}), in terms of a self-energy $\Sigma^{\alpha\beta}(\omega)$ which contains the effect of interaction, incoherent drive and dissipation and which is in general a functional of the non-Markovian bath kernel $\Delta^{\alpha\beta}(\omega)$. Hubbard-I consists in approximating the impurity self-energy by its value for ${\Delta}=0$, i.e. in absence of the bath, when it can be written as
\bea
\Sigma^{\alpha\beta}(\omega)\approx \left[g^{\alpha\beta}_0(\omega)\right]^{-1}-\left[G_0^{\alpha\beta}(\omega) \right]^{-1}\,.
\eea
Here $G^{\alpha\beta}_0(\omega)$ is the Green's function of the impurity site with interaction, incoherent drive and dissipation, but without the bath (the latter condition is indicated by the index $0$); it can be computed numerically~\cite{scarlatellaClerkSchiro2018}. In contrast $g^{\alpha\beta}_0(\omega)$ corresponds to the Green's function of the impurity site in absence of the bath and without interactions (lowercase letter), which is known analytically.  Plugging this self-energy back in the Dyson equation~(\ref{eqchdmft:impurityGreens}), and using the self-consistency condition on the Bethe Lattice,  we obtain a closed matrix equation for the Keldysh components of the local lattice Green's functions 
\be
\label{eq:hubbI}
\left[G^{\alpha\beta}(\omega) \right]^{-1}=\left[G_0^{\alpha\beta}(\omega) \right]^{-1}-\frac{J^2}{z}G^{\alpha\beta}(\omega) 
\ee
The expressions of the retarded and Keldysh components are given explicitly in appendix \ref{app:hubbI}. 
In the appendix, we also show that the Hubbard-I approximation, despite introducing a beyond mean-field hopping dependence of Green's functions, still yields the same phase diagram as mean-field, motivating the need for a more powerful solver.

\subsection{Super-Operator Hybridization Expansion and Non-Crossing Approximation}\label{sec:NCA}

\begin{figure}[t]
\begin{center}
\epsfig{figure=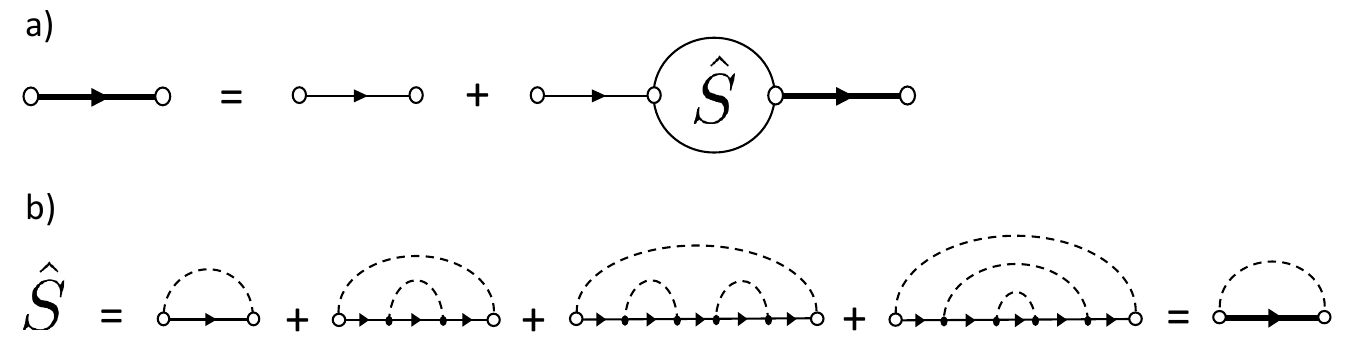,width=1\linewidth}
\caption{
a) Diagrammatic representation of the Dyson series in Eq.~(\ref{eq:dysonNCA}). The bold (thin) solid lines represent the full, non-Markovian (bare, Markovian) impurity super-operator $\hat{\mcV}$ ($\hat{\mcV}_0$), while the dashed lines correspond to the non-Markovian bath $\bold{\Delta}$. 
b) Expression of the self-energy $\hat{S}$ in the NCA approximation, where only diagrams with non-crossing $\bold{\Delta}$ lines are retained whose sum can be explicitly evaluated, see Eq.~(\ref{eq:selfEnergyExplicit}).
}
\label{fig:1PI_diagrams}
\end{center}
\end{figure}

To go beyond the Hubbard-I approximation, we build upon the method recently formulated in Ref.~\cite{Scarlatella2019}  and applied so far only to a simple toy model fermionic system, to develop a DMFT/NCA impurity solver for bosonic Markovian systems.
The idea is to perform a diagrammatic expansion in powers of the non-Markovian bath $\bold{\Delta}$ and to resum an infinite set of diagrams by solving a self-consistent Dyson-type equation. We remark that this expansion is carried out around an interacting problem, the single-site Markovian impurity, hence it is not based on Wick's theorem as in weak-coupling perturbation theories. 
As such, working directly with Green's functions is not convenient and the more natural formulation is in terms of evolution super-operators, that we will denote in the following with a hat~\cite{Scarlatella2019}. We start by defining the evolution super-operator $ \hat{\mcV}$ of the reduced density matrix of the impurity
\be
\label{eq:vMap}
\rho_{\rm imp}(t) = \hat{\mcV}(t,0) \rho_{\rm imp}(0)
 \ee
 formally obtained by tracing out the bath degrees of freedom. We note that Eq.~(\ref{eq:vMap}) assumes that at time $t=0$ the non-Markovian bath is not entangled with the impurity site, i.e. in the original lattice problem the initial condition corresponds to the limit of decoupled sites. Since the bath degrees of freedom are treated as non-interacting, only the single-particle Green's function of the bath enters the reduced dynamics, the hybridization function $\bold{\Delta}$ introduced in Eq.~(\ref{eqn:S_eff}). Expanding the super-operator $\hat{\mcV}(t,0) $ in powers of $\bold{\Delta}$ we obtain a series which can be represented diagrammatically as shown in Fig. \ref{fig:1PI_diagrams}, where bold solid lines describe the propagator $\hat{\mcV}(t,0)$, dashed lines represent the hybridization function $\bold{\Delta}$ while simple solid lines  
represent the bare Markovian evolution super-operator $\hat{\mcV}_0(t,0) = T \exp \lp \int_0^t dt' \hat{\lind}_\eff (t') \rp$, where $T$ is the time-ordering and $\hat{\lind}_\eff (t)= \hat{\lind}_0 + i [ \hat{\bb}^\da \hat{\bold{\Phi}}(t), \bullet] $ the effective single site Lindblad super-operator with argument $\bullet$. 

This diagrammatic representation allows to define the self-energy $\hat{S}$ of the series as the sum of one-particle-irreducible (1PI) diagrams, which cannot be cut into two disconnected parts by removing a solid line, and thus to formally resum the series into the Dyson equation

\beq
\label{eq:dysonNCA} 
{
 \hat{\mcV} (t,t') =\hat{\mcV}_0(t,t') + \int_{t'}^t dt_1 \int_{t'}^{t_1} dt_2  \hat{\mcV}_0(t,t_1)   \hat{S} (t_1,t_2) \hat{\mcV} (t_2,t') }
\eeq 
We remark that $\hat{\mcV}_0$, $\hat{\mcV}$ and $\hat{S}$ here are super-operators and that the self-energy $\hat{S}$ is a functional of the  propagator $\hat{\mcV} $ whose closed form is not known in general. The resulting series~(\ref{eq:dysonNCA}) generalizes to the case of Markovian impurities the hybridization expansion obtained for unitary quantum impurity models \cite{SchoellerSchonPRB94,muhlbacherRabaniPRL2008,schiroFabrizioPRB2009, schiroPRB2009,Werner_Keldysh_09}. For the latter, exact resummation techniques based on diagrammatic Monte Carlo~\cite{Gull_RMP11} have been employed but generically suffer from the so called sign problem, especially out of equilibrium, limiting the propagation time.
Here instead we adopt a self-consistent approximation for the self-energy $\hat{S}$. This can be written in general as a systematic expansion in diagrams with an increasing number of crossing hybridization lines, an approach which has been extensively used  for unitary quantum impurity models~\cite{Bickers1987,Nordlander1999,Eckstein2010,rueggMillisPRB2013}.  The lowest order contribution is given by non-crossing diagrams, e.g. in Fig. \ref{fig:1PI_diagrams}, giving an explicit expression for the NCA self-energy 
\begin{widetext}
\begin{align}
\label{eq:selfEnergyExplicit}
&\hat{S} (t_1,t_2) =- \frac{i}{2} \underset{ab}{\sum_{\alpha \beta  }}    \alpha \beta  \lsq  \Delta_{b  a }^{\beta \alpha} (t_1, t_2 ) \hat{b}_{ \beta  b }^\da \hat{\mcV} (t_1,t_2) \hat{b}_{ \alpha a } +   \Delta_{ a b }^{\alpha \beta} (t_2,t_1 ) \hat{b}_{\beta b} \hat{\mcV} (t_1,t_2) \hat{b}^\da_{\alpha a}   \rsq 
\end{align}
\end{widetext}
In the above expression $\alpha, \beta=\pm$ are Keldysh indices and $a,b=\left\{1,2\right\}$ are Nambu indices. Thus $\Delta^{\alpha\beta}_{ab}$ is a given component of the bath hybridization function introduced in Eq.~(\ref{eqn:S_eff}) , i.e. ${\Delta}^{\alpha\beta}_{ab}=\left(\bold{\Delta}^{\alpha\beta}\right)_{ab}$. We also introduce the super-operators analogues of the Nambu fields of Eq.~(\ref{eqn:Nambu}), that we define as 
\begin{eqnarray}\label{eqn:NambuSuper}
\hat{\bb}^\da_{\alpha} =
\lp
\begin{array}{l}
\hat{b}^{\dagger}_{\alpha} \; \hat{b}_{\alpha}
\end{array}\rp
\,\;\;
\hat{\bb}_{\alpha}=\lp
\begin{array}{l}
\hat{b}_{\alpha}\\
\hat{b}^{\dagger}_{\alpha} 
\end{array}\rp
\end{eqnarray}
and denote their $a$ Nambu component as $\hat{b}_{\alpha a}$ in Eq.~(\ref{eq:selfEnergyExplicit}). The Keldysh index $\alpha=\pm$ for a super-operator specifies whether it should act from the left or the right of its argument, i.e.
 \beq \hat{b}_{ + } = {b} \bullet \qquad
  \hat{b}_{ - } = \bullet{b}
  \eeq
  and similarly for $\hat{b}^{\dagger}_{\alpha}$.
We notice that the self-energy depends on the full propagator $\hat{\mcV}$, rather than on the bare one $\hat{\mcV}_0$, thus containing diagrams to all orders in $\bold{\Delta}$. Corrections to the NCA can be obtained systematically including self-energy diagrams with higher number of crossings, although the resulting computational cost increases. In this work, where we focus on the normal phase and its instability, we will limit ourselves to the NCA scheme, while we expect that to access the superfluid phase or for lower values of the connectivity higher order corrections would become important. We note in fact that self-energy corrections including higher number of crossings diagrams come with higher powers of the DMFT bath, which for bosons is of order $J^2/z$ (see Eq.~(\ref{eq:selfConsDelta})) and therefore subleading at least for large to moderate values of the connectivity.

Once the self-energy $\hat{S}$ is known in closed form, the propagator $\hat{\mcV}$ can be obtained numerically by solving Eqs.~\eqref{eq:dysonNCA} and \eqref{eq:selfEnergyExplicit}. 
To use this NCA impurity solver in our DMFT approach, we need to compute the one-particle Green's functions of the impurity, Eq.~(\ref{eqn:GreenFunction}). This can be obtained by taking the functional derivative  with respect to $\bold{\Delta}$ of the partition function $Z =\tr \lsq\rho_{\rm imp}(\infty) \rsq$ in Eq.~(\ref{eq:vMap}) and using the Dyson equation for $\hat{\mcV}$ (see Appendix~\ref{app:ncaGreenFunctions}). The final result reads
\beq
\label{eq:NCAGreen}
\es{
G^{\alpha \beta}_{a b}(t,t') = -i \Big\lbrace & \tr \lsq \hat{b}_{\alpha a} \hat{\mcV}(t,t')\hat{b}_{\beta b}^\da \rho_{\rm imp}(t')\rsq \theta(t-t') + \\ + & \tr \lsq \hat{b}_{\beta b}^\da  \hat{\mcV}(t',t) \hat{b}_{\alpha a}\rho_{\rm imp}(t) \rsq \theta(t'-t) \Big\rbrace +\\
+ &
i \Phi_{\alpha a}(t) \Phi_{\beta b}^\da(t') 
}
\eeq
where as before we have written explicitly both the  Keldysh indices $\alpha,\beta$ and the Nambu ones $a,b$ and where
$ \Phi_{\alpha a}(t)=\tr\lsq \hat{b}_{\alpha a}\rho_{\rm imp}(t)\rsq$. We notice that this result, which resembles a quantum regression theorem~\cite{breuerPetruccione2010} for the non-Markovian map $\hat{\mcV}(t,t')$ is only valid within NCA, while including higher order diagrams into the self-energy would lead to further terms which can be interpreted as vertex corrections.

Finally, we conclude by emphasizing that the solver introduced in this section is different from other NCA approaches to  quantum impurity models with or without dissipation~\cite{Eckstein2010,strandWernerPRX2015,Chen2016,PeronaciEtAlPRB2020,erpenbeck2021revealing}, which treat at the non-crossing level all couplings to the baths. Here, by formulating the hybridization expansion at the super-operator level, we are able to  fully capture the underlying local Markovian dynamics, resorting to an NCA approximation only for the non-Markovian DMFT bath. This introduces several differences with respect to the NCA literature, including the way the Green's functions are evaluated (see Eq.~(\ref{eq:NCAGreen})) and in the way the stationary-state theory is constructed, as we discuss further in the next section.

\subsubsection{Stationary state DMFT/NCA}\label{sec:stationarystateNCA}

While the formalism introduced so far allows to compute the whole transient dynamics, in this section we show how to directly address the stationary state properties of the system within our DMFT/NCA approach. At stationarity we expect the local Green's functions \eqref{eqn:GreenFunction} and, through the self-consistent condition~(\ref{eq:selfConsDelta}), the bath hybridization function $\bold{\Delta}^{\alpha\beta}$ to depend only on time differences.  We can then solve the NCA Dyson equation \eqref{eq:dysonNCA} for the stationary state propagator $\hat{\mcV}(t-t')$ which also depends only on time differences. This allows to significantly reduce the computational cost  for time-propagating this equation from $O(t_{\rm max}^3)$ to $O(t_{\rm max}^2)$, where $t_{\rm max}$ is the maximum integration time.

A complete steady-state DMFT/NCA procedure requires to compute, in addition to the stationary state propagator, also the steady-state density matrix of the impurity $\rho_s \equiv \hat{\mcV}(\infty,0)\rho_{\rm imp}(0)$, which is needed to evaluate the impurity Green's functions (See Eq.~(\ref{eq:NCAGreen})). While in principle this would require to perform the full transient dynamics from an arbitrary initial condition, here we show how to obtain $\rho_s$ directly from the stationary state propagator $\hat{\mcV}(t-t')$. We note that for Markovian open quantum systems the stationary-state density matrix can be directly obtained as zero-eigenvalue of the Lindblad supeoperator generating the dynamics. This argument however does not directly apply to the present case, since the DMFT bath makes the map $\hat{\mcV}(t,t')$ non-Markovian. A generalized stationarity condition for the non-Markovian map~(\ref{eq:dysonNCA}) can be obtained~ \cite{Scarlatella2019}  by requiring the derivative of Eq.~(\ref{eq:dysonNCA}) to vanish at long times, i.e. $\mbox{lim}_{t\rightarrow
\infty}\partial_t\hat{\mcV}(t,0)\rho_0=0$. This equation however still requires the knowledge of the full transient propagator.
A major simplification arises in DMFT if the system reaches a stationary state becoming time-translational invariant. Then the condition for the impurity density matrix simplifies to (see Appendix~(\ref{sec:rho_ss}))
\beq\label{eqchnca:ss_equation}
\lp \hat{\mathcal{L}}_\eff(\infty) +  \int_{0}^\infty d\tau \hat{S}(\tau) \rp \rho_{s} = 0 
\eeq
where the self-energy $\hat{S}(\tau)$ depends only on the steady-state propagator $\hat{\mcV}(\tau)$ and not on the transient dynamics, allowing to compute $\rho_s$ in a steady-state DMFT procedure. Equation~(\ref{eqchnca:ss_equation}) is analogous to the well known condition for the stationary state of Markovian master equations, with an additional contribution of the non-Markovian bath given by the time-integral of the NCA self-energy.
In practice, to solve DMFT/NCA for the stationary state, we solve the Dyson equation \eqref{eq:dysonNCA} for $\hat{\mcV}(t)$ starting from an initial ansaz for $\bold{\Delta}(t-t'), \bold{\Phi}(t')$ and $\rho_s$. As an initial condition we usually compute these quantities from the steady-state solution of the single-site problem.
Then we compute the updated stationary density matrix $\rho_s$ using Eq. \eqref{eqchnca:ss_equation} and the updated $\bold{\Delta}(t-t'), \bold{\Phi}(t')$ from Eqs. \eqref{eq:phieff},\eqref{eq:selfConsDelta} and iterate until convergence is reached. 
We conclude by noting that in principle the stationary state approximation could break down, leading to oscillatory behaviors. It is therefore important to study the stability of the steady state, which is encoded in the retarded Green's function as we discuss more in detail in Sec.~\ref{sec:DMFT_phaseDiag}.

\section{DMFT Results for a Driven-Dissipative Bose-Hubbard Lattice}
\label{sec:results}

In this section we discuss our results for the driven-dissipative Bose-Hubbard model introduced in Sec.~\ref{sec:model},  comparing different impurity solvers (NCA and Hubbard-I approximation) and highlighting the effect of introducing fluctuations beyond Gutzwiller mean-field due to the finite lattice connectivity. 
We start by discussing the properties of the normal phase at low hopping as encoded in its local spectral function (Sec.~\ref{sec:spectFunctionNorm}).  We then move on to occupation properties of the nonequilibrium normal phase (Sec.~\ref{sec:occupation}) from the point of view of the local density and populations of the stationary-state reduced density matrix.
In Sec.~\ref{sec:DMFT_phaseDiag} we discuss the finite-frequency instability of the normal phase, leading to the DMFT/NCA phase diagram, and provide a physical interpretation based on hopping-induced dissipation for the large reduction of the ordered phase found in DMFT with respect to the Gutzwiller mean-field result.
In Sec.~\ref{sec:synch} we connect the phase transition in our driven-dissipative Bose-Hubbard model to the physics of an array of quantum Van der Pol oscillators, in particular to the onset of many body synchronization and limit cycles and discuss their fate at finite lattice connectivity. Finally, in Sec.~\ref{sec:quantumzeno} we discuss the regime of large two-body losses, where Quantum Zeno physics emerges and the qualitative differences between Gutzwiller and DMFT/NCA results appear even more clearly.

Unless stated otherwise, we work in the regime where the interaction strength dominates the dissipation scale, i.e. we fix $\eta/U=0.02$, and study the model as a functions of the pump/loss ratio $r$ and the hopping to interaction ratio $J/U$. We set $U=5$ and $\omega_0=1$, although we note that this latter scale only sets the zero of energy and can be eliminated by going to a rotating frame, so it does not play any role in the physics.

We introduce a cutoff on the local Hilbert space $dim_H$, whose value will be specified for each result. 
We solve DMFT for the normal phase, where $\bold{\Phi}=0$ and the anomalous (Nambu) Green's function components vanish so that the self-consistent bath only retains Keldysh indexes $\Delta^{\alpha\beta}$. The NCA propagator in the stationary regime $\mcV(t)$ is obtained, as described in Sec. \ref{sec:stationarystateNCA}, by propagating in time the derivative of the Dyson equation \eqref{eq:dysonNCA} assuming time-translational invariance
\beq
\label{eq:dysonNCAdiff} 
{
\pt \hat{\mcV} (t) = \hat{\lind}_0 \hat{\mcV}(t) + \int_{0}^t dt_1   \hat{S} (t-t_1) \hat{\mcV} (t_1) }
\eeq 
with an implicit second-order Runge-Kutta scheme \cite{aokiWernerRMP2014}, a propagation time $t_{max}=10$ and a time step $dt=0.004$. We note that in the regime under consideration in this work the dynamics of the Dyson equation is dominated by the non-Markovian bath rather than by the two particle losses and therefore a $t_{max}=10=1/\eta$ is sufficient to reach convergence. Convergence of the implicit Runge-Kutta at each time step is assumed to be reached when $1/(dim_H)^4 \sum_{jk} |\mcV_{jk}^{(i)}(t) - \mcV_{jk}^{(i-1)}(t)| < 10^{-5}$, being $i$ the iteration index. The convergence of the DMFT scheme is assessed by checking that $ 1/(2 t_{max}) \sum_{\alpha} \int_{0}^{t_{max}} | \lp {\Delta}^{\alpha,-\alpha}\rp^{(i)}(t)  - \lp {\Delta}^{\alpha,-\alpha}\rp^{(i-1)}(t) | < 10^{-5}$, being $i$ the index of the DMFT iteration. 
We have checked that our results essentially do not change by decreasing those thresholds or increasing the Hilbert space cutoff.

\subsection{Spectral Function in the Normal Phase}
\label{sec:spectFunctionNorm}

\begin{figure}[t]
\begin{center}
\epsfig{figure=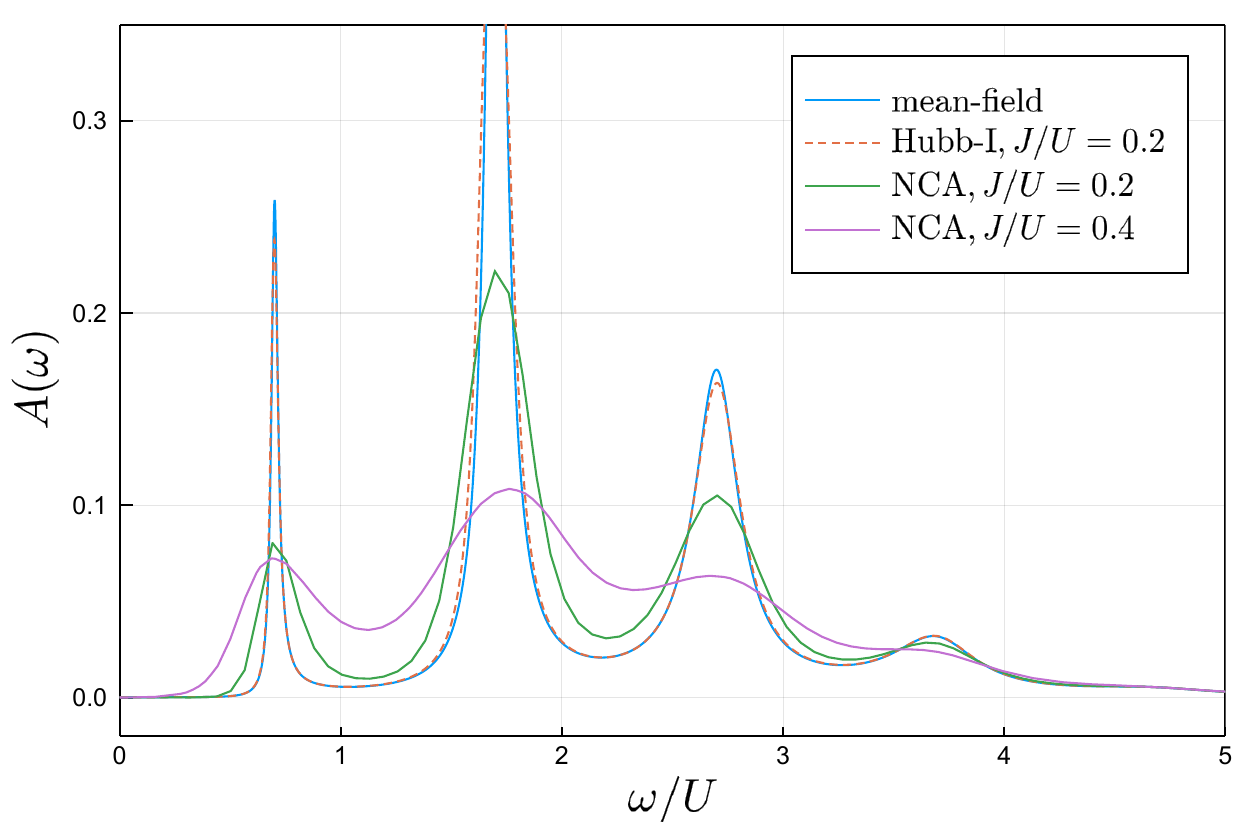,width=1\linewidth}
\caption{
Local spectral function $A(\omega)$ for different values of $J/U$, as computed within Gutzwiller mean-field theory (blue line), DMFT with Hubbard-I impurity solver (dashed red line) and NCA impurity solver (green and violet full lines), for fixed $z=6$ and $r=0.6$. Gutzwiller results show a series of narrow peaks broadened by the local dissipation only. DMFT instead is able to capture additional broadening processes, already evident for $J/U=0.2$ within Hubbard-I, which however largely underestimates the effect of the non-Markovian bath as confirmed by the comparison with the more accurate NCA.
Parameters: 
$dim_H=10$. 
}
\label{fig:specFunc_smalldrive}
\end{center}
\end{figure}

To characterize the properties of the system, we first focus on the local retarded Green's function defined in Eq.~(\ref{eqn:GRAK}). Since in the normal phase  all anomalous (Nambu) Green's function components vanish as well as the average of the order parameter, we have only one independent Nambu component $G^R(t)=-i\theta(t)\langle [b(t),b^{\dagger}(0)]\rangle$. Its imaginary part defines the local spectral function
\be 
\label{eq:specFunc}
A(\w)=-\frac{1}{\pi}\mbox{Im} G^R(\w)
\ee
In Fig.~\ref{fig:specFunc_smalldrive} we plot the local spectral function in the low pump regime, $r=0.6$, for different values of $J/U$ and compare the DMFT/NCA results with those obtained with Hubbard-I impurity solver and Gutzwiller mean-field.

The Gutzwiller mean-field spectral function shows a series of narrow peaks, whose broadening is controlled only by the local dissipation. We remark that in this approach, corresponding to the infinite coordination number limit $z=\infty$, all properties of the normal phase are independent on the hopping and coincide with the single-site $J=0$ limit. Indeed as we have discussed in Sec.~\ref{sec:mean_field_conn}, for $z=\infty$ the only feedback from neighboring sites comes through the order parameter $\bold{\Phi}$, which vanishes in this the normal phase.

DMFT instead is able to capture the effect of coherent hopping processes, resulting in a further broadening of the resonances. This finite hopping correction to the spectral function reflects the fact that the stationary density matrix in the normal  phase is \emph{not} a tensor product of single-site density matrices, as predicted by Gutzwiller, but rather includes correlations among neighboring sites encoded within DMFT in the non-Markovian bath.

A comparison between Hubbard-I and NCA, shows that the former largely underestimates the effect of the bath. Indeed within NCA the sharp peaks of the isolated single site problem are largely broadened already for a moderate value of the hopping rate $J/U=0.2$, a trend that further increases for larger values of $J/U$. At the same time the location of the poles is found to be weakly dependent on the hopping rate and, at least for $J/U=0.2$, essentially captured already by Hubbard-I and Gutzwiller mean-field.
This difference can be understood by noticing that there are two main sources of resonance broadening within our DMFT approach, one coming from the bare non-Markovian bath $\Delta(\omega)$, the other coming from the Markovian interacting single-site problem, encoded in the self-energy $\Sigma(\omega)$ (see Sec.~\ref{sec:lattice}). Within Hubbard-I the latter is independent of $J$, and only set by pump and losses. NCA on the other hand accounts for many body scattering channels \emph{mediated} by the bath and results in an imaginary part of $\Sigma(\omega)$, also scaling with the hopping strength and responsible for the larger broadening. 
We emphasize that while the main effect of the DMFT bath in this regime is to broaden the resonances, this broadening is not uniform in frequency, i.e. it could not be reproduced by treating the DMFT bath with a Markov approximation. In fact, the self-consistent condition Eq.~(\ref{eq:selfConsDelta}) implies that the spectrum of the DMFT bath is given by the local spectral function itself, namely it comes with a rich multi-peak structure in frequency which prevents the use of a simple Markovian approximation.
Overall the spectral function in this low-drive, low-hopping normal region is very reminiscent of an equilibrium Bose Hubbard model in the Mott insulating phase~\cite{strandWernerPRA2015}, with Hubbard bands, describing doublon/holons and multiparticle excitations, which are partially filled by incoherent pump and dissipation. As we show in the next section, increasing the pump strength reveals a spectral feature which is instead unique to interacting driven-dissipative systems.

\begin{figure}[t]
\begin{center}
\epsfig{figure=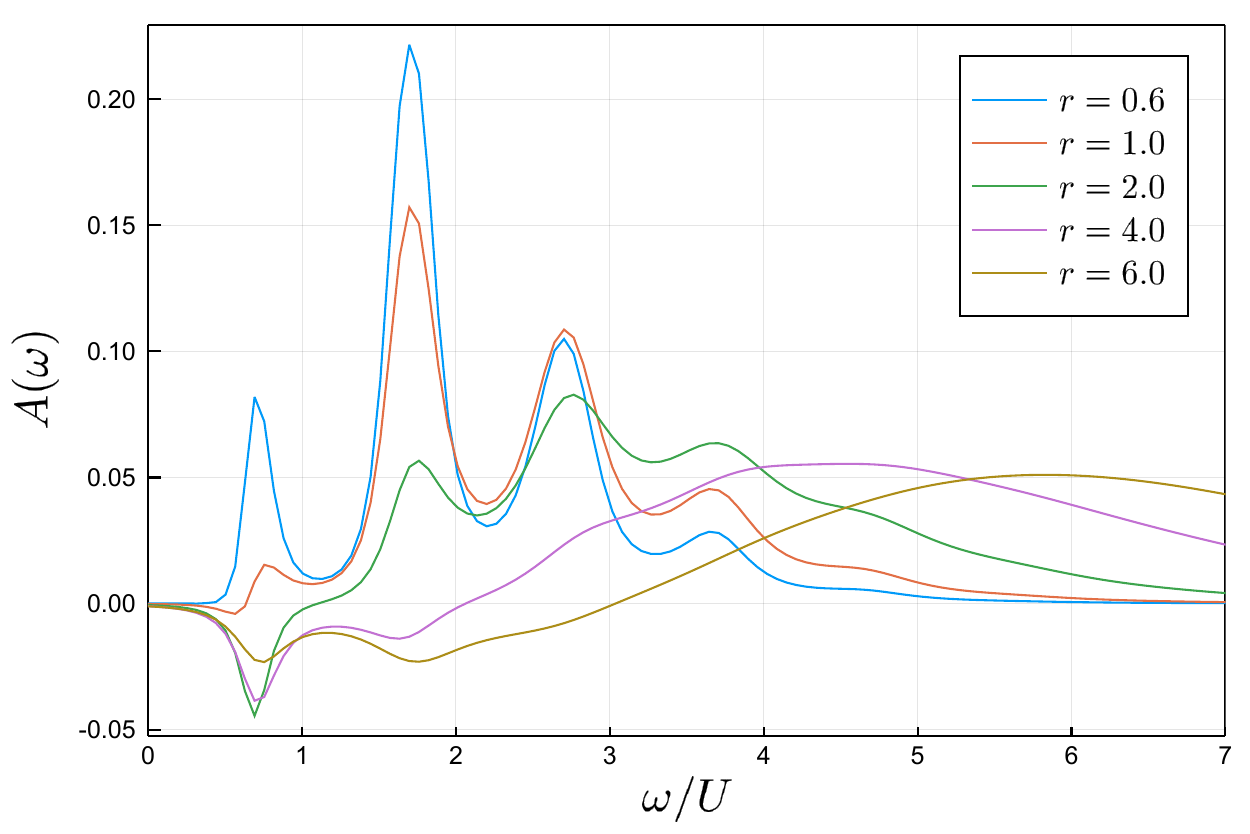,width=1\linewidth}
\caption{Local spectral function  $A(\omega)$ obtained from DMFT/NCA for different values of the pump/loss ratio $r$ at fixed $J/U=0.2$ and $z=6$.  Upon increasing $r$ the lowest Hubbard band flips sign and a region of Negative Density of States (NDoS) emerge at positive frequencies, up to $\omega=\Omega_0(r)$ where the spectral function vanishes, $A(\Omega_0)=0$. The spectral range of NDoS increases with $r$.
Parameters: 
$dim_H=14$.}
\label{fig:specFunc_largedrive}
\end{center}
\end{figure}

\subsubsection{Negative Density of States}
\label{sec:NDoS}

We now discuss how the spectral features of the normal phase evolve upon increasing the strength of the drive/loss ratio $r$. While in the low pump regime all the peaks of the spectral function are positive, see Fig. \ref{fig:specFunc_smalldrive}, a novel effect appears at large drives. Above a threshold $r_{\rm ndos}$  the lowest Hubbard band flips sign and a region of \emph{Negative Density of States} (NDoS) appears in a positive frequency range~\footnote{The spectral function obeys the frequency sum rule $\int_{-\infty}^\infty d\w A(\w) = A(t=0)= 1$, which is exactly enforced by our NCA approximation of Green functions \eqref{eq:NCAGreen}. This is easily verified by using the identity $\hat{\mathcal{V}}(t=0) = \hat{\id}$. In the numerics, the truncation of the bosonic Hilbert space spoils this exact identity, which we therefore use to assess that the Hilbert space cut-off we consider is sufficiently large.}.

We show this in Fig. \ref{fig:specFunc_largedrive} where we plot the spectral function obtained within NCA/DMFT for different values of drive/loss ratio $r$, at fixed $J/U=0.2$. The region of NDoS  extends up to $\omega=\Omega_0$, a frequency at which the imaginary part of the retarded Green's function linearly vanishes, i.e. we have
\be\label{eqn:Omega0}
A(\omega)=\gamma (\omega-\Omega_0)\;\;\;\;\mbox{for}\;\omega\simeq \Omega_0
\ee
with $\gamma>0$, while for $\omega>\Omega_0$ the conventional positive sign is recovered. As we show in figure~\ref{fig:specFunc_largedrive} the 
spectral range of NDoS increases with the drive $r$ and so does the frequency $\Omega_0(r)$.  
We stress that a negative spectral function at positive frequency is a genuine nonequilibrium phenomenon that cannot happen for closed systems in thermal equilibrium~\cite{scarlatellaClerkSchiro2018}. It has a direct physical consequences on the response of the system to a weak local coherent drive oscillating at frequency $\omega$, $V(t)=\sum_i \left( v^*_i(t) b_i+\mbox{hc}\right)$ with $v_i^*(t)=v_0\delta_{i,0}\,e^{i\omega t}$. Indeed for an open system the power absorbed from the perturbation, defined as~\cite{Alicki_1979,FeldmannEtAlPRE03,RivasEtAlPRL20}  $\dot{W}=\mbox{Tr}\rho(t)\dot{V}$ can be written within linear response theory as (see appendix~\ref{app:Response})
\be\label{eqn:Power}
\dot{W}=v_0^2\omega A(\omega) 
\ee
This expression highlights how the spectral function at frequency $\omega$ controls the power absorbed by the system under an external drive. A change in sign of this quantity, i.e. a negative absorbed power, signals the onset of energy emission and gain, a condition which is generally associated to optical amplification and lasing~\cite{ScovilEtAlPRL59,BoukobzaEtAlPRA062,BoukobzaEtAlPRL07,Liu285}. As we are going to discuss in Sec~\ref{sec:DMFT_phaseDiag} the NDoS effect and the frequency $\Omega_0$ will play a crucial role in the nonequilibrium phase transition from the normal to the superfluid phase.

\begin{figure}[t]
\begin{center}
\epsfig{figure=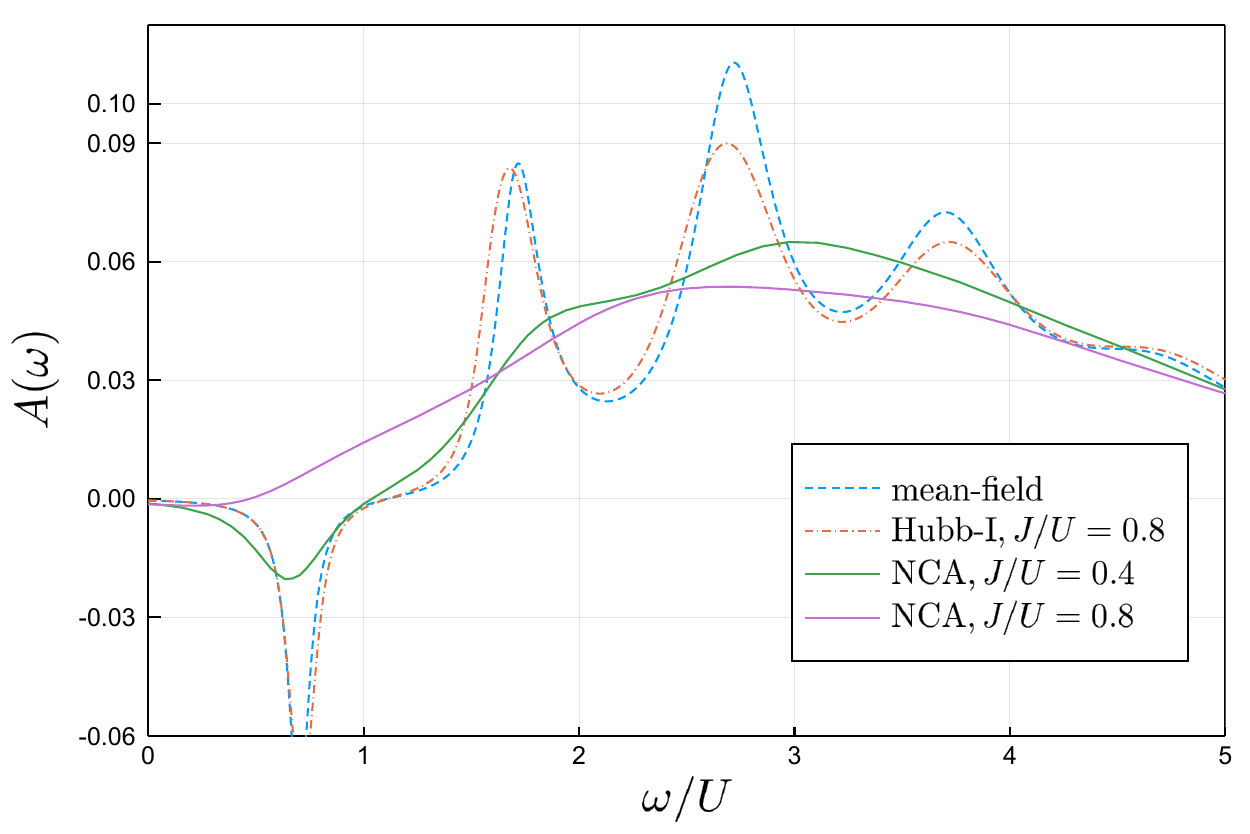,width=1\linewidth}
\caption{Local spectral function $A(\omega)$ for different values of $J/U$, as computed within Gutzwiller mean-field theory (dashed blue line), DMFT with Hubbard-I impurity solver (dashed red line) and NCA impurity solver (green and pink full lines), for fixed $z=6$ and $r=2$.  Within NCA we see that increasing the hopping $J$ changes qualitatively the structure of the low frequency spectrum, washing out the NDoS and restoring a positive spectrum at $\omega>0$. This effect of the DMFT bath is completely missed by Gutzwiller mean-field as well as Hubbard-I approximations. Parameters: 
$dim_H=10$}
\label{fig:specFunc_NDOS}
\end{center}
\end{figure}
We emphasize that the NDoS effect arises already in the single-site problem, i.e for $J=0$ in our model,  above a threshold pump $r_{\rm ndos}$ which depends on the strength of Kerr nonlinearity, as discussed in Ref.~\cite{scarlatellaClerkSchiro2018}. As a result, it naturally appears at large drive in the normal-phase spectral function of our lattice model calculated within Gutzwiller mean-field theory as well as DMFT/Hubbard-I, both built out of the exact solution of the single site problem. 

We now discuss the dependence of the NDoS effect from the hopping $J$. Clearly such a question goes beyond Gutzwiller mean-field theory, which as we stressed cannot capture any effect due to coherent hopping within the normal phase. In figure~\ref{fig:specFunc_NDOS} we plot the spectral function obtained with DMFT/Hubbard-I and NCA, for increasing values of $J/U$ and compare with the results obtained from Gutzwiller. We find that the NCA spectral function is strongly affected by the hopping, which broadens
the sharp high-energy peaks and decreases the strength of the negative peak around  $\Omega_0$, up to a value of $J/U\simeq 0.8$ at which this peak turns back to positive, washing away the NDoS effect. In other words NCA is able to capture a renormalization of the scale $\Omega_0$ from the hopping $J$. 
This is surprising at first since $J$ is a purely coherent energy scale while we have seen in Figure~\ref{fig:specFunc_largedrive} that the strength of the peaks and the NDoS is controlled by the dissipative scales, i.e. the pump to loss ratio. We interpret this effect as a first example of hopping-induced losses, a mechanism that is unique to open quantum systems and that plays a key role in the physics of our model.  Importantly, this effect is completely missed by the simple impurity solver Hubbard-I, whose spectral function, also shown in Figure~\ref{fig:specFunc_NDOS}, changes very little with respect to the Gutzwiller mean-field one~
\footnote{In fact one can show analytically, from the expression of the retarded Green function in appendix~\ref{app:hubbI}, that $\Omega_0$ in Hubbard-I is independent of the hopping and equivalent to the single-site and mean-field value.}.

To summarize, we have seen that changing drive and hopping largely affects the spectral properties of the normal phase. In particular we have identified for positive frequencies $0<\omega<\Omega_0(r,J/U)$ a region of NDoS emerging above a threshold drive strength $r>r_{\rm ndos}(J/U)$. Both these quantities depend within DMFT/NCA from the hopping-to-interaction ratio $J/U$, an effect which is completely missed by Gutzwiller mean field as well as by Hubbard-I. As we are going to discuss in Sec.~\ref{sec:DMFT_phaseDiag}, these dependencies of the critical frequency $\Omega_0$ and of the threshold drive $r_{\rm ndos}$ from the hopping strength $J$ will have direct consequences on the phase diagram of the model.

\subsection{Steady State Local Density Matrix and Population Inversion} 
\label{sec:occupation}

We now discuss the occupation properties of the stationary state distribution in the normal phase. For a lattice problem computing the full many-body density matrix can be done only for very small systems. Nevertheless, within our DMFT/NCA approach, describing the thermodynamic limit of infinitely-many sites, we can compute the reduced steady-state density matrix of a given site of the lattice, say site $i=0$, obtained by performing a partial trace on all other sites, namely $\rho_s = \tr_{j\neq0} \rho_{\rm latt,s}$.
This corresponds to the steady-state density matrix of the DMFT self-consistent quantum impurity model \eqref{eqn:S_eff} and thus of the non-Markovian map $\mathcal{\hat{V}}$ \eqref{eq:vMap} and it obeys Eq. (\ref{eqchnca:ss_equation}).
This reduced on-site stationary density matrix allows to study the change of the local populations of bosons due to hopping processes, which is completely missed by Gutzwiller mean-field theory. Also, for open systems these hopping processes enable new, effective dissipative channels. For example a particle can be injected from the Markovian environment on one site, hop to another site and escape the system, rather than just being created and annihilated on the same site. Those processes are captured by our DMFT approach and mimicked by the non-Markovian environment $\Delta$ and unlock interesting new physics which we are going to discuss here.

\subsubsection{Local Occupation vs J}

\begin{figure}[t]
\begin{center}
\epsfig{figure=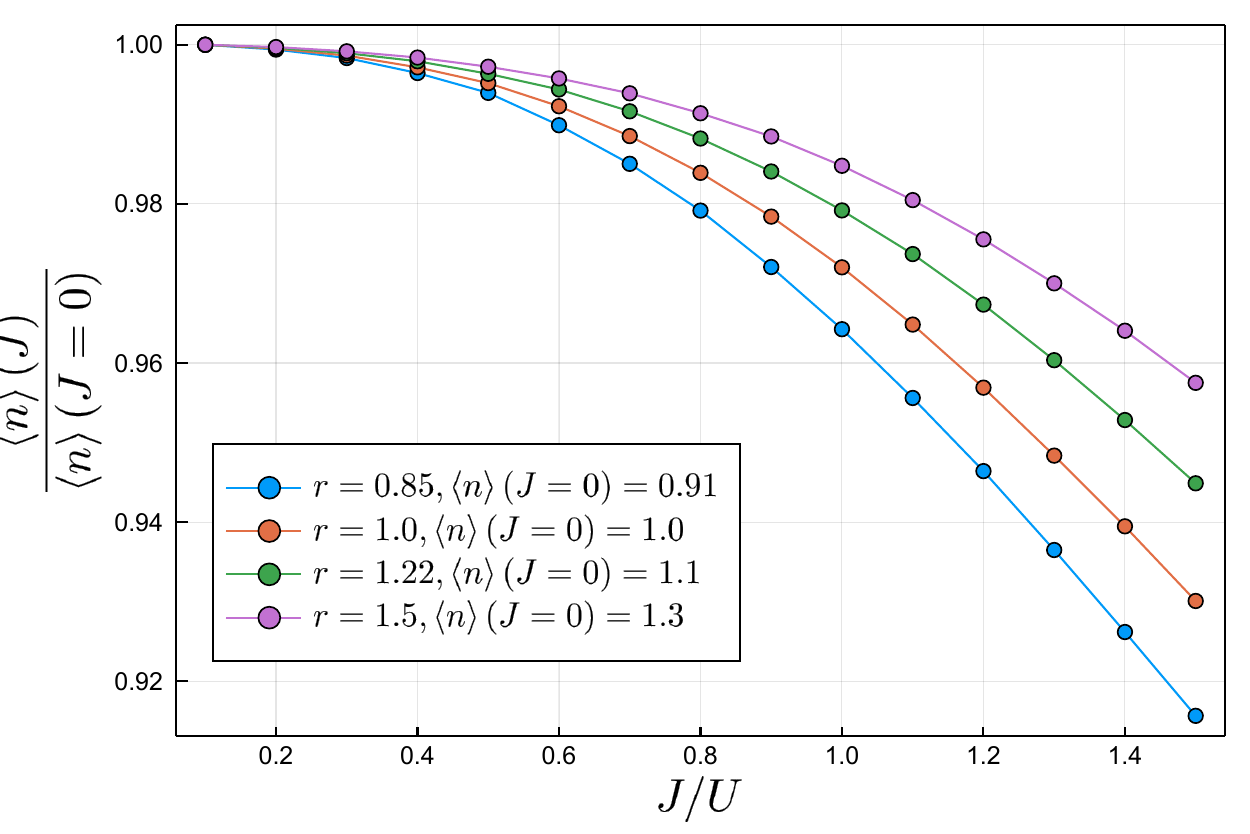,width=1\linewidth}
\caption{Local density of particles as a function of the hopping to interaction ratio $J/U$, within DMFT/NCA, for different values of the pump to loss ratio $r$, normalized to the values of the single site problem. We see that the density decreases with $J/U$, a specific feature of driven-dissipative lattices with two-body losses (see main text), which is captured by DMFT. 
Parameters: $z=50$, 
$dim_H=10$. The drive values used are marked on the y axis of Fig. \ref{fig:phasediag_vsNDOS}.}
\label{fig:nVsJ}
\end{center}
\end{figure}

From the knowledge of the on-site reduced stationary state density matrix $\rho_s$ we can obtain the average local density $\langle n \rangle = \tr \lp b^\da b \rho_s \rp$. 
We notice that the local density can be also obtained from the Green's functions, in particular from the Keldysh component at equal times 
\beq\label{eqn:GK}
 G^{K}(t,t) = -i \aver{ \lbr b(t),b^\dagger(t) \rbr } = -i \lp 2 \aver{ b^\dagger(t) b(t) } +1 \rp
\eeq
which gives consistently the same result in our NCA approach.

Within DMFT the local density acquires a dependence from the hopping $J$, which is obviously missing in Gutzwiller mean field. 
In Fig.~\ref{fig:nVsJ} we plot the density as a function of $J/U$ at $z=6$ and for different values of the drive, normalised to the mean-field value ($z=\infty$).  We see that quite generically the density decreases smoothly upon increasing the hopping within the normal phase, i.e. for $J<J_c$. This can be understood as an interplay of two particle losses and coherent hopping between neighboring sites, an effect that will be further explained by discussing the stationary-state populations in the next section. Interestingly the rate of decrease of the density with hopping changes quite strongly with the strength of the pump $r$ and in particular we notice in Figure~\ref{fig:nVsJ} that a large drive seems to make the density more pinned to the single-site value.

The result in Fig.~\ref{fig:nVsJ} turns out to be a specific feature of dissipative lattices with two-particle losses. In fact one can generically prove that for a driven-dissipative Bose-Hubbard model with only single particle losses and single particle drive the stationary state density matrix is independent of any Hamiltonian parameter~\cite{Lebreuilly2016}, leading to a density of particles independent of $J$ (although not necessarily integer, as it would be in the equilibrium Mott ground state of the Bose-Hubbard model ) and only set by drive/loss balance. In App. \ref{app:bench} we show that this effect is correctly captured by our DMFT/NCA approach, a highly non-trivial benchmark for its validity.

\subsubsection{Steady-State Populations and Population Inversion}
\label{sec:Inv_vs_NDOS}

\begin{figure}[t]
\begin{center}
\epsfig{figure=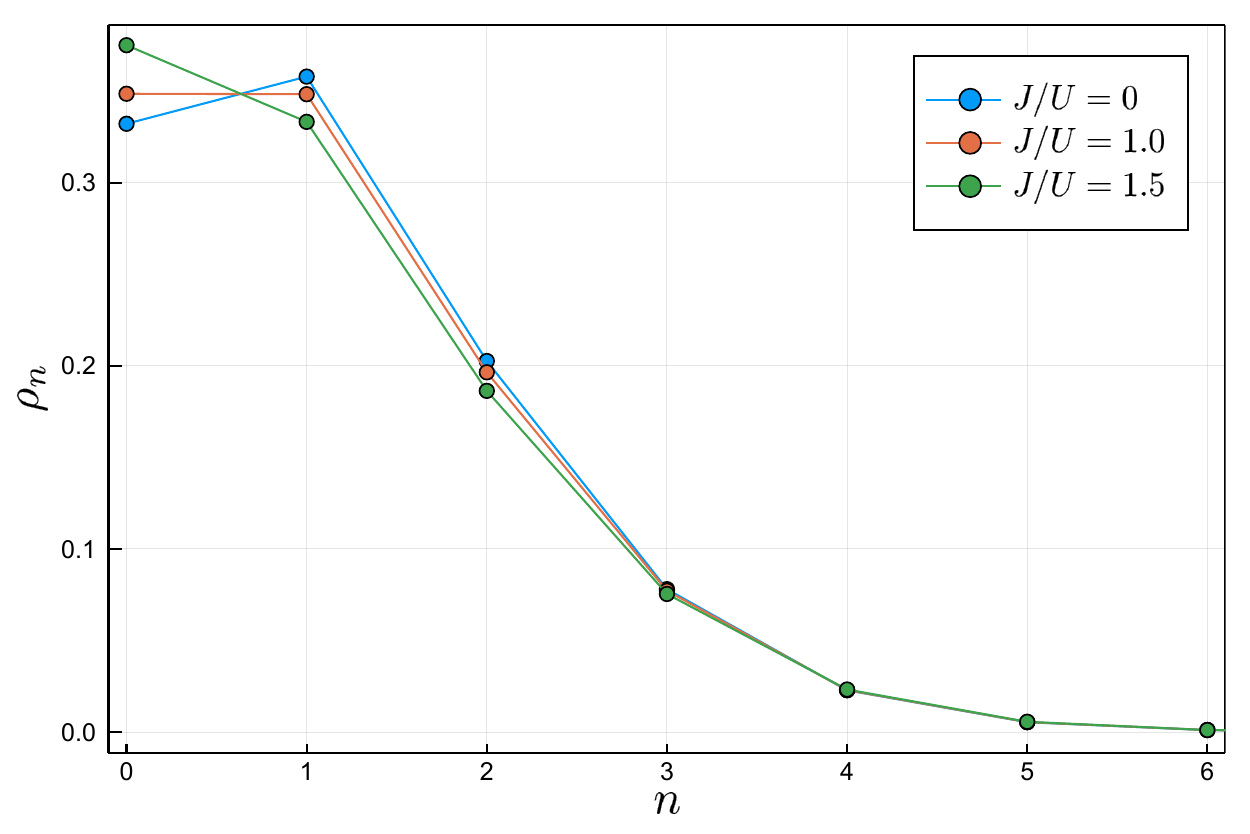,width=1\linewidth}
\caption{
Steady state populations of the reduced density matrix $\rho_n=\langle n\vert \rho_s\vert n\rangle $, within DMFT/NCA, for different values of the hopping to interaction ratio $J/U$ and for $r=1.22$. We see that increasing $J/U$ changes the populations at low values of $n$ and ultimately washes away the population inversion found in the single site limit $J/U=0$.
Parameters: $z=50$, 
$dim_H=10$. The drive value used is marked on the y axis of Fig. \ref{fig:phasediag_vsNDOS}.}
\label{fig:popVsJ}
\end{center}
\end{figure}

In this section we discuss the effect of coherent hopping processes on the steady state reduced density matrix, which as we show exhibits richer physics than the local occupancy. In the normal phase this quantity is diagonal in the basis of Fock states with $n$ bosons per site and the steady state populations $\rho_n$  are shown in Fig. \ref{fig:popVsJ} for different hopping values, $z=6$ and $r=3$.

First we observe that the single-site model, corresponding to $J=0$, shows a non-monotonic behavior of the populations as a function of the number of bosons per site, for drive/loss ratio $r>1$ and any value of the Kerr non-linearity $U$ (which in fact does not affect the stationary state as it is has been long known~\cite{dykman1978}). This \emph{population inversion} at $J=0$ appears clearly in Fig.~\ref{fig:popVsJ}, where the probability of finding $n$ bosons per site is maximum at $n=1$ despite the fact that a finite bosonic occupation costs energy  $E_n=\omega_0 n+Un^2/2 \sim U$ and should be therefore thermodynamically suppressed. 

Increasing the hopping changes the populations at low occupancy while leaving essentially unaffected the tail at large $n$. In particular, the coherent hopping from and into the neighboring sites increases the probability of having an empty site at expenses of finite  occupation. This is a genuine feature of our dissipative many body lattice problem with local two body losses: starting from a state with average filling $n\sim 1$, hopping processes towards neighboring sites creates double occupations which escape at a rate $\eta$, reducing the total occupation. This trend goes on upon further increasing $J$,  ultimately suppressing the population inversion above a threshold hopping. This mechanism also explains more in detail the observed overall decrease of average occupation with $J$, Fig. \ref{fig:nVsJ}, which we already discussed in the previous section. 

An interesting question concerns the relation between the NDoS effect discussed in Sec.~\ref{sec:spectFunctionNorm} and the population inversion in the reduced stationary density matrix. In closed quantum systems described by unitary evolution the two concepts are directly related, namely a NDoS could only emerge in presence of a inversion of populations where higher energy states are more occupied than lower energy ones.  For open quantum systems the situation is more subtle and the two concepts are not in one-to-one correspondence~\cite{scarlatellaClerkSchiro2018}. 
In Figure~\ref{fig:phasediag_vsNDOS}  we plot the behavior of the threshold for population inversion $r_{\rm inv}$ and for NDoS  $r_{\rm ndos}$ as a function of $J/U$. We notice that those thresholds are independent from the hopping within Gutzwiller mean field (see dashed lines, which coincide with the $J=0$ values of DMFT) while they are substantially renormalized in DMFT. In particular the two scales $r_{\rm ndos}<r_{\rm inv}$  further deviates from each other as the hopping is increased.  We note that $r_{\rm inv}$ increases monotonically with the hopping strength $J$ in DMFT. Based on closed systems arguments, this  \textit{hopping-induced suppression} of population inversion would suggest that the NDoS is also always suppressed by hopping, as for example Fig. \ref{fig:specFunc_NDOS} shows.
Surprisingly, this is not always the case. Figure~\ref{fig:phasediag_vsNDOS} shows that $r_{\rm ndos}$ has a non-monotonic behavior with the hopping rate $J$, namely its behavior changes from small and large hopping values. 
While for large values of $J$ the NDoS threshold $r_{\rm ndos}$ indeed increases following the behavior of the $r_{\rm inv}$ threshold, as expected from closed system arguments, for small values of $J$ it is actually reduced below the $J=0$ threshold, corresponding to the single site. Namely, for small values of $J$, the non-Markovian bath $\Delta$ actually generates a NDoS, even in a regime where the single site model would not present any signature of this effect. This is a unique feature of dissipative quantum systems, where an NDoS can be generated even in absence of population inversion. For Markovian systems it has been shown that this can be traced back to the structure of excitations on top of the stationary state, which come with characteristic complex weights, leading to anti-lorentzian lineshapes \cite{scarlatellaClerkSchiro2018}.

\subsection{Finite Frequency Instability of the Normal Phase}
\label{sec:DMFT_phaseDiag}

In this section we discuss how the peculiar spectral and occupation properties of the normal phase contribute to an instability towards a spontaneous breaking of $U(1)$ symmetry. We show that the conventional static superfluid transition of the equilibrium Bose Hubbard model as a function of the hopping to interaction ratio $J/U$ is pushed to finite frequency as a result of drive and dissipation, leading to an order parameter oscillating in time. We emphasize the role of the NDoS for the onset of the phase transition and compare the DMFT/NCA and Gutzwiller phase boundaries. We argue that the effect of finite-connectivity fluctuations is not only quantitative, but rather underlines a qualitatively new physical mechanism for the onset of an ordered phase in open quantum lattices with two-body losses, which cannot be simply interpreted as the destruction of an ordered phase by thermal fluctuations in an effective equilibrium problem.

\subsubsection{DMFT Phase Boundary}

Within our DMFT approach, we can derive an equation for the phase boundary separating the normal and the broken symmetry phases.
We assume to be in the early symmetry-broken phase, where the order parameter $\bold{\Phi}(t) = \aver{\bold{b}(t)}$ has just formed and it is small. This implies a small external field $\bold{\Phi}_\eff(t)$  \eqref{eq:phieff} in the DMFT effective action \eqref{eqn:S_eff}.
We also assume to be in a stationary regime at long times, such that two point correlators depend only on time differences, and move to Fourier space.  The average value of the bosonic field $\bold{\Phi} (\w) \equiv \aver{\bold{b}(\w) }$ is, to linear order in $\bold{\Phi}_\eff$ 
\beq
\label{eqapp:linResp}
\bold{\Phi} (\w)= -     \bold{G}^R (\w)  \bold{\Phi}_\eff (\w) 
\eeq
where we used the fact that $ \aver{\bold{b}(\w) }_{\bold{\Phi}_\eff = 0}= 0 $. A key point now is that at finite $z$ the effective field $ \bold{\Phi}_\eff (\w) $ in DMFT has two contributions, one from the local order parameter itself and the other from neighboring sites encoded in the non-Markovian bath, see Eq.~(\ref{eq:phieff}) which now reads (using $\bold{\Phi}_+ = \bold{\Phi}_{-}$ as well as $\bold{\Phi}_{\eff +} = \bold{\Phi}_{\eff -}$ )
\begin{align}
\label{eqapp:phieffRet}
\bold{\Phi}_\eff (\w) &=  J \bold{\Phi} (\w) + \bold{\Delta}^R (\w) \bold{\Phi}(\w)
\end{align}
Plugging \eqref{eqapp:phieffRet} into \eqref{eqapp:linResp} and using the DMFT self-consistency on the Bethe lattice  \eqref{eq:selfConsDelta} one finally gets 
\beq 
\bold{\Phi}(\w)  = \left(- J   \bold{G}^R (\w)- \frac{J^2}{z}  \bold{G} ^R(\w) \bold{G} ^R(\w) \right)\bold{\Phi}(\w) 
\eeq 
The critical coupling $J_c$ and critical frequency $\Omega_c$ needed for a self-consistent broken-symmetry solution, $\bold{\Phi}(\Omega_c) \neq0$ corresponding to an order parameter whose phase oscillates in time $\langle b(t)\rangle\sim e^{-i\Omega_c t}$ for $J>J_c$, are given by
\beq
\label{eq:dmft_cp}
\frac{1}{J_c}+ G ^R  (\Omega_c,J_c)  + \frac{J_c}{z} \lsq G^R   (\Omega_c,J_c )  \rsq^2  = 0
\eeq

\begin{figure}[t]
\begin{center}
\epsfig{figure=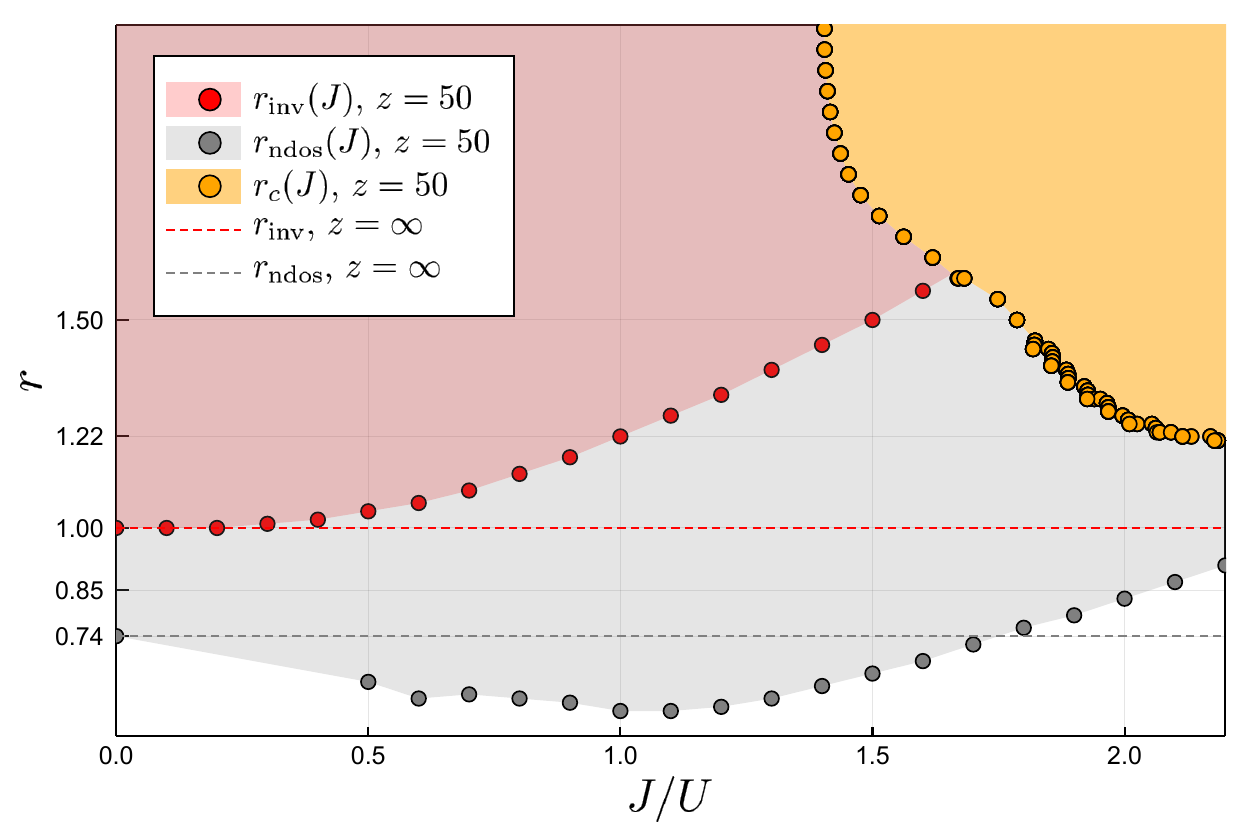,width=1\linewidth}
\caption{Threshold drives for NDoS $r_{\rm ndos}(J)$ and for population inversion in the density matrix $r_{\rm inv}(J)$, as a function of the hopping and compared with the critical drive $r_{\rm c}(J)$ (yellow points) for the finite-frequency superfluid transition. 
The ticks of the y axis correspond to the values of $r$ used for Figures \ref{fig:popVsJ} and \ref{fig:nVsJ}. We see that for small value of $r$ the phase transition occurs even in absence of a population inversion, i.e. it is the NDoS the key effect leading to the instability of the normal phase.
Parameters: $z=50$. 
Red curve: $t_{max}=10$, $dim_H=14$. Gray curve: $t_{max}=20$, $dim_H=10$.}
\label{fig:phasediag_vsNDOS}
\end{center}
\end{figure}

Equation \eqref{eq:dmft_cp}, which to the best of our knowledge is an original result of this paper, is generic for bosonic DMFT theories on the Bethe lattice and it holds also for equilibrium problems. Its solution, leading to the phase boundary in Figures~\ref{fig:phaseDiag} and \ref{fig:phasediag_vsNDOS}, strongly depends on the driven-dissipative nature of the problem, as we are going to discuss now. First, Eq.~(\ref{eq:dmft_cp}) has real and imaginary parts, which both need to vanish simultaneously, resulting in the two conditions

\begin{align}
\label{eq:dmft_cp_im}
\im G^R(\Omega_c,J_c) &= 0 \\
\label{eq:dmft_cp_re}
\frac{1}{J_c} + \re G^R(\Omega_c,J_c) + \frac{J_c}{z} \lsq \re G^R(\Omega_c,J_c)\rsq ^2 &= 0 
\end{align}
We remark that there is another solution possible, where $\im G^R(\Omega_c,J_c)\neq0$, but this is never realized in our simulations.
In thermal equilibrium the first condition Eq.~\eqref{eq:dmft_cp_im} can be only satisfied at zero frequency, where fluctuation-dissipation theorem constraints the imaginary part of a bosonic retarded Green's function to vanish, thus allowing for static symmetry breaking patterns (as in equilibrium superfluids).
Far from equilibrium this does not need to be the case~\cite{Keeling2010,ScarlatellaFazioSchiroPRB19} and indeed we have seen that the normal phase shows, above a threshold drive $r_{\rm ndos}$, a spectral function vanishing at a positive frequency, corresponding to the formation of a NDoS and the onset of gain in the system. The critical frequency $\Omega_c$ solving Eq. \eqref{eq:dmft_cp_im} corresponds therefore to the frequency at which the local spectral function of the normal phase vanishes 
\be\label{eqn:Omegac}
\Omega_c=\Omega_0(r,J_c(r)) 
\ee
for a critical value of hopping $J_c$ determined by jointly solving Eq. \eqref{eq:dmft_cp_re}. The energy scale $\Omega_0$ for the NDoS is therefore a precursor of the mode that will become unstable at the transition.  We conclude that the NDoS effect discussed in previous section is a key, necessary condition for a phase transition into the nonequilibrium superfluid phase.
This is clearly shown in Figure~\ref{fig:phasediag_vsNDOS}, where we plot the threshold pump $r_{\rm ndos}$ for NDoS and the critical drive $r_c$ obtained from solving Eq.~ \eqref{eq:dmft_cp} with DMFT/NCA as a function of $J/U$. We see that generically $r_{\rm ndos}<r_c$, namely the system first develops gain and then becomes truly unstable towards $U(1)$ symmetry breaking. On the other hand, from Figure~\ref{fig:phasediag_vsNDOS} we see that one can obtain an instability even in absence of population inversion.

\subsubsection{Role of finite-connectivity fluctuations and comparison with Gutzwiller}\label{sec:desynch}

\begin{figure}[t]
\begin{center}
\epsfig{figure=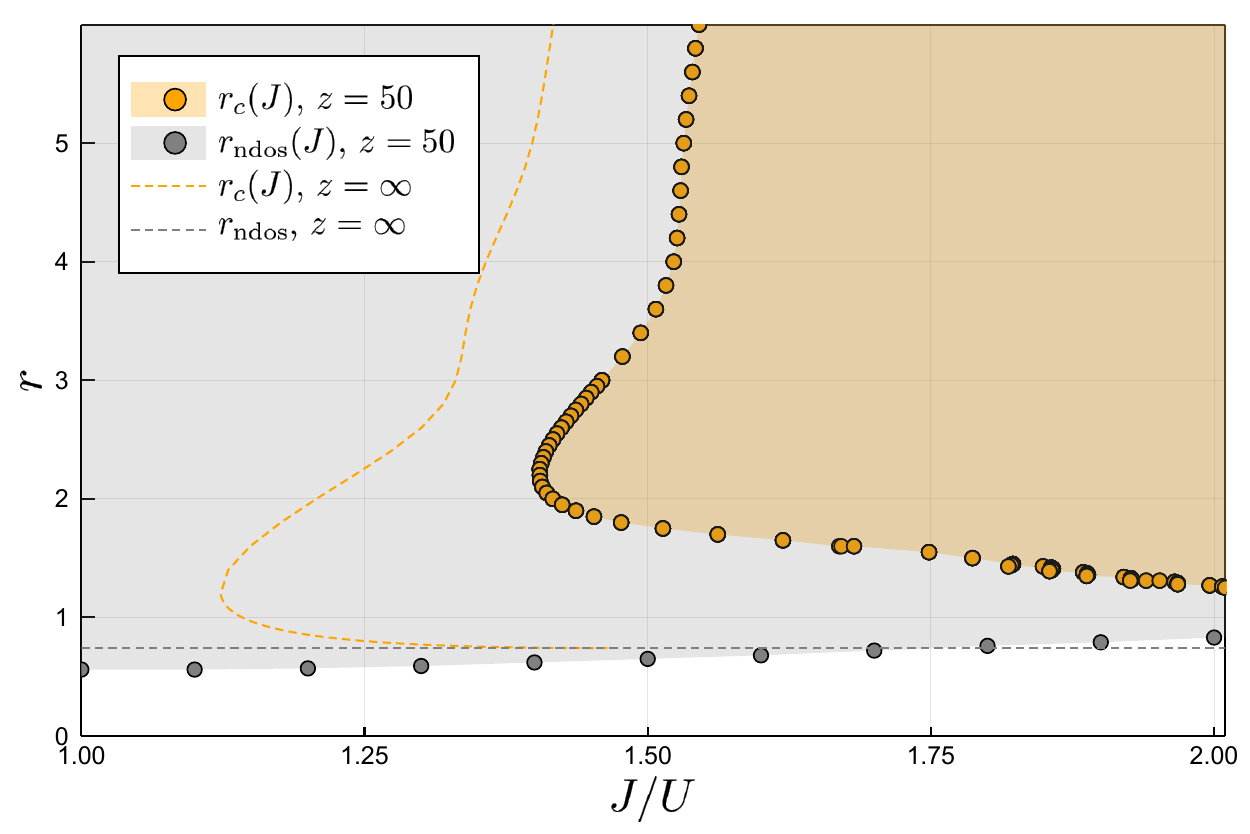,width=1\linewidth}
\caption{Phase diagram in the pump versus hopping plane obtained by DMFT/NCA for $z=50$ (yellow points) and Gutzwiller mean field theory ($z=\infty$, yellow dashed line). We further plots the thresholds for NDoS obtained within the two approaches. We see that fluctuations due to finite connectivity reduce the broken symmetry phase, pushing it towards higher values of hopping and drive. We interpret this effect as a signature of hopping-induced losses, which reduce the local gain and prevents the system to become unstable at finite frequency. Parameters :$dim_H=10$
}
\label{fig:phaseDiag_vs_negSpec}
\end{center}
\end{figure}

We now go back to an important aspect mentioned at the beginning of our paper, namely the large renormalisation to the phase boundary obtained within DMFT/NCA upon decreasing the connectivity $z$. As we show in Figure~\ref{fig:phaseDiag} and Figure~\ref{fig:phaseDiag_vs_negSpec}, fluctuations due to the finite-number of neighbors shift the phase boundary towards larger values of the hopping $J$ and pump/loss ratio $r$. We now provide a physical interpretation of this effect based on the properties of the normal phase discussed so far.

We start by considering the condition for the normal phase instability obtained within Gutzwiller mean-field theory, which corresponds to the $z\rightarrow\infty$ limit of the Eq.~(\ref{eq:dmft_cp}). In this limit the problem reduces to a quantum single site, while the feedback from neighboring sites is treated at a purely classical level~(See Sec.~\ref{sec:mean_field_conn}), in terms of a self-consistent coherent field which reads  $\Phi_{\rm eff}=J_c\langle \bb(t)\rangle\sim J e^{i\Omega_c t}$ near the instability. As such, if we repeat the argument of the previous section we can obtain a condition for the Gutzwiller phase boundary, which reads
\beq
\label{eq:mf_cp}
\frac{1}{J_{c}(z=\infty)}+ G_0^R  (\Omega_c(z=\infty)) = 0
\eeq
where the first term is the effective field contribution and $G_0^R(\omega) $ is the retarded Green's function of the isolated single-site problem and is therefore independent from the hopping. Within Gutzwiller mean-field theory the hopping $J$ has only the role of triggering the symmetry breaking, through the self-consistent field $\Phi_{\rm eff}$, while the onset of gain is controlled by the pump to loss ratio $r$. Indeed we known that the single site spectral function develops a NDoS  above a constant threshold pump $r_{\rm ndos}$ (see gray dashed line in Figure~\ref{fig:phaseDiag_vs_negSpec}). The  feedback from the neighboring sites acts as a seed for a single site which is on the verge of energy emission (negative absorbed power at $\Omega_c$, see Eq.~(\ref{eqn:Power})) and leads, above a threshold hopping $J_c$ shown in Figure~\ref{fig:phaseDiag_vs_negSpec}, to amplification of the local coherent field at frequency $\Omega_c$ and a spontaneous breaking of $U(1)$ and time-translation symmetry. Interestingly, the Gutzwiller phase boundary is very close to the line $r_{\rm ndos}$ (see Figure~\ref{fig:phaseDiag_vs_negSpec}) suggesting that at large hopping as soon as the system develops gain the symmetry breaking occurs.

DMFT/NCA on the other hand accounts for a more subtle effect of neighboring sites, which are encoded in the non-Markovian quantum bath, in addition to the classical coherent field. As we know this provides an increased effective dissipation, due to hopping processes from lossy neighboring sites, which is responsible for wiping out the NDoS region in the local spectral function, a necessary condition for the onset of the instability. This is clearly shown in Figure~\ref{fig:phasediag_vsNDOS} where for large values of $J/U$  the threshold for NDoS  eventually becomes larger than the Gutzwiller mean-field one (see gray dashed line in Figure~\ref{fig:phasediag_vsNDOS}), leaving a normal phase which would be superfluid at $z=\infty$. 
These \emph{hopping-induced losses} provide therefore a clear physical mechanism behind the finite-connectivity renormalization of the phase boundary. The picture that emerges from DMFT is the one of a single site on the verge of energy emission, coupled to an oscillating seed field which would favor optical amplification and embedded in a non-Markovian bath which is instead able to absorb part of the emitted power from the system, thus reducing the effective gain and requiring a stronger value of pump to trigger the instability.  This mechanism is presumably also effective at intermediate value of the hopping where,  as we see in Figure~\ref{fig:phaseDiag_vs_negSpec}, the DMFT/NCA threshold for NDoS remains well below the critical drive responsible for the true many-body instability at intermediate coupling, while approaching it at large hopping.
 
We conclude that this \emph{hopping-induced} dissipation is a qualitatively new mechanism for the destruction of an ordered phase, which is unique to open systems settings and one of the hallmark of our DMFT/NCA approach. Quite interestingly this mechanism is not only completely missed by the Gutzwiller mean field approach, but also by a perturbative solver such as the Hubbard-I approximation. In fact we have discussed (see in figure~\ref{fig:specFunc_NDOS}) how the NDoS is changed very little within this scheme. In appendix \ref{app:hubbI} we also show that the Hubbard-I phase diagram, obtained using the DMFT equation for the critical point \eqref{eq:dmft_cp}, still reduces to the Gutzwiller mean-field one. This further highlights the non-perturbative nature of the fluctuations responsible for the observed renormalization of the NDoS and the importance of using a self-consistent scheme such as our NCA approach.

\subsubsection{Thermal vs. Non-Thermal Origin of Finite Connectivity Fluctuations}\label{sec:Teff}

A different perspective on the role of finite-connectivity fluctuations on the phase boundary can be obtained by looking at the occupation of single particle modes at finite frequency, describing excitations on top of the stationary state and their effective thermal character. 
In fact, in an effective equilibrium picture, one could expect heating to provide an efficient mechanism for the reduction of a broken symmetry phase.
To investigate this physics we look at the Keldysh Green's function, defined in Eq~(\ref{eqn:GK}), which heuristically describes the fluctuations of the observable ${b}$.  If the system was in true thermal equilibrium, the quantum fluctuation-dissipation theorem (FDT) would constrain the Keldysh and the retarded components to obey the relation \cite{kamenev2011field}
\beq
\label{eq:fdt}
\frac{G^{K}(\w)}{-2 \pi i A(\w)} \equiv F_{\rm eq}(\omega)= \coth \lp \frac{ \w }{2 T } \rp 
\eeq
where $T$ is the system temperature. At low frequency or high temperatures, $\omega\ll T$, one has $F_{eq}(\omega)\sim T/\omega$.
In a non-equilibrium system on the contrary there is no well-defined temperature and the FDT does not hold in general.  Nonetheless, it is useful to use the left-hand side of the FDT relation in Eq.~(\ref{eq:fdt}) to {\it define} an effective distribution function 
$$
F_{\rm neq}(\omega)=\frac{iG^{K}(\w)}{2 \pi A(\w) }
$$ 
and to study its frequency dependence.
\begin{figure}[t]
\begin{center}
\epsfig{figure=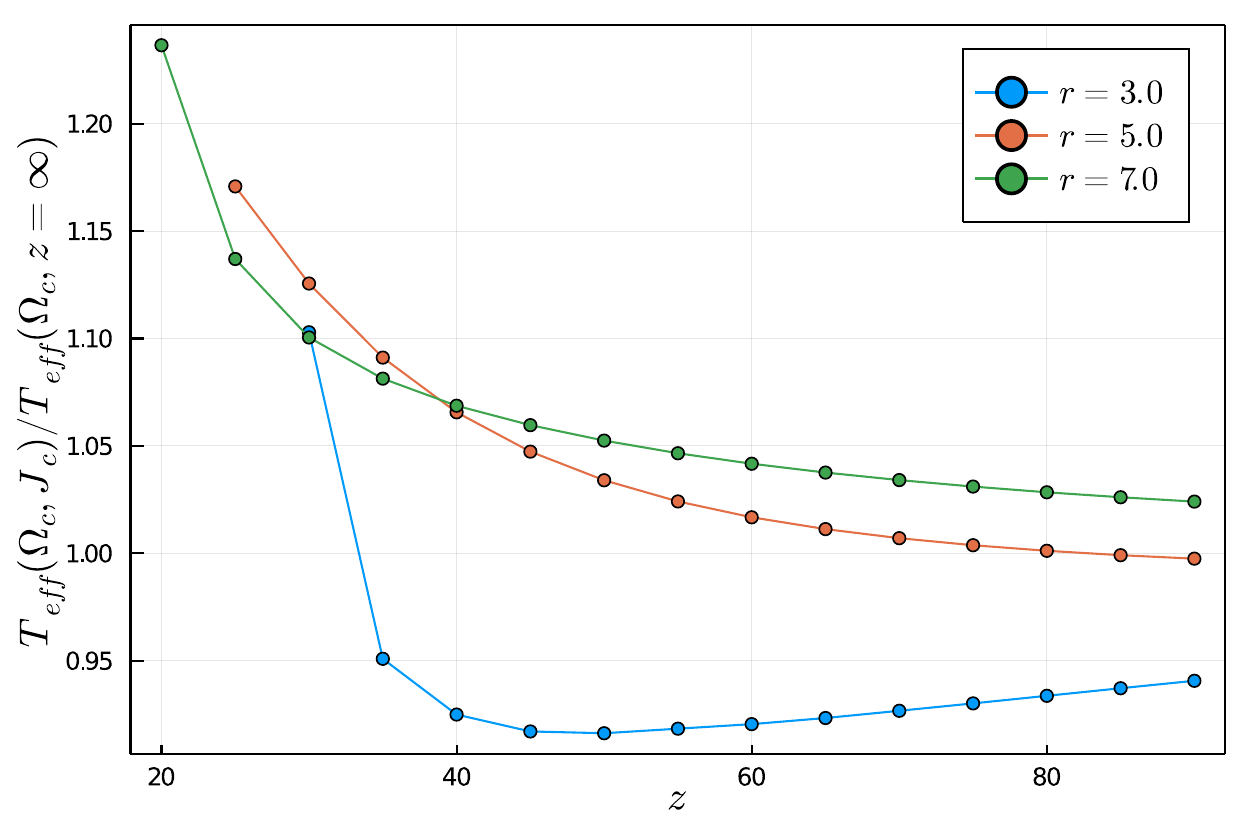,width=1\linewidth}
\caption{Change of the effective temperature $T_{\rm eff}$ at the critical point $(\Omega_c, J_c)$ as a function of lattice connectivity $z$, with respect to its mean-field value at $z=\infty$, for different values of the pump to loss ratio $r$. 
We see that the relative change in $T_{\rm eff}$ remains of the order of few percents up to connectivity $z=50$ and slightly increases up to $20\%$ for the lowest connectivity $z=20$. This should be compared to the DMFT renormalization of the critical hopping for similar values of drive and connectivity, which is instead much stronger as shown in Figure~\ref{fig:phaseDiag}. In particular for $z=30$ and $r=3$ the critical hopping is pushed to infinity corresponding to a complete destruction of the normal phase.
Parameters : $dim_H=14$.  
}
\label{fig:TeffvsZ}
\end{center}
\end{figure}
Within the DMFT and Gutzwiller normal phases, for pump above the threshold for NDoS $r>r_{\rm ndos}$, the spectral function  $A(\omega) $ vanishes at frequency $\Omega_0$, with linear corrections (see Eq.~(\ref{eqn:Omega0} ). On the other hand we find that the Keldysh component has a finite non-zero value at $\Omega_0$, which gives a distribution function of pseudo-equilibrium form at least for the modes around $\Omega_0$
\beq \label{eq:effTemp}
 F_{\rm neq}(\omega)\simeq \frac{T_{\rm eff}}{\omega-\Omega_0}
\eeq
From this expression we can therefore identify an effective temperature $T_{\rm eff}$, which emerges quite ubiquitously in nonequilibrium quantum systems~\cite{MitraEtAlPRL06,ClerkRMP2010,FoiniCugliandoloGambassi_PRB11,DallaTorreetalPRB2012,SchiroMitraPRL14,SiebererRepProgPhys2016}. 
At $z=\infty$, when the normal phase is described as a collection of independent sites, an effective temperature emerges due to the interplay of local drive and Kerr interaction~\cite{scarlatellaClerkSchiro2018}. 
Within DMFT/NCA one could in principle expect important corrections due to the non-Markvovian bath, as we have seen for the NDoS. To assess this point we sit at the phase boundary, $J_c(r)$, choose the corresponding critical frequency $\Omega_0(J_c,r)=\Omega_c$  and plot the behavior of $T_{\rm eff}$, scaled with respect to the $z=\infty$ value,  as a function of the lattice connectivity $z$ and for different values of the pump-to-loss ratio $r$.  We see that at large values of $r$ the effective temperature slightly increases upon decreasing $z$, while at smaller pumps, i.e. $r=3$, it shows a weak non-monotonic behavior with $z$. Overall, the relative variation of $T_{\rm eff}$ with respect to the Gutzwiller mean-field value remains rather moderate. On the other hand, in the same range of variation of $z$, the DMFT critical hopping shows instead strong renormalizations. This is particularly true for the small drive regime, where already for $r=3$ and $z=30$ the ordered phase is completely washed out. This observation suggests that effective thermal fluctuations and heating are not enough to explain the large renormalization of the phase diagram observed in DMFT, which is instead mainly driven by the reduction of local gain by virtual hopping processes, through the mechanism of hopping-induced losses.

\subsection{Nonequilibrium  Superfluidity, Lasing and Many Body Synchronization of Van der Pol Arrays}
\label{sec:synch}

We now discuss how the nonequilibrium superfluid transition in our Bose-Hubbard model is connected to other dynamical phenomena associated with breaking of time-translation symmetry, such as nonequilibrium Bose-Einstein condensation, lasing and synchronization of Van der Pol oscillators. To appreciate this point it is useful to start from the semiclassical limit of our model (See Appendix~\ref{app:semiclassics}) where non-linear quantum fields contributions) are disregarded. The dynamics of the bosonic field at site $j$ takes the form of a Langevin equation
\be\label{eqn:EOMcl}
i\dot{b}_{j\rm cl}=\left(\tilde{\omega}_0( b_{j\rm cl})+i\gamma( b_{j\rm cl})\right) b_{j\rm cl}
+\frac{J}{z}\sum_{\langle j'\rangle}
b_{j'\rm cl}+\xi_j(t)=0
\ee
with an effective frequency and damping terms which depend non-linearly on the field itself, i.e. $\tilde{\omega}_0( b_{j\rm cl})=\omega_0+U\vert b_{j\rm cl}\vert^2/2$  and $\gamma(b_{j\rm cl})=f/2-\eta\vert b_{j\rm cl}\vert^2/2$, and 
where $\xi_j(t)$ is a zero average white noise $\langle \xi_i(t)\xi_j(t')\rangle=f \delta(t-t')\delta_{ij}$. In the continuum limit this reduces to a complex Gross-Pitaevski equation~\cite{Pitaevskii:2143198,aranson2002theworld} with pump and non-linear losses which describes a variety of non-linear phenomena from exciton polariton condensates to multi-mode lasers~\cite{keeling2020bose}. In absence of any noise the spatially uniform stationary state admits a stable limit cycle, i.e. $\beta(t)=\vert\beta\vert e^{-i\omega_{\rm vdp}t}$ for $r>0$ and any $J$ with frequency $\omega_{\rm vdp}$
\be\label{eqn:omegavdp}
 \omega_{\rm vdp}=\omega_0-J+U\vert\beta\vert^2/2
\ee
and amplitude set by the incoherent drive, $\vert\beta\vert=\sqrt{r}$.Equation~(\ref{eqn:omegavdp}) is therefore the semiclassical version of our condition for finite-frequency instability in the normal phase. A major difference exists between the lasing threshold and our case, namely that the threshold for the onset of gain and the one for development of full coherence is well separated, while it coincides in the usual lasing regime~\cite{Keeling2010}. This can be seen easily by looking at the Green's function which develops a NDoS as soon as the system becomes unstable.

\begin{figure}[t]
\begin{center}
\epsfig{figure=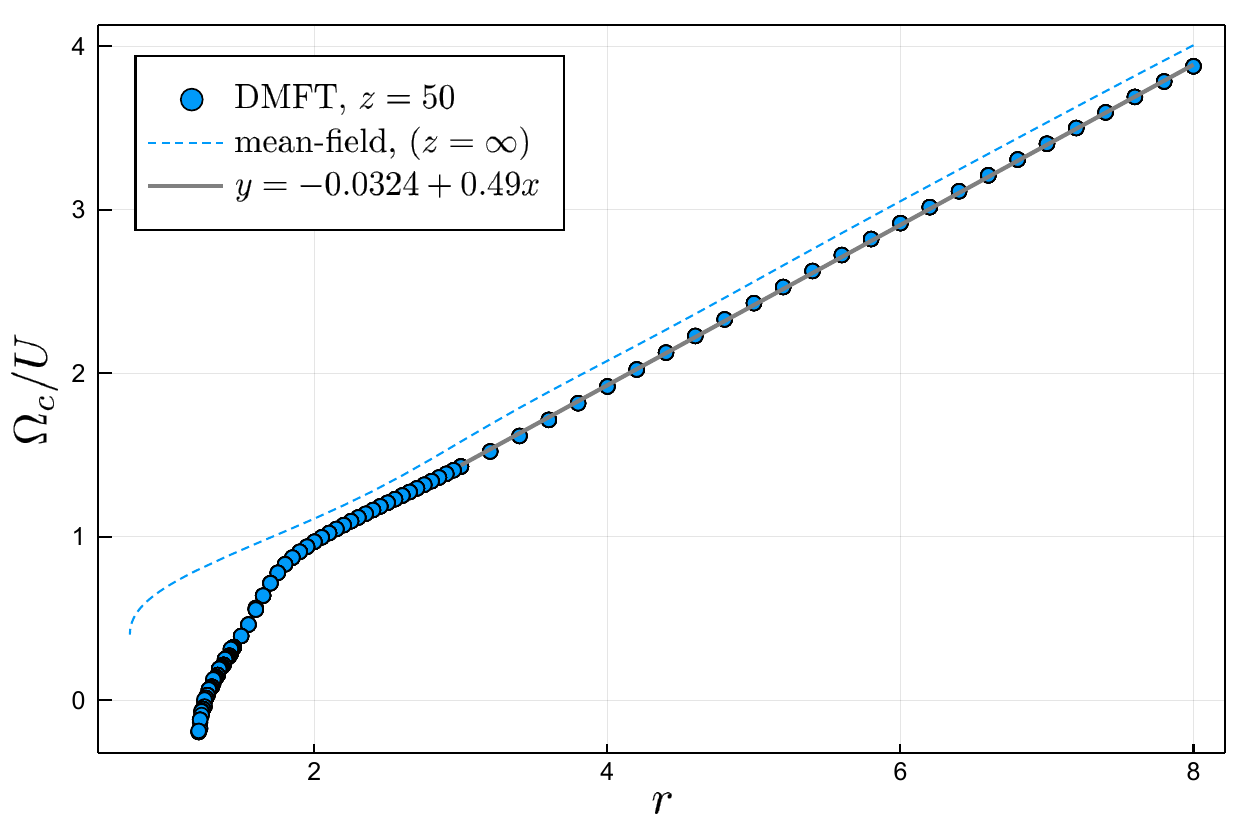,width=1\linewidth}
\caption{ Critical frequency for quantum synchronization $\Omega_c$  as a function of the pump/loss ratio $r$, within DMFT/NCA for connectivity $z=50$ and Gutzwiller mean field theory ($z=\infty$). In the region of large drive the fit shows that $\Omega_c/U \sim r/2$, in agreement with the semiclassical result for the limit cycle in the VdP array. At lower drives, in the regimes of few bosons per sites, quantum fluctuations become relevant and the critical frequency is strongly renormalized. Parameters: 
$dim_H=14$. 
}
\label{fig:wcVzR}
\end{center}
\end{figure}

The equation above describe also an array of coupled classical Van der Pol oscillators. Our driven-dissipative Bose-Hubbard model can be therefore also seen as a quantum many-body version of the VdP array.   From this perspective, the onset of finite-frequency oscillations at $J_c(r)$ described in the last section can be seen as a signature of a quantum synchronization~~\cite{lee2013,lorch2016,walter2014,walter2015,roulet2018a,roulet2018,dutta2019,giorgi2012,manzano2013,qiao2018,SonarEtalPRL18,
jaseem2019,tindall2020}, where above a certain coupling $J$ all quantum VdP oscillators enter into a collective limit-cycle phase. As we are going to discuss next, these oscillations share qualitative features with the semiclassical solution at least at large drive values, while deviate significantly for smaller drives where quantum fluctuations are important.


Despite the similarities the semiclassical limit described above is rather different from the large connectivity limit: in the former case one has a deterministic non-linear equation and desynchronization can only appear due to the noise. In the latter instead, corresponding to DMFT, one reduces to a quantum VdP oscillator coupled to a self-consistent field and a quantum bath. This is responsible for the large separation between the onset of gain and the true instability.

\subsubsection{Critical Frequency versus  Drive/Loss Ratio and Kerr Non-Linearity}
\label{sec:Kerr}

In this section we discuss the behavior of the critical frequency $\Omega_c$, signaling the onset of a quantum synchronized phase, as a function of pump/loss ratio $r$ and Kerr non-linearity $U$. 
%
In Figure \ref{fig:wcVzR} we plot this frequency as a function of the incoherent drive amplitude $r$, both for DMFT (for $z=50$) and for Gutzwiller mean field theory ($z=\infty$). We see that $\Omega_c$ scales linearly with the drive at large values of $r$ (see fit in Fig. \ref{fig:wcVzR}), a result which is in agreement with the semiclassical result obtained for the VdP array Eq.~(\ref{eqn:omegavdp}) where $\omega_{\rm vdp}/U \sim r/2$. 

As the pump is reduced and the number of bosons per site decreases one expects quantum fluctuations to become more important. Indeed we see significant deviations from the semiclassical result at small $r$, already captured by Gutzwiller mean field but more pronounced for DMFT.
\begin{figure}[t]
\begin{center}
\epsfig{figure=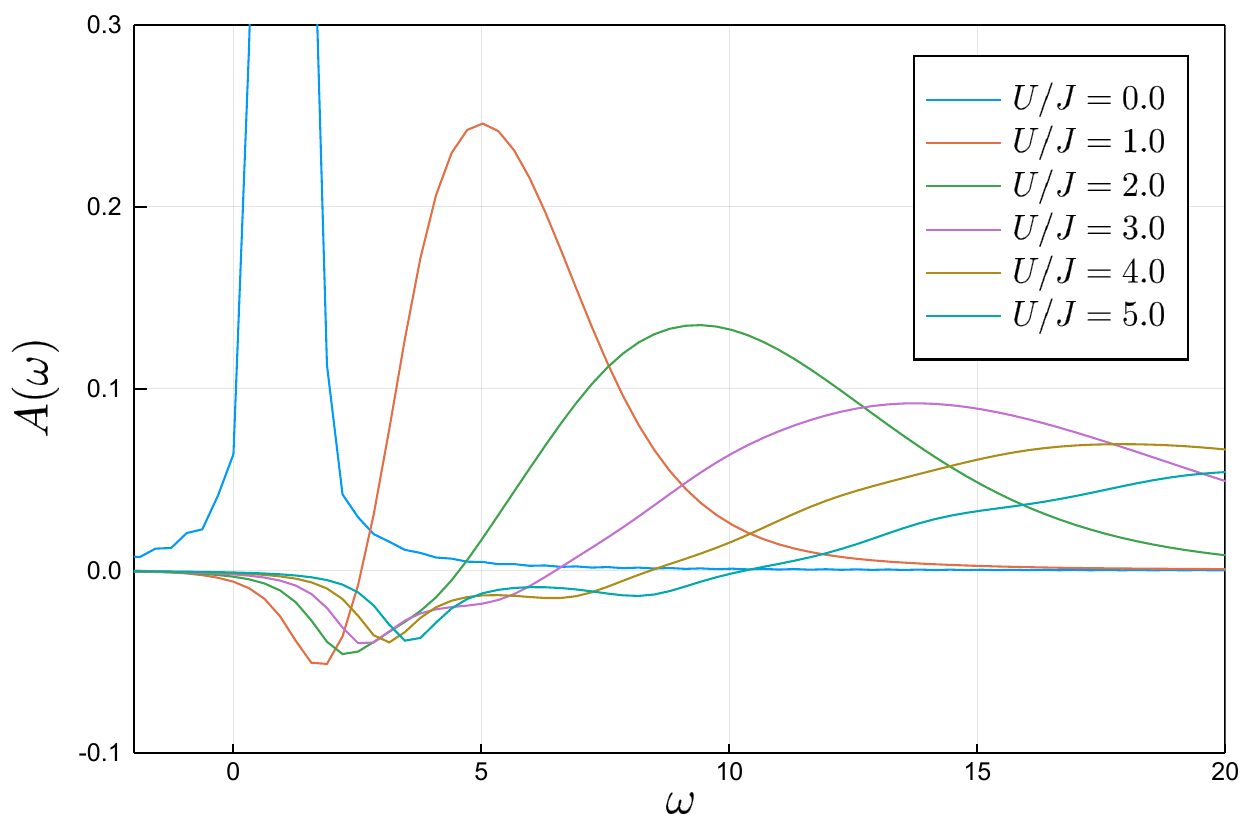,width=1\linewidth}
\caption{Local spectral function $A(\omega)$ for increasing values of the Kerr non-linearity $U$, at fixed value of hopping $J=\omega_0$ and drive-to-loss ratio $r=3$ within DMFT/NCA. We see that the frequency $\Omega_0$ at which the density of states vanishes, decreases with $U$ and eventually disappears for small enough values of $U$.
Parameters: $z=6$, $dim_H=14$.
}
\label{fig:specvsU}
\end{center}
\end{figure}
Another interesting aspect is the role played by the Kerr non-linearity. In Fig.~\ref{fig:specvsU} we plot the local density of states for fixed value of hopping and drive/loss ratio and for different values of the interaction $U$. A first interesting observation is that the frequency $\Omega_0$ at which NDoS emerges,  related to the mode becoming critical at the synchronization transition (see Eq.~(\ref{eqn:Omegac}), decreases upon reducing the Kerr non-linearity. A detailed analysis shows $\Omega_c\sim U$. 
We notice the analogy with the semiclassical result~(\ref{eqn:omegavdp}), nevertheless we stress that in this picture the critical frequency is proportional to the modulus square of the order parameter $ \vert  \langle b\rangle\vert^2\neq 0$, while within DMFT we find $\Omega_c\sim U $ \emph{at the critical point}, where by definition $ \langle b\rangle= 0$ . This further highlights the quantum nature of the synchronization transition considered in this work. Indeed we have seen how $\Omega_c$ is smoothly connected to the frequency $\Omega_0$ where NDoS emerges, a scale that exists already well inside the normal incoherent phase and that is a genuine feature of the quantum impurity model. Furthermore this implies that the Kerr non linearity is crucial in order to push the transition at finite frequency, implying undamped oscillations of the order parameter. We indeed observe that for a related model of all-to-all coupled quantum VdP oscillators recent Gutzwiller mean-field analysis reported a static (first-order) transition in absence of any Kerr non-linearity~\cite{lee2013,Davis_Tilley_2018}.

We conclude by noting that an oscillating phase in the order parameter could be in principle gauged away by going to an appropriate rotating frame. This however could not be done a priori, since as we have seen the critical frequency $\Omega_c$ itself depends from the many-body physics of the problem.  We are considering here only the instability of the normal phase, rather than the dynamics in the full broken symmetry phase which could show a more complex dynamical behavior,  whose description goes beyond the scope of the stationary-state oriented approach (see Sec. \ref{sec:stationarystateNCA}) used here.

\subsubsection{Limit Cycles at Finite Connectivity}

\begin{figure}[t]
\begin{center}
\epsfig{figure=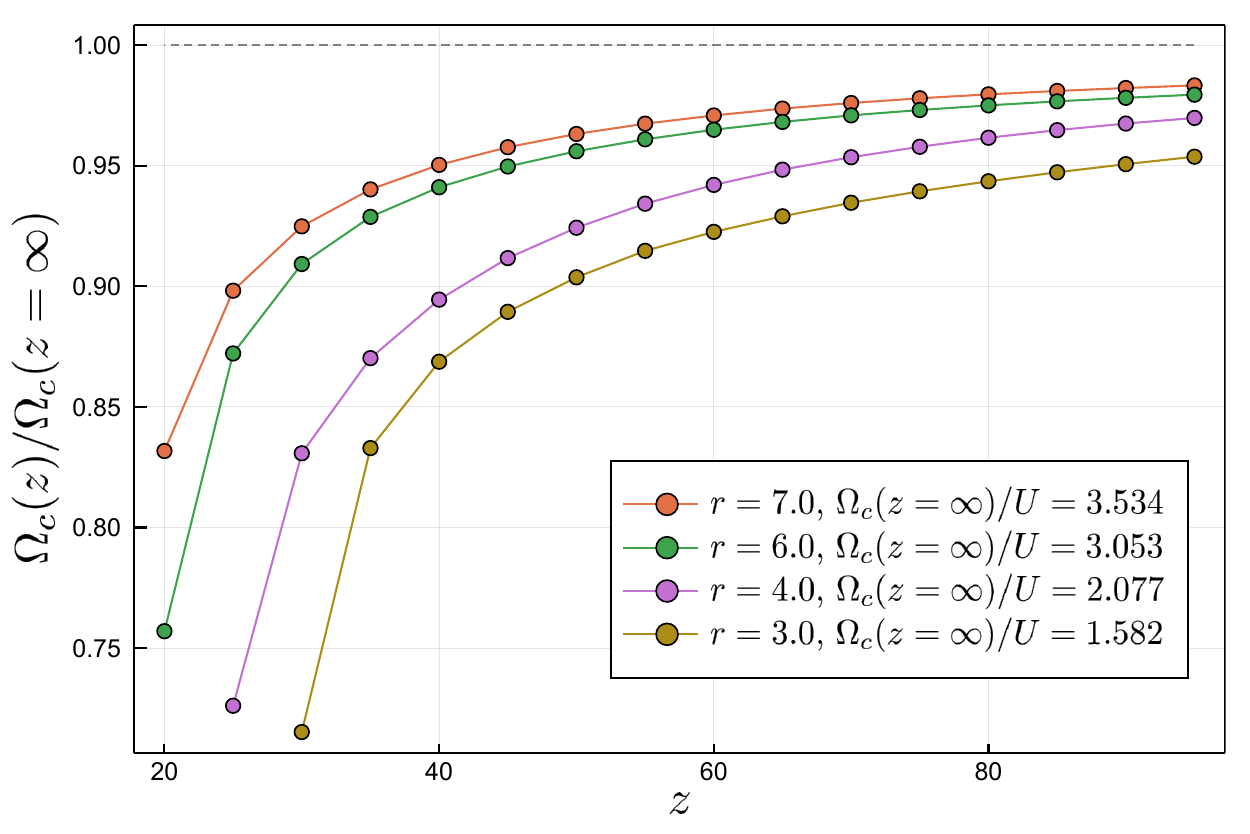,width=1\linewidth}
\caption{Critical Frequency for quantum synchronization, $\Omega_c$, within DMFT/NCA, as a function of the lattice connectivity $z$, for different values of the drive-to-loss ratio $r$ and hopping fixed on the phase boundary $J/U=(J/U)_c$. For clarity we normalize $\Omega_c$ to the mean field value obtained for $z=\infty$. We see that decreasing $z$ renormalizes down the frequency, but does not destroy the limit cycle. Parameters: 
$dim_H=14$. 
}
\label{fig:wcVzZ}
\end{center}
\end{figure}

%
Limit-cycles emerge ubiquitously within mean field approaches~\cite{lee_antiferromagnetic_2011,chan_limit-cycle_2015,wilson_collective_2016,SchiroPRL16}, and indeed even in the present problem the Gutzwiller solution at $z=\infty$ predicts one. The role of fluctuations on their stability has been discussed before, in particular in the context of a coherently driven anisotropic Heisenberg model with spontaneous decay ~\cite{Owen2018}. There an approach based on self-consistent Mori projection~\cite{degenfeld-schonburg2014}  and cluster mean field~\cite{JinEtAlPRX16} predicts that limit cycles disappear as the coordination number $z$ is decreased below a threshold value $z^*$ which depends on the system parameters. 

On the basis of these results, it is particularly interesting to study the fate of our synchronization transition beyond Gutzwiller mean-field theory.  In Fig. \ref{fig:wcVzZ} we plot the behavior of the critical frequency $\Omega_c$ obtained from DMFT/NCA, as a function of $z$ for different value of $r$. We observe that finite $z$ corrections tend to reduce the value of $\Omega_c$ with respect to the mean field value, which nevertheless remains finite down to the lowest value of connectivity at which, for a given value of drive, a synchronization transition exists, consistently with the  phase boundary moving to higher values of $r$ for decreasing $z$ (see Fig.~\ref{fig:phaseDiag}). In other words we find that, within our treatment, finite connectivity does not destroy the limit cycle phase which is only pushed at higher values of the incoherent drive.
Indeed in Fig. \ref{fig:wcVzZ} we show  that for drive $r=7$, the highest value that we can numerically access given the constraints on the local Hilbert space truncation, a limit cycle would exists down to $z=20$. We therefore expect that at higher drives the synchronized phase would survive down to low connectivity values. The regime of strong drive is however difficult to access within NCA and we leave this question open for future works. We remark nevertheless that our model is different from the one of Ref.~\cite{Owen2018}. In the present case the existence of a limit cycle phase is tightly related to a $U(1)$ symmetry present in the original Lindblad problem and spontaneously broken at the transition~\cite{ScarlatellaFazioSchiroPRB19}. 


\begin{figure}[b]
\begin{center}
\epsfig{figure=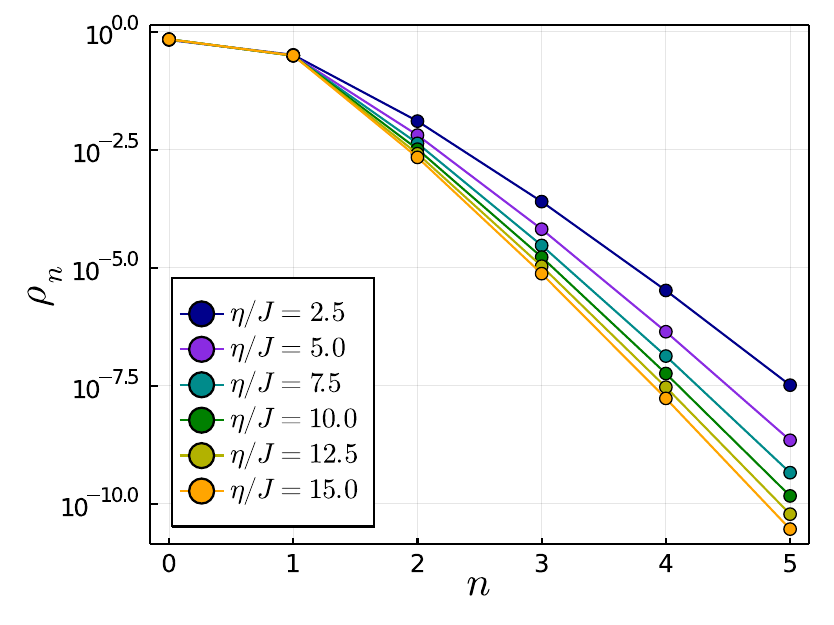,width=1\linewidth}
\caption{Quantum Zeno regime. Occupation probabilities of theDMFT/NCA on-site reduced density matrix $\rho_n=\langle n\vert \rho_s\vert n\rangle $,  as a function of the number of number of particles per site and for different values of two-particle-loss rates $\eta/J$.  For large $\eta/J$ we see that probabilities $\rho_{n\geq 2} $ are suppressed exponentially (note the log scale), with a rate that increases with $\eta/J$, while the states with $n=0,1$ bosons, not affected by the losses, retain a population of order one.}
\label{fig:Zenorhon}
\end{center}
\end{figure}

\begin{figure*}[t]
\begin{center}
\epsfig{figure=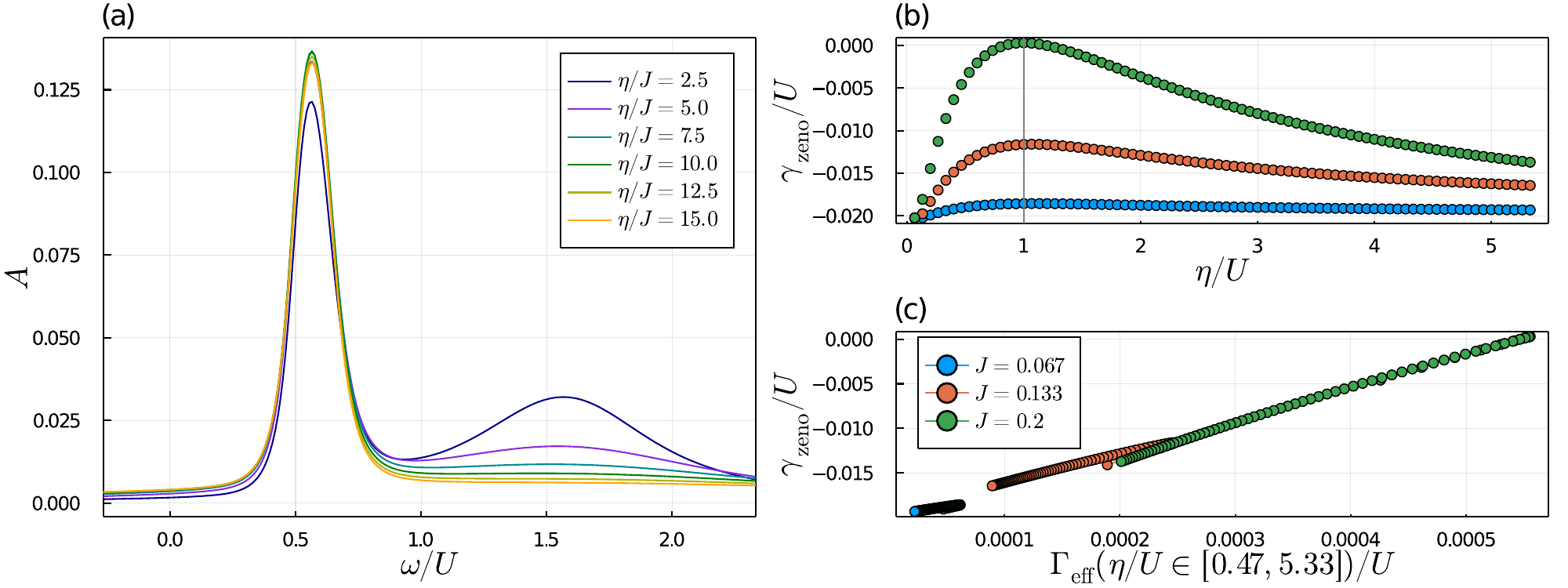,width=1\linewidth}
\caption{Quantum Zeno regime in the spectral function, increasing two-particle loss rate $\eta$ at fixed drive rate $f/U = r \eta/U =0.013$. Parameters: $t_{\rm max} =20$, $dt=0.002$, ${\rm dimH}=6$. (a) Spectral function at increasing $\eta/J$. The first peak corresponds to transitions between states with on-site occupation $n=0$ and $n=1$, while higher energy peaks are suppressed at large $\eta/J$. We fit the first peak as described in the main text in order to extract a measure of its lifetime $\gamma$.  (b) Partial lifetime $\gamma_{\rm zeno}/U$ of the first peak of the spectral function obtained by removing the zero hopping and zero drive/dissipation contributions. We show that this quantity has a non-monotonic behaviour, analogous to the populations in Fig.~(\ref{fig:zeno_highlight}), manifesting a quantum Zeno behaviour in the lifetime of excitations of the steady-state. (c) $\gamma_{\rm zeno}/U$ depends only on the dissipative scale $\Gamma_{\rm eff }/U$, analogously to the occupation probability $\rho_1$ as shown in the inset of Fig.~(\ref{fig:zeno_highlight}).
}
\label{fig:zeno1}
\end{center}
\end{figure*}

\subsection{Quantum-Zeno Regime}\label{sec:quantumzeno}

In the previous sections we have seen that the key feature of DMFT is the ability to capture hopping-induced dissipative processes in the normal phase, which are missed by Gutzwiller mean-field theory and ultimately responsible for the large corrections to the phase boundary due to finite connectivity. In this section we discuss a different parameter regime of our model, corresponding to large two-particle losses $\eta \gg J $, which alllows us to highlight even more the qualitative difference between DMFT and Gutzwiller predictions.
In the large two-particle losses limit, and in absence of any external drive, a perturbative analysis has shown that the dissipative dynamics of the Bose-Hubbard model effectively takes place in the subspace with zero and one boson per site, which are dark states of the local dissipator~\cite{garcia-ripoll2009}. These emergent hard-core bosons are subject to next-neighbor losses controlled by the scale
\beq
\label{eq:effDiss}
\Gamma_{\rm eff} = \frac{2 (J/z)^2}{U^2 + \eta^2} \eta 
\eeq
This effective dissipation is hopping-mediated -- namely it is non-zero only at finite hopping $J$ -- and remarkably it shows a non-monotonic behaviour as a function of the physical dissipation $\eta$: for $\eta \ll U$, it increases linearly with $\eta$. Instead, when  $\eta \gg U$, the effective dissipation $\Gamma_{\rm eff} $ is suppressed by increasing the physical dissipation $\eta$. The latter is a quantum-Zeno regime \cite{misra1977a,beige2000}, in which the coherent hopping dynamics bringing the system outside of the hard-core bosons subspace is suppressed due to the large coupling to the environment \cite{garcia-ripoll2009}.
In absence of an external pump the effective loss rate~(\ref{eq:effDiss}) ultimately leads the system to a zero density state and the Zeno scale~\ref{eq:effDiss} manifests itself in the transient dynamics~\cite{garcia-ripoll2009,rossini2020b}. Here instead we demonstrate how, in presence of a small pump, the Quantum Zeno regime emerges in the stationary state properties of the system and how this effect is completely missed by Gutzwiller, even at a qualitative level, while captured by our DMFT/NCA approach.

We consider the regime of  $\eta \gg J $ and a parametrically small pump rate $ f \ll \eta$.  In this case we expect the local occupation of all the states with more than one boson per site to be largely suppressed upon increasing the two-particle losses, since the small residual pump is not sufficient to counterbalance the losses. This is indeed shown in Fig. \ref{fig:Zenorhon} where we plot the occupation probabilities of the on-site reduced density matrix obtained from DMFT/NCA and see that only the occupation of the states $n=0,1$ remain of order one, while $\rho_{n\geq 1}$ is exponentially suppressed. We emphasize that in this case the long-time limit of the problem remains non-trivial within this subspace, due to the interplay between the small residual pump and the coherent dynamics generated by the Hamiltonian.
In Figure~\ref{fig:zeno_highlight} we have shown that our DMFT/NCA approach is able to capture the emergence of a stationary Zeno regime, through a non-monotonic behavior of the steady-state probability of having exactly one bosons per site, namely $\rho_1$, versus $\eta/U$, showing a universal collapse when plotted against $\Gamma_{\rm eff}$. This is remarkable, since the dissipative scale  in Eq.~(\ref{eq:effDiss}) describes particle losses which are non-local in space~\cite{garcia-ripoll2009} which a priori could have been expected to be beyond reach for a local approach such as DMFT. Instead, the non-Markovian bath is able to capture this hopping-induced dissipative processes. We emphasize that DMFT goes beyond the effective hard-core model of Ref.~\onlinecite{garcia-ripoll2009} in that the full crossover from normal to Zeno phase is captured as $\eta/J$ is increased. In this sense we can see DMFT as providing a non-perturbative solution of the effective hard-core model in the limit of large lattice connectivity. 

The presence of Zeno scale $\Gamma_{\rm eff}$ in the system not only affects the steady state populations, but is also expected to influence the lifetime of excitations of the steady-state, which can be extracted from the Green functions computed in DMFT.
In Fig. \ref{fig:zeno1} (a) we show the behaviour of the spectral function, defined in Eq. \eqref{eq:specFunc}, for increasing $\eta/J$. 
We work in a regime in which $U$ is large so that different resonances in the spectrum, corresponding to single-particle transitions between states with occupation $n$ and $n+1$, are well separated~\cite{scarlatellaClerkSchiro2018}.
By increasing $\eta/J$ all the peaks but the first one decrease in amplitude, as the former are proportional to the occupation of states with $n\geq 2$, which decreases with $\eta$ (see Fig.~\ref{fig:Zenorhon}). The amplitude of the lowest frequency peak, instead, increases with $\eta$ due to the increased probability of being in the subspace with occupation $n=0,1$. We concentrate on this peak in order to extract its  width. In order to do so, we fit the peak with a generalized Lorentzian $A(\omega) =a \frac{(\omega-\omega_0) b + 1}{(\omega-\omega_0)^2 + \gamma^2}$, with fitting parameters $a,b,\gamma$, which exactly describes the shape of the peak at zero hopping and which approximates it at small hopping, whose lifetime is given by the fitting parameter $\gamma$. 

We remark that the width $\gamma$ of the first peak has three contributions, $\gamma =  \gamma_{\rm{diss}} +\gamma_{\rm{coh}} +\gamma_{\rm {zeno}} $. The first term comes from dissipation and pump at zero hopping $\gamma_{\rm{diss}}= \gamma(J=0,\eta,f,U)$ and it can be well estimated with perturbation theory~\cite{scarlatellaClerkSchiro2018}, which shows it does not depend on $\eta$, as the transition $n=0\rightarrow n=1$ is unaffected by the two-particle losses at $J=0$. The second contribution to the lifetime is given by the hybridization of the energy levels of different sites at finite hopping, which is a purely coherent effect, independent of drive and dissipation. For example, this effect is responsible for the formation of the holon and doublon bands in the local spectral functions of the isolated Bose-Hubbard model, see e.g. \cite{strandWernerPRA2015}. We call this term $\gamma_{\rm{coh}} = \gamma(J,\eta=0,f=0, U)$. 

Finally, we expect a contribution coming from the hopping-induced decay, which we call $\gamma_{{\rm zeno}}$, which is present only for finite hopping and dissipation/drive.
In Figure Fig. \ref{fig:zeno1} (b), we show that the latter contribution follows a similar non-monotonic behaviour as $\Gamma_{\rm eff}$ as a function of $\eta/U$, witnessing the presence of quantum-Zeno behaviour in steady-state excitations. Analogously to populations, in Fig. \ref{fig:zeno1} (c) we show that plotting $\gamma_{{\rm zeno}}/U$ as a function of $\Gamma_{\rm eff}/U$, the data points at different values of $J/U$ all collapse on the same curve, demonstrating that $\gamma_{{\rm zeno}}$ only depends on $J$ and $\eta$ through $\Gamma_{\rm eff}$.

\section{Conclusions}
\label{sec:conclusions}


In this paper we formulated the Dynamical Mean Field Theory for bosonic open quantum many-body systems described by a Lindblad master equation on the lattice. This method is based on a systematic expansion in the limit of large lattice connectivity $z$. Within DMFT, fluctuations due to finite lattice connectivity are treated non-perturbatively through the solution of a self-consistent quantum impurity model, which in our case amounts to a Markovian single-site problem coupled to a coherent field and to a non-Markovian quantum bath mimicking the rest of the lattice. The non-Markovian bath contains the key new ingredient of DMFT, which makes it different from the Gutzwiller mean-field theory, to which it reduces in the $z=\infty$ limit. In particular, with respect to the former, DMFT is able to capture  both coherent and dissipative processes arising from the neighboring sites, the former playing a particularly crucial role in open quantum systems with correlated jump operators.

Using DMFT, together with a non-perturbative bosonic impurity solver based on a super-operator hybridization expansion truncated at the NCA level, we solved a Bose-Hubbard model in presence of two-particle losses and single particle pump, relevant for dissipative ultracold atoms as well as for arrays of superconducting circuits.  We have shown that this model features a dissipative phase transition from an incoherent normal phase to a nonequilibrium superfluid, which occurs above a critical hopping or pump strength. 

Within DMFT/NCA the phase boundary is strongly renormalized with respect to the Gutzwiller one, and pushed towards higher values of the couplings, leading to an increase of the normal phase. We have identified a new mechanism for this reduction of the broken symmetry phase which is unique to open quantum systems and arises from the suppression of local gain due to hopping-induced losses. This is a key feature brought forth by NCA, which treats the non-Markovian DMFT bath non-perturbatively, and it is instead missed by a simpler perturbative DMFT solver based on the Hubbard-I approximation.
We have further discussed how the increased effective dissipation due to the finite number of lossy neighbors affects all the unusual properties of the DMFT/NCA normal phase, from the renormalization of the gain (NDOS) threshold and steady-state populations due to hopping, to the emergence of a stationary-state quantum Zeno regime for large two-body losses. These effects are all qualitatively missed by Gutzwiller mean-field theory.

Finally, we have shown that the transition into the nonequilibrium superfluid phase occurs at  finite-frequency, corresponding to a local order parameter that oscillates in time at a frequency $\Omega_*$. Within DMFT/NCA this scale depends on both coherent and dissipative couplings, another important aspect  which is missed by Gutzwiller mean-field theory. Drawing from the physics of quantum VdP oscillators we interpreted this phenomenon as the onset of quantum many body synchronization and limit cycles at finite lattice connectivity.

Our DMFT holds the promise to be applied to a variety of driven-dissipative quantum many body problems.   Different bosonic models or driving schemes can be considered and readily studied with the NCA approach developed here, such as the recently introduced quadratically driven Kerr resonator~\cite{RotaEtAlPRL19}, the coherently driven case explored in the context of quantum bistability~\cite{LeBoiteEtAlPRL13,biondi_nonequilibrium_2017,foss-feig_emergent_2017,VicentiniEtAPRA18,LandaEtAlPRL20} or models relevant for the dissipatively stabilised Mott insulators of photons~\cite{ma2019}. Interesting directions for the future involve the development of other quantum impurity solvers, made possible by the Markovian structure of the quantum impurity problem. In particular, one could go beyond the Non-Crossing Approximation by including higher-order diagrams or resumming the full hybridization expansion~\cite{Scarlatella2019} with real-time diagrammatic Monte Carlo methods~\cite{schiroFabrizioPRB2009,Werner_Keldysh_09}. Alternatively, one could map back the non-Markovian Keldysh action into a Lindbladian problem with a finite number of bath levels~\cite{arrigoni2013} which could then be solved by exact diagonalisation~\cite{secli2021signatures}, quantum trajectories, through a matrix product operator representations of the density matrix~\cite{strathearn2018efficient,kilda2019fluorescence} or using an extension of the numerical renormalization group~\cite{lotem2020renormalized}. Further extensions of this approach could include the development of a cluster version of DMFT building upon recent developments for Markovian problems~\cite{JinEtAlPRX16}. 

Finally, we notice that a similar DMFT/NCA impurity solver, could be developed for driven-dissipative fermionic problems~\cite{ArrigoniEtAlPRL13,panas2019densitywave} or quantum spins~\cite{otsuki2013dynamical}, which are also being actively investigated and relevant for different experimental platforms~\cite{BarreiroEtAlNature11,LeeEtAlPRL13,HallerEtAlNatPhys15,YamamotoEtAlPRL19,NakagawaEtAlPRL20}.

\section*{Acknowledgments}

This work was supported by the University of Chicago through a FACCTS grant (“Franceand  Chicago  Collaborating  in  The  Sciences”),  by  the CNRS  through  the  PICS-USA-147504,  by  a  grant “Investissements  d’Avenir”  from  LabEx  PALM  (ANR-10-LABX-0039-PALM) and from the ANR grant "NonEQuMat" (ANR-19-CE47-0001). AC acknowledges support from the Air Force Office of Scientific Research MURI program, under Grant No. FA9550-19-1-0399.

\appendix
\section{Deriving DMFT for Open Markovian Quantum Systems}
\label{app:derivDMFT}

\begin{figure}[t]
\includegraphics[width=\columnwidth]{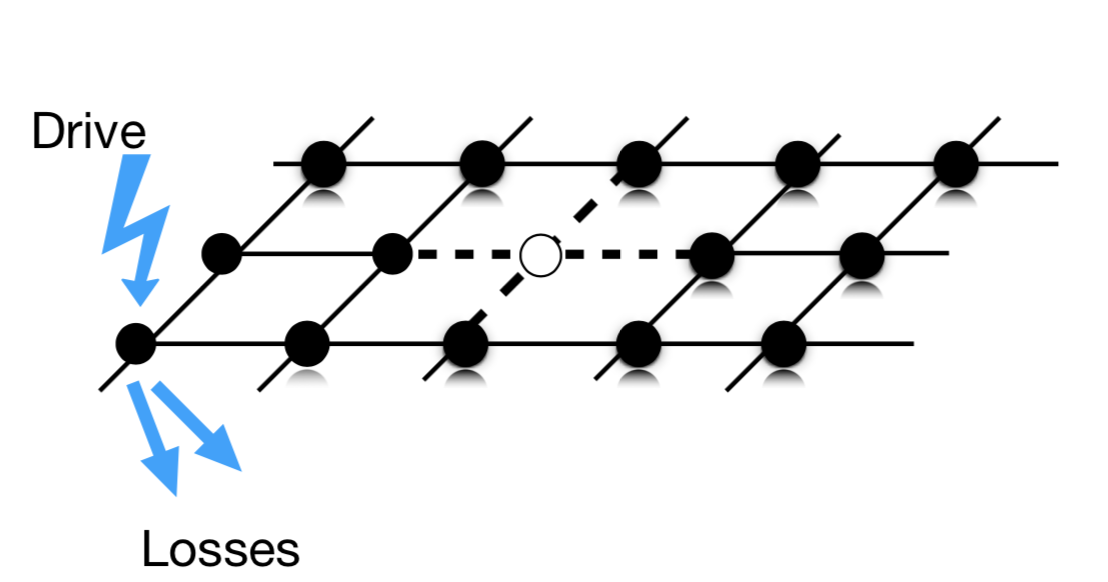}
\caption{Sketch of the quantum cavity method to derive the effective action of DMFT for Markovian lattice problems (see text).
A given lattice site is singled out (white dot in the figure) and all the remaining degrees of freedom are integrated out. Due to the coupling between the site and its $2z$ neighbors (dashed lines) this integration of degrees of freedom reduces to evaluate correlations function of a lattice with a cavity, containing all the remaining sites.}
\label{fig:DMFTcavity}
\end{figure}

We present here a sketch of the derivation of the DMFT action and self-consistency conditions, using the quantum cavity method and following and extending Ref~\cite{strandWernerPRX2015} to the open case. The main idea is to single out a given site of the lattice,  $i=0$ in the following, and to write down its effective Keldysh action obtained after integrating out all the other sites.
 \be
i\mathcal{S}_{\eff}[\bar{b}_0,b_0]= \mbox{log}\int \prod_{i\neq 0}D \lsq \bar{b}_i,b_i \rsq \,e^{i\mathcal{S}}
\ee
This effective single-site action describes, in principle, all the effects on site $i=0$ due to the coupling to the other sites by the hopping $J$, which, for our assumption of local jump operators and interactions (Sec.~\ref{sec:summary}), is the only term responsible for the coupling between different sites. To proceed, we notice that the full action Eq.~(\ref{eqn:actionFromME}) can be divided in three therms, 
$$
\mathcal{S}=\mathcal{S}_0+\delta\mathcal{S}+\mathcal{S}^{(0)}_{cav}\,$$ 
respectively $\mathcal{S}_0=\mathcal{S}[\bar{b}_0,b_0]$ containing only the terms in Eqns.~(\ref{eqn:actionFromME}-\ref{eqn:actionFromME2}) involving fields at site $i=0$, 
a term $\delta\mathcal{S}$ containing the hopping terms to the $2z$ fields ($\bar{b}_j,b_j$) at the neighboring sites 
\be
\delta\mathcal{S}=\frac{J}{z}\int dt \sum_{\alpha}\alpha\,\sum_{\langle 0j\rangle} \left(\bar{b}_{0\alpha}b_{j\alpha}+hc\right)
\ee
and a term $\mathcal{S}^{(0)}_{cav}$ describing a lattice with a cavity, namely including all the degrees of freedom except those at site $i=0$, see Figure~\ref{fig:DMFTcavity}. For an interacting many-body problem on a finite dimensional lattice integrating over the neighboring sites can only be done formally, as a cumulant expansion in $\delta S$, and leads to an effective action containing arbitrary powers of the local fields $\bar{b}_0,b_0$,  with coefficients given by the multipoints correlation functions of the cavity problem \cite{Review_DMFT_96}.  In the large connectivity limit, $z\gg 1$, one can formally organize this expansion in power of $1/z$ and obtain
\beq\label{eqn:S_eff_0}
\es{
\mathcal{S}_{\eff}&[\bb^\da_{0\alpha},\bb_{0\alpha}]=S_0
+\int dt \sum_{\alpha=\pm}\alpha\bold{\Phi}_{\eff\alpha}^\da (t) \bb_{0\alpha}(t)+\\
& 
+\frac{i^2}{2}\int dtdt\sum_{\alpha,\beta=\pm}\alpha\beta\,\bb^\da_{0\alpha}(t)\bold{\Delta}^{\alpha\beta}(t,t')\bb_{0\beta}(t') 
}
\eeq
where, in order to allow for condensed phases, we have introduced the Nambu-spinor notation 
$$
\bb^\da_0 =
\lp
\begin{array}{l}
b_0 \; \bar{b}_0
\end{array}\rp
\,\;\;
\bb_0=\lp
\begin{array}{l}
b_0\\
\bar{b}_0
\end{array}\rp
$$ 
The coefficients entering the effective action~(\ref{eqn:S_eff_0}) are related to quantum averages over the cavity action $\mathcal{S}^{(0)}_{cav}$ 
\bea 
\label{eq:sc_phi_implic}
\bold{\Phi}_{\eff\,\alpha}^\da (t)=J/z\sum_{\langle j0\rangle}  \langle \bb^{\dagger}_{j\alpha}(t)\rangle_{cav}^{(0)}
\\
\label{eq:sc_delta_implic}
\bold{\Delta}^{\alpha\beta}(t,t')=J^2/z^2 \sum_{\langle j0\rangle\langle k0\rangle} \langle \bb_{k\alpha}(t) \bb^{\dagger}_{j\beta}(t')\rangle^{(0)}_{cav}
\eea
The final step, in order to obtain the self-consistent conditions, is to relate the average of bosonic fields on the cavity action to those evaluated to the effective action~(\ref{eqn:S_eff_0}). For what concerns the non-Markovian bath in Eq.~(\ref{eq:sc_delta_implic}) this depends on the lattice geometry: it becomes particularly transparent for a Bethe lattice, a lattice with no loops such that once a cavity is created two neighbors $j,k$ are completely disconnected, where one gets 
\bea
\langle \bb_{k\alpha}(t) \bb^{\dagger}_{j\beta}(t')\rangle^{(0)}_{cav}= \delta_{kj}\langle \bb_{j\alpha}(t) \bb^{\dagger}_{k\beta}(t')\rangle^{(0)}_{cav}=\nonumber \\
= \delta_{kj}\langle \bb_{j\alpha}(t) \bb^{\dagger}_{j\beta}(t')\rangle_{S} = \delta_{kj}\langle \bb_{0\alpha}(t) \bb^{\dagger}_{0\beta}(t')\rangle_{S_{\rm eff}} 
\eea
where we have used the property of the Bethe lattice in the first equality, the fact that in thermodynamic limit the local property of the cavity action and the original action must be the same in the second equality and translational invariance in the last step. Plugging this into Eq.~(\ref{eq:sc_delta_implic}) gives the self-consistency condition~(\ref{eq:selfConsDelta}). We notice that simular arguments can be used for a different lattice, to relate averages over cavity and effective action, the only difference would be a more complicated self-consistency relation between bath and local Green's function~\cite{Review_DMFT_96}.
Finally, the average of the bosonic field taken on the cavity action can be related to the one on the effective action~(\ref{eqn:S_eff_0}) as~\cite{strandWernerPRX2015}
\bea
 \langle \bb^{\dagger}_{j\alpha}\rangle_{cav}^{(0)}= \langle \bb^{\dagger}_{0\alpha}\rangle_{S_{\rm eff}} +\\
  \int \dt '\sum_{\beta=\pm}\beta\langle \bb^{\dagger}_{0\beta}(t')\rangle_{S_{\rm eff}} 
\langle \bb_{k\beta}(t') \bb^{\dagger}_{j\alpha}(t)\rangle^{(0)}_{cav}
\eea
which can be plugged in Eq.~(\ref{eq:sc_phi_implic}) to give the second self-consistent condition~(\ref{eq:phieff}).


\section{NCA Benchmark: Local Occupation with Single-Particle Losses}
\label{app:bench}

\begin{figure}[t]
\begin{center}
\epsfig{figure=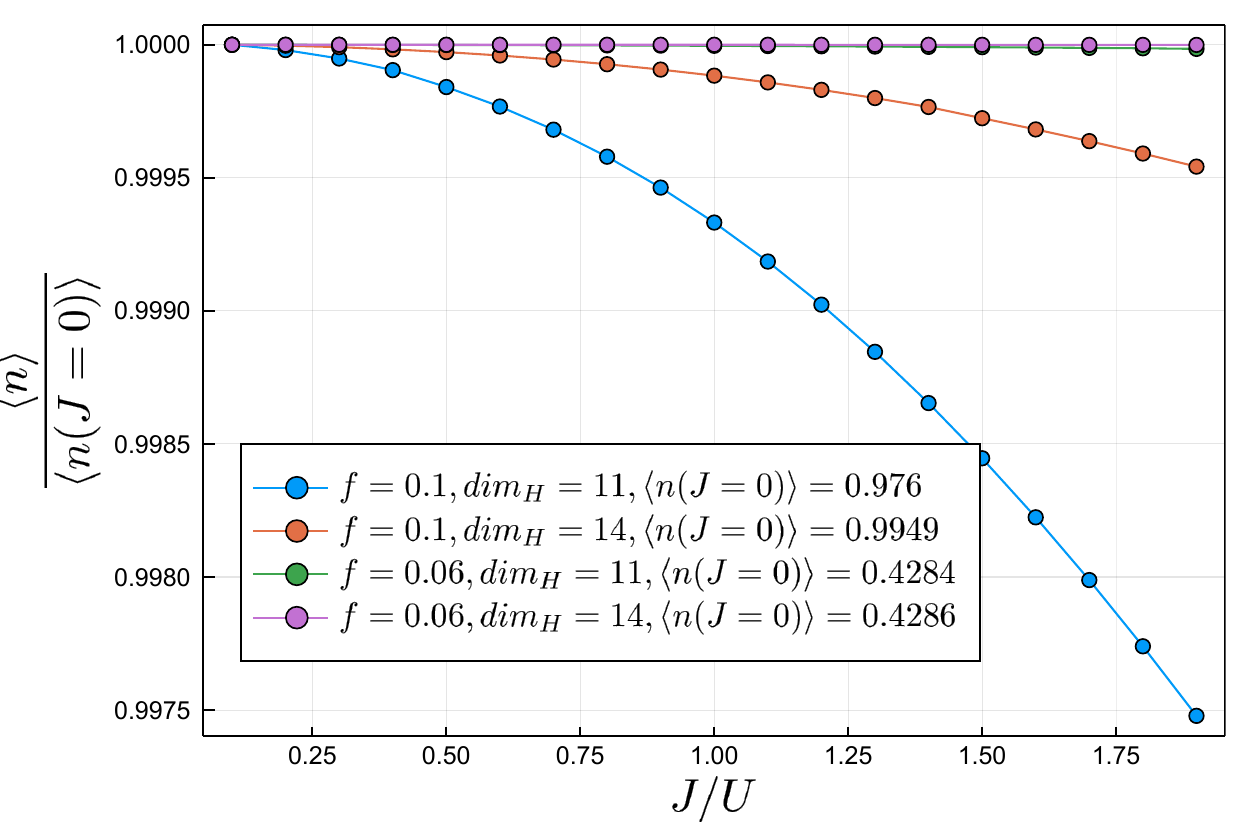,width=1\linewidth}
\caption{Local density of particles for the BH model with single-particle losses, as a function of the hopping to interaction ratio $J/U$, for different values of the drive amplitude $f$, normalized to the values of the single site problem. Parameters: $z=30$, $\kappa=0.2$, $dt=0.005$. }
\label{fig:1plBenchmark}
\end{center}
\end{figure}

In this appendix we report a  benchmark our DMFT/NCA approach for driven-dissipative many-body master equations defined by Eqs. \eqref{eq:masterEquation2} and \eqref{eq:BH}. 
We consider a different model from the main text, namely the driven-dissipative Bose-Hubbard model with single-particle losses and single-particle drive. This is specified by the same Bose-Hubbard Hamiltonian Eqs. \eqref{eq:BH} and \eqref{eq:locHam} as in the main text, but with the jump operators
\bea
\hat{L}_{i, l}=\sqrt{\kappa}\,\hat{b}_i \\
\hat{L}_{i, p}=\sqrt{f}\hat{b}_i^{\dagger}
\eea
where there is single-particle losses instead of the two-particle losses of the main text.

One can prove~\cite{Lebreuilly2016} that the stationary state density matrix of this model is independent from any Hamiltonian parameter, thus, for example, the on-site occupation is constant with the hopping rate $J$.
Obtaining this constant occupation with $J$ is a highly non-trivial benchmark for our DMFT/NCA approach. Figure \ref{fig:1plBenchmark} shows that this behavior is correctly reproduced by our approach. 
Small deviations from constant occupation show that this property is not exactly enforced by our numerical scheme and thus is a good test to validate our approach.
Deviations from constant occupation are mostly a result of local Hilbert space truncation, which become more severe increasing the drive, but they are reduced by increasing the cutoff $\dim_H$.

\section{Keldysh Field Theory and Semiclassical Limit}
\label{app:semiclassics}

In this section we write down the Keldysh field-theory associated to the Lindblad master equation for our driven-dissipative Bose-Hubbard model in Eq.~(\ref{eq:BH}-\ref{eqn:jumps}). This is done by writing a coherent path-integral representation of the trace over the density matrix~\cite{kamenev2011field,SiebererRepProgPhys2016}
$$
Z=\mbox{Tr}\rho(t)=\int \prod_i D\bar{b}_{i\rm cl}Db_{i\rm cl}
D\bar{b}_{i\rm q}Db_{i\rm q}\,\exp\left(i\mathcal{S}\right)
$$
in terms of bosonic fields in the classical-quantum basis $b_{j\rm cl/q}=\left(b_{j+}\pm b_{j-}\right)/\sqrt{2}$, where the Keldysh action reads
\begin{widetext}
\begin{align}
S = \int dt \sum_j \left[\bar{b}_{j\rm q}\left(i\partial_t-\omega_0-i \chi_-\right)b_{j\rm cl} +\mbox{hc}\right]
-\frac{1}{2}\int dt \sum_j \left[\left(U+i\eta\right)\left(\bar{b}_{j\rm cl}^2b_{j\rm cl}b_{j\rm q}+\bar{b}_{j\rm q}^2b_{j\rm cl}b_{j\rm q}\right)+hc\right]+\\
+\int dt \sum_j\left(2i\chi_+-i\eta\vert b_{j\rm cl}\vert^2\right)\bar{b}_{j\rm q}b_{j\rm q}
-\frac{J}{z}\int dt\sum_{\langle jj'\rangle}
\left([\bar{b}_{j\rm q}b_{j'\rm cl}+hc \right)
\end{align}
\end{widetext}
where the first line describes the local non-interacting contribution, including the dissipative couplings $\chi_{\pm}=\left(\gamma_p\pm\gamma_l\right)/2$ which in the case considered in the main text, where $\gamma_l=0$ and $\gamma_p=f=r\eta$, reduce to $\chi_{\pm}=f/2$. The second line includes non-linearities due to interaction and two-body losses while the last term account for the hopping between neighboring sites. 
The semiclassical limit is obtained by disregarding terms with more than two quantum fields, which leaves the action at most quadratic in the quantum fields $b_{j\rm q}$~\cite{SiebererRepProgPhys2016}. Those quadratic terms can be decoupled using a classical stochastic field~\cite{kamenev2011field} which allow to write the action as
\begin{align}
S &= \int dt \sum_j \left[\bar{b}_{j\rm q}\left(i\partial_t-\omega_0-i \chi_-+\xi_j(t) \right)
b_{j\rm cl}+\mbox{hc} \right]+\nonumber\\
&+\left(U-i\eta\right)\int dt \sum_j \bar{b}_{j\rm q}\vert b_{j\rm cl}\vert^2 b_{j\rm cl}+\mbox{hc}\nonumber\\
&-\frac{J}{z}\int dt\sum_{\langle jj'\rangle}
\left([\bar{b}_{j\rm q}b_{j'\rm cl}+\mbox{hc} \right)
\end{align}
We can now take the saddle point conditions
$$
\frac{\delta S}{\delta\bar{b}_{j\rm q}}=0=\frac{\delta S}{\delta\bar{b}_{j\rm cl}}
$$
and obtain the Langevin equation given in Eq.~(\ref{eqn:EOMcl}) of the main text. We note that the non-interacting retarded Green's function of the problem reads 
$$
G^R_0(\omega)=\frac{1}{\omega-\omega_0-i\chi_-}
$$
which has NDoS as soon as $\chi_-\ge0$, i.e. $r>0$ which is in the condition for the instability towards an oscillating solution.

\section{NDoS and Response to a  Weak Coherent Drive }
\label{app:Response}

In this appendix we show that a direct consequence of the NDoS effect is that the average power absorbed from a weak coherent drive becomes negative, indicating the onset of gain and  energy emission. We consider our lattice model in presence of a time dependent perturbation of the Hamiltonian describing a weak coherent drive with frequency $\omega$
\be
H(t)=H+V(t)\equiv H+\sum_i \left(v_i^* e^{i\omega t} b_i +v_i e^{-i\omega t} b^{\dagger}_i\right)
\ee
where $H$ is the Bose Hubbard Hamiltonian defined in Eq.(\ref{eq:BH}-\ref{eq:locHam}). The evolution of the density matrix of the system is described by the Lindblad master equation in presence of $H(t)$
\be
\partial_t \hat{\rho} = - i[H(t),\hat{\rho}]+\sum_{i\mu} \left(\hat{L}_{i\mu} \hat{\rho} \hat{L}_{i\mu}^{\dagger}
-\frac{1}{2}\left\{\hat{L}^{\dagger}_{i\mu} \hat{L}_{i\mu},\hat{\rho} \right\}\right) 
\ee
We notice that the time derivative of the internal energy $E(t)=\langle H(t)\rangle$ in an open quantum system has two contributions, respectively coming from the derivative of the system Hamiltonian and from the time derivative of the density matrix
\be 
\dot{E}(t)=\mbox{Tr}\rho(t) \dot{H}(t)+\mbox{Tr}\dot{\rho}(t) H(t)\,
\ee
the latter being identified as heat flow which would be zero in a closed system evolving with unitary dynamics. We are interested in the first term which measures the absorbed power from the external perturbation~\cite{RivasEtAlPRL20} and can be written as
\be\label{eqn:power}
\dot{W}(t)= \mbox{Tr}\rho(t)\dot{H}(t)=\sum_i
\left(i\omega v^*_i\,e^{i\omega t}\langle b_i\rangle +\mbox{hc}\right)
\ee
Within linear response theory the average of the bosonic field can be written as
\bea
\langle b_i\rangle =\sum_j \int dt' G^R_{ij}(t-t')v_j(t')
\eea
where $G^R_{ij}(t-t')=-i\theta(t-t')\langle [b_i(t),b^{\dagger}_j(t')]\rangle$. Plugging this expression into Eq.~(\ref{eqn:power}) we obtain the average absorbed power 
\be 
\dot{W}=\sum_{ij}v^*_iv_j\omega A_{ij}(\omega)
\ee
written in terms of the spectral function $A_{ij}(\omega)=(-1/\pi)\mbox{Im}G^R_{ij}(\omega)$. For a localized perturbation, $v_i=v_0\delta_{i,0}$ we obtain the result given in the main text, showing how a NDoS implies a negative absorption rate, namely the system transfer some of the energy from the perturbation to the emitted radiation.

\section{Hubbard-I Impurity Solver} 
\label{app:hubbI}
In this appendix we compute the impurity Green's functions in the simple Hubbard-I approximation discussed in Sec.~\ref{sec:HubbardI} and show that the phase diagram computed in Hubbard-I coincides with the mean-field one.  We restrict to study the symmetric phase, i.e. $\bold{\Phi}=0$, where equations involve one Nambu component and to the stationary regime where convolutions turn into product under Fourier transform. 

Equation \eqref{eq:hubbI} is a closed second-order algebraic equation for $G^{\alpha\beta}(\omega)$, which is easily solved in the classical/quantum basis for the retarded and Keldysh components. The retarded Green's function is simply given by taking its retarded component, yielding
\beq
\label{eqn:GretH1}
{J^2}/{z} \, G^R(\w) =  \left[G _{0}^R(\w)\right]^{-1} -\left[G^R(\w)\right]^{-1} 
\eeq
which can be solved and gives
\beq
\label{eqapp:retHubbI}
G^R(\w) = \frac{z}{2 J^2} { G^R _{0}(\omega) } ^\mo  \lp 1- \sqrt{1-\frac{4 J^2}{z} { G^R _{0}(\omega) } ^2  } \rp
\eeq
The inverse Keldysh Green's function is given by $ \lsq G^\mo \rsq^K = \lsq G_{0}^\mo \rsq^K - \frac{J^2}{ z} G^K $, that we invert with the standard relation $G^K = 	-  G^R  \lsq G^\mo \rsq^K G^A $ giving
\beq
\label{eq:keldyshGreensFunction}
G^K(\w) =  \frac{   \abs{G^R(\w)}^2 G_{0}^K(\w) }{ \abs{G_{0}^R(\w)}^2 \lp 1  -\frac{J^2}{z} \abs{G^R(\w)}^2 \rp}
\eeq

We finally show that the Hubbard-I impurity solver gives the same phase boundary and critical frequency as Gutzwiller mean-field. In fact if we plug the equation for the Hubbard-I retarded Green's function, Eq.~(\ref{eqn:GretH1}), into the equation for the finite-frequency transition in DMFT, Eq.~\eqref{eq:dmft_cp} that we rewrite for convenience here 
\beq 
\frac{1}{J_c}+ G ^R  (\Omega_c,J_c)  + \frac{J_c}{z} \lsq G^R   (\Omega_c,J_c )  \rsq^2  = 0
\eeq
we obtain
\beq
G ^R  (\Omega_c,J_c)\left(J_c+\left[G _{0}^R(\Omega_c)\right]^{-1}\right)=0
\eeq
where the term in parenthesis is exactly the Gutzwiller mean-field condition for critical hopping and frequency (see Eq.~\eqref{eq:mf_cp} or Eq.~\eqref{eq:dmft_cp} for $z=\infty$). We conclude therefore that the reduction of the ordered phase discussed in the main text and the dependence of $\Omega_c$ from the hopping are key features of the NCA impurity solver.

\section{Stationary state in DMFT/NCA}\label{sec:rho_ss}

A complete steady-state DMFT/NCA procedure requires to compute the steady-state density matrix of the impurity, which is defined by $\rho_s = \hat{\mathcal{V}}(\infty,0) \rho_0 $, where we assume a unique steady-state.
While in principle this would require to perform the full transient dynamics from an arbitrary initial condition, here we show how to obtain $\rho_s$ directly from the stationary-state propagator $ \hat{\mathcal{V}}(t-t') $.
The steady-state density matrix satisfies the condition $\lim_{t\rightarrow \infty} \partial_t \rho(t) = \lim_{t\rightarrow \infty} \partial_t \hat{\mathcal{V}}(t,0) \rho_0 = 0$, where $\rho_0$ is an  arbitrary initial state. Using the Dyson equation~(\ref{eq:dysonNCA}) this condition translates into the equation
\begin{equation}\label{eqn:ssNCA}
\lim_{t\rightarrow \infty} \left[\hat{\mathcal{L}}_{\rm eff } (t)\hat{\mathcal{V}}(t,0) +  \int_{0}^{t} d\tau \hat{S}(t,t-\tau) \hat{\mathcal{V}}(t-\tau,0) \right] \rho_0=0 
\end{equation}
where we have performed a change of variable in the convolution integral. The NCA self-energy $\hat{S}(t,t-\tau)$ in the equation above depends on the hybridization function of the bath $\bold{\Delta}$ (see Eq.~\ref{eq:selfEnergyExplicit}), which is itself related to the local Green's function of the system by the DMFT self-consistency. Therefore the assumption that for $t\rightarrow\infty$ the system reaches a unique steady-state, forgetting about its initial condition, translates within DMFT in a requirement on the behavior of the NCA self-energy for large values of its argument. In particular $\hat{S}(t,t-\tau)$ must vanish for $ \tau\sim t\rightarrow\infty$, restricting  the time convolution in the Dyson equation to times of the order $\tau\sim \tau_*\ll t$. We can therefore safely take the long time limit $t\rightarrow\infty$ in Eq.~(\ref{eqn:ssNCA}), use the fact that in this limit  the self-energy becomes translational invariant, $\hat{S}(t,t-\tau) \rightarrow \hat{S}(\tau)$, and the state $\hat{\mathcal{V}}(t-\tau,0) \rho_0 $ approaches the stationary state $\rho_s$, and finally send the scale $\tau_*$ to infinity, yielding the eigenvalue condition
\begin{equation}
 \left( \hat{\mathcal{L}}_{\rm eff } (\infty)   +  \int_{0}^{\infty} d\tau \hat{S}(\tau) \right)\rho_s = 0 
\end{equation}
This condition depends only on the stationary state propagator $\hat{\mathcal{V}}(\tau)$ through the NCA self-energy $\hat{S}(\tau)$, allowing to compute the stationary density matrix $\rho_s$ in the steady-state DMFT procedure.


\widetext
\section{NCA Green functions Derivation}
\label{app:ncaGreenFunctions}
In this section we show how to compute impurity Green functions knowing $\mcV(t,t')$.
From the impurity action Eq. \eqref{eqn:S_eff}, the trace of the density matrix at infinite time is obtained by 
\beq
Z = \tr{\rho(\infty)} =  \int \mathcal{D} \lsq \bar{b}_\alpha b_\alpha \rsq e^{  iS_{loc}[ \bar{b}_{\alpha ,a} , b_{\beta ,b}]+i \int_0^\infty dt \underset{a}{\underset{{\alpha}}{\sum}}\alpha\;\bar{\Phi}_{\eff\,\alpha a} (t) b_{\alpha a}(t) - \frac{i}{2} \int_0^\infty  \int_0^\infty  dt_1 dt_2 \underset{ab}{\underset{{\alpha \beta  }}{\sum}}   \bar{b}_{\alpha a} (t_1) \alpha \beta \Delta^{\alpha \beta}_{a b}(t_1, t_2) b_{\beta b}(t_2) }
\eeq
where we explicited both Keldysh $\alpha, \beta \in \lbr +, - \rbr $ and Nambu $a,b$ indices, as in Sec. \ref{sec:NCA}.
One-particle Green functions can be obtained as functional derivatives of $Z$
\beq
G^{\alpha \beta}_{a b} \lp t, t' \rp - i \phi_{\alpha a}(t) \phi^\da_{\beta b} (t')= - i \aver{b_{\alpha a} (t ) \bar{b}_{\beta b} (t') }  =2 \alpha \beta \frac{\delta  Z }{\delta \Delta^{\beta \alpha}_{b a} (t',t)} 
\eeq
We remark that the derivative over $\Delta$ gives both the connected components $G^{\alpha \beta}_{a b} $, as defined in the main text Eq. \eqref{eqn:GreenFunction}, and the disconnected ones $- i \phi_{\alpha a}(t) \phi^\da_{\beta b} (t')$. 
The density matrix trace can equally be expressed in terms of the evolution superoperator: $ Z = \tr \lsq \rho (\infty) \rsq = \tr \lsq \hat{\mcV}(\infty,0) \rho (0) \rsq $. 
Using the Dyson equation \eqref{eq:dysonNCA} for $\hat{\mcV}(\infty,0)$, we get 
\beq
\label{eq:greensFuncInterm}
\es{
- i \aver{b_{\alpha a} (t ) \bar{b}_{\beta b} (t') } = 2 \alpha \beta \tr \lbr  \int_{0}^\infty dt_1 \int_{0}^{t_1} dt_2 \lsq \hat{\mcV}_0(\infty ,t_1)  \frac{\delta \hat{S}(t_1,t_2)}{\delta \Delta_{b a}^{\beta \alpha} ( t',t)}   \hat{\mcV} (t_2,0) +  \hat{\mcV}_0(\infty ,t_1)  \hat{S}(t_1,t_2)  \frac{\delta \hat{\mcV} (t_2,0)  }{\delta \Delta_{b a}^{\beta \alpha} ( t',t)}  \rsq\rho(0) \rbr
}
\eeq
In the last expression we first drop all $\hat{\mcV}_0$s as they leave the trace unchanged. 
The second contribution to the trace vanishes because $ \tr{ \lp  \hat{S} (t_1,t_2)  \bullet  \rp = 0}$ for $t_1 >t_2$, as dictated by the time integrals, as we now show. In order to to do this, we report here the NCA self-energy Eq. \eqref{eq:selfEnergyExplicit} for simplicity:
\begin{align}
&\hat{S} (t_1,t_2) =-\frac{i}{2} \underset{ab}{\sum_{\alpha \beta  }}    \alpha \beta  \lsq  \Delta_{b  a }^{\beta \alpha} (t_1, t_2 ) \hat{b}_{ \beta  b }^\da \hat{\mcV} (t_1,t_2) \hat{b}_{ \alpha a } +   \Delta_{ a b }^{\alpha \beta} (t_2,t_1 ) \hat{b}_{\beta b} \hat{\mcV} (t_1,t_2) \hat{b}^\da_{\alpha a}   \rsq 
\end{align}
We remark that, for the cyclic property of the trace, it holds that $\tr \lsq  \hat{X}_+ \bullet \rsq =\tr \lsq  \hat{X}_- \bullet \rsq  $, $\hat{X}_\beta$ being a generic super-operator. 
Then we can fix the $ \hat{b}^\da_\beta,  \hat{b}_\beta$ super-operators to be $ \hat{b}^\da_+,  \hat{b}_+$ under the trace, getting 
\beq
\label{eq:traceSVanishes}
\es{
 &\tr \lsq  \hat{S}(t_1,t_2)  \bullet \rsq = \\ 
 = &\sum_{a , b} \sum_{\alpha \beta \in \{ + , - \}  }   - \alpha \beta \, \frac{i}{2} \,  \tr \lsq   \Delta_{b  a }^{\beta \alpha} (t_1, t_2 ) \hat{b}_{ \beta  b }^\da \hat{\mcV}(t_1,t_2) \hat{b}_{ \alpha a } \bullet +    \Delta_{ a b }^{\alpha \beta} (t_2,t_1 ) \hat{b}_{\beta b} \hat{\mcV}(t_1,t_2) \hat{b}^\da_{\alpha a}    \bullet \rsq  = \\ 
  = &\sum_{a , b} \sum_{\alpha  \in \{ + , - \}  }   - \alpha  \, \frac{i}{2} \,  \tr \lbr  \sum_{\beta  \in \{ + , - \}  }  \lsq \beta \Delta_{b  a }^{\beta \alpha} (t_1, t_2 ) \rsq \hat{b}_{ +  b }^\da \hat{\mcV}(t_1,t_2) \hat{b}_{ \alpha a }  \bullet +   \sum_{\beta  \in \{ + , - \}  }  \lsq \beta \Delta_{ a b }^{\alpha \beta} (t_2,t_1 ) \rsq \hat{b}_{+ b} \hat{\mcV}(t_1,t_2) \hat{b}^\da_{\alpha a}    \bullet \rbr  
 }
 \eeq
Finally, we remark that the two sums over $\beta$ in the last line vanish respectively because of the identities on the Keldysh components $ \Delta_{ba}^{++}(t_1,t_2) =  \Delta_{ba}^{-+}(t_1,t_2) $, $ \Delta_{ba}^{+-}(t_1,t_2) =  \Delta_{ba}^{--}(t_1,t_2) $ and $ \Delta_{ab}^{++}(t_2,t_1)=  \Delta_{ab}^{+-}(t_2,t_1)$, $ \Delta_{ab}^{-+}(t_2,t_1)=  \Delta_{ab}^{--}(t_2,t_1)$ which hold for $t_1>t_2$ according to the definition of $ \Delta$ in terms of bath Green functions: schematically its Keldysh components are given by $ \Delta^{\beta \alpha}( {t_1},{t_2}) \sim \sum \dots  \aver{T_C c(t_1,\beta) c^\da (t_2,\alpha)} $ with $c, c^\da$ bath operators and $T_C$ the contour-time-ordering operator.
Going back to Eq. \eqref{eq:greensFuncInterm}, from its first term we get
\beq
G^{\alpha \beta}_{a b } (t,t' ) - i \phi_{\alpha a}(t) \phi^\da_{\beta b} (t') = 2 \alpha \beta  \tr \lbr  \int_{0}^\infty dt_1 \int_{0}^{t_1} dt_2  
 \frac{\delta \hat{S}(t_1,t_2)}{\delta \Delta_{ba}^{\beta \alpha} ( t',t)}  \hat{ \mcV } (t_2,0)  \rho(0) \rbr
\eeq
Taking derivatives of the NCA self-energy, we get two contributions of the form $\simeq \tr \lp \frac{\delta \Delta }{\delta \Delta} X_{1} \mathcal{V} X_2 + \Delta X_{1} \frac{\delta \mathcal{V} }{\delta \Delta} X_2 \rp $. The second one vanishes and the proof is analogous to Eq. \eqref{eq:traceSVanishes}, where we showed that the trace of the NCA self-energy vanishes. 
For the first term, we need remembering that $t_1> t_2$ and thus that we have to distinguish the two cases $t>t'$ and $t<t'$, getting 
\beq
\label{eq:derivativeSigma}
   \frac{\delta \hat{S}(t_1,t_2)}{\delta \Delta_{b a}^{\beta \alpha} ( t',t)} =  
- \frac{i}{2} \alpha  \beta \lsq  \hat{b}_{\alpha a } \hat{\mcV}(t,t') \hat{b}_{\beta b}^\da \theta(t-t')  \delta(t_1-t) \delta(t_2 - t') + 
  \hat{b}_{\beta b} ^\da \hat{\mcV}(t',t) \hat{b}_{\alpha a} \theta(t'-t)\delta(t_1-t') \delta(t_2 - t) \rsq
\eeq
and the Green functions are eventually given by 
\beq
 G_{a b }^{\alpha \beta}(t,t')  - i \phi_{\alpha a}(t) \phi^\da_{\beta b} (t') = -i  \lbr \tr \lsq \hat{b}_{\alpha a} \hat{\mcV}(t,t') \hat{b}_{\beta b}^\da \hat{\mcV}(t',0) \rho(0) \rsq \theta(t-t') + \tr \lsq \hat{b}_{\beta b}^\da  \hat{\mcV}(t',t) \hat{b}_{\alpha a} \hat{\mcV}(t,0) \rho(0) \rsq \theta(t'-t) \rbr
\eeq
that is equivalent to Eq. \eqref{eq:NCAGreen} of the main text. 
This equation has the same form of the quantum regression formulae for one-particle Green functions for Markovian systems and shows that these generalize to the case of non-Markovian baths within the NCA approximation.

\twocolumngrid

%
%
%
%

%

%


\begin{thebibliography}{173}%
\makeatletter
\providecommand \@ifxundefined [1]{%
 \@ifx{#1\undefined}
}%
\providecommand \@ifnum [1]{%
 \ifnum #1\expandafter \@firstoftwo
 \else \expandafter \@secondoftwo
 \fi
}%
\providecommand \@ifx [1]{%
 \ifx #1\expandafter \@firstoftwo
 \else \expandafter \@secondoftwo
 \fi
}%
\providecommand \natexlab [1]{#1}%
\providecommand \enquote  [1]{``#1''}%
\providecommand \bibnamefont  [1]{#1}%
\providecommand \bibfnamefont [1]{#1}%
\providecommand \citenamefont [1]{#1}%
\providecommand \href@noop [0]{\@secondoftwo}%
\providecommand \href [0]{\begingroup \@sanitize@url \@href}%
\providecommand \@href[1]{\@@startlink{#1}\@@href}%
\providecommand \@@href[1]{\endgroup#1\@@endlink}%
\providecommand \@sanitize@url [0]{\catcode `\\12\catcode `\$12\catcode
  `\&12\catcode `\#12\catcode `\^12\catcode `\_12\catcode `\%12\relax}%
\providecommand \@@startlink[1]{}%
\providecommand \@@endlink[0]{}%
\providecommand \url  [0]{\begingroup\@sanitize@url \@url }%
\providecommand \@url [1]{\endgroup\@href {#1}{\urlprefix }}%
\providecommand \urlprefix  [0]{URL }%
\providecommand \Eprint [0]{\href }%
\providecommand \doibase [0]{http://dx.doi.org/}%
\providecommand \selectlanguage [0]{\@gobble}%
\providecommand \bibinfo  [0]{\@secondoftwo}%
\providecommand \bibfield  [0]{\@secondoftwo}%
\providecommand \translation [1]{[#1]}%
\providecommand \BibitemOpen [0]{}%
\providecommand \bibitemStop [0]{}%
\providecommand \bibitemNoStop [0]{.\EOS\space}%
\providecommand \EOS [0]{\spacefactor3000\relax}%
\providecommand \BibitemShut  [1]{\csname bibitem#1\endcsname}%
\let\auto@bib@innerbib\@empty
\bibitem [{\citenamefont {Cirac}\ and\ \citenamefont
  {Zoller}(1995)}]{CiracZollerPRL95}%
  \BibitemOpen
  \bibfield  {author} {\bibinfo {author} {\bibfnamefont {J.~I.}\ \bibnamefont
  {Cirac}}\ and\ \bibinfo {author} {\bibfnamefont {P.}~\bibnamefont {Zoller}},\
  }\bibfield  {title} {\enquote {\bibinfo {title} {Quantum computations with
  cold trapped ions},}\ }\href {\doibase 10.1103/PhysRevLett.74.4091}
  {\bibfield  {journal} {\bibinfo  {journal} {Phys. Rev. Lett.}\ }\textbf
  {\bibinfo {volume} {74}},\ \bibinfo {pages} {4091--4094} (\bibinfo {year}
  {1995})}\BibitemShut {NoStop}%
\bibitem [{\citenamefont {Raimond}\ \emph {et~al.}(2001)\citenamefont
  {Raimond}, \citenamefont {Brune},\ and\ \citenamefont
  {Haroche}}]{Haroche_RMP}%
  \BibitemOpen
  \bibfield  {author} {\bibinfo {author} {\bibfnamefont {J.~M.}\ \bibnamefont
  {Raimond}}, \bibinfo {author} {\bibfnamefont {M.}~\bibnamefont {Brune}}, \
  and\ \bibinfo {author} {\bibfnamefont {S.}~\bibnamefont {Haroche}},\
  }\bibfield  {title} {\enquote {\bibinfo {title} {Manipulating quantum
  entanglement with atoms and photons in a cavity},}\ }\href {\doibase
  10.1103/RevModPhys.73.565} {\bibfield  {journal} {\bibinfo  {journal} {Rev.
  Mod. Phys.}\ }\textbf {\bibinfo {volume} {73}},\ \bibinfo {pages} {565--582}
  (\bibinfo {year} {2001})}\BibitemShut {NoStop}%
\bibitem [{\citenamefont {Schoelkopf}\ and\ \citenamefont
  {Girvin}(2008)}]{Schoelkopf2008}%
  \BibitemOpen
  \bibfield  {author} {\bibinfo {author} {\bibfnamefont {R.~J.}\ \bibnamefont
  {Schoelkopf}}\ and\ \bibinfo {author} {\bibfnamefont {S.~M.}\ \bibnamefont
  {Girvin}},\ }\bibfield  {title} {\enquote {\bibinfo {title} {Wiring up
  quantum systems},}\ }\href {\doibase 10.1038/451664a} {\bibfield  {journal}
  {\bibinfo  {journal} {Nature}\ }\textbf {\bibinfo {volume} {451}},\ \bibinfo
  {pages} {664--669} (\bibinfo {year} {2008})}\BibitemShut {NoStop}%
\bibitem [{\citenamefont {Aspelmeyer}\ \emph {et~al.}(2014)\citenamefont
  {Aspelmeyer}, \citenamefont {Kippenberg},\ and\ \citenamefont
  {Marquardt}}]{aspelmeyer2014}%
  \BibitemOpen
  \bibfield  {author} {\bibinfo {author} {\bibfnamefont {Markus}\ \bibnamefont
  {Aspelmeyer}}, \bibinfo {author} {\bibfnamefont {Tobias~J.}\ \bibnamefont
  {Kippenberg}}, \ and\ \bibinfo {author} {\bibfnamefont {Florian}\
  \bibnamefont {Marquardt}},\ }\bibfield  {title} {\enquote {\bibinfo {title}
  {Cavity optomechanics},}\ }\href {\doibase 10.1103/RevModPhys.86.1391}
  {\bibfield  {journal} {\bibinfo  {journal} {Reviews of Modern Physics}\
  }\textbf {\bibinfo {volume} {86}},\ \bibinfo {pages} {1391--1452} (\bibinfo
  {year} {2014})}\BibitemShut {NoStop}%
\bibitem [{\citenamefont {Bloch}\ \emph {et~al.}(2012)\citenamefont {Bloch},
  \citenamefont {Dalibard},\ and\ \citenamefont
  {Nascimb{\`e}ne}}]{BlochDalibardNascimbeneNatPhys12}%
  \BibitemOpen
  \bibfield  {author} {\bibinfo {author} {\bibfnamefont {Immanuel}\
  \bibnamefont {Bloch}}, \bibinfo {author} {\bibfnamefont {Jean}\ \bibnamefont
  {Dalibard}}, \ and\ \bibinfo {author} {\bibfnamefont {Sylvain}\ \bibnamefont
  {Nascimb{\`e}ne}},\ }\bibfield  {title} {\enquote {\bibinfo {title} {Quantum
  simulations with ultracold quantum gases},}\ }\href
  {http://dx.doi.org/10.1038/nphys2259} {\bibfield  {journal} {\bibinfo
  {journal} {Nature Physics}\ }\textbf {\bibinfo {volume} {8}},\ \bibinfo
  {pages} {267 EP --} (\bibinfo {year} {2012})}\BibitemShut {NoStop}%
\bibitem [{\citenamefont {Browaeys}\ and\ \citenamefont
  {Lahaye}(2020)}]{BrowaeysNatPhys}%
  \BibitemOpen
  \bibfield  {author} {\bibinfo {author} {\bibfnamefont {Antoine}\ \bibnamefont
  {Browaeys}}\ and\ \bibinfo {author} {\bibfnamefont {Thierry}\ \bibnamefont
  {Lahaye}},\ }\bibfield  {title} {\enquote {\bibinfo {title} {Many-body
  physics with individually controlled rydberg atoms},}\ }\href {\doibase
  10.1038/s41567-019-0733-z} {\bibfield  {journal} {\bibinfo  {journal} {Nature
  Physics}\ }\textbf {\bibinfo {volume} {16}},\ \bibinfo {pages} {132--142}
  (\bibinfo {year} {2020})}\BibitemShut {NoStop}%
\bibitem [{\citenamefont {Blatt}\ and\ \citenamefont
  {Roos}(2012)}]{BlattRoosNatPhys12}%
  \BibitemOpen
  \bibfield  {author} {\bibinfo {author} {\bibfnamefont {R.}~\bibnamefont
  {Blatt}}\ and\ \bibinfo {author} {\bibfnamefont {C.~F.}\ \bibnamefont
  {Roos}},\ }\bibfield  {title} {\enquote {\bibinfo {title} {Quantum
  simulations with trapped ions},}\ }\href
  {http://dx.doi.org/10.1038/nphys2252} {\bibfield  {journal} {\bibinfo
  {journal} {Nature Physics}\ }\textbf {\bibinfo {volume} {8}},\ \bibinfo
  {pages} {277 EP --} (\bibinfo {year} {2012})}\BibitemShut {NoStop}%
\bibitem [{\citenamefont {Houck}\ \emph {et~al.}(2012)\citenamefont {Houck},
  \citenamefont {T{\"u}reci},\ and\ \citenamefont {Koch}}]{Houck2012}%
  \BibitemOpen
  \bibfield  {author} {\bibinfo {author} {\bibfnamefont {Andrew~A.}\
  \bibnamefont {Houck}}, \bibinfo {author} {\bibfnamefont {Hakan~E.}\
  \bibnamefont {T{\"u}reci}}, \ and\ \bibinfo {author} {\bibfnamefont {Jens}\
  \bibnamefont {Koch}},\ }\bibfield  {title} {\enquote {\bibinfo {title}
  {On-chip quantum simulation with superconducting circuits},}\ }\href
  {\doibase 10.1038/nphys2251} {\bibfield  {journal} {\bibinfo  {journal}
  {Nature Physics}\ }\textbf {\bibinfo {volume} {8}},\ \bibinfo {pages}
  {292--299} (\bibinfo {year} {2012})}\BibitemShut {NoStop}%
\bibitem [{\citenamefont {Salath{\'e}}\ \emph {et~al.}(2015)\citenamefont
  {Salath{\'e}}, \citenamefont {Mondal}, \citenamefont {Oppliger},
  \citenamefont {Heinsoo}, \citenamefont {Kurpiers}, \citenamefont {Poto{\v
  c}nik}, \citenamefont {Mezzacapo}, \citenamefont {Las~Heras}, \citenamefont
  {Lamata}, \citenamefont {Solano}, \citenamefont {Filipp},\ and\ \citenamefont
  {Wallraff}}]{Salathe2015}%
  \BibitemOpen
  \bibfield  {author} {\bibinfo {author} {\bibfnamefont {Y.}~\bibnamefont
  {Salath{\'e}}}, \bibinfo {author} {\bibfnamefont {M.}~\bibnamefont {Mondal}},
  \bibinfo {author} {\bibfnamefont {M.}~\bibnamefont {Oppliger}}, \bibinfo
  {author} {\bibfnamefont {J.}~\bibnamefont {Heinsoo}}, \bibinfo {author}
  {\bibfnamefont {P.}~\bibnamefont {Kurpiers}}, \bibinfo {author}
  {\bibfnamefont {A.}~\bibnamefont {Poto{\v c}nik}}, \bibinfo {author}
  {\bibfnamefont {A.}~\bibnamefont {Mezzacapo}}, \bibinfo {author}
  {\bibfnamefont {U.}~\bibnamefont {Las~Heras}}, \bibinfo {author}
  {\bibfnamefont {L.}~\bibnamefont {Lamata}}, \bibinfo {author} {\bibfnamefont
  {E.}~\bibnamefont {Solano}}, \bibinfo {author} {\bibfnamefont
  {S.}~\bibnamefont {Filipp}}, \ and\ \bibinfo {author} {\bibfnamefont
  {A.}~\bibnamefont {Wallraff}},\ }\bibfield  {title} {\enquote {\bibinfo
  {title} {Digital quantum simulation of spin models with circuit quantum
  electrodynamics},}\ }\href {\doibase 10.1103/PhysRevX.5.021027} {\bibfield
  {journal} {\bibinfo  {journal} {Physical Review X}\ }\textbf {\bibinfo
  {volume} {5}},\ \bibinfo {pages} {21027} (\bibinfo {year}
  {2015})}\BibitemShut {NoStop}%
\bibitem [{\citenamefont {Puertas~Mart{\'i}nez}\ \emph
  {et~al.}(2019)\citenamefont {Puertas~Mart{\'i}nez}, \citenamefont
  {L{\'e}ger}, \citenamefont {Gheeraert}, \citenamefont {Dassonneville},
  \citenamefont {Planat}, \citenamefont {Foroughi}, \citenamefont {Krupko},
  \citenamefont {Buisson}, \citenamefont {Naud}, \citenamefont
  {{Hasch-Guichard}}, \citenamefont {Florens}, \citenamefont {Snyman},\ and\
  \citenamefont {Roch}}]{puertasmartinez2019}%
  \BibitemOpen
  \bibfield  {author} {\bibinfo {author} {\bibfnamefont {Javier}\ \bibnamefont
  {Puertas~Mart{\'i}nez}}, \bibinfo {author} {\bibfnamefont {S{\'e}bastien}\
  \bibnamefont {L{\'e}ger}}, \bibinfo {author} {\bibfnamefont {Nicolas}\
  \bibnamefont {Gheeraert}}, \bibinfo {author} {\bibfnamefont {R{\'e}my}\
  \bibnamefont {Dassonneville}}, \bibinfo {author} {\bibfnamefont {Luca}\
  \bibnamefont {Planat}}, \bibinfo {author} {\bibfnamefont {Farshad}\
  \bibnamefont {Foroughi}}, \bibinfo {author} {\bibfnamefont {Yuriy}\
  \bibnamefont {Krupko}}, \bibinfo {author} {\bibfnamefont {Olivier}\
  \bibnamefont {Buisson}}, \bibinfo {author} {\bibfnamefont {C{\'e}cile}\
  \bibnamefont {Naud}}, \bibinfo {author} {\bibfnamefont {Wiebke}\ \bibnamefont
  {{Hasch-Guichard}}}, \bibinfo {author} {\bibfnamefont {Serge}\ \bibnamefont
  {Florens}}, \bibinfo {author} {\bibfnamefont {Izak}\ \bibnamefont {Snyman}},
  \ and\ \bibinfo {author} {\bibfnamefont {Nicolas}\ \bibnamefont {Roch}},\
  }\bibfield  {title} {\enquote {\bibinfo {title} {A tunable {{Josephson}}
  platform to explore many-body quantum optics in circuit-{{QED}}},}\ }\href
  {\doibase 10.1038/s41534-018-0104-0} {\bibfield  {journal} {\bibinfo
  {journal} {npj Quantum Information}\ }\textbf {\bibinfo {volume} {5}},\
  \bibinfo {pages} {19} (\bibinfo {year} {2019})}\BibitemShut {NoStop}%
\bibitem [{\citenamefont {Carusotto}\ \emph {et~al.}(2020)\citenamefont
  {Carusotto}, \citenamefont {Houck}, \citenamefont {Koll{\'a}r}, \citenamefont
  {Roushan}, \citenamefont {Schuster},\ and\ \citenamefont
  {Simon}}]{CarusottoEtAlNatPhys2020}%
  \BibitemOpen
  \bibfield  {author} {\bibinfo {author} {\bibfnamefont {Iacopo}\ \bibnamefont
  {Carusotto}}, \bibinfo {author} {\bibfnamefont {Andrew~A.}\ \bibnamefont
  {Houck}}, \bibinfo {author} {\bibfnamefont {Alicia~J.}\ \bibnamefont
  {Koll{\'a}r}}, \bibinfo {author} {\bibfnamefont {Pedram}\ \bibnamefont
  {Roushan}}, \bibinfo {author} {\bibfnamefont {David~I.}\ \bibnamefont
  {Schuster}}, \ and\ \bibinfo {author} {\bibfnamefont {Jonathan}\ \bibnamefont
  {Simon}},\ }\bibfield  {title} {\enquote {\bibinfo {title} {Photonic
  materials in circuit quantum electrodynamics},}\ }\href {\doibase
  10.1038/s41567-020-0815-y} {\bibfield  {journal} {\bibinfo  {journal} {Nature
  Physics}\ }\textbf {\bibinfo {volume} {16}},\ \bibinfo {pages} {268--279}
  (\bibinfo {year} {2020})}\BibitemShut {NoStop}%
\bibitem [{\citenamefont {Fink}\ \emph {et~al.}(2017)\citenamefont {Fink},
  \citenamefont {Dombi}, \citenamefont {Vukics}, \citenamefont {Wallraff},\
  and\ \citenamefont {Domokos}}]{FinkEtAlPRX17}%
  \BibitemOpen
  \bibfield  {author} {\bibinfo {author} {\bibfnamefont {J.~M.}\ \bibnamefont
  {Fink}}, \bibinfo {author} {\bibfnamefont {A.}~\bibnamefont {Dombi}},
  \bibinfo {author} {\bibfnamefont {A.}~\bibnamefont {Vukics}}, \bibinfo
  {author} {\bibfnamefont {A.}~\bibnamefont {Wallraff}}, \ and\ \bibinfo
  {author} {\bibfnamefont {P.}~\bibnamefont {Domokos}},\ }\bibfield  {title}
  {\enquote {\bibinfo {title} {Observation of the {{Photon}}-{{Blockade
  Breakdown Phase Transition}}},}\ }\href {\doibase 10.1103/PhysRevX.7.011012}
  {\bibfield  {journal} {\bibinfo  {journal} {Phys. Rev. X}\ }\textbf {\bibinfo
  {volume} {7}},\ \bibinfo {pages} {11012} (\bibinfo {year}
  {2017})}\BibitemShut {NoStop}%
\bibitem [{\citenamefont {Fitzpatrick}\ \emph {et~al.}(2017)\citenamefont
  {Fitzpatrick}, \citenamefont {Sundaresan}, \citenamefont {Li}, \citenamefont
  {Koch},\ and\ \citenamefont {Houck}}]{FitzpatrickEtAlPRX17}%
  \BibitemOpen
  \bibfield  {author} {\bibinfo {author} {\bibfnamefont {Mattias}\ \bibnamefont
  {Fitzpatrick}}, \bibinfo {author} {\bibfnamefont {Neereja~M.}\ \bibnamefont
  {Sundaresan}}, \bibinfo {author} {\bibfnamefont {Andy C.Y.~Y}\ \bibnamefont
  {Li}}, \bibinfo {author} {\bibfnamefont {Jens}\ \bibnamefont {Koch}}, \ and\
  \bibinfo {author} {\bibfnamefont {Andrew~A.}\ \bibnamefont {Houck}},\
  }\bibfield  {title} {\enquote {\bibinfo {title} {Observation of a
  {{Dissipative Phase Transition}} in a {{One}}-{{Dimensional Circuit QED
  Lattice}}},}\ }\href {\doibase 10.1103/PhysRevX.7.011016} {\bibfield
  {journal} {\bibinfo  {journal} {Phys. Rev. X}\ }\textbf {\bibinfo {volume}
  {7}},\ \bibinfo {pages} {11016} (\bibinfo {year} {2017})}\BibitemShut
  {NoStop}%
\bibitem [{\citenamefont {Fink}\ \emph {et~al.}(2018)\citenamefont {Fink},
  \citenamefont {Schade}, \citenamefont {H{\"o}fling}, \citenamefont
  {Schneider},\ and\ \citenamefont {Imamoglu}}]{Fink2018}%
  \BibitemOpen
  \bibfield  {author} {\bibinfo {author} {\bibfnamefont {Thomas}\ \bibnamefont
  {Fink}}, \bibinfo {author} {\bibfnamefont {Anne}\ \bibnamefont {Schade}},
  \bibinfo {author} {\bibfnamefont {Sven}\ \bibnamefont {H{\"o}fling}},
  \bibinfo {author} {\bibfnamefont {Christian}\ \bibnamefont {Schneider}}, \
  and\ \bibinfo {author} {\bibfnamefont {Ata{\c c}}\ \bibnamefont {Imamoglu}},\
  }\bibfield  {title} {\enquote {\bibinfo {title} {Signatures of a dissipative
  phase transition in photon correlation measurements},}\ }\href {\doibase
  10.1038/s41567-017-0020-9} {\bibfield  {journal} {\bibinfo  {journal} {Nature
  Physics}\ }\textbf {\bibinfo {volume} {14}},\ \bibinfo {pages} {365--369}
  (\bibinfo {year} {2018})}\BibitemShut {NoStop}%
\bibitem [{\citenamefont {Raftery}\ \emph {et~al.}(2014)\citenamefont
  {Raftery}, \citenamefont {Sadri}, \citenamefont {Schmidt}, \citenamefont
  {T{\"u}reci}, \citenamefont {Houck}, \citenamefont {Tureci},\ and\
  \citenamefont {Houck}}]{Raftery2014}%
  \BibitemOpen
  \bibfield  {author} {\bibinfo {author} {\bibfnamefont {J.}~\bibnamefont
  {Raftery}}, \bibinfo {author} {\bibfnamefont {D.}~\bibnamefont {Sadri}},
  \bibinfo {author} {\bibfnamefont {S.}~\bibnamefont {Schmidt}}, \bibinfo
  {author} {\bibfnamefont {H~E}\ \bibnamefont {T{\"u}reci}}, \bibinfo {author}
  {\bibfnamefont {A.~A.}\ \bibnamefont {Houck}}, \bibinfo {author}
  {\bibfnamefont {H.~E.}\ \bibnamefont {Tureci}}, \ and\ \bibinfo {author}
  {\bibfnamefont {A.~A.}\ \bibnamefont {Houck}},\ }\bibfield  {title} {\enquote
  {\bibinfo {title} {Observation of a dissipation-induced classical to quantum
  transition},}\ }\href {\doibase 10.1103/PhysRevX.4.031043} {\bibfield
  {journal} {\bibinfo  {journal} {Physical Review X}\ }\textbf {\bibinfo
  {volume} {4}},\ \bibinfo {pages} {031043} (\bibinfo {year}
  {2014})}\BibitemShut {NoStop}%
\bibitem [{\citenamefont {Ma}\ \emph {et~al.}(2019)\citenamefont {Ma},
  \citenamefont {Saxberg}, \citenamefont {Owens}, \citenamefont {Leung},
  \citenamefont {Lu}, \citenamefont {Simon},\ and\ \citenamefont
  {Schuster}}]{ma2019}%
  \BibitemOpen
  \bibfield  {author} {\bibinfo {author} {\bibfnamefont {Ruichao}\ \bibnamefont
  {Ma}}, \bibinfo {author} {\bibfnamefont {Brendan}\ \bibnamefont {Saxberg}},
  \bibinfo {author} {\bibfnamefont {Clai}\ \bibnamefont {Owens}}, \bibinfo
  {author} {\bibfnamefont {Nelson}\ \bibnamefont {Leung}}, \bibinfo {author}
  {\bibfnamefont {Yao}\ \bibnamefont {Lu}}, \bibinfo {author} {\bibfnamefont
  {Jonathan}\ \bibnamefont {Simon}}, \ and\ \bibinfo {author} {\bibfnamefont
  {David~I.}\ \bibnamefont {Schuster}},\ }\bibfield  {title} {\enquote
  {\bibinfo {title} {A dissipatively stabilized {{Mott}} insulator of
  photons},}\ }\href {\doibase 10.1038/s41586-019-0897-9} {\bibfield  {journal}
  {\bibinfo  {journal} {Nature}\ }\textbf {\bibinfo {volume} {566}},\ \bibinfo
  {pages} {51--57} (\bibinfo {year} {2019})}\BibitemShut {NoStop}%
\bibitem [{\citenamefont {Syassen}\ \emph {et~al.}(2008)\citenamefont
  {Syassen}, \citenamefont {Bauer}, \citenamefont {Lettner}, \citenamefont
  {Volz}, \citenamefont {Dietze}, \citenamefont {Garc{\'\i}a-Ripoll},
  \citenamefont {Cirac}, \citenamefont {Rempe},\ and\ \citenamefont
  {D{\"u}rr}}]{Syassen1329}%
  \BibitemOpen
  \bibfield  {author} {\bibinfo {author} {\bibfnamefont {N.}~\bibnamefont
  {Syassen}}, \bibinfo {author} {\bibfnamefont {D.~M.}\ \bibnamefont {Bauer}},
  \bibinfo {author} {\bibfnamefont {M.}~\bibnamefont {Lettner}}, \bibinfo
  {author} {\bibfnamefont {T.}~\bibnamefont {Volz}}, \bibinfo {author}
  {\bibfnamefont {D.}~\bibnamefont {Dietze}}, \bibinfo {author} {\bibfnamefont
  {J.~J.}\ \bibnamefont {Garc{\'\i}a-Ripoll}}, \bibinfo {author} {\bibfnamefont
  {J.~I.}\ \bibnamefont {Cirac}}, \bibinfo {author} {\bibfnamefont
  {G.}~\bibnamefont {Rempe}}, \ and\ \bibinfo {author} {\bibfnamefont
  {S.}~\bibnamefont {D{\"u}rr}},\ }\bibfield  {title} {\enquote {\bibinfo
  {title} {Strong dissipation inhibits losses and induces correlations in cold
  molecular gases},}\ }\href {\doibase 10.1126/science.1155309} {\bibfield
  {journal} {\bibinfo  {journal} {Science}\ }\textbf {\bibinfo {volume}
  {320}},\ \bibinfo {pages} {1329--1331} (\bibinfo {year} {2008})}\BibitemShut {NoStop}%
\bibitem [{\citenamefont {Tomita}\ \emph {et~al.}(2017)\citenamefont {Tomita},
  \citenamefont {Nakajima}, \citenamefont {Danshita}, \citenamefont {Takasu},\
  and\ \citenamefont {Takahashi}}]{TomitaEtAlScienceAdv17}%
  \BibitemOpen
  \bibfield  {author} {\bibinfo {author} {\bibfnamefont {Takafumi}\
  \bibnamefont {Tomita}}, \bibinfo {author} {\bibfnamefont {Shuta}\
  \bibnamefont {Nakajima}}, \bibinfo {author} {\bibfnamefont {Ippei}\
  \bibnamefont {Danshita}}, \bibinfo {author} {\bibfnamefont {Yosuke}\
  \bibnamefont {Takasu}}, \ and\ \bibinfo {author} {\bibfnamefont {Yoshiro}\
  \bibnamefont {Takahashi}},\ }\bibfield  {title} {\enquote {\bibinfo {title}
  {Observation of the mott insulator to superfluid crossover of a
  driven-dissipative bose-hubbard system},}\ }\href {\doibase 10.1126/sciadv.1701513} {\bibfield  {journal} {\bibinfo  {journal} {Science
  Advances}\ }\textbf {\bibinfo {volume} {3}} (\bibinfo {year} {2017}),\
  10.1126/sciadv.1701513}\BibitemShut {NoStop}%
\bibitem [{\citenamefont {L\"uschen}\ \emph {et~al.}(2017)\citenamefont
  {L\"uschen}, \citenamefont {Bordia}, \citenamefont {Hodgman}, \citenamefont
  {Schreiber}, \citenamefont {Sarkar}, \citenamefont {Daley}, \citenamefont
  {Fischer}, \citenamefont {Altman}, \citenamefont {Bloch},\ and\ \citenamefont
  {Schneider}}]{LuschenEtAlPRX17}%
  \BibitemOpen
  \bibfield  {author} {\bibinfo {author} {\bibfnamefont {Henrik~P.}\
  \bibnamefont {L\"uschen}}, \bibinfo {author} {\bibfnamefont {Pranjal}\
  \bibnamefont {Bordia}}, \bibinfo {author} {\bibfnamefont {Sean~S.}\
  \bibnamefont {Hodgman}}, \bibinfo {author} {\bibfnamefont {Michael}\
  \bibnamefont {Schreiber}}, \bibinfo {author} {\bibfnamefont {Saubhik}\
  \bibnamefont {Sarkar}}, \bibinfo {author} {\bibfnamefont {Andrew~J.}\
  \bibnamefont {Daley}}, \bibinfo {author} {\bibfnamefont {Mark~H.}\
  \bibnamefont {Fischer}}, \bibinfo {author} {\bibfnamefont {Ehud}\
  \bibnamefont {Altman}}, \bibinfo {author} {\bibfnamefont {Immanuel}\
  \bibnamefont {Bloch}}, \ and\ \bibinfo {author} {\bibfnamefont {Ulrich}\
  \bibnamefont {Schneider}},\ }\bibfield  {title} {\enquote {\bibinfo {title}
  {Signatures of many-body localization in a controlled open quantum system},}\
  }\href {\doibase 10.1103/PhysRevX.7.011034} {\bibfield  {journal} {\bibinfo
  {journal} {Phys. Rev. X}\ }\textbf {\bibinfo {volume} {7}},\ \bibinfo {pages}
  {011034} (\bibinfo {year} {2017})}\BibitemShut {NoStop}%
\bibitem [{\citenamefont {Bouganne}\ \emph {et~al.}(2020)\citenamefont
  {Bouganne}, \citenamefont {Bosch~Aguilera}, \citenamefont {Ghermaoui},
  \citenamefont {Beugnon},\ and\ \citenamefont {Gerbier}}]{bouganne2020}%
  \BibitemOpen
  \bibfield  {author} {\bibinfo {author} {\bibfnamefont {Rapha{\"e}l}\
  \bibnamefont {Bouganne}}, \bibinfo {author} {\bibfnamefont {Manel}\
  \bibnamefont {Bosch~Aguilera}}, \bibinfo {author} {\bibfnamefont {Alexis}\
  \bibnamefont {Ghermaoui}}, \bibinfo {author} {\bibfnamefont {J{\'e}r{\^o}me}\
  \bibnamefont {Beugnon}}, \ and\ \bibinfo {author} {\bibfnamefont {Fabrice}\
  \bibnamefont {Gerbier}},\ }\bibfield  {title} {\enquote {\bibinfo {title}
  {Anomalous decay of coherence in a dissipative many-body system},}\ }\href
  {\doibase 10.1038/s41567-019-0678-2} {\bibfield  {journal} {\bibinfo
  {journal} {Nature Physics}\ }\textbf {\bibinfo {volume} {16}},\ \bibinfo
  {pages} {21--25} (\bibinfo {year} {2020})}\BibitemShut {NoStop}%
\bibitem [{\citenamefont {Carusotto}\ and\ \citenamefont
  {Ciuti}(2013)}]{CarusottoCiutiRMP13}%
  \BibitemOpen
  \bibfield  {author} {\bibinfo {author} {\bibfnamefont {Iacopo}\ \bibnamefont
  {Carusotto}}\ and\ \bibinfo {author} {\bibfnamefont {Cristiano}\ \bibnamefont
  {Ciuti}},\ }\bibfield  {title} {\enquote {\bibinfo {title} {Quantum fluids of
  light},}\ }\href {\doibase 10.1103/RevModPhys.85.299} {\bibfield  {journal}
  {\bibinfo  {journal} {Rev. Mod. Phys.}\ }\textbf {\bibinfo {volume} {85}},\
  \bibinfo {pages} {299--366} (\bibinfo {year} {2013})}\BibitemShut {NoStop}%
\bibitem [{\citenamefont {Ritsch}\ \emph {et~al.}(2013)\citenamefont {Ritsch},
  \citenamefont {Domokos}, \citenamefont {Brennecke},\ and\ \citenamefont
  {Esslinger}}]{ritsch2013}%
  \BibitemOpen
  \bibfield  {author} {\bibinfo {author} {\bibfnamefont {Helmut}\ \bibnamefont
  {Ritsch}}, \bibinfo {author} {\bibfnamefont {Peter}\ \bibnamefont {Domokos}},
  \bibinfo {author} {\bibfnamefont {Ferdinand}\ \bibnamefont {Brennecke}}, \
  and\ \bibinfo {author} {\bibfnamefont {Tilman}\ \bibnamefont {Esslinger}},\
  }\bibfield  {title} {\enquote {\bibinfo {title} {Cold atoms in
  cavity-generated dynamical optical potentials},}\ }\href {\doibase
  10.1103/RevModPhys.85.553} {\bibfield  {journal} {\bibinfo  {journal}
  {Reviews of Modern Physics}\ }\textbf {\bibinfo {volume} {85}},\ \bibinfo
  {pages} {553--601} (\bibinfo {year} {2013})}\BibitemShut {NoStop}%
\bibitem [{\citenamefont {Schmidt}\ and\ \citenamefont
  {Koch}(2013)}]{SchmidtKochAnnPhy13}%
  \BibitemOpen
  \bibfield  {author} {\bibinfo {author} {\bibfnamefont {Sebastian}\
  \bibnamefont {Schmidt}}\ and\ \bibinfo {author} {\bibfnamefont {Jens}\
  \bibnamefont {Koch}},\ }\bibfield  {title} {\enquote {\bibinfo {title}
  {Circuit {{QED}} lattices: {{Towards}} quantum simulation with
  superconducting circuits},}\ }\href {\doibase 10.1002/andp.201200261}
  {\bibfield  {journal} {\bibinfo  {journal} {Annalen der Physik}\ }\textbf
  {\bibinfo {volume} {525}},\ \bibinfo {pages} {395--412} (\bibinfo {year}
  {2013})}\BibitemShut {NoStop}%
\bibitem [{\citenamefont {Hur}\ \emph {et~al.}(2016)\citenamefont {Hur},
  \citenamefont {Henriet}, \citenamefont {Petrescu}, \citenamefont {Plekhanov},
  \citenamefont {Roux},\ and\ \citenamefont {Schir\'o}}]{LeHurReview16}%
  \BibitemOpen
  \bibfield  {author} {\bibinfo {author} {\bibfnamefont {Karyn~Le}\
  \bibnamefont {Hur}}, \bibinfo {author} {\bibfnamefont {Loïc}\ \bibnamefont
  {Henriet}}, \bibinfo {author} {\bibfnamefont {Alexandru}\ \bibnamefont
  {Petrescu}}, \bibinfo {author} {\bibfnamefont {Kirill}\ \bibnamefont
  {Plekhanov}}, \bibinfo {author} {\bibfnamefont {Guillaume}\ \bibnamefont
  {Roux}}, \ and\ \bibinfo {author} {\bibfnamefont {Marco}\ \bibnamefont
  {Schir\'o}},\ }\bibfield  {title} {\enquote {\bibinfo {title} {Many-body
  quantum electrodynamics networks: Non-equilibrium condensed matter physics
  with light},}\ }\href@noop {} {\bibfield  {journal} {\bibinfo  {journal}
  {Comptes Rendus Physique}\ }\textbf {\bibinfo {volume} {17}},\ \bibinfo
  {pages} {808 -- 835} (\bibinfo {year} {2016})}\BibitemShut {NoStop}%
\bibitem [{\citenamefont {Noh}\ and\ \citenamefont
  {Angelakis}(2017)}]{noh2017}%
  \BibitemOpen
  \bibfield  {author} {\bibinfo {author} {\bibfnamefont {Changsuk}\
  \bibnamefont {Noh}}\ and\ \bibinfo {author} {\bibfnamefont {Dimitris~G.}\
  \bibnamefont {Angelakis}},\ }\bibfield  {title} {\enquote {\bibinfo {title}
  {Quantum simulations and many-body physics with light},}\ }\href {\doibase
  10.1088/0034-4885/80/1/016401} {\bibfield  {journal} {\bibinfo  {journal}
  {Reports on Progress in Physics}\ }\textbf {\bibinfo {volume} {80}},\
  \bibinfo {pages} {16401} (\bibinfo {year} {2017})}\BibitemShut {NoStop}%
\bibitem [{\citenamefont {Hartmann}(2016)}]{Hartmann2016}%
  \BibitemOpen
  \bibfield  {author} {\bibinfo {author} {\bibfnamefont {Michael~J.}\
  \bibnamefont {Hartmann}},\ }\bibfield  {title} {\enquote {\bibinfo {title}
  {Quantum simulation with interacting photons},}\ }\href {\doibase
  10.1088/2040-8978/18/10/104005} {\bibfield  {journal} {\bibinfo  {journal}
  {Journal of Optics}\ }\textbf {\bibinfo {volume} {18}},\ \bibinfo {pages}
  {104005} (\bibinfo {year} {2016})}\BibitemShut {NoStop}%
\bibitem [{\citenamefont {Breuer}\ and\ \citenamefont
  {Petruccione}(2007)}]{breuerPetruccione2010}%
  \BibitemOpen
  \bibfield  {author} {\bibinfo {author} {\bibfnamefont {Heinz~Peter}\
  \bibnamefont {Breuer}}\ and\ \bibinfo {author} {\bibfnamefont {Francesco}\
  \bibnamefont {Petruccione}},\ }\href {\doibase
  10.1093/acprof:oso/9780199213900.001.0001} {\emph {\bibinfo {title} {The
  {{Theory}} of {{Open Quantum Systems}}}}},\ \bibinfo {edition} {1st}\ ed.,\
  Vol.\ \bibinfo {volume} {9780199213}\ (\bibinfo  {publisher} {{OUP Oxford}},\
  \bibinfo {year} {2007})\BibitemShut {NoStop}%
\bibitem [{\citenamefont {Diehl}\ \emph {et~al.}(2008)\citenamefont {Diehl},
  \citenamefont {Micheli}, \citenamefont {Kantian}, \citenamefont {Kraus},
  \citenamefont {B{\"u}chler},\ and\ \citenamefont
  {Zoller}}]{DiehlEtalNatPhys08}%
  \BibitemOpen
  \bibfield  {author} {\bibinfo {author} {\bibfnamefont {S.}~\bibnamefont
  {Diehl}}, \bibinfo {author} {\bibfnamefont {A.}~\bibnamefont {Micheli}},
  \bibinfo {author} {\bibfnamefont {A.}~\bibnamefont {Kantian}}, \bibinfo
  {author} {\bibfnamefont {B.}~\bibnamefont {Kraus}}, \bibinfo {author}
  {\bibfnamefont {H.~P.}\ \bibnamefont {B{\"u}chler}}, \ and\ \bibinfo {author}
  {\bibfnamefont {P.}~\bibnamefont {Zoller}},\ }\bibfield  {title} {\enquote
  {\bibinfo {title} {Quantum states and phases in driven open quantum systems
  with cold atoms},}\ }\href {\doibase 10.1038/nphys1073} {\bibfield  {journal}
  {\bibinfo  {journal} {Nature Physics}\ }\textbf {\bibinfo {volume} {4}},\
  \bibinfo {pages} {878--883} (\bibinfo {year} {2008})}\BibitemShut {NoStop}%
\bibitem [{\citenamefont {Verstraete}\ \emph {et~al.}(2009)\citenamefont
  {Verstraete}, \citenamefont {Wolf},\ and\ \citenamefont
  {Ignacio~Cirac}}]{Verstraete2009}%
  \BibitemOpen
  \bibfield  {author} {\bibinfo {author} {\bibfnamefont {Frank}\ \bibnamefont
  {Verstraete}}, \bibinfo {author} {\bibfnamefont {Michael~M.}\ \bibnamefont
  {Wolf}}, \ and\ \bibinfo {author} {\bibfnamefont {J.}~\bibnamefont
  {Ignacio~Cirac}},\ }\bibfield  {title} {\enquote {\bibinfo {title} {Quantum
  computation and quantum-state engineering driven by dissipation},}\ }\href
  {\doibase 10.1038/nphys1342} {\bibfield  {journal} {\bibinfo  {journal}
  {Nature Physics}\ }\textbf {\bibinfo {volume} {5}},\ \bibinfo {pages}
  {633--636} (\bibinfo {year} {2009})}\BibitemShut {NoStop}%
\bibitem [{\citenamefont {Murch}\ \emph {et~al.}(2012)\citenamefont {Murch},
  \citenamefont {Vool}, \citenamefont {Zhou}, \citenamefont {Weber},
  \citenamefont {Girvin},\ and\ \citenamefont
  {Siddiqi}}]{Siddiqi_quantum_bath_engineering}%
  \BibitemOpen
  \bibfield  {author} {\bibinfo {author} {\bibfnamefont {K.~W.}\ \bibnamefont
  {Murch}}, \bibinfo {author} {\bibfnamefont {U.}~\bibnamefont {Vool}},
  \bibinfo {author} {\bibfnamefont {D.}~\bibnamefont {Zhou}}, \bibinfo {author}
  {\bibfnamefont {S.~J.}\ \bibnamefont {Weber}}, \bibinfo {author}
  {\bibfnamefont {S.~M.}\ \bibnamefont {Girvin}}, \ and\ \bibinfo {author}
  {\bibfnamefont {I.}~\bibnamefont {Siddiqi}},\ }\bibfield  {title} {\enquote
  {\bibinfo {title} {Cavity-assisted quantum bath engineering},}\ }\href
  {\doibase 10.1103/PhysRevLett.109.183602} {\bibfield  {journal} {\bibinfo
  {journal} {Physical Review Letters}\ }\textbf {\bibinfo {volume} {109}},\
  \bibinfo {pages} {183602} (\bibinfo {year} {2012})}\BibitemShut {NoStop}%
\bibitem [{\citenamefont {Leghtas}\ \emph {et~al.}(2015)\citenamefont
  {Leghtas}, \citenamefont {Touzard}, \citenamefont {Pop}, \citenamefont {Kou},
  \citenamefont {Vlastakis}, \citenamefont {Petrenko}, \citenamefont {Sliwa},
  \citenamefont {Narla}, \citenamefont {Shankar}, \citenamefont {Hatridge},
  \citenamefont {Reagor}, \citenamefont {Frunzio}, \citenamefont {Schoelkopf},
  \citenamefont {Mirrahimi},\ and\ \citenamefont {Devoret}}]{Leghtas2015}%
  \BibitemOpen
  \bibfield  {author} {\bibinfo {author} {\bibfnamefont {Z.}~\bibnamefont
  {Leghtas}}, \bibinfo {author} {\bibfnamefont {S.}~\bibnamefont {Touzard}},
  \bibinfo {author} {\bibfnamefont {I.~M.}\ \bibnamefont {Pop}}, \bibinfo
  {author} {\bibfnamefont {A.}~\bibnamefont {Kou}}, \bibinfo {author}
  {\bibfnamefont {B.}~\bibnamefont {Vlastakis}}, \bibinfo {author}
  {\bibfnamefont {A.}~\bibnamefont {Petrenko}}, \bibinfo {author}
  {\bibfnamefont {K.~M.}\ \bibnamefont {Sliwa}}, \bibinfo {author}
  {\bibfnamefont {A.}~\bibnamefont {Narla}}, \bibinfo {author} {\bibfnamefont
  {S.}~\bibnamefont {Shankar}}, \bibinfo {author} {\bibfnamefont {M.~J.}\
  \bibnamefont {Hatridge}}, \bibinfo {author} {\bibfnamefont {M.}~\bibnamefont
  {Reagor}}, \bibinfo {author} {\bibfnamefont {L.}~\bibnamefont {Frunzio}},
  \bibinfo {author} {\bibfnamefont {R.~J.}\ \bibnamefont {Schoelkopf}},
  \bibinfo {author} {\bibfnamefont {M.}~\bibnamefont {Mirrahimi}}, \ and\
  \bibinfo {author} {\bibfnamefont {M.~H.}\ \bibnamefont {Devoret}},\
  }\bibfield  {title} {\enquote {\bibinfo {title} {Confining the state of light
  to a quantum manifold by engineered two-photon loss},}\ }\href {\doibase
  10.1126/science.aaa2085} {\bibfield  {journal} {\bibinfo  {journal}
  {Science}\ }\textbf {\bibinfo {volume} {347}},\ \bibinfo {pages} {853--857}
  (\bibinfo {year} {2015})}\BibitemShut {NoStop}%
\bibitem [{\citenamefont {Diehl}\ \emph {et~al.}(2010)\citenamefont {Diehl},
  \citenamefont {Tomadin}, \citenamefont {Micheli}, \citenamefont {Fazio},\
  and\ \citenamefont {Zoller}}]{Tomadin_prl10}%
  \BibitemOpen
  \bibfield  {author} {\bibinfo {author} {\bibfnamefont {Sebastian}\
  \bibnamefont {Diehl}}, \bibinfo {author} {\bibfnamefont {Andrea}\
  \bibnamefont {Tomadin}}, \bibinfo {author} {\bibfnamefont {Andrea}\
  \bibnamefont {Micheli}}, \bibinfo {author} {\bibfnamefont {Rosario}\
  \bibnamefont {Fazio}}, \ and\ \bibinfo {author} {\bibfnamefont {Peter}\
  \bibnamefont {Zoller}},\ }\bibfield  {title} {\enquote {\bibinfo {title}
  {Dynamical phase transitions and instabilities in open atomic many-body
  systems},}\ }\href {\doibase 10.1103/PhysRevLett.105.015702} {\bibfield
  {journal} {\bibinfo  {journal} {Phys. Rev. Lett.}\ }\textbf {\bibinfo
  {volume} {105}},\ \bibinfo {pages} {015702} (\bibinfo {year}
  {2010})}\BibitemShut {NoStop}%
\bibitem [{\citenamefont {Poletti}\ \emph {et~al.}(2012)\citenamefont
  {Poletti}, \citenamefont {Bernier}, \citenamefont {Georges},\ and\
  \citenamefont {Kollath}}]{PolettiEtAlPRL12}%
  \BibitemOpen
  \bibfield  {author} {\bibinfo {author} {\bibfnamefont {Dario}\ \bibnamefont
  {Poletti}}, \bibinfo {author} {\bibfnamefont {Jean-S\'ebastien}\ \bibnamefont
  {Bernier}}, \bibinfo {author} {\bibfnamefont {Antoine}\ \bibnamefont
  {Georges}}, \ and\ \bibinfo {author} {\bibfnamefont {Corinna}\ \bibnamefont
  {Kollath}},\ }\bibfield  {title} {\enquote {\bibinfo {title}
  {Interaction-induced impeding of decoherence and anomalous diffusion},}\
  }\href {\doibase 10.1103/PhysRevLett.109.045302} {\bibfield  {journal}
  {\bibinfo  {journal} {Phys. Rev. Lett.}\ }\textbf {\bibinfo {volume} {109}},\
  \bibinfo {pages} {045302} (\bibinfo {year} {2012})}\BibitemShut {NoStop}%
\bibitem [{\citenamefont {Poletti}\ \emph {et~al.}(2013)\citenamefont
  {Poletti}, \citenamefont {Barmettler}, \citenamefont {Georges},\ and\
  \citenamefont {Kollath}}]{PolettiEtAlPRL13}%
  \BibitemOpen
  \bibfield  {author} {\bibinfo {author} {\bibfnamefont {Dario}\ \bibnamefont
  {Poletti}}, \bibinfo {author} {\bibfnamefont {Peter}\ \bibnamefont
  {Barmettler}}, \bibinfo {author} {\bibfnamefont {Antoine}\ \bibnamefont
  {Georges}}, \ and\ \bibinfo {author} {\bibfnamefont {Corinna}\ \bibnamefont
  {Kollath}},\ }\bibfield  {title} {\enquote {\bibinfo {title} {Emergence of
  glasslike dynamics for dissipative and strongly interacting bosons},}\ }\href
  {\doibase 10.1103/PhysRevLett.111.195301} {\bibfield  {journal} {\bibinfo
  {journal} {Phys. Rev. Lett.}\ }\textbf {\bibinfo {volume} {111}},\ \bibinfo
  {pages} {195301} (\bibinfo {year} {2013})}\BibitemShut {NoStop}%
\bibitem [{\citenamefont {Ludwig}\ and\ \citenamefont
  {Marquardt}(2013)}]{LudwigMarquardtPRL13}%
  \BibitemOpen
  \bibfield  {author} {\bibinfo {author} {\bibfnamefont {Max}\ \bibnamefont
  {Ludwig}}\ and\ \bibinfo {author} {\bibfnamefont {Florian}\ \bibnamefont
  {Marquardt}},\ }\bibfield  {title} {\enquote {\bibinfo {title} {Quantum
  many-body dynamics in optomechanical arrays},}\ }\href {\doibase
  10.1103/PhysRevLett.111.073603} {\bibfield  {journal} {\bibinfo  {journal}
  {Phys. Rev. Lett.}\ }\textbf {\bibinfo {volume} {111}},\ \bibinfo {pages}
  {073603} (\bibinfo {year} {2013})}\BibitemShut {NoStop}%
\bibitem [{\citenamefont {Sieberer}\ \emph {et~al.}(2013)\citenamefont
  {Sieberer}, \citenamefont {Huber}, \citenamefont {Altman},\ and\
  \citenamefont {Diehl}}]{SiebereHuberAltmanDiehlPRL13}%
  \BibitemOpen
  \bibfield  {author} {\bibinfo {author} {\bibfnamefont {L.~M.}\ \bibnamefont
  {Sieberer}}, \bibinfo {author} {\bibfnamefont {S.~D.}\ \bibnamefont {Huber}},
  \bibinfo {author} {\bibfnamefont {E.}~\bibnamefont {Altman}}, \ and\ \bibinfo
  {author} {\bibfnamefont {S.}~\bibnamefont {Diehl}},\ }\bibfield  {title}
  {\enquote {\bibinfo {title} {Dynamical {{Critical Phenomena}} in
  {{Driven}}-{{Dissipative Systems}}},}\ }\href {\doibase
  10.1103/PhysRevLett.110.195301} {\bibfield  {journal} {\bibinfo  {journal}
  {Physical Review Letters}\ }\textbf {\bibinfo {volume} {110}},\ \bibinfo
  {pages} {195301} (\bibinfo {year} {2013})}\BibitemShut {NoStop}%
\bibitem [{\citenamefont {Marino}\ and\ \citenamefont
  {Diehl}(2016)}]{MarinoDiehlPRL16}%
  \BibitemOpen
  \bibfield  {author} {\bibinfo {author} {\bibfnamefont {Jamir}\ \bibnamefont
  {Marino}}\ and\ \bibinfo {author} {\bibfnamefont {Sebastian}\ \bibnamefont
  {Diehl}},\ }\bibfield  {title} {\enquote {\bibinfo {title} {Driven markovian
  quantum criticality},}\ }\href@noop {} {\bibfield  {journal} {\bibinfo
  {journal} {Phys. Rev. Lett.}\ }\textbf {\bibinfo {volume} {116}},\ \bibinfo
  {pages} {070407} (\bibinfo {year} {2016})}\BibitemShut {NoStop}%
\bibitem [{\citenamefont {Schir\'o}\ \emph {et~al.}(2016)\citenamefont
  {Schir\'o}, \citenamefont {Joshi}, \citenamefont {Bordyuh}, \citenamefont
  {Fazio}, \citenamefont {Keeling},\ and\ \citenamefont
  {T\"ureci}}]{SchiroPRL16}%
  \BibitemOpen
  \bibfield  {author} {\bibinfo {author} {\bibfnamefont {M.}~\bibnamefont
  {Schir\'o}}, \bibinfo {author} {\bibfnamefont {C.}~\bibnamefont {Joshi}},
  \bibinfo {author} {\bibfnamefont {M.}~\bibnamefont {Bordyuh}}, \bibinfo
  {author} {\bibfnamefont {R.}~\bibnamefont {Fazio}}, \bibinfo {author}
  {\bibfnamefont {J.}~\bibnamefont {Keeling}}, \ and\ \bibinfo {author}
  {\bibfnamefont {H.~E.}\ \bibnamefont {T\"ureci}},\ }\bibfield  {title}
  {\enquote {\bibinfo {title} {Exotic attractors of the nonequilibrium
  rabi-hubbard model},}\ }\href@noop {} {\bibfield  {journal} {\bibinfo
  {journal} {Phys. Rev. Lett.}\ }\textbf {\bibinfo {volume} {116}},\ \bibinfo
  {pages} {143603} (\bibinfo {year} {2016})}\BibitemShut {NoStop}%
\bibitem [{\citenamefont {Minganti}\ \emph {et~al.}(2018)\citenamefont
  {Minganti}, \citenamefont {Biella}, \citenamefont {Bartolo},\ and\
  \citenamefont {Ciuti}}]{MingantiEtAlPRA18}%
  \BibitemOpen
  \bibfield  {author} {\bibinfo {author} {\bibfnamefont {Fabrizio}\
  \bibnamefont {Minganti}}, \bibinfo {author} {\bibfnamefont {Alberto}\
  \bibnamefont {Biella}}, \bibinfo {author} {\bibfnamefont {Nicola}\
  \bibnamefont {Bartolo}}, \ and\ \bibinfo {author} {\bibfnamefont {Cristiano}\
  \bibnamefont {Ciuti}},\ }\bibfield  {title} {\enquote {\bibinfo {title}
  {Spectral theory of liouvillians for dissipative phase transitions},}\ }\href
  {\doibase 10.1103/PhysRevA.98.042118} {\bibfield  {journal} {\bibinfo
  {journal} {Phys. Rev. A}\ }\textbf {\bibinfo {volume} {98}},\ \bibinfo
  {pages} {042118} (\bibinfo {year} {2018})}\BibitemShut {NoStop}%
\bibitem [{\citenamefont {Rota}\ \emph
  {et~al.}(2019{\natexlab{a}})\citenamefont {Rota}, \citenamefont {Minganti},
  \citenamefont {Ciuti},\ and\ \citenamefont {Savona}}]{RotaEtAlPRL19}%
  \BibitemOpen
  \bibfield  {author} {\bibinfo {author} {\bibfnamefont {Riccardo}\
  \bibnamefont {Rota}}, \bibinfo {author} {\bibfnamefont {Fabrizio}\
  \bibnamefont {Minganti}}, \bibinfo {author} {\bibfnamefont {Cristiano}\
  \bibnamefont {Ciuti}}, \ and\ \bibinfo {author} {\bibfnamefont {Vincenzo}\
  \bibnamefont {Savona}},\ }\bibfield  {title} {\enquote {\bibinfo {title}
  {Quantum critical regime in a quadratically driven nonlinear photonic
  lattice},}\ }\href {\doibase 10.1103/PhysRevLett.122.110405} {\bibfield
  {journal} {\bibinfo  {journal} {Phys. Rev. Lett.}\ }\textbf {\bibinfo
  {volume} {122}},\ \bibinfo {pages} {110405} (\bibinfo {year}
  {2019}{\natexlab{a}})}\BibitemShut {NoStop}%
\bibitem [{\citenamefont {Young}\ \emph {et~al.}(2020)\citenamefont {Young},
  \citenamefont {Gorshkov}, \citenamefont {Foss-Feig},\ and\ \citenamefont
  {Maghrebi}}]{YoungEtAlPRX20}%
  \BibitemOpen
  \bibfield  {author} {\bibinfo {author} {\bibfnamefont {Jeremy~T.}\
  \bibnamefont {Young}}, \bibinfo {author} {\bibfnamefont {Alexey~V.}\
  \bibnamefont {Gorshkov}}, \bibinfo {author} {\bibfnamefont {Michael}\
  \bibnamefont {Foss-Feig}}, \ and\ \bibinfo {author} {\bibfnamefont
  {Mohammad~F.}\ \bibnamefont {Maghrebi}},\ }\bibfield  {title} {\enquote
  {\bibinfo {title} {Nonequilibrium fixed points of coupled ising models},}\
  }\href {\doibase 10.1103/PhysRevX.10.011039} {\bibfield  {journal} {\bibinfo
  {journal} {Phys. Rev. X}\ }\textbf {\bibinfo {volume} {10}},\ \bibinfo
  {pages} {011039} (\bibinfo {year} {2020})}\BibitemShut {NoStop}%
\bibitem [{\citenamefont {Daley}(2014)}]{DaleyAdvPhys2014}%
  \BibitemOpen
  \bibfield  {author} {\bibinfo {author} {\bibfnamefont {Andrew~J.}\
  \bibnamefont {Daley}},\ }\bibfield  {title} {\enquote {\bibinfo {title}
  {Quantum trajectories and open many-body quantum systems},}\ }\href {\doibase 10.1080/00018732.2014.933502} {\bibfield  {journal} {\bibinfo  {journal}
  {Advances in Physics}\ }\textbf {\bibinfo {volume} {63}},\ \bibinfo {pages}
  {77--149} (\bibinfo {year} {2014})} \BibitemShut {NoStop}%
\bibitem [{\citenamefont {Verstraete}\ \emph {et~al.}(2004)\citenamefont
  {Verstraete}, \citenamefont {Garc\'{\i}a-Ripoll},\ and\ \citenamefont
  {Cirac}}]{VerstraeteEtAlPRL04}%
  \BibitemOpen
  \bibfield  {author} {\bibinfo {author} {\bibfnamefont {F.}~\bibnamefont
  {Verstraete}}, \bibinfo {author} {\bibfnamefont {J.~J.}\ \bibnamefont
  {Garc\'{\i}a-Ripoll}}, \ and\ \bibinfo {author} {\bibfnamefont {J.~I.}\
  \bibnamefont {Cirac}},\ }\bibfield  {title} {\enquote {\bibinfo {title}
  {Matrix product density operators: Simulation of finite-temperature and
  dissipative systems},}\ }\href {\doibase 10.1103/PhysRevLett.93.207204}
  {\bibfield  {journal} {\bibinfo  {journal} {Phys. Rev. Lett.}\ }\textbf
  {\bibinfo {volume} {93}},\ \bibinfo {pages} {207204} (\bibinfo {year}
  {2004})}\BibitemShut {NoStop}%
\bibitem [{\citenamefont {Zwolak}\ and\ \citenamefont
  {Vidal}(2004)}]{ZwolakVidalPRL04}%
  \BibitemOpen
  \bibfield  {author} {\bibinfo {author} {\bibfnamefont {Michael}\ \bibnamefont
  {Zwolak}}\ and\ \bibinfo {author} {\bibfnamefont {Guifr\'e}\ \bibnamefont
  {Vidal}},\ }\bibfield  {title} {\enquote {\bibinfo {title} {Mixed-state
  dynamics in one-dimensional quantum lattice systems: A time-dependent
  superoperator renormalization algorithm},}\ }\href {\doibase
  10.1103/PhysRevLett.93.207205} {\bibfield  {journal} {\bibinfo  {journal}
  {Phys. Rev. Lett.}\ }\textbf {\bibinfo {volume} {93}},\ \bibinfo {pages}
  {207205} (\bibinfo {year} {2004})}\BibitemShut {NoStop}%
\bibitem [{\citenamefont {Kilda}\ and\ \citenamefont
  {Keeling}(2019{\natexlab{a}})}]{KildaEtAlPRL19}%
  \BibitemOpen
  \bibfield  {author} {\bibinfo {author} {\bibfnamefont {Dainius}\ \bibnamefont
  {Kilda}}\ and\ \bibinfo {author} {\bibfnamefont {Jonathan}\ \bibnamefont
  {Keeling}},\ }\bibfield  {title} {\enquote {\bibinfo {title} {Fluorescence
  spectrum and thermalization in a driven coupled cavity array},}\ }\href
  {\doibase 10.1103/PhysRevLett.122.043602} {\bibfield  {journal} {\bibinfo
  {journal} {Phys. Rev. Lett.}\ }\textbf {\bibinfo {volume} {122}},\ \bibinfo
  {pages} {043602} (\bibinfo {year} {2019}{\natexlab{a}})}\BibitemShut
  {NoStop}%
\bibitem [{\citenamefont {Kshetrimayum}\ \emph {et~al.}(2017)\citenamefont
  {Kshetrimayum}, \citenamefont {Weimer},\ and\ \citenamefont
  {Or{\'u}s}}]{KshetrimayumNatComm17}%
  \BibitemOpen
  \bibfield  {author} {\bibinfo {author} {\bibfnamefont {Augustine}\
  \bibnamefont {Kshetrimayum}}, \bibinfo {author} {\bibfnamefont {Hendrik}\
  \bibnamefont {Weimer}}, \ and\ \bibinfo {author} {\bibfnamefont {Rom{\'a}n}\
  \bibnamefont {Or{\'u}s}},\ }\bibfield  {title} {\enquote {\bibinfo {title} {A
  simple tensor network algorithm for two-dimensional steady states},}\ }\href
  {\doibase 10.1038/s41467-017-01511-6} {\bibfield  {journal} {\bibinfo
  {journal} {Nature Communications}\ }\textbf {\bibinfo {volume} {8}},\
  \bibinfo {pages} {1291} (\bibinfo {year} {2017})}\BibitemShut {NoStop}%
\bibitem [{\citenamefont {Landa}\ \emph {et~al.}(2020)\citenamefont {Landa},
  \citenamefont {Schir\'o},\ and\ \citenamefont {Misguich}}]{LandaEtAlPRL20}%
  \BibitemOpen
  \bibfield  {author} {\bibinfo {author} {\bibfnamefont {Haggai}\ \bibnamefont
  {Landa}}, \bibinfo {author} {\bibfnamefont {Marco}\ \bibnamefont {Schir\'o}},
  \ and\ \bibinfo {author} {\bibfnamefont {Gr\'egoire}\ \bibnamefont
  {Misguich}},\ }\bibfield  {title} {\enquote {\bibinfo {title} {Multistability
  of driven-dissipative quantum spins},}\ }\href {\doibase
  10.1103/PhysRevLett.124.043601} {\bibfield  {journal} {\bibinfo  {journal}
  {Phys. Rev. Lett.}\ }\textbf {\bibinfo {volume} {124}},\ \bibinfo {pages}
  {043601} (\bibinfo {year} {2020})}\BibitemShut {NoStop}%
\bibitem [{\citenamefont {Kilda}\ \emph {et~al.}(2021)\citenamefont {Kilda},
  \citenamefont {Biella}, \citenamefont {Schiró}, \citenamefont {Fazio},\ and\
  \citenamefont {Keeling}}]{kilda2021stability}%
  \BibitemOpen
  \bibfield  {author} {\bibinfo {author} {\bibfnamefont {Dainius}\ \bibnamefont
  {Kilda}}, \bibinfo {author} {\bibfnamefont {Alberto}\ \bibnamefont {Biella}},
  \bibinfo {author} {\bibfnamefont {Marco}\ \bibnamefont {Schiró}}, \bibinfo
  {author} {\bibfnamefont {Rosario}\ \bibnamefont {Fazio}}, \ and\ \bibinfo
  {author} {\bibfnamefont {Jonathan}\ \bibnamefont {Keeling}},\ }\href@noop {}
  {\enquote {\bibinfo {title} {On the stability of the infinite projected
  entangled pair operator ansatz for driven-dissipative 2d lattices},}\ }
  (\bibinfo {year} {2021}),\ \Eprint {http://arxiv.org/abs/2012.03095}
  {arXiv:2012.03095 [cond-mat.other]} \BibitemShut {NoStop}%
\bibitem [{\citenamefont {Finazzi}\ \emph {et~al.}(2015)\citenamefont
  {Finazzi}, \citenamefont {Le~Boit\'e}, \citenamefont {Storme}, \citenamefont
  {Baksic},\ and\ \citenamefont {Ciuti}}]{FinazziEtAlPRL15}%
  \BibitemOpen
  \bibfield  {author} {\bibinfo {author} {\bibfnamefont {S.}~\bibnamefont
  {Finazzi}}, \bibinfo {author} {\bibfnamefont {A.}~\bibnamefont {Le~Boit\'e}},
  \bibinfo {author} {\bibfnamefont {F.}~\bibnamefont {Storme}}, \bibinfo
  {author} {\bibfnamefont {A.}~\bibnamefont {Baksic}}, \ and\ \bibinfo {author}
  {\bibfnamefont {C.}~\bibnamefont {Ciuti}},\ }\bibfield  {title} {\enquote
  {\bibinfo {title} {Corner-space renormalization method for driven-dissipative
  two-dimensional correlated systems},}\ }\href {\doibase
  10.1103/PhysRevLett.115.080604} {\bibfield  {journal} {\bibinfo  {journal}
  {Phys. Rev. Lett.}\ }\textbf {\bibinfo {volume} {115}},\ \bibinfo {pages}
  {080604} (\bibinfo {year} {2015})}\BibitemShut {NoStop}%
\bibitem [{\citenamefont {Weimer}(2015)}]{WeimerPRL15}%
  \BibitemOpen
  \bibfield  {author} {\bibinfo {author} {\bibfnamefont {Hendrik}\ \bibnamefont
  {Weimer}},\ }\bibfield  {title} {\enquote {\bibinfo {title} {Variational
  principle for steady states of dissipative quantum many-body systems},}\
  }\href@noop {} {\bibfield  {journal} {\bibinfo  {journal} {Phys. Rev. Lett.}\
  }\textbf {\bibinfo {volume} {114}},\ \bibinfo {pages} {040402} (\bibinfo
  {year} {2015})}\BibitemShut {NoStop}%
\bibitem [{\citenamefont {Jin}\ \emph {et~al.}(2016)\citenamefont {Jin},
  \citenamefont {Biella}, \citenamefont {Viyuela}, \citenamefont {Mazza},
  \citenamefont {Keeling}, \citenamefont {Fazio},\ and\ \citenamefont
  {Rossini}}]{JinEtAlPRX16}%
  \BibitemOpen
  \bibfield  {author} {\bibinfo {author} {\bibfnamefont {Jiasen}\ \bibnamefont
  {Jin}}, \bibinfo {author} {\bibfnamefont {Alberto}\ \bibnamefont {Biella}},
  \bibinfo {author} {\bibfnamefont {Oscar}\ \bibnamefont {Viyuela}}, \bibinfo
  {author} {\bibfnamefont {Leonardo}\ \bibnamefont {Mazza}}, \bibinfo {author}
  {\bibfnamefont {Jonathan}\ \bibnamefont {Keeling}}, \bibinfo {author}
  {\bibfnamefont {Rosario}\ \bibnamefont {Fazio}}, \ and\ \bibinfo {author}
  {\bibfnamefont {Davide}\ \bibnamefont {Rossini}},\ }\bibfield  {title}
  {\enquote {\bibinfo {title} {Cluster mean-field approach to the steady-state
  phase diagram of dissipative spin systems},}\ }\href {\doibase
  10.1103/PhysRevX.6.031011} {\bibfield  {journal} {\bibinfo  {journal} {Phys.
  Rev. X}\ }\textbf {\bibinfo {volume} {6}},\ \bibinfo {pages} {031011}
  (\bibinfo {year} {2016})}\BibitemShut {NoStop}%
\bibitem [{\citenamefont {Sieberer}\ \emph {et~al.}(2016)\citenamefont
  {Sieberer}, \citenamefont {Buchhold},\ and\ \citenamefont
  {Diehl}}]{SiebererRepProgPhys2016}%
  \BibitemOpen
  \bibfield  {author} {\bibinfo {author} {\bibfnamefont {L~M}\ \bibnamefont
  {Sieberer}}, \bibinfo {author} {\bibfnamefont {M}~\bibnamefont {Buchhold}}, \
  and\ \bibinfo {author} {\bibfnamefont {S}~\bibnamefont {Diehl}},\ }\bibfield
  {title} {\enquote {\bibinfo {title} {Keldysh field theory for driven open
  quantum systems},}\ }\href@noop {} {\bibfield  {journal} {\bibinfo  {journal}
  {Reports on Progress in Physics}\ }\textbf {\bibinfo {volume} {79}},\
  \bibinfo {pages} {096001} (\bibinfo {year} {2016})}\BibitemShut {NoStop}%
\bibitem [{\citenamefont {Biondi}\ \emph
  {et~al.}(2017{\natexlab{a}})\citenamefont {Biondi}, \citenamefont {Lienhard},
  \citenamefont {Blatter}, \citenamefont {Türeci},\ and\ \citenamefont
  {Schmidt}}]{Biondi_2017}%
  \BibitemOpen
  \bibfield  {author} {\bibinfo {author} {\bibfnamefont {Matteo}\ \bibnamefont
  {Biondi}}, \bibinfo {author} {\bibfnamefont {Saskia}\ \bibnamefont
  {Lienhard}}, \bibinfo {author} {\bibfnamefont {Gianni}\ \bibnamefont
  {Blatter}}, \bibinfo {author} {\bibfnamefont {Hakan~E}\ \bibnamefont
  {Türeci}}, \ and\ \bibinfo {author} {\bibfnamefont {Sebastian}\ \bibnamefont
  {Schmidt}},\ }\bibfield  {title} {\enquote {\bibinfo {title} {Spatial
  correlations in driven-dissipative photonic lattices},}\ }\href {\doibase
  10.1088/1367-2630/aa99b2} {\bibfield  {journal} {\bibinfo  {journal} {New
  Journal of Physics}\ }\textbf {\bibinfo {volume} {19}},\ \bibinfo {pages}
  {125016} (\bibinfo {year} {2017}{\natexlab{a}})}\BibitemShut {NoStop}%
\bibitem [{\citenamefont {Vicentini}\ \emph {et~al.}(2018)\citenamefont
  {Vicentini}, \citenamefont {Minganti}, \citenamefont {Rota}, \citenamefont
  {Orso},\ and\ \citenamefont {Ciuti}}]{VicentiniEtAPRA18}%
  \BibitemOpen
  \bibfield  {author} {\bibinfo {author} {\bibfnamefont {Filippo}\ \bibnamefont
  {Vicentini}}, \bibinfo {author} {\bibfnamefont {Fabrizio}\ \bibnamefont
  {Minganti}}, \bibinfo {author} {\bibfnamefont {Riccardo}\ \bibnamefont
  {Rota}}, \bibinfo {author} {\bibfnamefont {Giuliano}\ \bibnamefont {Orso}}, \
  and\ \bibinfo {author} {\bibfnamefont {Cristiano}\ \bibnamefont {Ciuti}},\
  }\bibfield  {title} {\enquote {\bibinfo {title} {Critical slowing down in
  driven-dissipative bose-hubbard lattices},}\ }\href {\doibase
  10.1103/PhysRevA.97.013853} {\bibfield  {journal} {\bibinfo  {journal} {Phys.
  Rev. A}\ }\textbf {\bibinfo {volume} {97}},\ \bibinfo {pages} {013853}
  (\bibinfo {year} {2018})}\BibitemShut {NoStop}%
\bibitem [{\citenamefont {{Weimer}}\ \emph {et~al.}(2019)\citenamefont
  {{Weimer}}, \citenamefont {{Kshetrimayum}},\ and\ \citenamefont
  {{Or{\'u}s}}}]{WeimerEtAlArxiv19}%
  \BibitemOpen
  \bibfield  {author} {\bibinfo {author} {\bibfnamefont {Hendrik}\ \bibnamefont
  {{Weimer}}}, \bibinfo {author} {\bibfnamefont {Augustine}\ \bibnamefont
  {{Kshetrimayum}}}, \ and\ \bibinfo {author} {\bibfnamefont {Rom{\'a}n}\
  \bibnamefont {{Or{\'u}s}}},\ }\bibfield  {title} {\enquote {\bibinfo {title}
  {{Simulation methods for open quantum many-body systems}},}\ }\href {\doibase 10.1103/RevModPhys.93.015008} {\bibfield
  {journal} {\bibinfo  {journal} {Rev. Mod. Phys.}\ }\textbf {\bibinfo
  {volume} {93}},\ \bibinfo {pages} {015008} (\bibinfo {year}
  {2021})}\BibitemShut {NoStop}%
\bibitem [{\citenamefont {Vicentini}\ \emph {et~al.}(2019)\citenamefont
  {Vicentini}, \citenamefont {Biella}, \citenamefont {Regnault},\ and\
  \citenamefont {Ciuti}}]{VicentiniEtAlPRL19}%
  \BibitemOpen
  \bibfield  {author} {\bibinfo {author} {\bibfnamefont {Filippo}\ \bibnamefont
  {Vicentini}}, \bibinfo {author} {\bibfnamefont {Alberto}\ \bibnamefont
  {Biella}}, \bibinfo {author} {\bibfnamefont {Nicolas}\ \bibnamefont
  {Regnault}}, \ and\ \bibinfo {author} {\bibfnamefont {Cristiano}\
  \bibnamefont {Ciuti}},\ }\bibfield  {title} {\enquote {\bibinfo {title}
  {Variational neural-network ansatz for steady states in open quantum
  systems},}\ }\href {\doibase 10.1103/PhysRevLett.122.250503} {\bibfield
  {journal} {\bibinfo  {journal} {Phys. Rev. Lett.}\ }\textbf {\bibinfo
  {volume} {122}},\ \bibinfo {pages} {250503} (\bibinfo {year}
  {2019})}\BibitemShut {NoStop}%
\bibitem [{\citenamefont {Yoshioka}\ and\ \citenamefont
  {Hamazaki}(2019)}]{YoshiokaHamazakiPRB19}%
  \BibitemOpen
  \bibfield  {author} {\bibinfo {author} {\bibfnamefont {Nobuyuki}\
  \bibnamefont {Yoshioka}}\ and\ \bibinfo {author} {\bibfnamefont {Ryusuke}\
  \bibnamefont {Hamazaki}},\ }\bibfield  {title} {\enquote {\bibinfo {title}
  {Constructing neural stationary states for open quantum many-body systems},}\
  }\href {\doibase 10.1103/PhysRevB.99.214306} {\bibfield  {journal} {\bibinfo
  {journal} {Phys. Rev. B}\ }\textbf {\bibinfo {volume} {99}},\ \bibinfo
  {pages} {214306} (\bibinfo {year} {2019})}\BibitemShut {NoStop}%
\bibitem [{\citenamefont {Nagy}\ and\ \citenamefont
  {Savona}(2019)}]{NagySavonaPRL19}%
  \BibitemOpen
  \bibfield  {author} {\bibinfo {author} {\bibfnamefont {Alexandra}\
  \bibnamefont {Nagy}}\ and\ \bibinfo {author} {\bibfnamefont {Vincenzo}\
  \bibnamefont {Savona}},\ }\bibfield  {title} {\enquote {\bibinfo {title}
  {Variational quantum monte carlo method with a neural-network ansatz for open
  quantum systems},}\ }\href {\doibase 10.1103/PhysRevLett.122.250501}
  {\bibfield  {journal} {\bibinfo  {journal} {Phys. Rev. Lett.}\ }\textbf
  {\bibinfo {volume} {122}},\ \bibinfo {pages} {250501} (\bibinfo {year}
  {2019})}\BibitemShut {NoStop}%
\bibitem [{\citenamefont {Hartmann}\ and\ \citenamefont
  {Carleo}(2019)}]{HartmannCarleoPRL19}%
  \BibitemOpen
  \bibfield  {author} {\bibinfo {author} {\bibfnamefont {Michael~J.}\
  \bibnamefont {Hartmann}}\ and\ \bibinfo {author} {\bibfnamefont {Giuseppe}\
  \bibnamefont {Carleo}},\ }\bibfield  {title} {\enquote {\bibinfo {title}
  {Neural-network approach to dissipative quantum many-body dynamics},}\ }\href
  {\doibase 10.1103/PhysRevLett.122.250502} {\bibfield  {journal} {\bibinfo
  {journal} {Phys. Rev. Lett.}\ }\textbf {\bibinfo {volume} {122}},\ \bibinfo
  {pages} {250502} (\bibinfo {year} {2019})}\BibitemShut {NoStop}%
\bibitem [{\citenamefont {Metzner}\ and\ \citenamefont
  {Vollhardt}(1989)}]{Metzner1989}%
  \BibitemOpen
  \bibfield  {author} {\bibinfo {author} {\bibfnamefont {Walter}\ \bibnamefont
  {Metzner}}\ and\ \bibinfo {author} {\bibfnamefont {Dieter}\ \bibnamefont
  {Vollhardt}},\ }\bibfield  {title} {\enquote {\bibinfo {title} {Correlated
  {{Lattice Fermions}} in {{Infinite Dimensions}}},}\ }\href {\doibase
  10.1103/PhysRevLett.62.324} {\bibfield  {journal} {\bibinfo  {journal}
  {Physical Review Letters}\ }\textbf {\bibinfo {volume} {62}},\ \bibinfo
  {pages} {324--327} (\bibinfo {year} {1989})}\BibitemShut {NoStop}%
\bibitem [{\citenamefont {Georges}\ and\ \citenamefont
  {Kotliar}(1992)}]{Georges_Kotliar_PRB92}%
  \BibitemOpen
  \bibfield  {author} {\bibinfo {author} {\bibfnamefont {Antoine}\ \bibnamefont
  {Georges}}\ and\ \bibinfo {author} {\bibfnamefont {Gabriel}\ \bibnamefont
  {Kotliar}},\ }\bibfield  {title} {\enquote {\bibinfo {title} {Hubbard model
  in infinite dimensions},}\ }\href {\doibase 10.1103/PhysRevB.45.6479}
  {\bibfield  {journal} {\bibinfo  {journal} {Physical Review B}\ }\textbf
  {\bibinfo {volume} {45}},\ \bibinfo {pages} {6479--6483} (\bibinfo {year}
  {1992})}\BibitemShut {NoStop}%
\bibitem [{\citenamefont {Georges}\ \emph {et~al.}(1996)\citenamefont
  {Georges}, \citenamefont {Kotliar}, \citenamefont {Krauth},\ and\
  \citenamefont {Rozenberg}}]{Review_DMFT_96}%
  \BibitemOpen
  \bibfield  {author} {\bibinfo {author} {\bibfnamefont {Antoine}\ \bibnamefont
  {Georges}}, \bibinfo {author} {\bibfnamefont {Gabriel}\ \bibnamefont
  {Kotliar}}, \bibinfo {author} {\bibfnamefont {Werner}\ \bibnamefont
  {Krauth}}, \ and\ \bibinfo {author} {\bibfnamefont {Marcelo~J.}\ \bibnamefont
  {Rozenberg}},\ }\bibfield  {title} {\enquote {\bibinfo {title} {Dynamical
  mean-field theory of strongly correlated fermion systems and the limit of
  infinite dimensions},}\ }\href {\doibase 10.1103/RevModPhys.68.13} {\bibfield
   {journal} {\bibinfo  {journal} {Reviews of Modern Physics}\ }\textbf
  {\bibinfo {volume} {68}},\ \bibinfo {pages} {13--125} (\bibinfo {year}
  {1996})}\BibitemShut {NoStop}%
\bibitem [{\citenamefont {Byczuk}\ and\ \citenamefont
  {Vollhardt}(2008)}]{ByczukVollhardtPRB08}%
  \BibitemOpen
  \bibfield  {author} {\bibinfo {author} {\bibfnamefont {Krzysztof}\
  \bibnamefont {Byczuk}}\ and\ \bibinfo {author} {\bibfnamefont {Dieter}\
  \bibnamefont {Vollhardt}},\ }\bibfield  {title} {\enquote {\bibinfo {title}
  {Correlated bosons on a lattice: Dynamical mean-field theory for
  bose-einstein condensed and normal phases},}\ }\href {\doibase
  10.1103/PhysRevB.77.235106} {\bibfield  {journal} {\bibinfo  {journal} {Phys.
  Rev. B}\ }\textbf {\bibinfo {volume} {77}},\ \bibinfo {pages} {235106}
  (\bibinfo {year} {2008})}\BibitemShut {NoStop}%
\bibitem [{\citenamefont {Anders}\ \emph {et~al.}(2010)\citenamefont {Anders},
  \citenamefont {Gull}, \citenamefont {Pollet}, \citenamefont {Troyer},\ and\
  \citenamefont {Werner}}]{AndersEtAlPRL10}%
  \BibitemOpen
  \bibfield  {author} {\bibinfo {author} {\bibfnamefont {Peter}\ \bibnamefont
  {Anders}}, \bibinfo {author} {\bibfnamefont {Emanuel}\ \bibnamefont {Gull}},
  \bibinfo {author} {\bibfnamefont {Lode}\ \bibnamefont {Pollet}}, \bibinfo
  {author} {\bibfnamefont {Matthias}\ \bibnamefont {Troyer}}, \ and\ \bibinfo
  {author} {\bibfnamefont {Philipp}\ \bibnamefont {Werner}},\ }\bibfield
  {title} {\enquote {\bibinfo {title} {Dynamical mean field solution of the
  bose-hubbard model},}\ }\href {\doibase 10.1103/PhysRevLett.105.096402}
  {\bibfield  {journal} {\bibinfo  {journal} {Phys. Rev. Lett.}\ }\textbf
  {\bibinfo {volume} {105}},\ \bibinfo {pages} {096402} (\bibinfo {year}
  {2010})}\BibitemShut {NoStop}%
\bibitem [{\citenamefont {Anders}\ \emph {et~al.}(2011)\citenamefont {Anders},
  \citenamefont {Gull}, \citenamefont {Pollet}, \citenamefont {Troyer},\ and\
  \citenamefont {Werner}}]{andersWernerNJP2011}%
  \BibitemOpen
  \bibfield  {author} {\bibinfo {author} {\bibfnamefont {Peter}\ \bibnamefont
  {Anders}}, \bibinfo {author} {\bibfnamefont {Emanuel}\ \bibnamefont {Gull}},
  \bibinfo {author} {\bibfnamefont {Lode}\ \bibnamefont {Pollet}}, \bibinfo
  {author} {\bibfnamefont {Matthias}\ \bibnamefont {Troyer}}, \ and\ \bibinfo
  {author} {\bibfnamefont {Philipp}\ \bibnamefont {Werner}},\ }\bibfield
  {title} {\enquote {\bibinfo {title} {Dynamical mean-field theory for
  bosons},}\ }\href {\doibase 10.1088/1367-2630/13/7/075013} {\bibfield
  {journal} {\bibinfo  {journal} {New Journal of Physics}\ }\textbf {\bibinfo
  {volume} {13}},\ \bibinfo {pages} {075013} (\bibinfo {year}
  {2011})}\BibitemShut {NoStop}%
\bibitem [{\citenamefont {Aoki}\ \emph {et~al.}(2014)\citenamefont {Aoki},
  \citenamefont {Tsuji}, \citenamefont {Eckstein}, \citenamefont {Kollar},
  \citenamefont {Oka},\ and\ \citenamefont {Werner}}]{aokiWernerRMP2014}%
  \BibitemOpen
  \bibfield  {author} {\bibinfo {author} {\bibfnamefont {Hideo}\ \bibnamefont
  {Aoki}}, \bibinfo {author} {\bibfnamefont {Naoto}\ \bibnamefont {Tsuji}},
  \bibinfo {author} {\bibfnamefont {Martin}\ \bibnamefont {Eckstein}}, \bibinfo
  {author} {\bibfnamefont {Marcus}\ \bibnamefont {Kollar}}, \bibinfo {author}
  {\bibfnamefont {Takashi}\ \bibnamefont {Oka}}, \ and\ \bibinfo {author}
  {\bibfnamefont {Philipp}\ \bibnamefont {Werner}},\ }\bibfield  {title}
  {\enquote {\bibinfo {title} {Nonequilibrium dynamical mean-field theory and
  its applications},}\ }\href {\doibase 10.1103/RevModPhys.86.779} {\bibfield
  {journal} {\bibinfo  {journal} {Reviews of Modern Physics}\ }\textbf
  {\bibinfo {volume} {86}},\ \bibinfo {pages} {779--837} (\bibinfo {year}
  {2014})}\BibitemShut {NoStop}%
\bibitem [{\citenamefont {Panas}\ \emph {et~al.}(2019)\citenamefont {Panas},
  \citenamefont {Pasek}, \citenamefont {Dhar}, \citenamefont {Qin},
  \citenamefont {Gei\ss{}ler}, \citenamefont {Hafez-Torbati}, \citenamefont
  {Sorantin}, \citenamefont {Titvinidze},\ and\ \citenamefont
  {Hofstetter}}]{panas2019densitywave}%
  \BibitemOpen
  \bibfield  {author} {\bibinfo {author} {\bibfnamefont {Jaromir}\ \bibnamefont
  {Panas}}, \bibinfo {author} {\bibfnamefont {Michael}\ \bibnamefont {Pasek}},
  \bibinfo {author} {\bibfnamefont {Arya}\ \bibnamefont {Dhar}}, \bibinfo
  {author} {\bibfnamefont {Tao}\ \bibnamefont {Qin}}, \bibinfo {author}
  {\bibfnamefont {Andreas}\ \bibnamefont {Gei\ss{}ler}}, \bibinfo {author}
  {\bibfnamefont {Mohsen}\ \bibnamefont {Hafez-Torbati}}, \bibinfo {author}
  {\bibfnamefont {Max~E.}\ \bibnamefont {Sorantin}}, \bibinfo {author}
  {\bibfnamefont {Irakli}\ \bibnamefont {Titvinidze}}, \ and\ \bibinfo {author}
  {\bibfnamefont {Walter}\ \bibnamefont {Hofstetter}},\ }\bibfield  {title}
  {\enquote {\bibinfo {title} {Density-wave steady-state phase of dissipative
  ultracold fermions with nearest-neighbor interactions},}\ }\href {\doibase
  10.1103/PhysRevB.99.115125} {\bibfield  {journal} {\bibinfo  {journal} {Phys.
  Rev. B}\ }\textbf {\bibinfo {volume} {99}},\ \bibinfo {pages} {115125}
  (\bibinfo {year} {2019})}\BibitemShut {NoStop}%
\bibitem [{\citenamefont {Strand}\ \emph
  {et~al.}(2015{\natexlab{a}})\citenamefont {Strand}, \citenamefont
  {Eckstein},\ and\ \citenamefont {Werner}}]{strandWernerPRX2015}%
  \BibitemOpen
  \bibfield  {author} {\bibinfo {author} {\bibfnamefont {Hugo U.~R.}\
  \bibnamefont {Strand}}, \bibinfo {author} {\bibfnamefont {Martin}\
  \bibnamefont {Eckstein}}, \ and\ \bibinfo {author} {\bibfnamefont {Philipp}\
  \bibnamefont {Werner}},\ }\bibfield  {title} {\enquote {\bibinfo {title}
  {Nonequilibrium dynamical mean-field theory for bosonic lattice models},}\
  }\href {\doibase 10.1103/PhysRevX.5.011038} {\bibfield  {journal} {\bibinfo
  {journal} {Physical Review X}\ }\textbf {\bibinfo {volume} {5}},\ \bibinfo
  {pages} {11038} (\bibinfo {year} {2015}{\natexlab{a}})}\BibitemShut {NoStop}%
\bibitem [{\citenamefont {Schiro}\ and\ \citenamefont
  {Scarlatella}(2019)}]{Scarlatella2019}%
  \BibitemOpen
  \bibfield  {author} {\bibinfo {author} {\bibfnamefont {Marco}\ \bibnamefont
  {Schiro}}\ and\ \bibinfo {author} {\bibfnamefont {Orazio}\ \bibnamefont
  {Scarlatella}},\ }\bibfield  {title} {\enquote {\bibinfo {title} {Quantum
  impurity models coupled to {{Markovian}} and non-{{Markovian}} baths},}\
  }\href {\doibase 10.1063/1.5100157} {\bibfield  {journal} {\bibinfo
  {journal} {The Journal of Chemical Physics}\ }\textbf {\bibinfo {volume}
  {151}},\ \bibinfo {pages} {044102} (\bibinfo {year} {2019})}\BibitemShut
  {NoStop}%
\bibitem [{\citenamefont {Lee}\ and\ \citenamefont
  {Sadeghpour}(2013)}]{lee2013}%
  \BibitemOpen
  \bibfield  {author} {\bibinfo {author} {\bibfnamefont {Tony~E.}\ \bibnamefont
  {Lee}}\ and\ \bibinfo {author} {\bibfnamefont {H.~R.}\ \bibnamefont
  {Sadeghpour}},\ }\bibfield  {title} {\enquote {\bibinfo {title} {Quantum
  {{Synchronization}} of {{Quantum}} van der {{Pol Oscillators}} with {{Trapped
  Ions}}},}\ }\href {\doibase 10.1103/PhysRevLett.111.234101} {\bibfield
  {journal} {\bibinfo  {journal} {Physical Review Letters}\ }\textbf {\bibinfo
  {volume} {111}},\ \bibinfo {pages} {234101} (\bibinfo {year}
  {2013})}\BibitemShut {NoStop}%
\bibitem [{\citenamefont {L{\"o}rch}\ \emph {et~al.}(2016)\citenamefont
  {L{\"o}rch}, \citenamefont {Amitai}, \citenamefont {Nunnenkamp},\ and\
  \citenamefont {Bruder}}]{lorch2016}%
  \BibitemOpen
  \bibfield  {author} {\bibinfo {author} {\bibfnamefont {Niels}\ \bibnamefont
  {L{\"o}rch}}, \bibinfo {author} {\bibfnamefont {Ehud}\ \bibnamefont
  {Amitai}}, \bibinfo {author} {\bibfnamefont {Andreas}\ \bibnamefont
  {Nunnenkamp}}, \ and\ \bibinfo {author} {\bibfnamefont {Christoph}\
  \bibnamefont {Bruder}},\ }\bibfield  {title} {\enquote {\bibinfo {title}
  {Genuine {{Quantum Signatures}} in {{Synchronization}} of {{Anharmonic
  Self}}-{{Oscillators}}},}\ }\href {\doibase 10.1103/PhysRevLett.117.073601}
  {\bibfield  {journal} {\bibinfo  {journal} {Physical Review Letters}\
  }\textbf {\bibinfo {volume} {117}},\ \bibinfo {pages} {073601} (\bibinfo
  {year} {2016})}\BibitemShut {NoStop}%
\bibitem [{\citenamefont {Walter}\ \emph {et~al.}(2014)\citenamefont {Walter},
  \citenamefont {Nunnenkamp},\ and\ \citenamefont {Bruder}}]{walter2014}%
  \BibitemOpen
  \bibfield  {author} {\bibinfo {author} {\bibfnamefont {Stefan}\ \bibnamefont
  {Walter}}, \bibinfo {author} {\bibfnamefont {Andreas}\ \bibnamefont
  {Nunnenkamp}}, \ and\ \bibinfo {author} {\bibfnamefont {Christoph}\
  \bibnamefont {Bruder}},\ }\bibfield  {title} {\enquote {\bibinfo {title}
  {Quantum {{Synchronization}} of a {{Driven Self}}-{{Sustained
  Oscillator}}},}\ }\href {\doibase 10.1103/PhysRevLett.112.094102} {\bibfield
  {journal} {\bibinfo  {journal} {Physical Review Letters}\ }\textbf {\bibinfo
  {volume} {112}},\ \bibinfo {pages} {094102} (\bibinfo {year}
  {2014})}\BibitemShut {NoStop}%
\bibitem [{\citenamefont {Walter}\ \emph {et~al.}(2015)\citenamefont {Walter},
  \citenamefont {Nunnenkamp},\ and\ \citenamefont {Bruder}}]{walter2015}%
  \BibitemOpen
  \bibfield  {author} {\bibinfo {author} {\bibfnamefont {Stefan}\ \bibnamefont
  {Walter}}, \bibinfo {author} {\bibfnamefont {Andreas}\ \bibnamefont
  {Nunnenkamp}}, \ and\ \bibinfo {author} {\bibfnamefont {Christoph}\
  \bibnamefont {Bruder}},\ }\bibfield  {title} {\enquote {\bibinfo {title}
  {Quantum synchronization of two {{Van}} der {{Pol}} oscillators},}\ }\href
  {\doibase 10.1002/andp.201400144} {\bibfield  {journal} {\bibinfo  {journal}
  {Annalen der Physik}\ }\textbf {\bibinfo {volume} {527}},\ \bibinfo {pages}
  {131--138} (\bibinfo {year} {2015})}\BibitemShut {NoStop}%
\bibitem [{\citenamefont {Roulet}\ and\ \citenamefont
  {Bruder}(2018{\natexlab{a}})}]{roulet2018a}%
  \BibitemOpen
  \bibfield  {author} {\bibinfo {author} {\bibfnamefont {Alexandre}\
  \bibnamefont {Roulet}}\ and\ \bibinfo {author} {\bibfnamefont {Christoph}\
  \bibnamefont {Bruder}},\ }\bibfield  {title} {\enquote {\bibinfo {title}
  {Synchronizing the {{Smallest Possible System}}},}\ }\href {\doibase
  10.1103/PhysRevLett.121.053601} {\bibfield  {journal} {\bibinfo  {journal}
  {Physical Review Letters}\ }\textbf {\bibinfo {volume} {121}},\ \bibinfo
  {pages} {053601} (\bibinfo {year} {2018}{\natexlab{a}})}\BibitemShut
  {NoStop}%
\bibitem [{\citenamefont {Roulet}\ and\ \citenamefont
  {Bruder}(2018{\natexlab{b}})}]{roulet2018}%
  \BibitemOpen
  \bibfield  {author} {\bibinfo {author} {\bibfnamefont {Alexandre}\
  \bibnamefont {Roulet}}\ and\ \bibinfo {author} {\bibfnamefont {Christoph}\
  \bibnamefont {Bruder}},\ }\bibfield  {title} {\enquote {\bibinfo {title}
  {Quantum {{Synchronization}} and {{Entanglement Generation}}},}\ }\href
  {\doibase 10.1103/PhysRevLett.121.063601} {\bibfield  {journal} {\bibinfo
  {journal} {Physical Review Letters}\ }\textbf {\bibinfo {volume} {121}},\
  \bibinfo {pages} {063601} (\bibinfo {year} {2018}{\natexlab{b}})}\BibitemShut
  {NoStop}%
\bibitem [{\citenamefont {Dutta}\ and\ \citenamefont
  {Cooper}(2019)}]{dutta2019}%
  \BibitemOpen
  \bibfield  {author} {\bibinfo {author} {\bibfnamefont {Shovan}\ \bibnamefont
  {Dutta}}\ and\ \bibinfo {author} {\bibfnamefont {Nigel~R.}\ \bibnamefont
  {Cooper}},\ }\bibfield  {title} {\enquote {\bibinfo {title} {Critical
  {{Response}} of a {{Quantum}} van der {{Pol Oscillator}}},}\ }\href {\doibase
  10.1103/PhysRevLett.123.250401} {\bibfield  {journal} {\bibinfo  {journal}
  {Physical Review Letters}\ }\textbf {\bibinfo {volume} {123}},\ \bibinfo
  {pages} {250401} (\bibinfo {year} {2019})}\BibitemShut {NoStop}%
\bibitem [{\citenamefont {Giorgi}\ \emph {et~al.}(2012)\citenamefont {Giorgi},
  \citenamefont {Galve}, \citenamefont {Manzano}, \citenamefont {Colet},\ and\
  \citenamefont {Zambrini}}]{giorgi2012}%
  \BibitemOpen
  \bibfield  {author} {\bibinfo {author} {\bibfnamefont {Gian~Luca}\
  \bibnamefont {Giorgi}}, \bibinfo {author} {\bibfnamefont {Fernando}\
  \bibnamefont {Galve}}, \bibinfo {author} {\bibfnamefont {Gonzalo}\
  \bibnamefont {Manzano}}, \bibinfo {author} {\bibfnamefont {Pere}\
  \bibnamefont {Colet}}, \ and\ \bibinfo {author} {\bibfnamefont {Roberta}\
  \bibnamefont {Zambrini}},\ }\bibfield  {title} {\enquote {\bibinfo {title}
  {Quantum correlations and mutual synchronization},}\ }\href {\doibase
  10.1103/PhysRevA.85.052101} {\bibfield  {journal} {\bibinfo  {journal}
  {Physical Review A}\ }\textbf {\bibinfo {volume} {85}},\ \bibinfo {pages}
  {052101} (\bibinfo {year} {2012})}\BibitemShut {NoStop}%
\bibitem [{\citenamefont {Manzano}\ \emph {et~al.}(2013)\citenamefont
  {Manzano}, \citenamefont {Galve}, \citenamefont {Giorgi}, \citenamefont
  {{Hern{\'a}ndez-Garc{\'i}a}},\ and\ \citenamefont {Zambrini}}]{manzano2013}%
  \BibitemOpen
  \bibfield  {author} {\bibinfo {author} {\bibfnamefont {Gonzalo}\ \bibnamefont
  {Manzano}}, \bibinfo {author} {\bibfnamefont {Fernando}\ \bibnamefont
  {Galve}}, \bibinfo {author} {\bibfnamefont {Gian~Luca}\ \bibnamefont
  {Giorgi}}, \bibinfo {author} {\bibfnamefont {Emilio}\ \bibnamefont
  {{Hern{\'a}ndez-Garc{\'i}a}}}, \ and\ \bibinfo {author} {\bibfnamefont
  {Roberta}\ \bibnamefont {Zambrini}},\ }\bibfield  {title} {\enquote {\bibinfo
  {title} {Synchronization, quantum correlations and entanglement in oscillator
  networks},}\ }\href {\doibase 10.1038/srep01439} {\bibfield  {journal}
  {\bibinfo  {journal} {Scientific Reports}\ }\textbf {\bibinfo {volume} {3}},\
  \bibinfo {pages} {1--6} (\bibinfo {year} {2013})}\BibitemShut {NoStop}%
\bibitem [{\citenamefont {Qiao}\ \emph {et~al.}(2018)\citenamefont {Qiao},
  \citenamefont {Gao}, \citenamefont {Liu},\ and\ \citenamefont
  {Yi}}]{qiao2018}%
  \BibitemOpen
  \bibfield  {author} {\bibinfo {author} {\bibfnamefont {Guo-jian}\
  \bibnamefont {Qiao}}, \bibinfo {author} {\bibfnamefont {Hui-xia}\
  \bibnamefont {Gao}}, \bibinfo {author} {\bibfnamefont {Hao-di}\ \bibnamefont
  {Liu}}, \ and\ \bibinfo {author} {\bibfnamefont {X.~X.}\ \bibnamefont {Yi}},\
  }\bibfield  {title} {\enquote {\bibinfo {title} {Quantum synchronization of
  two mechanical oscillators in coupled optomechanical systems with {{Kerr}}
  nonlinearity},}\ }\href {\doibase 10.1038/s41598-018-33903-z} {\bibfield
  {journal} {\bibinfo  {journal} {Scientific Reports}\ }\textbf {\bibinfo
  {volume} {8}},\ \bibinfo {pages} {1--11} (\bibinfo {year}
  {2018})}\BibitemShut {NoStop}%
\bibitem [{\citenamefont {Sonar}\ \emph {et~al.}(2018)\citenamefont {Sonar},
  \citenamefont {Hajdu\ifmmode~\check{s}\else \v{s}\fi{}ek}, \citenamefont
  {Mukherjee}, \citenamefont {Fazio}, \citenamefont {Vedral}, \citenamefont
  {Vinjanampathy},\ and\ \citenamefont {Kwek}}]{SonarEtalPRL18}%
  \BibitemOpen
  \bibfield  {author} {\bibinfo {author} {\bibfnamefont {Sameer}\ \bibnamefont
  {Sonar}}, \bibinfo {author} {\bibfnamefont {Michal}\ \bibnamefont
  {Hajdu\ifmmode~\check{s}\else \v{s}\fi{}ek}}, \bibinfo {author}
  {\bibfnamefont {Manas}\ \bibnamefont {Mukherjee}}, \bibinfo {author}
  {\bibfnamefont {Rosario}\ \bibnamefont {Fazio}}, \bibinfo {author}
  {\bibfnamefont {Vlatko}\ \bibnamefont {Vedral}}, \bibinfo {author}
  {\bibfnamefont {Sai}\ \bibnamefont {Vinjanampathy}}, \ and\ \bibinfo {author}
  {\bibfnamefont {Leong-Chuan}\ \bibnamefont {Kwek}},\ }\bibfield  {title}
  {\enquote {\bibinfo {title} {Squeezing enhances quantum synchronization},}\
  }\href {\doibase 10.1103/PhysRevLett.120.163601} {\bibfield  {journal}
  {\bibinfo  {journal} {Phys. Rev. Lett.}\ }\textbf {\bibinfo {volume} {120}},\
  \bibinfo {pages} {163601} (\bibinfo {year} {2018})}\BibitemShut {NoStop}%
\bibitem [{\citenamefont {Jaseem}\ \emph {et~al.}(2020)\citenamefont {Jaseem},
  \citenamefont {Hajdu\ifmmode~\check{s}\else \v{s}\fi{}ek}, \citenamefont
  {Vedral}, \citenamefont {Fazio}, \citenamefont {Kwek},\ and\ \citenamefont
  {Vinjanampathy}}]{jaseem2019}%
  \BibitemOpen
  \bibfield  {author} {\bibinfo {author} {\bibfnamefont {Noufal}\ \bibnamefont
  {Jaseem}}, \bibinfo {author} {\bibfnamefont {Michal}\ \bibnamefont
  {Hajdu\ifmmode~\check{s}\else \v{s}\fi{}ek}}, \bibinfo {author}
  {\bibfnamefont {Vlatko}\ \bibnamefont {Vedral}}, \bibinfo {author}
  {\bibfnamefont {Rosario}\ \bibnamefont {Fazio}}, \bibinfo {author}
  {\bibfnamefont {Leong-Chuan}\ \bibnamefont {Kwek}}, \ and\ \bibinfo {author}
  {\bibfnamefont {Sai}\ \bibnamefont {Vinjanampathy}},\ }\bibfield  {title}
  {\enquote {\bibinfo {title} {Quantum synchronization in nanoscale heat
  engines},}\ }\href {\doibase 10.1103/PhysRevE.101.020201} {\bibfield
  {journal} {\bibinfo  {journal} {Phys. Rev. E}\ }\textbf {\bibinfo {volume}
  {101}},\ \bibinfo {pages} {020201} (\bibinfo {year} {2020})}\BibitemShut
  {NoStop}%
\bibitem [{\citenamefont {Tindall}\ \emph {et~al.}(2020)\citenamefont
  {Tindall}, \citenamefont {Mu{\~n}oz}, \citenamefont {Bu{\v c}a},\ and\
  \citenamefont {Jaksch}}]{tindall2020}%
  \BibitemOpen
  \bibfield  {author} {\bibinfo {author} {\bibfnamefont {J.}~\bibnamefont
  {Tindall}}, \bibinfo {author} {\bibfnamefont {C.~S{\'a}nchez}\ \bibnamefont
  {Mu{\~n}oz}}, \bibinfo {author} {\bibfnamefont {B.}~\bibnamefont {Bu{\v
  c}a}}, \ and\ \bibinfo {author} {\bibfnamefont {D.}~\bibnamefont {Jaksch}},\
  }\bibfield  {title} {\enquote {\bibinfo {title} {Quantum synchronisation
  enabled by dynamical symmetries and dissipation},}\ }\href {\doibase
  10.1088/1367-2630/ab60f5} {\bibfield  {journal} {\bibinfo  {journal} {New
  Journal of Physics}\ }\textbf {\bibinfo {volume} {22}},\ \bibinfo {pages}
  {013026} (\bibinfo {year} {2020})}\BibitemShut {NoStop}%
\bibitem [{\citenamefont {Fisher}\ \emph {et~al.}(1989)\citenamefont {Fisher},
  \citenamefont {Weichman}, \citenamefont {Grinstein},\ and\ \citenamefont
  {Fisher}}]{fisherFisherPRB1989}%
  \BibitemOpen
  \bibfield  {author} {\bibinfo {author} {\bibfnamefont {Matthew~P.A.}\
  \bibnamefont {Fisher}}, \bibinfo {author} {\bibfnamefont {Peter~B.}\
  \bibnamefont {Weichman}}, \bibinfo {author} {\bibfnamefont {G.}~\bibnamefont
  {Grinstein}}, \ and\ \bibinfo {author} {\bibfnamefont {Daniel~S.}\
  \bibnamefont {Fisher}},\ }\bibfield  {title} {\enquote {\bibinfo {title}
  {Boson localization and the superfluid-insulator transition},}\ }\href
  {\doibase 10.1103/PhysRevB.40.546} {\bibfield  {journal} {\bibinfo  {journal}
  {Physical Review B}\ }\textbf {\bibinfo {volume} {40}},\ \bibinfo {pages}
  {546--570} (\bibinfo {year} {1989})}\BibitemShut {NoStop}%
\bibitem [{\citenamefont {Joura}\ \emph {et~al.}(2008)\citenamefont {Joura},
  \citenamefont {Freericks},\ and\ \citenamefont
  {Pruschke}}]{joura2008steadystate}%
  \BibitemOpen
  \bibfield  {author} {\bibinfo {author} {\bibfnamefont {A.~V.}\ \bibnamefont
  {Joura}}, \bibinfo {author} {\bibfnamefont {J.~K.}\ \bibnamefont
  {Freericks}}, \ and\ \bibinfo {author} {\bibfnamefont {Th.}\ \bibnamefont
  {Pruschke}},\ }\bibfield  {title} {\enquote {\bibinfo {title} {Steady-state
  nonequilibrium density of states of driven strongly correlated lattice models
  in infinite dimensions},}\ }\href {\doibase 10.1103/PhysRevLett.101.196401}
  {\bibfield  {journal} {\bibinfo  {journal} {Phys. Rev. Lett.}\ }\textbf
  {\bibinfo {volume} {101}},\ \bibinfo {pages} {196401} (\bibinfo {year}
  {2008})}\BibitemShut {NoStop}%
\bibitem [{\citenamefont {Li}\ \emph {et~al.}(2015)\citenamefont {Li},
  \citenamefont {Aron}, \citenamefont {Kotliar},\ and\ \citenamefont
  {Han}}]{li2015electricfield}%
  \BibitemOpen
  \bibfield  {author} {\bibinfo {author} {\bibfnamefont {Jiajun}\ \bibnamefont
  {Li}}, \bibinfo {author} {\bibfnamefont {Camille}\ \bibnamefont {Aron}},
  \bibinfo {author} {\bibfnamefont {Gabriel}\ \bibnamefont {Kotliar}}, \ and\
  \bibinfo {author} {\bibfnamefont {Jong~E.}\ \bibnamefont {Han}},\ }\bibfield
  {title} {\enquote {\bibinfo {title} {Electric-field-driven resistive
  switching in the dissipative hubbard model},}\ }\href {\doibase
  10.1103/PhysRevLett.114.226403} {\bibfield  {journal} {\bibinfo  {journal}
  {Phys. Rev. Lett.}\ }\textbf {\bibinfo {volume} {114}},\ \bibinfo {pages}
  {226403} (\bibinfo {year} {2015})}\BibitemShut {NoStop}%
\bibitem [{\citenamefont {Arrigoni}\ \emph
  {et~al.}(2013{\natexlab{a}})\citenamefont {Arrigoni}, \citenamefont {Knap},\
  and\ \citenamefont {Von Der~Linden}}]{arrigoni2013}%
  \BibitemOpen
  \bibfield  {author} {\bibinfo {author} {\bibfnamefont {Enrico}\ \bibnamefont
  {Arrigoni}}, \bibinfo {author} {\bibfnamefont {Michael}\ \bibnamefont
  {Knap}}, \ and\ \bibinfo {author} {\bibfnamefont {Wolfgang}\ \bibnamefont
  {Von Der~Linden}},\ }\bibfield  {title} {\enquote {\bibinfo {title}
  {Nonequilibrium dynamical mean-field theory: {{An}} auxiliary quantum master
  equation approach},}\ }\href {\doibase 10.1103/PhysRevLett.110.086403}
  {\bibfield  {journal} {\bibinfo  {journal} {Physical Review Letters}\
  }\textbf {\bibinfo {volume} {110}},\ \bibinfo {pages} {86403} (\bibinfo
  {year} {2013}{\natexlab{a}})}\BibitemShut {NoStop}%
\bibitem [{\citenamefont {Titvinidze}\ \emph {et~al.}(2018)\citenamefont
  {Titvinidze}, \citenamefont {Sorantin}, \citenamefont {Dorda}, \citenamefont
  {von~der Linden},\ and\ \citenamefont {Arrigoni}}]{titvinidze2018charge}%
  \BibitemOpen
  \bibfield  {author} {\bibinfo {author} {\bibfnamefont {Irakli}\ \bibnamefont
  {Titvinidze}}, \bibinfo {author} {\bibfnamefont {Max~E.}\ \bibnamefont
  {Sorantin}}, \bibinfo {author} {\bibfnamefont {Antonius}\ \bibnamefont
  {Dorda}}, \bibinfo {author} {\bibfnamefont {Wolfgang}\ \bibnamefont {von~der
  Linden}}, \ and\ \bibinfo {author} {\bibfnamefont {Enrico}\ \bibnamefont
  {Arrigoni}},\ }\bibfield  {title} {\enquote {\bibinfo {title} {Charge
  redistribution in correlated heterostuctures within nonequilibrium real-space
  dynamical mean-field theory},}\ }\href {\doibase 10.1103/PhysRevB.98.035146}
  {\bibfield  {journal} {\bibinfo  {journal} {Phys. Rev. B}\ }\textbf {\bibinfo
  {volume} {98}},\ \bibinfo {pages} {035146} (\bibinfo {year}
  {2018})}\BibitemShut {NoStop}%
\bibitem [{\citenamefont {Matthies}\ \emph {et~al.}(2018)\citenamefont
  {Matthies}, \citenamefont {Li},\ and\ \citenamefont
  {Eckstein}}]{mattheis2018control}%
  \BibitemOpen
  \bibfield  {author} {\bibinfo {author} {\bibfnamefont {Anne}\ \bibnamefont
  {Matthies}}, \bibinfo {author} {\bibfnamefont {Jiajun}\ \bibnamefont {Li}}, \
  and\ \bibinfo {author} {\bibfnamefont {Martin}\ \bibnamefont {Eckstein}},\
  }\bibfield  {title} {\enquote {\bibinfo {title} {Control of competing
  superconductivity and charge order by nonequilibrium currents},}\ }\href
  {\doibase 10.1103/PhysRevB.98.180502} {\bibfield  {journal} {\bibinfo
  {journal} {Phys. Rev. B}\ }\textbf {\bibinfo {volume} {98}},\ \bibinfo
  {pages} {180502} (\bibinfo {year} {2018})}\BibitemShut {NoStop}%
\bibitem [{\citenamefont {Murakami}\ and\ \citenamefont
  {Werner}(2018)}]{murakami2018nonequilibrium}%
  \BibitemOpen
  \bibfield  {author} {\bibinfo {author} {\bibfnamefont {Yuta}\ \bibnamefont
  {Murakami}}\ and\ \bibinfo {author} {\bibfnamefont {Philipp}\ \bibnamefont
  {Werner}},\ }\bibfield  {title} {\enquote {\bibinfo {title} {Nonequilibrium
  steady states of electric field driven mott insulators},}\ }\href {\doibase
  10.1103/PhysRevB.98.075102} {\bibfield  {journal} {\bibinfo  {journal} {Phys.
  Rev. B}\ }\textbf {\bibinfo {volume} {98}},\ \bibinfo {pages} {075102}
  (\bibinfo {year} {2018})}\BibitemShut {NoStop}%
\bibitem [{\citenamefont {Tsuji}\ \emph {et~al.}(2008)\citenamefont {Tsuji},
  \citenamefont {Oka},\ and\ \citenamefont {Aoki}}]{tsuji2008correlated}%
  \BibitemOpen
  \bibfield  {author} {\bibinfo {author} {\bibfnamefont {Naoto}\ \bibnamefont
  {Tsuji}}, \bibinfo {author} {\bibfnamefont {Takashi}\ \bibnamefont {Oka}}, \
  and\ \bibinfo {author} {\bibfnamefont {Hideo}\ \bibnamefont {Aoki}},\
  }\bibfield  {title} {\enquote {\bibinfo {title} {Correlated electron systems
  periodically driven out of equilibrium: $\text{Floquet}+\text{DMFT}$
  formalism},}\ }\href {\doibase 10.1103/PhysRevB.78.235124} {\bibfield
  {journal} {\bibinfo  {journal} {Phys. Rev. B}\ }\textbf {\bibinfo {volume}
  {78}},\ \bibinfo {pages} {235124} (\bibinfo {year} {2008})}\BibitemShut
  {NoStop}%
\bibitem [{\citenamefont {Murakami}\ \emph {et~al.}(2017)\citenamefont
  {Murakami}, \citenamefont {Tsuji}, \citenamefont {Eckstein},\ and\
  \citenamefont {Werner}}]{murakami2017nonequilibrium}%
  \BibitemOpen
  \bibfield  {author} {\bibinfo {author} {\bibfnamefont {Yuta}\ \bibnamefont
  {Murakami}}, \bibinfo {author} {\bibfnamefont {Naoto}\ \bibnamefont {Tsuji}},
  \bibinfo {author} {\bibfnamefont {Martin}\ \bibnamefont {Eckstein}}, \ and\
  \bibinfo {author} {\bibfnamefont {Philipp}\ \bibnamefont {Werner}},\
  }\bibfield  {title} {\enquote {\bibinfo {title} {Nonequilibrium steady states
  and transient dynamics of conventional superconductors under phonon
  driving},}\ }\href {\doibase 10.1103/PhysRevB.96.045125} {\bibfield
  {journal} {\bibinfo  {journal} {Phys. Rev. B}\ }\textbf {\bibinfo {volume}
  {96}},\ \bibinfo {pages} {045125} (\bibinfo {year} {2017})}\BibitemShut
  {NoStop}%
\bibitem [{\citenamefont {Qin}\ and\ \citenamefont
  {Hofstetter}(2017)}]{qin2017spectral}%
  \BibitemOpen
  \bibfield  {author} {\bibinfo {author} {\bibfnamefont {Tao}\ \bibnamefont
  {Qin}}\ and\ \bibinfo {author} {\bibfnamefont {Walter}\ \bibnamefont
  {Hofstetter}},\ }\bibfield  {title} {\enquote {\bibinfo {title} {Spectral
  functions of a time-periodically driven falicov-kimball model: Real-space
  floquet dynamical mean-field theory study},}\ }\href {\doibase
  10.1103/PhysRevB.96.075134} {\bibfield  {journal} {\bibinfo  {journal} {Phys.
  Rev. B}\ }\textbf {\bibinfo {volume} {96}},\ \bibinfo {pages} {075134}
  (\bibinfo {year} {2017})}\BibitemShut {NoStop}%
\bibitem [{\citenamefont {Li}\ and\ \citenamefont
  {Eckstein}(2021)}]{li2021nonequilibrium}%
  \BibitemOpen
  \bibfield  {author} {\bibinfo {author} {\bibfnamefont {Jiajun}\ \bibnamefont
  {Li}}\ and\ \bibinfo {author} {\bibfnamefont {Martin}\ \bibnamefont
  {Eckstein}},\ }\bibfield  {title} {\enquote {\bibinfo {title} {Nonequilibrium
  steady-state theory of photodoped mott insulators},}\ }\href {\doibase
  10.1103/PhysRevB.103.045133} {\bibfield  {journal} {\bibinfo  {journal}
  {Phys. Rev. B}\ }\textbf {\bibinfo {volume} {103}},\ \bibinfo {pages}
  {045133} (\bibinfo {year} {2021})}\BibitemShut {NoStop}%
\bibitem [{\citenamefont {Eckstein}\ and\ \citenamefont
  {Werner}(2013)}]{eckstein2013photoinduced}%
  \BibitemOpen
  \bibfield  {author} {\bibinfo {author} {\bibfnamefont {Martin}\ \bibnamefont
  {Eckstein}}\ and\ \bibinfo {author} {\bibfnamefont {Philipp}\ \bibnamefont
  {Werner}},\ }\bibfield  {title} {\enquote {\bibinfo {title} {Photoinduced
  states in a mott insulator},}\ }\href {\doibase
  10.1103/PhysRevLett.110.126401} {\bibfield  {journal} {\bibinfo  {journal}
  {Phys. Rev. Lett.}\ }\textbf {\bibinfo {volume} {110}},\ \bibinfo {pages}
  {126401} (\bibinfo {year} {2013})}\BibitemShut {NoStop}%
\bibitem [{\citenamefont {Gole\ifmmode~\check{z}\else \v{z}\fi{}}\ \emph
  {et~al.}(2015)\citenamefont {Gole\ifmmode~\check{z}\else \v{z}\fi{}},
  \citenamefont {Eckstein},\ and\ \citenamefont {Werner}}]{golez2015dynamics}%
  \BibitemOpen
  \bibfield  {author} {\bibinfo {author} {\bibfnamefont {Denis}\ \bibnamefont
  {Gole\ifmmode~\check{z}\else \v{z}\fi{}}}, \bibinfo {author} {\bibfnamefont
  {Martin}\ \bibnamefont {Eckstein}}, \ and\ \bibinfo {author} {\bibfnamefont
  {Philipp}\ \bibnamefont {Werner}},\ }\bibfield  {title} {\enquote {\bibinfo
  {title} {Dynamics of screening in photodoped mott insulators},}\ }\href
  {\doibase 10.1103/PhysRevB.92.195123} {\bibfield  {journal} {\bibinfo
  {journal} {Phys. Rev. B}\ }\textbf {\bibinfo {volume} {92}},\ \bibinfo
  {pages} {195123} (\bibinfo {year} {2015})}\BibitemShut {NoStop}%
\bibitem [{\citenamefont {Chen}\ \emph {et~al.}(2016)\citenamefont {Chen},
  \citenamefont {Cohen}, \citenamefont {Millis},\ and\ \citenamefont
  {Reichman}}]{Chen2016}%
  \BibitemOpen
  \bibfield  {author} {\bibinfo {author} {\bibfnamefont {Hsing-Ta~Ta}\
  \bibnamefont {Chen}}, \bibinfo {author} {\bibfnamefont {Guy}\ \bibnamefont
  {Cohen}}, \bibinfo {author} {\bibfnamefont {Andrew~J.}\ \bibnamefont
  {Millis}}, \ and\ \bibinfo {author} {\bibfnamefont {David~R.}\ \bibnamefont
  {Reichman}},\ }\bibfield  {title} {\enquote {\bibinfo {title}
  {Anderson-{{Holstein}} model in two flavors of the noncrossing
  approximation},}\ }\href {\doibase 10.1103/PhysRevB.93.174309} {\bibfield
  {journal} {\bibinfo  {journal} {Physical Review B}\ }\textbf {\bibinfo
  {volume} {93}},\ \bibinfo {pages} {174309} (\bibinfo {year}
  {2016})}\BibitemShut {NoStop}%
\bibitem [{\citenamefont {Bittner}\ \emph {et~al.}(2018)\citenamefont
  {Bittner}, \citenamefont {Gole\ifmmode~\check{z}\else \v{z}\fi{}},
  \citenamefont {Strand}, \citenamefont {Eckstein},\ and\ \citenamefont
  {Werner}}]{bittner2018coupled}%
  \BibitemOpen
  \bibfield  {author} {\bibinfo {author} {\bibfnamefont {Nikolaj}\ \bibnamefont
  {Bittner}}, \bibinfo {author} {\bibfnamefont {Denis}\ \bibnamefont
  {Gole\ifmmode~\check{z}\else \v{z}\fi{}}}, \bibinfo {author} {\bibfnamefont
  {Hugo U.~R.}\ \bibnamefont {Strand}}, \bibinfo {author} {\bibfnamefont
  {Martin}\ \bibnamefont {Eckstein}}, \ and\ \bibinfo {author} {\bibfnamefont
  {Philipp}\ \bibnamefont {Werner}},\ }\bibfield  {title} {\enquote {\bibinfo
  {title} {Coupled charge and spin dynamics in a photoexcited doped mott
  insulator},}\ }\href {\doibase 10.1103/PhysRevB.97.235125} {\bibfield
  {journal} {\bibinfo  {journal} {Phys. Rev. B}\ }\textbf {\bibinfo {volume}
  {97}},\ \bibinfo {pages} {235125} (\bibinfo {year} {2018})}\BibitemShut
  {NoStop}%
\bibitem [{\citenamefont {Peronaci}\ \emph {et~al.}(2020)\citenamefont
  {Peronaci}, \citenamefont {Parcollet},\ and\ \citenamefont
  {Schir\'o}}]{PeronaciEtAlPRB2020}%
  \BibitemOpen
  \bibfield  {author} {\bibinfo {author} {\bibfnamefont {Francesco}\
  \bibnamefont {Peronaci}}, \bibinfo {author} {\bibfnamefont {Olivier}\
  \bibnamefont {Parcollet}}, \ and\ \bibinfo {author} {\bibfnamefont {Marco}\
  \bibnamefont {Schir\'o}},\ }\bibfield  {title} {\enquote {\bibinfo {title}
  {Enhancement of local pairing correlations in periodically driven mott
  insulators},}\ }\href {\doibase 10.1103/PhysRevB.101.161101} {\bibfield
  {journal} {\bibinfo  {journal} {Phys. Rev. B}\ }\textbf {\bibinfo {volume}
  {101}},\ \bibinfo {pages} {161101} (\bibinfo {year} {2020})}\BibitemShut
  {NoStop}%
\bibitem [{\citenamefont {Rota}\ \emph
  {et~al.}(2019{\natexlab{b}})\citenamefont {Rota}, \citenamefont {Minganti},
  \citenamefont {Ciuti},\ and\ \citenamefont {Savona}}]{rota2019quantum}%
  \BibitemOpen
  \bibfield  {author} {\bibinfo {author} {\bibfnamefont {Riccardo}\
  \bibnamefont {Rota}}, \bibinfo {author} {\bibfnamefont {Fabrizio}\
  \bibnamefont {Minganti}}, \bibinfo {author} {\bibfnamefont {Cristiano}\
  \bibnamefont {Ciuti}}, \ and\ \bibinfo {author} {\bibfnamefont {Vincenzo}\
  \bibnamefont {Savona}},\ }\bibfield  {title} {\enquote {\bibinfo {title}
  {Quantum critical regime in a quadratically driven nonlinear photonic
  lattice},}\ }\href {\doibase 10.1103/PhysRevLett.122.110405} {\bibfield
  {journal} {\bibinfo  {journal} {Phys. Rev. Lett.}\ }\textbf {\bibinfo
  {volume} {122}},\ \bibinfo {pages} {110405} (\bibinfo {year}
  {2019}{\natexlab{b}})}\BibitemShut {NoStop}%
\bibitem [{\citenamefont {Lescanne}\ \emph {et~al.}(2020)\citenamefont
  {Lescanne}, \citenamefont {Villiers}, \citenamefont {Peronnin}, \citenamefont
  {Sarlette}, \citenamefont {Delbecq}, \citenamefont {Huard}, \citenamefont
  {Kontos}, \citenamefont {Mirrahimi},\ and\ \citenamefont
  {Leghtas}}]{lescanne2020exponential}%
  \BibitemOpen
  \bibfield  {author} {\bibinfo {author} {\bibfnamefont {Rapha{\"e}l}\
  \bibnamefont {Lescanne}}, \bibinfo {author} {\bibfnamefont {Marius}\
  \bibnamefont {Villiers}}, \bibinfo {author} {\bibfnamefont {Th{\'e}au}\
  \bibnamefont {Peronnin}}, \bibinfo {author} {\bibfnamefont {Alain}\
  \bibnamefont {Sarlette}}, \bibinfo {author} {\bibfnamefont {Matthieu}\
  \bibnamefont {Delbecq}}, \bibinfo {author} {\bibfnamefont {Benjamin}\
  \bibnamefont {Huard}}, \bibinfo {author} {\bibfnamefont {Takis}\ \bibnamefont
  {Kontos}}, \bibinfo {author} {\bibfnamefont {Mazyar}\ \bibnamefont
  {Mirrahimi}}, \ and\ \bibinfo {author} {\bibfnamefont {Zaki}\ \bibnamefont
  {Leghtas}},\ }\bibfield  {title} {\enquote {\bibinfo {title} {Exponential
  suppression of bit-flips in a qubit encoded in an oscillator},}\ }\href
  {\doibase 10.1038/s41567-020-0824-x} {\bibfield  {journal} {\bibinfo
  {journal} {Nature Physics}\ }\textbf {\bibinfo {volume} {16}},\ \bibinfo
  {pages} {509--513} (\bibinfo {year} {2020})}\BibitemShut {NoStop}%
\bibitem [{\citenamefont {Hartmann}\ \emph {et~al.}(2006)\citenamefont
  {Hartmann}, \citenamefont {Brand{\~a}o},\ and\ \citenamefont
  {Plenio}}]{HartmannEtAlNatPhys06}%
  \BibitemOpen
  \bibfield  {author} {\bibinfo {author} {\bibfnamefont {Michael~J.}\
  \bibnamefont {Hartmann}}, \bibinfo {author} {\bibfnamefont {Fernando G.
  S.~L.}\ \bibnamefont {Brand{\~a}o}}, \ and\ \bibinfo {author} {\bibfnamefont
  {Martin~B.}\ \bibnamefont {Plenio}},\ }\bibfield  {title} {\enquote {\bibinfo
  {title} {Strongly interacting polaritons in coupled arrays of cavities},}\
  }\href {\doibase 10.1038/nphys462} {\bibfield  {journal} {\bibinfo  {journal}
  {Nature Physics}\ }\textbf {\bibinfo {volume} {2}},\ \bibinfo {pages}
  {849--855} (\bibinfo {year} {2006})}\BibitemShut {NoStop}%
\bibitem [{\citenamefont {Angelakis}\ \emph {et~al.}(2007)\citenamefont
  {Angelakis}, \citenamefont {Santos},\ and\ \citenamefont
  {Bose}}]{AngelakisEtAlPRA07}%
  \BibitemOpen
  \bibfield  {author} {\bibinfo {author} {\bibfnamefont {Dimitris~G.}\
  \bibnamefont {Angelakis}}, \bibinfo {author} {\bibfnamefont {Marcelo~Franca}\
  \bibnamefont {Santos}}, \ and\ \bibinfo {author} {\bibfnamefont {Sougato}\
  \bibnamefont {Bose}},\ }\bibfield  {title} {\enquote {\bibinfo {title}
  {Photon-blockade-induced mott transitions and $xy$ spin models in coupled
  cavity arrays},}\ }\href {\doibase 10.1103/PhysRevA.76.031805} {\bibfield
  {journal} {\bibinfo  {journal} {Phys. Rev. A}\ }\textbf {\bibinfo {volume}
  {76}},\ \bibinfo {pages} {031805} (\bibinfo {year} {2007})}\BibitemShut
  {NoStop}%
\bibitem [{\citenamefont {Hartmann}(2010)}]{HartmannPRL10}%
  \BibitemOpen
  \bibfield  {author} {\bibinfo {author} {\bibfnamefont {Michael~J.}\
  \bibnamefont {Hartmann}},\ }\bibfield  {title} {\enquote {\bibinfo {title}
  {Polariton crystallization in driven arrays of lossy nonlinear resonators},}\
  }\href {\doibase 10.1103/PhysRevLett.104.113601} {\bibfield  {journal}
  {\bibinfo  {journal} {Phys. Rev. Lett.}\ }\textbf {\bibinfo {volume} {104}},\
  \bibinfo {pages} {113601} (\bibinfo {year} {2010})}\BibitemShut {NoStop}%
\bibitem [{\citenamefont {Jin}\ \emph {et~al.}(2013)\citenamefont {Jin},
  \citenamefont {Rossini}, \citenamefont {Fazio}, \citenamefont {Leib},\ and\
  \citenamefont {Hartmann}}]{JinEtAlPRL13}%
  \BibitemOpen
  \bibfield  {author} {\bibinfo {author} {\bibfnamefont {Jiasen}\ \bibnamefont
  {Jin}}, \bibinfo {author} {\bibfnamefont {Davide}\ \bibnamefont {Rossini}},
  \bibinfo {author} {\bibfnamefont {Rosario}\ \bibnamefont {Fazio}}, \bibinfo
  {author} {\bibfnamefont {Martin}\ \bibnamefont {Leib}}, \ and\ \bibinfo
  {author} {\bibfnamefont {Michael~J.}\ \bibnamefont {Hartmann}},\ }\bibfield
  {title} {\enquote {\bibinfo {title} {Photon solid phases in driven arrays of
  nonlinearly coupled cavities},}\ }\href {\doibase
  10.1103/PhysRevLett.110.163605} {\bibfield  {journal} {\bibinfo  {journal}
  {Phys. Rev. Lett.}\ }\textbf {\bibinfo {volume} {110}},\ \bibinfo {pages}
  {163605} (\bibinfo {year} {2013})}\BibitemShut {NoStop}%
\bibitem [{\citenamefont {Le~Boit\'e}\ \emph {et~al.}(2013)\citenamefont
  {Le~Boit\'e}, \citenamefont {Orso},\ and\ \citenamefont
  {Ciuti}}]{LeBoiteEtAlPRL13}%
  \BibitemOpen
  \bibfield  {author} {\bibinfo {author} {\bibfnamefont {Alexandre}\
  \bibnamefont {Le~Boit\'e}}, \bibinfo {author} {\bibfnamefont {Giuliano}\
  \bibnamefont {Orso}}, \ and\ \bibinfo {author} {\bibfnamefont {Cristiano}\
  \bibnamefont {Ciuti}},\ }\bibfield  {title} {\enquote {\bibinfo {title}
  {Steady-state phases and tunneling-induced instabilities in the driven
  dissipative bose-hubbard model},}\ }\href {\doibase
  10.1103/PhysRevLett.110.233601} {\bibfield  {journal} {\bibinfo  {journal}
  {Phys. Rev. Lett.}\ }\textbf {\bibinfo {volume} {110}},\ \bibinfo {pages}
  {233601} (\bibinfo {year} {2013})}\BibitemShut {NoStop}%
\bibitem [{\citenamefont {Lebreuilly}\ \emph {et~al.}(2017)\citenamefont
  {Lebreuilly}, \citenamefont {Biella}, \citenamefont {Storme}, \citenamefont
  {Rossini}, \citenamefont {Fazio}, \citenamefont {Ciuti},\ and\ \citenamefont
  {Carusotto}}]{LebreuillyEtAlPRA17}%
  \BibitemOpen
  \bibfield  {author} {\bibinfo {author} {\bibfnamefont {Jos\'e}\ \bibnamefont
  {Lebreuilly}}, \bibinfo {author} {\bibfnamefont {Alberto}\ \bibnamefont
  {Biella}}, \bibinfo {author} {\bibfnamefont {Florent}\ \bibnamefont
  {Storme}}, \bibinfo {author} {\bibfnamefont {Davide}\ \bibnamefont
  {Rossini}}, \bibinfo {author} {\bibfnamefont {Rosario}\ \bibnamefont
  {Fazio}}, \bibinfo {author} {\bibfnamefont {Cristiano}\ \bibnamefont
  {Ciuti}}, \ and\ \bibinfo {author} {\bibfnamefont {Iacopo}\ \bibnamefont
  {Carusotto}},\ }\bibfield  {title} {\enquote {\bibinfo {title} {Stabilizing
  strongly correlated photon fluids with non-markovian reservoirs},}\ }\href
  {\doibase 10.1103/PhysRevA.96.033828} {\bibfield  {journal} {\bibinfo
  {journal} {Phys. Rev. A}\ }\textbf {\bibinfo {volume} {96}},\ \bibinfo
  {pages} {033828} (\bibinfo {year} {2017})}\BibitemShut {NoStop}%
\bibitem [{\citenamefont {Foss-Feig}\ \emph {et~al.}(2017)\citenamefont
  {Foss-Feig}, \citenamefont {Niroula}, \citenamefont {Young}, \citenamefont
  {Hafezi}, \citenamefont {Gorshkov}, \citenamefont {Wilson},\ and\
  \citenamefont {Maghrebi}}]{foss-feig_emergent_2017}%
  \BibitemOpen
  \bibfield  {author} {\bibinfo {author} {\bibfnamefont {M.}~\bibnamefont
  {Foss-Feig}}, \bibinfo {author} {\bibfnamefont {P.}~\bibnamefont {Niroula}},
  \bibinfo {author} {\bibfnamefont {J.~T.}\ \bibnamefont {Young}}, \bibinfo
  {author} {\bibfnamefont {M.}~\bibnamefont {Hafezi}}, \bibinfo {author}
  {\bibfnamefont {A.~V.}\ \bibnamefont {Gorshkov}}, \bibinfo {author}
  {\bibfnamefont {R.~M.}\ \bibnamefont {Wilson}}, \ and\ \bibinfo {author}
  {\bibfnamefont {M.~F.}\ \bibnamefont {Maghrebi}},\ }\bibfield  {title}
  {\enquote {\bibinfo {title} {Emergent equilibrium in many-body optical
  bistability},}\ }\href {\doibase 10.1103/PhysRevA.95.043826} {\bibfield
  {journal} {\bibinfo  {journal} {Phys. Rev. A}\ }\textbf {\bibinfo {volume}
  {95}},\ \bibinfo {pages} {043826} (\bibinfo {year} {2017})}\BibitemShut
  {NoStop}%
\bibitem [{\citenamefont {Biondi}\ \emph
  {et~al.}(2017{\natexlab{b}})\citenamefont {Biondi}, \citenamefont {Blatter},
  \citenamefont {Türeci},\ and\ \citenamefont
  {Schmidt}}]{biondi_nonequilibrium_2017}%
  \BibitemOpen
  \bibfield  {author} {\bibinfo {author} {\bibfnamefont {Matteo}\ \bibnamefont
  {Biondi}}, \bibinfo {author} {\bibfnamefont {Gianni}\ \bibnamefont
  {Blatter}}, \bibinfo {author} {\bibfnamefont {Hakan~E.}\ \bibnamefont
  {Türeci}}, \ and\ \bibinfo {author} {\bibfnamefont {Sebastian}\ \bibnamefont
  {Schmidt}},\ }\bibfield  {title} {\enquote {\bibinfo {title} {Nonequilibrium
  gas-liquid transition in the driven-dissipative photonic lattice},}\ }\href
  {\doibase 10.1103/PhysRevA.96.043809} {\bibfield  {journal} {\bibinfo
  {journal} {Phys. Rev. A}\ }\textbf {\bibinfo {volume} {96}},\ \bibinfo
  {pages} {043809} (\bibinfo {year} {2017}{\natexlab{b}})}\BibitemShut
  {NoStop}%
\bibitem [{\citenamefont {Scarlatella}\ \emph {et~al.}(2019)\citenamefont
  {Scarlatella}, \citenamefont {Fazio},\ and\ \citenamefont
  {Schir\'o}}]{ScarlatellaFazioSchiroPRB19}%
  \BibitemOpen
  \bibfield  {author} {\bibinfo {author} {\bibfnamefont {Orazio}\ \bibnamefont
  {Scarlatella}}, \bibinfo {author} {\bibfnamefont {Rosario}\ \bibnamefont
  {Fazio}}, \ and\ \bibinfo {author} {\bibfnamefont {Marco}\ \bibnamefont
  {Schir\'o}},\ }\bibfield  {title} {\enquote {\bibinfo {title} {Emergent
  finite frequency criticality of driven-dissipative correlated lattice
  bosons},}\ }\href {\doibase 10.1103/PhysRevB.99.064511} {\bibfield  {journal}
  {\bibinfo  {journal} {Phys. Rev. B}\ }\textbf {\bibinfo {volume} {99}},\
  \bibinfo {pages} {064511} (\bibinfo {year} {2019})}\BibitemShut {NoStop}%
\bibitem [{\citenamefont {Dodonov}\ and\ \citenamefont
  {Mizrahi}(1997)}]{dodonovMizrahiJPAMathGen1997}%
  \BibitemOpen
  \bibfield  {author} {\bibinfo {author} {\bibfnamefont {V.~V.}\ \bibnamefont
  {Dodonov}}\ and\ \bibinfo {author} {\bibfnamefont {S.~S.}\ \bibnamefont
  {Mizrahi}},\ }\bibfield  {title} {\enquote {\bibinfo {title} {Exact
  stationary photon distributions due to competition between one-and two-photon
  absorption and emission},}\ }\href {\doibase 10.1088/0305-4470/30/16/010}
  {\bibfield  {journal} {\bibinfo  {journal} {Journal of Physics A:
  Mathematical and General}\ }\textbf {\bibinfo {volume} {30}},\ \bibinfo
  {pages} {5657--5667} (\bibinfo {year} {1997})}\BibitemShut {NoStop}%
\bibitem [{\citenamefont {Dykman}(1978)}]{dykman1978}%
  \BibitemOpen
  \bibfield  {author} {\bibinfo {author} {\bibfnamefont {M.}~\bibnamefont
  {Dykman}},\ }\bibfield  {title} {\enquote {\bibinfo {title} {Heating and
  cooling of local and quasilocal vibrations by a nonresonance field.pdf},}\
  }\href@noop {} {\bibfield  {journal} {\bibinfo  {journal} {Sov. phys. Solid
  State}\ }\textbf {\bibinfo {volume} {20}},\ \bibinfo {pages} {1306--1311}
  (\bibinfo {year} {1978})}\BibitemShut {NoStop}%
\bibitem [{Note1()}]{Note1}%
  \BibitemOpen
  \bibinfo {note} {We are not limited to the coordination numbers we spanned in
  Fig. \ref {fig:phaseDiag} and we can run DMFT equations for smaller values of
  $z$, but computing the phase diagram becomes particularly challenging, as it
  moves towards higher pump-to-loss $r$ values, as shown in Fig. \ref
  {fig:phaseDiag}, which are numerically hard to access requiring bigger
  Hilbert space sizes.}\BibitemShut {Stop}%
\bibitem [{\citenamefont {Scarlatella}\ \emph {et~al.}(2018)\citenamefont
  {Scarlatella}, \citenamefont {Clerk},\ and\ \citenamefont
  {Schir{\`o}}}]{scarlatellaClerkSchiro2018}%
  \BibitemOpen
  \bibfield  {author} {\bibinfo {author} {\bibfnamefont {Orazio}\ \bibnamefont
  {Scarlatella}}, \bibinfo {author} {\bibfnamefont {Aashish~A.}\ \bibnamefont
  {Clerk}}, \ and\ \bibinfo {author} {\bibfnamefont {Marco}\ \bibnamefont
  {Schir{\`o}}},\ }\bibfield  {title} {\enquote {\bibinfo {title} {Spectral
  functions and negative density of states of a driven-dissipative nonlinear
  quantum resonator},}\ }\href {\doibase 10.1088/1367-2630/ab0ce9} {\bibfield
  {journal} {\bibinfo  {journal} {New Journal of Physics}\ }\textbf {\bibinfo
  {volume} {21}},\ \bibinfo {pages} {043040} (\bibinfo {year}
  {2018})}\BibitemShut {NoStop}%
\bibitem [{\citenamefont {Bu{\v c}a}\ \emph {et~al.}(2019)\citenamefont {Bu{\v
  c}a}, \citenamefont {Tindall},\ and\ \citenamefont
  {Jaksch}}]{bucaJaksch2018}%
  \BibitemOpen
  \bibfield  {author} {\bibinfo {author} {\bibfnamefont {Berislav}\
  \bibnamefont {Bu{\v c}a}}, \bibinfo {author} {\bibfnamefont {Joseph}\
  \bibnamefont {Tindall}}, \ and\ \bibinfo {author} {\bibfnamefont {Dieter}\
  \bibnamefont {Jaksch}},\ }\bibfield  {title} {\enquote {\bibinfo {title}
  {Non-stationary coherent quantum many-body dynamics through dissipation},}\
  }\href {\doibase 10.1038/s41467-019-09757-y} {\bibfield  {journal} {\bibinfo
  {journal} {Nature Communications}\ }\textbf {\bibinfo {volume} {10}},\
  \bibinfo {pages} {1804.06744} (\bibinfo {year} {2019})}\BibitemShut {NoStop}%
\bibitem [{\citenamefont {Iemini}\ \emph {et~al.}(2018)\citenamefont {Iemini},
  \citenamefont {Russomanno}, \citenamefont {Keeling}, \citenamefont
  {Schir\`o}, \citenamefont {Dalmonte},\ and\ \citenamefont
  {Fazio}}]{IeminiEtAlPRL18}%
  \BibitemOpen
  \bibfield  {author} {\bibinfo {author} {\bibfnamefont {F.}~\bibnamefont
  {Iemini}}, \bibinfo {author} {\bibfnamefont {A.}~\bibnamefont {Russomanno}},
  \bibinfo {author} {\bibfnamefont {J.}~\bibnamefont {Keeling}}, \bibinfo
  {author} {\bibfnamefont {M.}~\bibnamefont {Schir\`o}}, \bibinfo {author}
  {\bibfnamefont {M.}~\bibnamefont {Dalmonte}}, \ and\ \bibinfo {author}
  {\bibfnamefont {R.}~\bibnamefont {Fazio}},\ }\bibfield  {title} {\enquote
  {\bibinfo {title} {Boundary time crystals},}\ }\href {\doibase
  10.1103/PhysRevLett.121.035301} {\bibfield  {journal} {\bibinfo  {journal}
  {Phys. Rev. Lett.}\ }\textbf {\bibinfo {volume} {121}},\ \bibinfo {pages}
  {035301} (\bibinfo {year} {2018})}\BibitemShut {NoStop}%
\bibitem [{\citenamefont {Strogatz}\ and\ \citenamefont
  {Mirollo}(1991)}]{strogatz1991}%
  \BibitemOpen
  \bibfield  {author} {\bibinfo {author} {\bibfnamefont {Steven~H.}\
  \bibnamefont {Strogatz}}\ and\ \bibinfo {author} {\bibfnamefont {Renato~E.}\
  \bibnamefont {Mirollo}},\ }\bibfield  {title} {\enquote {\bibinfo {title}
  {Stability of incoherence in a population of coupled oscillators},}\ }\href
  {\doibase 10.1007/BF01029202} {\bibfield  {journal} {\bibinfo  {journal}
  {Journal of Statistical Physics}\ }\textbf {\bibinfo {volume} {63}},\
  \bibinfo {pages} {613--635} (\bibinfo {year} {1991})}\BibitemShut {NoStop}%
\bibitem [{\citenamefont {Matthews}\ \emph {et~al.}(1991)\citenamefont
  {Matthews}, \citenamefont {Mirollo},\ and\ \citenamefont
  {Strogatz}}]{matthews1991}%
  \BibitemOpen
  \bibfield  {author} {\bibinfo {author} {\bibfnamefont {Paul~C.}\ \bibnamefont
  {Matthews}}, \bibinfo {author} {\bibfnamefont {Renato~E.}\ \bibnamefont
  {Mirollo}}, \ and\ \bibinfo {author} {\bibfnamefont {Steven~H.}\ \bibnamefont
  {Strogatz}},\ }\bibfield  {title} {\enquote {\bibinfo {title} {Dynamics of a
  large system of coupled nonlinear oscillators},}\ }\href {\doibase
  10.1016/0167-2789(91)90129-W} {\bibfield  {journal} {\bibinfo  {journal}
  {Physica D: Nonlinear Phenomena}\ }\textbf {\bibinfo {volume} {52}},\
  \bibinfo {pages} {293--331} (\bibinfo {year} {1991})}\BibitemShut {NoStop}%
\bibitem [{\citenamefont {Cross}\ \emph {et~al.}(2004)\citenamefont {Cross},
  \citenamefont {Zumdieck}, \citenamefont {Lifshitz},\ and\ \citenamefont
  {Rogers}}]{cross2004}%
  \BibitemOpen
  \bibfield  {author} {\bibinfo {author} {\bibfnamefont {M.~C.}\ \bibnamefont
  {Cross}}, \bibinfo {author} {\bibfnamefont {A.}~\bibnamefont {Zumdieck}},
  \bibinfo {author} {\bibfnamefont {Ron}\ \bibnamefont {Lifshitz}}, \ and\
  \bibinfo {author} {\bibfnamefont {J.~L.}\ \bibnamefont {Rogers}},\ }\bibfield
   {title} {\enquote {\bibinfo {title} {Synchronization by {{Nonlinear
  Frequency Pulling}}},}\ }\href {\doibase 10.1103/PhysRevLett.93.224101}
  {\bibfield  {journal} {\bibinfo  {journal} {Physical Review Letters}\
  }\textbf {\bibinfo {volume} {93}},\ \bibinfo {pages} {224101} (\bibinfo
  {year} {2004})}\BibitemShut {NoStop}%
\bibitem [{\citenamefont {Davis-Tilley}\ \emph {et~al.}(2018)\citenamefont
  {Davis-Tilley}, \citenamefont {Teoh},\ and\ \citenamefont
  {Armour}}]{Davis_Tilley_2018}%
  \BibitemOpen
  \bibfield  {author} {\bibinfo {author} {\bibfnamefont {C}~\bibnamefont
  {Davis-Tilley}}, \bibinfo {author} {\bibfnamefont {C~K}\ \bibnamefont
  {Teoh}}, \ and\ \bibinfo {author} {\bibfnamefont {A~D}\ \bibnamefont
  {Armour}},\ }\bibfield  {title} {\enquote {\bibinfo {title} {Dynamics of
  many-body quantum synchronisation},}\ }\href {\doibase
  10.1088/1367-2630/aae947} {\bibfield  {journal} {\bibinfo  {journal} {New
  Journal of Physics}\ }\textbf {\bibinfo {volume} {20}},\ \bibinfo {pages}
  {113002} (\bibinfo {year} {2018})}\BibitemShut {NoStop}%
\bibitem [{\citenamefont {{Garc{\'i}a-Ripoll}}\ \emph
  {et~al.}(2009)\citenamefont {{Garc{\'i}a-Ripoll}}, \citenamefont {D{\"u}rr},
  \citenamefont {Syassen}, \citenamefont {Bauer}, \citenamefont {Lettner},
  \citenamefont {Rempe},\ and\ \citenamefont {Cirac}}]{garcia-ripoll2009}%
  \BibitemOpen
  \bibfield  {author} {\bibinfo {author} {\bibfnamefont {J.~J.}\ \bibnamefont
  {{Garc{\'i}a-Ripoll}}}, \bibinfo {author} {\bibfnamefont {S.}~\bibnamefont
  {D{\"u}rr}}, \bibinfo {author} {\bibfnamefont {N.}~\bibnamefont {Syassen}},
  \bibinfo {author} {\bibfnamefont {D.~M.}\ \bibnamefont {Bauer}}, \bibinfo
  {author} {\bibfnamefont {M.}~\bibnamefont {Lettner}}, \bibinfo {author}
  {\bibfnamefont {G.}~\bibnamefont {Rempe}}, \ and\ \bibinfo {author}
  {\bibfnamefont {J.~I.}\ \bibnamefont {Cirac}},\ }\bibfield  {title} {\enquote
  {\bibinfo {title} {Dissipation-induced hard-core boson gas in an optical
  lattice},}\ }\href@noop {} {\bibfield  {journal} {\bibinfo  {journal} {New
  Journal of Physics}\ }\textbf {\bibinfo {volume} {11}},\ \bibinfo {pages}
  {013053} (\bibinfo {year} {2009})}\BibitemShut {NoStop}%
\bibitem [{\citenamefont {Misra}\ and\ \citenamefont
  {Sudarshan}(1977)}]{misra1977a}%
  \BibitemOpen
  \bibfield  {author} {\bibinfo {author} {\bibfnamefont {B.}~\bibnamefont
  {Misra}}\ and\ \bibinfo {author} {\bibfnamefont {E.~C.~G.}\ \bibnamefont
  {Sudarshan}},\ }\bibfield  {title} {\enquote {\bibinfo {title} {The
  {{Zeno}}'s paradox in quantum theory},}\ }\href@noop {} {\bibfield  {journal}
  {\bibinfo  {journal} {Journal of Mathematical Physics}\ }\textbf {\bibinfo
  {volume} {18}},\ \bibinfo {pages} {756--763} (\bibinfo {year}
  {1977})}\BibitemShut {NoStop}%
\bibitem [{\citenamefont {Beige}\ \emph {et~al.}(2000)\citenamefont {Beige},
  \citenamefont {Braun}, \citenamefont {Tregenna},\ and\ \citenamefont
  {Knight}}]{beige2000}%
  \BibitemOpen
  \bibfield  {author} {\bibinfo {author} {\bibfnamefont {Almut}\ \bibnamefont
  {Beige}}, \bibinfo {author} {\bibfnamefont {Daniel}\ \bibnamefont {Braun}},
  \bibinfo {author} {\bibfnamefont {Ben}\ \bibnamefont {Tregenna}}, \ and\
  \bibinfo {author} {\bibfnamefont {Peter~L.}\ \bibnamefont {Knight}},\
  }\bibfield  {title} {\enquote {\bibinfo {title} {Quantum {{Computing Using
  Dissipation}} to {{Remain}} in a {{Decoherence}}-{{Free Subspace}}},}\
  }\href@noop {} {\bibfield  {journal} {\bibinfo  {journal} {Physical Review
  Letters}\ }\textbf {\bibinfo {volume} {85}},\ \bibinfo {pages} {1762--1765}
  (\bibinfo {year} {2000})}\BibitemShut {NoStop}%
\bibitem [{\citenamefont {Rossini}\ \emph {et~al.}(2020)\citenamefont
  {Rossini}, \citenamefont {Ghermaoui}, \citenamefont {Aguilera}, \citenamefont
  {Vatr{\'e}}, \citenamefont {Bouganne}, \citenamefont {Beugnon}, \citenamefont
  {Gerbier},\ and\ \citenamefont {Mazza}}]{rossini2020b}%
  \BibitemOpen
  \bibfield  {author} {\bibinfo {author} {\bibfnamefont {Davide}\ \bibnamefont
  {Rossini}}, \bibinfo {author} {\bibfnamefont {Alexis}\ \bibnamefont
  {Ghermaoui}}, \bibinfo {author} {\bibfnamefont {Manel~Bosch}\ \bibnamefont
  {Aguilera}}, \bibinfo {author} {\bibfnamefont {R{\'e}my}\ \bibnamefont
  {Vatr{\'e}}}, \bibinfo {author} {\bibfnamefont {Rapha{\"e}l}\ \bibnamefont
  {Bouganne}}, \bibinfo {author} {\bibfnamefont {J{\'e}r{\^o}me}\ \bibnamefont
  {Beugnon}}, \bibinfo {author} {\bibfnamefont {Fabrice}\ \bibnamefont
  {Gerbier}}, \ and\ \bibinfo {author} {\bibfnamefont {Leonardo}\ \bibnamefont
  {Mazza}},\ }\bibfield  {title} {\enquote {\bibinfo {title} {Strong
  correlations in lossy one-dimensional quantum gases: From the quantum
  {{Zeno}} effect to the generalized {{Gibbs}} ensemble},}\ }\href@noop {}
  {\bibfield  {journal} {\bibinfo  {journal} {arXiv:2011.04318 [cond-mat]}\ }
  (\bibinfo {year} {2020})}\BibitemShut {NoStop}%
\bibitem [{\citenamefont {Lemonde}\ \emph {et~al.}(2013)\citenamefont
  {Lemonde}, \citenamefont {Didier},\ and\ \citenamefont
  {Clerk}}]{LemondePRL2013}%
  \BibitemOpen
  \bibfield  {author} {\bibinfo {author} {\bibfnamefont {Marc~Antoine}\
  \bibnamefont {Lemonde}}, \bibinfo {author} {\bibfnamefont {Nicolas}\
  \bibnamefont {Didier}}, \ and\ \bibinfo {author} {\bibfnamefont {Aashish~A.}\
  \bibnamefont {Clerk}},\ }\bibfield  {title} {\enquote {\bibinfo {title}
  {Nonlinear interaction effects in a strongly driven optomechanical cavity},}\
  }\href {\doibase 10.1103/PhysRevLett.111.053602} {\bibfield  {journal}
  {\bibinfo  {journal} {Physical Review Letters}\ }\textbf {\bibinfo {volume}
  {111}},\ \bibinfo {pages} {53602} (\bibinfo {year} {2013})}\BibitemShut
  {NoStop}%
\bibitem [{\citenamefont {Levitan}\ \emph {et~al.}(2016)\citenamefont
  {Levitan}, \citenamefont {Metelmann},\ and\ \citenamefont
  {Clerk}}]{lavitanClerkNJP2016}%
  \BibitemOpen
  \bibfield  {author} {\bibinfo {author} {\bibfnamefont {B.~A.}\ \bibnamefont
  {Levitan}}, \bibinfo {author} {\bibfnamefont {A.}~\bibnamefont {Metelmann}},
  \ and\ \bibinfo {author} {\bibfnamefont {A.~A.}\ \bibnamefont {Clerk}},\
  }\bibfield  {title} {\enquote {\bibinfo {title} {Optomechanics with
  two-phonon driving},}\ }\href {\doibase 10.1088/1367-2630/18/9/093014}
  {\bibfield  {journal} {\bibinfo  {journal} {New Journal of Physics}\ }\textbf
  {\bibinfo {volume} {18}},\ \bibinfo {pages} {93014} (\bibinfo {year}
  {2016})}\BibitemShut {NoStop}%
\bibitem [{\citenamefont {Kamenev}(2011)}]{kamenev2011field}%
  \BibitemOpen
  \bibfield  {author} {\bibinfo {author} {\bibfnamefont {Alex}\ \bibnamefont
  {Kamenev}},\ }\href@noop {} {\emph {\bibinfo {title} {Field theory of
  non-equilibrium systems}}}\ (\bibinfo  {publisher} {Cambridge University
  Press},\ \bibinfo {year} {2011})\BibitemShut {NoStop}%
\bibitem [{\citenamefont {Strand}\ \emph
  {et~al.}(2015{\natexlab{b}})\citenamefont {Strand}, \citenamefont
  {Eckstein},\ and\ \citenamefont {Werner}}]{strandWernerPRA2015}%
  \BibitemOpen
  \bibfield  {author} {\bibinfo {author} {\bibfnamefont {Hugo U.~R.}\
  \bibnamefont {Strand}}, \bibinfo {author} {\bibfnamefont {Martin}\
  \bibnamefont {Eckstein}}, \ and\ \bibinfo {author} {\bibfnamefont {Philipp}\
  \bibnamefont {Werner}},\ }\bibfield  {title} {\enquote {\bibinfo {title}
  {Beyond the {{Hubbard}} bands in strongly correlated lattice bosons},}\
  }\href {\doibase 10.1103/PhysRevA.92.063602} {\bibfield  {journal} {\bibinfo
  {journal} {Physical Review A - Atomic, Molecular, and Optical Physics}\
  }\textbf {\bibinfo {volume} {92}},\ \bibinfo {pages} {63602} (\bibinfo {year}
  {2015}{\natexlab{b}})}\BibitemShut {NoStop}%
\bibitem [{\citenamefont {Schoeller}\ and\ \citenamefont
  {Sch\"on}(1994)}]{SchoellerSchonPRB94}%
  \BibitemOpen
  \bibfield  {author} {\bibinfo {author} {\bibfnamefont {Herbert}\ \bibnamefont
  {Schoeller}}\ and\ \bibinfo {author} {\bibfnamefont {Gerd}\ \bibnamefont
  {Sch\"on}},\ }\bibfield  {title} {\enquote {\bibinfo {title} {Mesoscopic
  quantum transport: Resonant tunneling in the presence of a strong coulomb
  interaction},}\ }\href {\doibase 10.1103/PhysRevB.50.18436} {\bibfield
  {journal} {\bibinfo  {journal} {Phys. Rev. B}\ }\textbf {\bibinfo {volume}
  {50}},\ \bibinfo {pages} {18436--18452} (\bibinfo {year} {1994})}\BibitemShut
  {NoStop}%
\bibitem [{\citenamefont {M{\"u}hlbacher}\ and\ \citenamefont
  {Rabani}(2008)}]{muhlbacherRabaniPRL2008}%
  \BibitemOpen
  \bibfield  {author} {\bibinfo {author} {\bibfnamefont {Lothar}\ \bibnamefont
  {M{\"u}hlbacher}}\ and\ \bibinfo {author} {\bibfnamefont {Eran}\ \bibnamefont
  {Rabani}},\ }\bibfield  {title} {\enquote {\bibinfo {title} {Real-time path
  integral approach to nonequilibrium many-body quantum systems},}\ }\href
  {\doibase 10.1103/PhysRevLett.100.176403} {\bibfield  {journal} {\bibinfo
  {journal} {Physical Review Letters}\ }\textbf {\bibinfo {volume} {100}},\
  \bibinfo {pages} {176403} (\bibinfo {year} {2008})}\BibitemShut {NoStop}%
\bibitem [{\citenamefont {Schir{\'o}}\ and\ \citenamefont
  {Fabrizio}(2009)}]{schiroFabrizioPRB2009}%
  \BibitemOpen
  \bibfield  {author} {\bibinfo {author} {\bibfnamefont {Marco}\ \bibnamefont
  {Schir{\'o}}}\ and\ \bibinfo {author} {\bibfnamefont {Michele}\ \bibnamefont
  {Fabrizio}},\ }\bibfield  {title} {\enquote {\bibinfo {title} {Real-time
  diagrammatic {{Monte Carlo}} for nonequilibrium quantum transport},}\ }\href
  {\doibase 10.1103/PhysRevB.79.153302} {\bibfield  {journal} {\bibinfo
  {journal} {Phys. Rev. B}\ }\textbf {\bibinfo {volume} {79}},\ \bibinfo
  {pages} {153302} (\bibinfo {year} {2009})}\BibitemShut {NoStop}%
\bibitem [{\citenamefont {Schir{\'o}}(2010)}]{schiroPRB2009}%
  \BibitemOpen
  \bibfield  {author} {\bibinfo {author} {\bibfnamefont {Marco}\ \bibnamefont
  {Schir{\'o}}},\ }\bibfield  {title} {\enquote {\bibinfo {title} {Real-time
  dynamics in quantum impurity models with diagrammatic {{Monte Carlo}}},}\
  }\href {\doibase 10.1103/PhysRevB.81.085126} {\bibfield  {journal} {\bibinfo
  {journal} {Physical Review B - Condensed Matter and Materials Physics}\
  }\textbf {\bibinfo {volume} {81}},\ \bibinfo {pages} {85126} (\bibinfo {year}
  {2010})}\BibitemShut {NoStop}%
\bibitem [{\citenamefont {Werner}\ \emph {et~al.}(2009)\citenamefont {Werner},
  \citenamefont {Oka},\ and\ \citenamefont {Millis}}]{Werner_Keldysh_09}%
  \BibitemOpen
  \bibfield  {author} {\bibinfo {author} {\bibfnamefont {Philipp}\ \bibnamefont
  {Werner}}, \bibinfo {author} {\bibfnamefont {Takashi}\ \bibnamefont {Oka}}, \
  and\ \bibinfo {author} {\bibfnamefont {Andrew~J.}\ \bibnamefont {Millis}},\
  }\bibfield  {title} {\enquote {\bibinfo {title} {Diagrammatic {{Monte Carlo}}
  simulation of nonequilibrium systems},}\ }\href {\doibase
  10.1103/PhysRevB.79.035320} {\bibfield  {journal} {\bibinfo  {journal}
  {Physical Review B - Condensed Matter and Materials Physics}\ }\textbf
  {\bibinfo {volume} {79}},\ \bibinfo {pages} {35320} (\bibinfo {year}
  {2009})}\BibitemShut {NoStop}%
\bibitem [{\citenamefont {Gull}\ \emph {et~al.}(2011)\citenamefont {Gull},
  \citenamefont {Millis}, \citenamefont {Lichtenstein}, \citenamefont
  {Rubtsov}, \citenamefont {Troyer},\ and\ \citenamefont
  {Werner}}]{Gull_RMP11}%
  \BibitemOpen
  \bibfield  {author} {\bibinfo {author} {\bibfnamefont {Emanuel}\ \bibnamefont
  {Gull}}, \bibinfo {author} {\bibfnamefont {Andrew~J.}\ \bibnamefont
  {Millis}}, \bibinfo {author} {\bibfnamefont {Alexander~I.}\ \bibnamefont
  {Lichtenstein}}, \bibinfo {author} {\bibfnamefont {Alexey~N.}\ \bibnamefont
  {Rubtsov}}, \bibinfo {author} {\bibfnamefont {Matthias}\ \bibnamefont
  {Troyer}}, \ and\ \bibinfo {author} {\bibfnamefont {Philipp}\ \bibnamefont
  {Werner}},\ }\bibfield  {title} {\enquote {\bibinfo {title} {Continuous-time
  monte~carlo methods for quantum impurity models},}\ }\href {\doibase
  10.1103/RevModPhys.83.349} {\bibfield  {journal} {\bibinfo  {journal} {Rev.
  Mod. Phys.}\ }\textbf {\bibinfo {volume} {83}},\ \bibinfo {pages} {349--404}
  (\bibinfo {year} {2011})}\BibitemShut {NoStop}%
\bibitem [{\citenamefont {Bickers}(1987)}]{Bickers1987}%
  \BibitemOpen
  \bibfield  {author} {\bibinfo {author} {\bibfnamefont {N.~E.}\ \bibnamefont
  {Bickers}},\ }\href {\doibase 10.1103/RevModPhys.59.845} {\emph {\bibinfo
  {title} {Review of Techniques in the Large-{{N}} Expansion for Dilute
  Magnetic Alloys}}},\ \bibinfo {type} {Tech. Rep.}\ (\bibinfo {year} {1987})\
  \bibinfo {note} {publication Title: Reviews of Modern Physics}\BibitemShut
  {NoStop}%
\bibitem [{\citenamefont {Nordlander}\ \emph {et~al.}(1999)\citenamefont
  {Nordlander}, \citenamefont {Pustilnik}, \citenamefont {Meir}, \citenamefont
  {Wingreen},\ and\ \citenamefont {Langreth}}]{Nordlander1999}%
  \BibitemOpen
  \bibfield  {author} {\bibinfo {author} {\bibfnamefont {Peter}\ \bibnamefont
  {Nordlander}}, \bibinfo {author} {\bibfnamefont {Michael}\ \bibnamefont
  {Pustilnik}}, \bibinfo {author} {\bibfnamefont {Yigal}\ \bibnamefont {Meir}},
  \bibinfo {author} {\bibfnamefont {Ned~S.}\ \bibnamefont {Wingreen}}, \ and\
  \bibinfo {author} {\bibfnamefont {David~C.}\ \bibnamefont {Langreth}},\
  }\bibfield  {title} {\enquote {\bibinfo {title} {How {{Long Does It Take}}
  for the {{Kondo Effect}} to {{Develop}}?}}\ }\href {\doibase
  10.1103/PhysRevLett.83.808} {\bibfield  {journal} {\bibinfo  {journal}
  {Physical Review Letters}\ }\textbf {\bibinfo {volume} {83}},\ \bibinfo
  {pages} {808--811} (\bibinfo {year} {1999})}\BibitemShut {NoStop}%
\bibitem [{\citenamefont {Eckstein}\ and\ \citenamefont
  {Werner}(2010)}]{Eckstein2010}%
  \BibitemOpen
  \bibfield  {author} {\bibinfo {author} {\bibfnamefont {Martin}\ \bibnamefont
  {Eckstein}}\ and\ \bibinfo {author} {\bibfnamefont {Philipp}\ \bibnamefont
  {Werner}},\ }\bibfield  {title} {\enquote {\bibinfo {title} {Nonequilibrium
  dynamical mean-field calculations based on the noncrossing approximation and
  its generalizations},}\ }\href {\doibase 10.1103/PhysRevB.82.115115}
  {\bibfield  {journal} {\bibinfo  {journal} {Physical Review B - Condensed
  Matter and Materials Physics}\ }\textbf {\bibinfo {volume} {82}},\ \bibinfo
  {pages} {115115} (\bibinfo {year} {2010})}\BibitemShut {NoStop}%
\bibitem [{\citenamefont {R{\"u}egg}\ \emph {et~al.}(2013)\citenamefont
  {R{\"u}egg}, \citenamefont {Gull}, \citenamefont {Fiete},\ and\ \citenamefont
  {Millis}}]{rueggMillisPRB2013}%
  \BibitemOpen
  \bibfield  {author} {\bibinfo {author} {\bibfnamefont {Andreas}\ \bibnamefont
  {R{\"u}egg}}, \bibinfo {author} {\bibfnamefont {Emanuel}\ \bibnamefont
  {Gull}}, \bibinfo {author} {\bibfnamefont {Gregory~A.}\ \bibnamefont
  {Fiete}}, \ and\ \bibinfo {author} {\bibfnamefont {Andrew~J.}\ \bibnamefont
  {Millis}},\ }\bibfield  {title} {\enquote {\bibinfo {title} {Sum rule
  violation in self-consistent hybridization expansions},}\ }\href {\doibase
  10.1103/PhysRevB.87.075124} {\bibfield  {journal} {\bibinfo  {journal}
  {Physical Review B - Condensed Matter and Materials Physics}\ }\textbf
  {\bibinfo {volume} {87}},\ \bibinfo {pages} {75124} (\bibinfo {year}
  {2013})}\BibitemShut {NoStop}%
\bibitem [{\citenamefont {Erpenbeck}\ \emph {et~al.}(2021)\citenamefont
  {Erpenbeck}, \citenamefont {Gull},\ and\ \citenamefont
  {Cohen}}]{erpenbeck2021revealing}%
  \BibitemOpen
  \bibfield  {author} {\bibinfo {author} {\bibfnamefont {A.}~\bibnamefont
  {Erpenbeck}}, \bibinfo {author} {\bibfnamefont {E.}~\bibnamefont {Gull}}, \
  and\ \bibinfo {author} {\bibfnamefont {G.}~\bibnamefont {Cohen}},\ }\bibfield
   {title} {\enquote {\bibinfo {title} {Revealing strong correlations in
  higher-order transport statistics: A noncrossing approximation approach},}\
  }\href {\doibase 10.1103/PhysRevB.103.125431} {\bibfield  {journal} {\bibinfo
   {journal} {Phys. Rev. B}\ }\textbf {\bibinfo {volume} {103}},\ \bibinfo
  {pages} {125431} (\bibinfo {year} {2021})}\BibitemShut {NoStop}%
\bibitem [{Note2()}]{Note2}%
  \BibitemOpen
  \bibinfo {note} {\protect \leavevmode {\protect \color {blue}The spectral
  function obeys the frequency sum rule $\DOTSI \intop \ilimits@ _{-\infty
  }^\infty d\omega A(\omega ) = A(t=0)= 1$, which is exactly enforced by our
  NCA approximation of Green functions \protect \textup {\hbox {\mathsurround
  \z@ \protect \normalfont (\ignorespaces \ref {eq:NCAGreen}\unskip
  \@@italiccorr )}}. This is easily verified by using the identity $\protect
  \hat {\protect \mathcal {V}}(t=0) = \protect \hat {\protect \mathbb {1}}$. In
  the numerics, the truncation of the bosonic Hilbert space spoils this exact
  identity, which we therefore use to assess that the Hilbert space cut-off we
  consider is sufficiently large.}}\BibitemShut {Stop}%
\bibitem [{\citenamefont {Alicki}(1979)}]{Alicki_1979}%
  \BibitemOpen
  \bibfield  {author} {\bibinfo {author} {\bibfnamefont {R}~\bibnamefont
  {Alicki}},\ }\bibfield  {title} {\enquote {\bibinfo {title} {The quantum open
  system as a model of the heat engine},}\ }\href {\doibase
  10.1088/0305-4470/12/5/007} {\bibfield  {journal} {\bibinfo  {journal}
  {Journal of Physics A: Mathematical and General}\ }\textbf {\bibinfo {volume}
  {12}},\ \bibinfo {pages} {L103--L107} (\bibinfo {year} {1979})}\BibitemShut
  {NoStop}%
\bibitem [{\citenamefont {Feldmann}\ and\ \citenamefont
  {Kosloff}(2003)}]{FeldmannEtAlPRE03}%
  \BibitemOpen
  \bibfield  {author} {\bibinfo {author} {\bibfnamefont {Tova}\ \bibnamefont
  {Feldmann}}\ and\ \bibinfo {author} {\bibfnamefont {Ronnie}\ \bibnamefont
  {Kosloff}},\ }\bibfield  {title} {\enquote {\bibinfo {title} {Quantum
  four-stroke heat engine: Thermodynamic observables in a model with intrinsic
  friction},}\ }\href {\doibase 10.1103/PhysRevE.68.016101} {\bibfield
  {journal} {\bibinfo  {journal} {Phys. Rev. E}\ }\textbf {\bibinfo {volume}
  {68}},\ \bibinfo {pages} {016101} (\bibinfo {year} {2003})}\BibitemShut
  {NoStop}%
\bibitem [{\citenamefont {Rivas}(2020)}]{RivasEtAlPRL20}%
  \BibitemOpen
  \bibfield  {author} {\bibinfo {author} {\bibfnamefont {\'Angel}\ \bibnamefont
  {Rivas}},\ }\bibfield  {title} {\enquote {\bibinfo {title} {Strong coupling
  thermodynamics of open quantum systems},}\ }\href {\doibase
  10.1103/PhysRevLett.124.160601} {\bibfield  {journal} {\bibinfo  {journal}
  {Phys. Rev. Lett.}\ }\textbf {\bibinfo {volume} {124}},\ \bibinfo {pages}
  {160601} (\bibinfo {year} {2020})}\BibitemShut {NoStop}%
\bibitem [{\citenamefont {Scovil}\ and\ \citenamefont
  {Schulz-DuBois}(1959)}]{ScovilEtAlPRL59}%
  \BibitemOpen
  \bibfield  {author} {\bibinfo {author} {\bibfnamefont {H.~E.~D.}\
  \bibnamefont {Scovil}}\ and\ \bibinfo {author} {\bibfnamefont {E.~O.}\
  \bibnamefont {Schulz-DuBois}},\ }\bibfield  {title} {\enquote {\bibinfo
  {title} {Three-level masers as heat engines},}\ }\href {\doibase
  10.1103/PhysRevLett.2.262} {\bibfield  {journal} {\bibinfo  {journal} {Phys.
  Rev. Lett.}\ }\textbf {\bibinfo {volume} {2}},\ \bibinfo {pages} {262--263}
  (\bibinfo {year} {1959})}\BibitemShut {NoStop}%
\bibitem [{\citenamefont {Boukobza}\ and\ \citenamefont
  {Tannor}(2006)}]{BoukobzaEtAlPRA062}%
  \BibitemOpen
  \bibfield  {author} {\bibinfo {author} {\bibfnamefont {E.}~\bibnamefont
  {Boukobza}}\ and\ \bibinfo {author} {\bibfnamefont {D.~J.}\ \bibnamefont
  {Tannor}},\ }\bibfield  {title} {\enquote {\bibinfo {title} {Thermodynamic
  analysis of quantum light amplification},}\ }\href {\doibase
  10.1103/PhysRevA.74.063822} {\bibfield  {journal} {\bibinfo  {journal} {Phys.
  Rev. A}\ }\textbf {\bibinfo {volume} {74}},\ \bibinfo {pages} {063822}
  (\bibinfo {year} {2006})}\BibitemShut {NoStop}%
\bibitem [{\citenamefont {Boukobza}\ and\ \citenamefont
  {Tannor}(2007)}]{BoukobzaEtAlPRL07}%
  \BibitemOpen
  \bibfield  {author} {\bibinfo {author} {\bibfnamefont {E.}~\bibnamefont
  {Boukobza}}\ and\ \bibinfo {author} {\bibfnamefont {D.~J.}\ \bibnamefont
  {Tannor}},\ }\bibfield  {title} {\enquote {\bibinfo {title} {Three-level
  systems as amplifiers and attenuators: A thermodynamic analysis},}\ }\href
  {\doibase 10.1103/PhysRevLett.98.240601} {\bibfield  {journal} {\bibinfo
  {journal} {Phys. Rev. Lett.}\ }\textbf {\bibinfo {volume} {98}},\ \bibinfo
  {pages} {240601} (\bibinfo {year} {2007})}\BibitemShut {NoStop}%
\bibitem [{\citenamefont {Liu}\ \emph {et~al.}(2015)\citenamefont {Liu},
  \citenamefont {Stehlik}, \citenamefont {Eichler}, \citenamefont {Gullans},
  \citenamefont {Taylor},\ and\ \citenamefont {Petta}}]{Liu285}%
  \BibitemOpen
  \bibfield  {author} {\bibinfo {author} {\bibfnamefont {Y.-Y.}\ \bibnamefont
  {Liu}}, \bibinfo {author} {\bibfnamefont {J.}~\bibnamefont {Stehlik}},
  \bibinfo {author} {\bibfnamefont {C.}~\bibnamefont {Eichler}}, \bibinfo
  {author} {\bibfnamefont {M.~J.}\ \bibnamefont {Gullans}}, \bibinfo {author}
  {\bibfnamefont {J.~M.}\ \bibnamefont {Taylor}}, \ and\ \bibinfo {author}
  {\bibfnamefont {J.~R.}\ \bibnamefont {Petta}},\ }\bibfield  {title} {\enquote
  {\bibinfo {title} {Semiconductor double quantum dot micromaser},}\ }\href
  {\doibase 10.1126/science.aaa2501} {\bibfield  {journal} {\bibinfo  {journal}
  {Science}\ }\textbf {\bibinfo {volume} {347}},\ \bibinfo {pages} {285--287}
  (\bibinfo {year} {2015})}\BibitemShut {NoStop}%
\bibitem [{Note3()}]{Note3}%
  \BibitemOpen
  \bibinfo {note} {\protect \leavevmode {\protect \color {blue}In fact one can
  show analytically, from the expression of the retarded Green function in
  appendix~\ref {app:hubbI}, that $\Omega _0$ in Hubbard-I is independent of
  the hopping and equivalent to the single-site and mean-field
  value.}}\BibitemShut {Stop}%
\bibitem [{\citenamefont {Lebreuilly}\ \emph {et~al.}(2016)\citenamefont
  {Lebreuilly}, \citenamefont {Wouters},\ and\ \citenamefont
  {Carusotto}}]{Lebreuilly2016}%
  \BibitemOpen
  \bibfield  {author} {\bibinfo {author} {\bibfnamefont {Jos{\'e}}\
  \bibnamefont {Lebreuilly}}, \bibinfo {author} {\bibfnamefont {Michiel}\
  \bibnamefont {Wouters}}, \ and\ \bibinfo {author} {\bibfnamefont {Iacopo}\
  \bibnamefont {Carusotto}},\ }\bibfield  {title} {\enquote {\bibinfo {title}
  {Towards strongly correlated photons in arrays of dissipative nonlinear
  cavities under a frequency-dependent incoherent pumping},}\ }\href {\doibase
  10.1016/j.crhy.2016.07.001} {\bibfield  {journal} {\bibinfo  {journal}
  {Comptes Rendus Physique}\ }\textbf {\bibinfo {volume} {17}},\ \bibinfo
  {pages} {836--860} (\bibinfo {year} {2016})}\BibitemShut {NoStop}%
\bibitem [{\citenamefont {Keeling}\ \emph {et~al.}(2010)\citenamefont
  {Keeling}, \citenamefont {Szyma{\'{n}}ska},\ and\ \citenamefont
  {Littlewood}}]{Keeling2010}%
  \BibitemOpen
  \bibfield  {author} {\bibinfo {author} {\bibfnamefont {Jonathan}\
  \bibnamefont {Keeling}}, \bibinfo {author} {\bibfnamefont {Marzena~H.}\
  \bibnamefont {Szyma{\'{n}}ska}}, \ and\ \bibinfo {author} {\bibfnamefont
  {Peter~B.}\ \bibnamefont {Littlewood}},\ }\enquote {\bibinfo {title} {Keldysh
  green's function approach to coherence in a non-equilibrium steady state:
  connecting bose-einstein condensation and lasing},}\ in\ \href {\doibase
  10.1007/978-3-642-12491-4_12} {\emph {\bibinfo {booktitle} {Optical
  Generation and Control of Quantum Coherence in Semiconductor
  Nanostructures}}},\ \bibinfo {editor} {edited by\ \bibinfo {editor}
  {\bibfnamefont {Gabriela}\ \bibnamefont {Slavcheva}}\ and\ \bibinfo {editor}
  {\bibfnamefont {Philippe}\ \bibnamefont {Roussignol}}}\ (\bibinfo
  {publisher} {Springer Berlin Heidelberg},\ \bibinfo {address} {Berlin,
  Heidelberg},\ \bibinfo {year} {2010})\ pp.\ \bibinfo {pages}
  {293--329}\BibitemShut {NoStop}%
\bibitem [{\citenamefont {Mitra}\ \emph {et~al.}(2006)\citenamefont {Mitra},
  \citenamefont {Takei}, \citenamefont {Kim},\ and\ \citenamefont
  {Millis}}]{MitraEtAlPRL06}%
  \BibitemOpen
  \bibfield  {author} {\bibinfo {author} {\bibfnamefont {Aditi}\ \bibnamefont
  {Mitra}}, \bibinfo {author} {\bibfnamefont {So}~\bibnamefont {Takei}},
  \bibinfo {author} {\bibfnamefont {Yong~Baek}\ \bibnamefont {Kim}}, \ and\
  \bibinfo {author} {\bibfnamefont {A.~J.}\ \bibnamefont {Millis}},\ }\bibfield
   {title} {\enquote {\bibinfo {title} {Nonequilibrium quantum criticality in
  open electronic systems},}\ }\href@noop {} {\bibfield  {journal} {\bibinfo
  {journal} {Phys. Rev. Lett.}\ }\textbf {\bibinfo {volume} {97}},\ \bibinfo
  {pages} {236808} (\bibinfo {year} {2006})}\BibitemShut {NoStop}%
\bibitem [{\citenamefont {Clerk}\ \emph {et~al.}(2010)\citenamefont {Clerk},
  \citenamefont {Devoret}, \citenamefont {Girvin}, \citenamefont {Marquardt},\
  and\ \citenamefont {Schoelkopf}}]{ClerkRMP2010}%
  \BibitemOpen
  \bibfield  {author} {\bibinfo {author} {\bibfnamefont {A.~A.}\ \bibnamefont
  {Clerk}}, \bibinfo {author} {\bibfnamefont {M.~H.}\ \bibnamefont {Devoret}},
  \bibinfo {author} {\bibfnamefont {S.~M.}\ \bibnamefont {Girvin}}, \bibinfo
  {author} {\bibfnamefont {Florian}\ \bibnamefont {Marquardt}}, \ and\ \bibinfo
  {author} {\bibfnamefont {R.~J.}\ \bibnamefont {Schoelkopf}},\ }\bibfield
  {title} {\enquote {\bibinfo {title} {Introduction to quantum noise,
  measurement, and amplification},}\ }\href {\doibase
  10.1103/RevModPhys.82.1155} {\bibfield  {journal} {\bibinfo  {journal}
  {Reviews of Modern Physics}\ }\textbf {\bibinfo {volume} {82}},\ \bibinfo
  {pages} {1155--1208} (\bibinfo {year} {2010})}\BibitemShut {NoStop}%
\bibitem [{\citenamefont {Foini}\ \emph {et~al.}(2011)\citenamefont {Foini},
  \citenamefont {Cugliandolo},\ and\ \citenamefont
  {Gambassi}}]{FoiniCugliandoloGambassi_PRB11}%
  \BibitemOpen
  \bibfield  {author} {\bibinfo {author} {\bibfnamefont {Laura}\ \bibnamefont
  {Foini}}, \bibinfo {author} {\bibfnamefont {Leticia~F.}\ \bibnamefont
  {Cugliandolo}}, \ and\ \bibinfo {author} {\bibfnamefont {Andrea}\
  \bibnamefont {Gambassi}},\ }\bibfield  {title} {\enquote {\bibinfo {title}
  {Fluctuation-dissipation relations and critical quenches in the transverse
  field ising chain},}\ }\href@noop {} {\bibfield  {journal} {\bibinfo
  {journal} {Phys. Rev. B}\ }\textbf {\bibinfo {volume} {84}},\ \bibinfo
  {pages} {212404} (\bibinfo {year} {2011})}\BibitemShut {NoStop}%
\bibitem [{\citenamefont {Dalla~Torre}\ \emph {et~al.}(2012)\citenamefont
  {Dalla~Torre}, \citenamefont {Demler}, \citenamefont {Giamarchi},\ and\
  \citenamefont {Altman}}]{DallaTorreetalPRB2012}%
  \BibitemOpen
  \bibfield  {author} {\bibinfo {author} {\bibfnamefont {Emanuele~G.}\
  \bibnamefont {Dalla~Torre}}, \bibinfo {author} {\bibfnamefont {Eugene}\
  \bibnamefont {Demler}}, \bibinfo {author} {\bibfnamefont {Thierry}\
  \bibnamefont {Giamarchi}}, \ and\ \bibinfo {author} {\bibfnamefont {Ehud}\
  \bibnamefont {Altman}},\ }\bibfield  {title} {\enquote {\bibinfo {title}
  {Dynamics and universality in noise-driven dissipative systems},}\
  }\href@noop {} {\bibfield  {journal} {\bibinfo  {journal} {Phys. Rev. B}\
  }\textbf {\bibinfo {volume} {85}},\ \bibinfo {pages} {184302} (\bibinfo
  {year} {2012})}\BibitemShut {NoStop}%
\bibitem [{\citenamefont {Schir\'o}\ and\ \citenamefont
  {Mitra}(2014)}]{SchiroMitraPRL14}%
  \BibitemOpen
  \bibfield  {author} {\bibinfo {author} {\bibfnamefont {Marco}\ \bibnamefont
  {Schir\'o}}\ and\ \bibinfo {author} {\bibfnamefont {Aditi}\ \bibnamefont
  {Mitra}},\ }\bibfield  {title} {\enquote {\bibinfo {title} {Transient
  orthogonality catastrophe in a time-dependent nonequilibrium environment},}\
  }\href@noop {} {\bibfield  {journal} {\bibinfo  {journal} {Phys. Rev. Lett.}\
  }\textbf {\bibinfo {volume} {112}},\ \bibinfo {pages} {246401} (\bibinfo
  {year} {2014})}\BibitemShut {NoStop}%
\bibitem [{\citenamefont {Pitaevskii}\ and\ \citenamefont
  {Stringari}(2016)}]{Pitaevskii:2143198}%
  \BibitemOpen
  \bibfield  {author} {\bibinfo {author} {\bibfnamefont {Lev}\ \bibnamefont
  {Pitaevskii}}\ and\ \bibinfo {author} {\bibfnamefont {Sandro}\ \bibnamefont
  {Stringari}},\ }\href {\doibase 10.1093/acprof:oso/9780198758884.001.0001}
  {\emph {\bibinfo {title} {{Bose-Einstein condensation and superfluidity}}}},\
  International series of monographs on physics\ (\bibinfo  {publisher} {Oxford
  University Press},\ \bibinfo {address} {Oxford},\ \bibinfo {year}
  {2016})\BibitemShut {NoStop}%
\bibitem [{\citenamefont {Aranson}\ and\ \citenamefont
  {Kramer}(2002)}]{aranson2002theworld}%
  \BibitemOpen
  \bibfield  {author} {\bibinfo {author} {\bibfnamefont {Igor~S.}\ \bibnamefont
  {Aranson}}\ and\ \bibinfo {author} {\bibfnamefont {Lorenz}\ \bibnamefont
  {Kramer}},\ }\bibfield  {title} {\enquote {\bibinfo {title} {The world of the
  complex ginzburg-landau equation},}\ }\href {\doibase
  10.1103/RevModPhys.74.99} {\bibfield  {journal} {\bibinfo  {journal} {Rev.
  Mod. Phys.}\ }\textbf {\bibinfo {volume} {74}},\ \bibinfo {pages} {99--143}
  (\bibinfo {year} {2002})}\BibitemShut {NoStop}%
\bibitem [{\citenamefont {Keeling}\ and\ \citenamefont
  {Kéna-Cohen}(2020)}]{keeling2020bose}%
  \BibitemOpen
  \bibfield  {author} {\bibinfo {author} {\bibfnamefont {Jonathan}\
  \bibnamefont {Keeling}}\ and\ \bibinfo {author} {\bibfnamefont {Stéphane}\
  \bibnamefont {Kéna-Cohen}},\ }\bibfield  {title} {\enquote {\bibinfo {title}
  {Bose–einstein condensation of exciton-polaritons in organic
  microcavities},}\ }\href {\doibase 10.1146/annurev-physchem-010920-102509}
  {\bibfield  {journal} {\bibinfo  {journal} {Annual Review of Physical
  Chemistry}\ }\textbf {\bibinfo {volume} {71}},\ \bibinfo {pages} {435--459}
  (\bibinfo {year} {2020})},\ \bibinfo {note} {pMID: 32126177} \BibitemShut
  {NoStop}%
\bibitem [{\citenamefont {Lee}\ \emph {et~al.}(2011)\citenamefont {Lee},
  \citenamefont {Häffner},\ and\ \citenamefont
  {Cross}}]{lee_antiferromagnetic_2011}%
  \BibitemOpen
  \bibfield  {author} {\bibinfo {author} {\bibfnamefont {Tony~E.}\ \bibnamefont
  {Lee}}, \bibinfo {author} {\bibfnamefont {H.}~\bibnamefont {Häffner}}, \
  and\ \bibinfo {author} {\bibfnamefont {M.~C.}\ \bibnamefont {Cross}},\
  }\bibfield  {title} {\enquote {\bibinfo {title} {Antiferromagnetic phase
  transition in a nonequilibrium lattice of {Rydberg} atoms},}\ }\href
  {\doibase 10.1103/PhysRevA.84.031402} {\bibfield  {journal} {\bibinfo
  {journal} {Phys. Rev. A}\ }\textbf {\bibinfo {volume} {84}},\ \bibinfo
  {pages} {031402} (\bibinfo {year} {2011})}\BibitemShut {NoStop}%
\bibitem [{\citenamefont {Chan}\ \emph {et~al.}(2015)\citenamefont {Chan},
  \citenamefont {Lee},\ and\ \citenamefont
  {Gopalakrishnan}}]{chan_limit-cycle_2015}%
  \BibitemOpen
  \bibfield  {author} {\bibinfo {author} {\bibfnamefont {Ching-Kit}\
  \bibnamefont {Chan}}, \bibinfo {author} {\bibfnamefont {Tony~E.}\
  \bibnamefont {Lee}}, \ and\ \bibinfo {author} {\bibfnamefont {Sarang}\
  \bibnamefont {Gopalakrishnan}},\ }\bibfield  {title} {\enquote {\bibinfo
  {title} {Limit-cycle phase in driven-dissipative spin systems},}\ }\href
  {\doibase 10.1103/PhysRevA.91.051601} {\bibfield  {journal} {\bibinfo
  {journal} {Phys. Rev. A}\ }\textbf {\bibinfo {volume} {91}},\ \bibinfo
  {pages} {051601} (\bibinfo {year} {2015})}\BibitemShut {NoStop}%
\bibitem [{\citenamefont {Wilson}\ \emph {et~al.}(2016)\citenamefont {Wilson},
  \citenamefont {Mahmud}, \citenamefont {Hu}, \citenamefont {Gorshkov},
  \citenamefont {Hafezi},\ and\ \citenamefont
  {Foss-Feig}}]{wilson_collective_2016}%
  \BibitemOpen
  \bibfield  {author} {\bibinfo {author} {\bibfnamefont {Ryan~M.}\ \bibnamefont
  {Wilson}}, \bibinfo {author} {\bibfnamefont {Khan~W.}\ \bibnamefont
  {Mahmud}}, \bibinfo {author} {\bibfnamefont {Anzi}\ \bibnamefont {Hu}},
  \bibinfo {author} {\bibfnamefont {Alexey~V.}\ \bibnamefont {Gorshkov}},
  \bibinfo {author} {\bibfnamefont {Mohammad}\ \bibnamefont {Hafezi}}, \ and\
  \bibinfo {author} {\bibfnamefont {Michael}\ \bibnamefont {Foss-Feig}},\
  }\bibfield  {title} {\enquote {\bibinfo {title} {Collective phases of
  strongly interacting cavity photons},}\ }\href {\doibase
  10.1103/PhysRevA.94.033801} {\bibfield  {journal} {\bibinfo  {journal} {Phys.
  Rev. A}\ }\textbf {\bibinfo {volume} {94}},\ \bibinfo {pages} {033801}
  (\bibinfo {year} {2016})}\BibitemShut {NoStop}%
\bibitem [{\citenamefont {Owen}\ \emph {et~al.}(2018)\citenamefont {Owen},
  \citenamefont {Jin}, \citenamefont {Rossini}, \citenamefont {Fazio},\ and\
  \citenamefont {Hartmann}}]{Owen2018}%
  \BibitemOpen
  \bibfield  {author} {\bibinfo {author} {\bibfnamefont {E.~T.}\ \bibnamefont
  {Owen}}, \bibinfo {author} {\bibfnamefont {J.}~\bibnamefont {Jin}}, \bibinfo
  {author} {\bibfnamefont {D.}~\bibnamefont {Rossini}}, \bibinfo {author}
  {\bibfnamefont {R.}~\bibnamefont {Fazio}}, \ and\ \bibinfo {author}
  {\bibfnamefont {M.~J.}\ \bibnamefont {Hartmann}},\ }\bibfield  {title}
  {\enquote {\bibinfo {title} {Quantum correlations and limit cycles in the
  driven-dissipative {{Heisenberg}} lattice},}\ }\href {\doibase
  10.1088/1367-2630/aab7d3} {\bibfield  {journal} {\bibinfo  {journal} {New
  Journal of Physics}\ }\textbf {\bibinfo {volume} {20}},\ \bibinfo {pages}
  {045004} (\bibinfo {year} {2018})}\BibitemShut {NoStop}%
\bibitem [{\citenamefont {{Degenfeld-Schonburg}}\ and\ \citenamefont
  {Hartmann}(2014)}]{degenfeld-schonburg2014}%
  \BibitemOpen
  \bibfield  {author} {\bibinfo {author} {\bibfnamefont {Peter}\ \bibnamefont
  {{Degenfeld-Schonburg}}}\ and\ \bibinfo {author} {\bibfnamefont {Michael~J.}\
  \bibnamefont {Hartmann}},\ }\bibfield  {title} {\enquote {\bibinfo {title}
  {Self-consistent projection operator theory for quantum many-body systems},}\
  }\href {\doibase 10.1103/PhysRevB.89.245108} {\bibfield  {journal} {\bibinfo
  {journal} {Physical Review B - Condensed Matter and Materials Physics}\
  }\textbf {\bibinfo {volume} {89}},\ \bibinfo {pages} {245108} (\bibinfo
  {year} {2014})}\BibitemShut {NoStop}%
\bibitem [{\citenamefont {Seclì}\ \emph {et~al.}(2021)\citenamefont {Seclì},
  \citenamefont {Capone},\ and\ \citenamefont {Schirò}}]{secli2021signatures}%
  \BibitemOpen
  \bibfield  {author} {\bibinfo {author} {\bibfnamefont {Matteo}\ \bibnamefont
  {Seclì}}, \bibinfo {author} {\bibfnamefont {Massimo}\ \bibnamefont
  {Capone}}, \ and\ \bibinfo {author} {\bibfnamefont {Marco}\ \bibnamefont
  {Schirò}},\ }\href@noop {} {\enquote {\bibinfo {title} {Signatures of
  self-trapping in the driven-dissipative bose-hubbard dimer},}\ } (\bibinfo
  {year} {2021}),\ \Eprint {http://arxiv.org/abs/2102.04076} {arXiv:2102.04076
  [quant-ph]} \BibitemShut {NoStop}%
\bibitem [{\citenamefont {Strathearn}\ \emph {et~al.}(2018)\citenamefont
  {Strathearn}, \citenamefont {Kirton}, \citenamefont {Kilda}, \citenamefont
  {Keeling},\ and\ \citenamefont {Lovett}}]{strathearn2018efficient}%
  \BibitemOpen
  \bibfield  {author} {\bibinfo {author} {\bibfnamefont {A.}~\bibnamefont
  {Strathearn}}, \bibinfo {author} {\bibfnamefont {P.}~\bibnamefont {Kirton}},
  \bibinfo {author} {\bibfnamefont {D.}~\bibnamefont {Kilda}}, \bibinfo
  {author} {\bibfnamefont {J.}~\bibnamefont {Keeling}}, \ and\ \bibinfo
  {author} {\bibfnamefont {B.~W.}\ \bibnamefont {Lovett}},\ }\bibfield  {title}
  {\enquote {\bibinfo {title} {Efficient non-markovian quantum dynamics using
  time-evolving matrix product operators},}\ }\href {\doibase
  10.1038/s41467-018-05617-3} {\bibfield  {journal} {\bibinfo  {journal}
  {Nature Communications}\ }\textbf {\bibinfo {volume} {9}},\ \bibinfo {pages}
  {3322} (\bibinfo {year} {2018})}\BibitemShut {NoStop}%
\bibitem [{\citenamefont {Kilda}\ and\ \citenamefont
  {Keeling}(2019{\natexlab{b}})}]{kilda2019fluorescence}%
  \BibitemOpen
  \bibfield  {author} {\bibinfo {author} {\bibfnamefont {Dainius}\ \bibnamefont
  {Kilda}}\ and\ \bibinfo {author} {\bibfnamefont {Jonathan}\ \bibnamefont
  {Keeling}},\ }\bibfield  {title} {\enquote {\bibinfo {title} {Fluorescence
  spectrum and thermalization in a driven coupled cavity array},}\ }\href
  {\doibase 10.1103/PhysRevLett.122.043602} {\bibfield  {journal} {\bibinfo
  {journal} {Phys. Rev. Lett.}\ }\textbf {\bibinfo {volume} {122}},\ \bibinfo
  {pages} {043602} (\bibinfo {year} {2019}{\natexlab{b}})}\BibitemShut
  {NoStop}%
\bibitem [{\citenamefont {Lotem}\ \emph {et~al.}(2020)\citenamefont {Lotem},
  \citenamefont {Weichselbaum}, \citenamefont {von Delft},\ and\ \citenamefont
  {Goldstein}}]{lotem2020renormalized}%
  \BibitemOpen
  \bibfield  {author} {\bibinfo {author} {\bibfnamefont {Matan}\ \bibnamefont
  {Lotem}}, \bibinfo {author} {\bibfnamefont {Andreas}\ \bibnamefont
  {Weichselbaum}}, \bibinfo {author} {\bibfnamefont {Jan}\ \bibnamefont {von
  Delft}}, \ and\ \bibinfo {author} {\bibfnamefont {Moshe}\ \bibnamefont
  {Goldstein}},\ }\bibfield  {title} {\enquote {\bibinfo {title} {Renormalized
  lindblad driving: A numerically exact nonequilibrium quantum impurity
  solver},}\ }\href {\doibase 10.1103/PhysRevResearch.2.043052} {\bibfield
  {journal} {\bibinfo  {journal} {Phys. Rev. Research}\ }\textbf {\bibinfo
  {volume} {2}},\ \bibinfo {pages} {043052} (\bibinfo {year}
  {2020})}\BibitemShut {NoStop}%
\bibitem [{\citenamefont {Arrigoni}\ \emph
  {et~al.}(2013{\natexlab{b}})\citenamefont {Arrigoni}, \citenamefont {Knap},\
  and\ \citenamefont {von~der Linden}}]{ArrigoniEtAlPRL13}%
  \BibitemOpen
  \bibfield  {author} {\bibinfo {author} {\bibfnamefont {Enrico}\ \bibnamefont
  {Arrigoni}}, \bibinfo {author} {\bibfnamefont {Michael}\ \bibnamefont
  {Knap}}, \ and\ \bibinfo {author} {\bibfnamefont {Wolfgang}\ \bibnamefont
  {von~der Linden}},\ }\bibfield  {title} {\enquote {\bibinfo {title}
  {Nonequilibrium dynamical mean-field theory: An auxiliary quantum master
  equation approach},}\ }\href {\doibase 10.1103/PhysRevLett.110.086403}
  {\bibfield  {journal} {\bibinfo  {journal} {Phys. Rev. Lett.}\ }\textbf
  {\bibinfo {volume} {110}},\ \bibinfo {pages} {086403} (\bibinfo {year}
  {2013}{\natexlab{b}})}\BibitemShut {NoStop}%
\bibitem [{\citenamefont {Otsuki}\ and\ \citenamefont
  {Kuramoto}(2013)}]{otsuki2013dynamical}%
  \BibitemOpen
  \bibfield  {author} {\bibinfo {author} {\bibfnamefont {Junya}\ \bibnamefont
  {Otsuki}}\ and\ \bibinfo {author} {\bibfnamefont {Yoshio}\ \bibnamefont
  {Kuramoto}},\ }\bibfield  {title} {\enquote {\bibinfo {title} {Dynamical
  mean-field theory for quantum spin systems: Test of solutions for
  magnetically ordered states},}\ }\href {\doibase 10.1103/PhysRevB.88.024427}
  {\bibfield  {journal} {\bibinfo  {journal} {Phys. Rev. B}\ }\textbf {\bibinfo
  {volume} {88}},\ \bibinfo {pages} {024427} (\bibinfo {year}
  {2013})}\BibitemShut {NoStop}%
\bibitem [{\citenamefont {Barreiro}\ \emph {et~al.}(2011)\citenamefont
  {Barreiro}, \citenamefont {M{\"u}ller}, \citenamefont {Schindler},
  \citenamefont {Nigg}, \citenamefont {Monz}, \citenamefont {Chwalla},
  \citenamefont {Hennrich}, \citenamefont {Roos}, \citenamefont {Zoller},\ and\
  \citenamefont {Blatt}}]{BarreiroEtAlNature11}%
  \BibitemOpen
  \bibfield  {author} {\bibinfo {author} {\bibfnamefont {Julio~T.}\
  \bibnamefont {Barreiro}}, \bibinfo {author} {\bibfnamefont {Markus}\
  \bibnamefont {M{\"u}ller}}, \bibinfo {author} {\bibfnamefont {Philipp}\
  \bibnamefont {Schindler}}, \bibinfo {author} {\bibfnamefont {Daniel}\
  \bibnamefont {Nigg}}, \bibinfo {author} {\bibfnamefont {Thomas}\ \bibnamefont
  {Monz}}, \bibinfo {author} {\bibfnamefont {Michael}\ \bibnamefont {Chwalla}},
  \bibinfo {author} {\bibfnamefont {Markus}\ \bibnamefont {Hennrich}}, \bibinfo
  {author} {\bibfnamefont {Christian~F.}\ \bibnamefont {Roos}}, \bibinfo
  {author} {\bibfnamefont {Peter}\ \bibnamefont {Zoller}}, \ and\ \bibinfo
  {author} {\bibfnamefont {Rainer}\ \bibnamefont {Blatt}},\ }\bibfield  {title}
  {\enquote {\bibinfo {title} {An open-system quantum simulator with trapped
  ions},}\ }\href {\doibase 10.1038/nature09801} {\bibfield  {journal}
  {\bibinfo  {journal} {Nature}\ }\textbf {\bibinfo {volume} {470}},\ \bibinfo
  {pages} {486--491} (\bibinfo {year} {2011})}\BibitemShut {NoStop}%
\bibitem [{\citenamefont {Lee}\ \emph {et~al.}(2013)\citenamefont {Lee},
  \citenamefont {Gopalakrishnan},\ and\ \citenamefont {Lukin}}]{LeeEtAlPRL13}%
  \BibitemOpen
  \bibfield  {author} {\bibinfo {author} {\bibfnamefont {Tony~E.}\ \bibnamefont
  {Lee}}, \bibinfo {author} {\bibfnamefont {Sarang}\ \bibnamefont
  {Gopalakrishnan}}, \ and\ \bibinfo {author} {\bibfnamefont {Mikhail~D.}\
  \bibnamefont {Lukin}},\ }\bibfield  {title} {\enquote {\bibinfo {title}
  {Unconventional magnetism via optical pumping of interacting spin systems},}\
  }\href {\doibase 10.1103/PhysRevLett.110.257204} {\bibfield  {journal}
  {\bibinfo  {journal} {Phys. Rev. Lett.}\ }\textbf {\bibinfo {volume} {110}},\
  \bibinfo {pages} {257204} (\bibinfo {year} {2013})}\BibitemShut {NoStop}%
\bibitem [{\citenamefont {Haller}\ \emph {et~al.}(2015)\citenamefont {Haller},
  \citenamefont {Hudson}, \citenamefont {Kelly}, \citenamefont {Cotta},
  \citenamefont {Peaudecerf}, \citenamefont {Bruce},\ and\ \citenamefont
  {Kuhr}}]{HallerEtAlNatPhys15}%
  \BibitemOpen
  \bibfield  {author} {\bibinfo {author} {\bibfnamefont {Elmar}\ \bibnamefont
  {Haller}}, \bibinfo {author} {\bibfnamefont {James}\ \bibnamefont {Hudson}},
  \bibinfo {author} {\bibfnamefont {Andrew}\ \bibnamefont {Kelly}}, \bibinfo
  {author} {\bibfnamefont {Dylan~A.}\ \bibnamefont {Cotta}}, \bibinfo {author}
  {\bibfnamefont {Bruno}\ \bibnamefont {Peaudecerf}}, \bibinfo {author}
  {\bibfnamefont {Graham~D.}\ \bibnamefont {Bruce}}, \ and\ \bibinfo {author}
  {\bibfnamefont {Stefan}\ \bibnamefont {Kuhr}},\ }\bibfield  {title} {\enquote
  {\bibinfo {title} {Single-atom imaging of fermions in a quantum-gas
  microscope},}\ }\href {\doibase 10.1038/nphys3403} {\bibfield  {journal}
  {\bibinfo  {journal} {Nature Physics}\ }\textbf {\bibinfo {volume} {11}},\
  \bibinfo {pages} {738--742} (\bibinfo {year} {2015})}\BibitemShut {NoStop}%
\bibitem [{\citenamefont {Yamamoto}\ \emph {et~al.}(2019)\citenamefont
  {Yamamoto}, \citenamefont {Nakagawa}, \citenamefont {Adachi}, \citenamefont
  {Takasan}, \citenamefont {Ueda},\ and\ \citenamefont
  {Kawakami}}]{YamamotoEtAlPRL19}%
  \BibitemOpen
  \bibfield  {author} {\bibinfo {author} {\bibfnamefont {Kazuki}\ \bibnamefont
  {Yamamoto}}, \bibinfo {author} {\bibfnamefont {Masaya}\ \bibnamefont
  {Nakagawa}}, \bibinfo {author} {\bibfnamefont {Kyosuke}\ \bibnamefont
  {Adachi}}, \bibinfo {author} {\bibfnamefont {Kazuaki}\ \bibnamefont
  {Takasan}}, \bibinfo {author} {\bibfnamefont {Masahito}\ \bibnamefont
  {Ueda}}, \ and\ \bibinfo {author} {\bibfnamefont {Norio}\ \bibnamefont
  {Kawakami}},\ }\bibfield  {title} {\enquote {\bibinfo {title} {Theory of
  non-hermitian fermionic superfluidity with a complex-valued interaction},}\
  }\href {\doibase 10.1103/PhysRevLett.123.123601} {\bibfield  {journal}
  {\bibinfo  {journal} {Phys. Rev. Lett.}\ }\textbf {\bibinfo {volume} {123}},\
  \bibinfo {pages} {123601} (\bibinfo {year} {2019})}\BibitemShut {NoStop}%
\bibitem [{\citenamefont {Nakagawa}\ \emph {et~al.}(2020)\citenamefont
  {Nakagawa}, \citenamefont {Tsuji}, \citenamefont {Kawakami},\ and\
  \citenamefont {Ueda}}]{NakagawaEtAlPRL20}%
  \BibitemOpen
  \bibfield  {author} {\bibinfo {author} {\bibfnamefont {Masaya}\ \bibnamefont
  {Nakagawa}}, \bibinfo {author} {\bibfnamefont {Naoto}\ \bibnamefont {Tsuji}},
  \bibinfo {author} {\bibfnamefont {Norio}\ \bibnamefont {Kawakami}}, \ and\
  \bibinfo {author} {\bibfnamefont {Masahito}\ \bibnamefont {Ueda}},\
  }\bibfield  {title} {\enquote {\bibinfo {title} {Dynamical sign reversal of
  magnetic correlations in dissipative hubbard models},}\ }\href {\doibase
  10.1103/PhysRevLett.124.147203} {\bibfield  {journal} {\bibinfo  {journal}
  {Phys. Rev. Lett.}\ }\textbf {\bibinfo {volume} {124}},\ \bibinfo {pages}
  {147203} (\bibinfo {year} {2020})}\BibitemShut {NoStop}%
\end{thebibliography}

%

\end{document}